\documentclass[aps,prd,reprint,superscriptaddress,nofootinbib, longbibliography]{revtex4-1}

\usepackage{amsmath,amssymb, amsthm,amstext}
\usepackage{natbib}
\usepackage{graphicx}
\usepackage{color}
\usepackage{array, enumerate}
\usepackage{bm}
\usepackage{multirow}
\usepackage[breaklinks,colorlinks,citecolor=blue]{hyperref}
\usepackage{braket}
\usepackage{txfonts}
\usepackage[normalem]{ulem}

\def\be{\begin{equation}}
\def\ee{\end{equation}}

\def\apjl{Astroph.J.Lett.}
\def\apjs{Astroph.J.S.}
\def\mnras{MNRAS}
\def\araa{ARA\&A}
\def\baas{Bulletin of the AAS}

\newcommand{\CS}{{\mbox{\tiny CS}}}
\newcommand{\FRB}{{\mbox{\tiny FRB}}}

\begin{document}
\title{
Repeating Fast Radio Bursts from Neutron Star Binaries: \\ Multi-band and Multi-messenger Opportunities }

\author{Zhen Pan}
\email{zpan@perimeterinstitute.ca}
\affiliation{Tsung-Dao Lee Institute, Shanghai Jiao-Tong University, Shanghai, 520 Shengrong Road, 201210, People’s Republic of China}
\affiliation{School of Physics \& Astronomy, Shanghai Jiao-Tong University, Shanghai, 800 Dongchuan Road, 200240, People’s Republic of China}
\affiliation{Perimeter Institute for Theoretical Physics, Ontario, N2L 2Y5, Canada}
\author{Huan Yang}
\email{hyang@perimeterinstitute.ca}
\affiliation{Perimeter Institute for Theoretical Physics, Ontario, N2L 2Y5, Canada}
\affiliation{University of Guelph, Guelph, Ontario N1G 2W1, Canada}
\author{Kent Yagi}
\email{ky5t@virginia.edu}
\affiliation{Department of Physics, University of Virginia, Charlottesville, Virginia 22904, USA}

\begin{abstract}
 Recent observations indicate that magnetars may  reside in merging compact binaries
 and at least part of fast radio bursts (FRBs) are sourced by magnetar activities. It is natural to speculate that a class of merging neutron star binaries may have FRB emitters.
  In this work, we study the observational aspects of these binaries - particularly those with FRB repeaters, which  are promising multi-band and multi-messenger observation targets of  radio telescopes and
 ground based gravitational wave detectors as the former telescopes can probe the systems at a much earlier stage in the inspiral than the latter.
  We show that observations of  FRB repeaters in compact binaries have a significant advantage in pinning down the binary spin dynamics, constraining neutron star equation of state, probing FRB production mechanisms, and testing beyond standard physics.
As a  proof of principle, we investigate several half-year long mock observations of FRB pulses originating from
pre-merger neutron star binaries, and we find that using the information of FRB arriving times alone,
the intrinsic parameters of this system (including the stellar masses,  spins, and quadrupole moments) can be measured with high precision, and
the angular dependence of the FRB emission pattern  can also be  well reconstructed.
The measurement of stellar masses (with an error of $\mathcal{O}(10^{-6}-10^{-5})$) and quadrupole moments (with an error of $\mathcal{O}(1\%-10\%)$) may be an unprecedented discriminator of nuclear equations of state in neutron stars. In addition, we find the multi-band and multi-messenger observations of this binary will be sensitive to alternative theories of gravity and beyond standard models, e.g., dynamical Chern-Simons gravity and axion field that is coupled to matter. For example, we find that the effect of gravitational parity violation can be probed much more accurately than existing observations/experiments by combining FRB observations with either gravitational observations or X-ray observations and applying universal relations for neutron stars that do not depend sensitively on the equations of state.

\end{abstract}
\maketitle

\section{Introduction}
The LIGO/Virgo/KAGRA Collaboration (LVKC)
has detected nearly a hundred  compact binary mergers,
including binary black holes (BBHs), binary neutron stars (BNSs) and black hole-neutron stars (BHNSs) \cite{LVC-GWTC1,LVC-GWTC2,LVC-GWTC2.1,LVK-GWTC3},
which have been used as novel probes to astrophysics, fundamental laws of gravity and cosmology \cite{LVK2021pop,LVK2021gr,LVK2021cos}.
Immediately after the first gravitational wave (GW) detection of a BBH merger event GW150914 \cite{LVC2016},
Sesana \cite{Sesana2016} pointed out the prospects of multi-band observations of BBHs by both space-borne and ground-based GW detectors, which offer promising opportunities in probing BBH astrophysics,
testing non-General Relativity (GR) theories and cosmology \cite{Sesana2016,Barausse:2016eii,Wong:2018uwb,Carson:2019rda,Carson:2019kkh,Gerosa:2019dbe,Jani:2019ffg,Perkins:2020tra,Gupta:2020lxa,Datta:2020vcj}.
In addition, the first multi-messenger observations of a BNS merger event GW170817 \cite{GW170817, LVC170817}
provides unique insights into the problems including confirming BNS mergers as a progenitor of short gamma-ray bursts (GRBs) and tightly constraining
the difference between the speed of gravity and the speed of light \cite{2017ApJ...848L..13A}.
 GW170817 has also been widely used as a test bed of neutron star (NS) structure and nuclear physics \cite[e.g.,][]{Margalit2017,Essick2020,Pan2020}, light particles \cite[e.g.,][]{Hook2018,Huang2019,Zhang2021},
and as a bright siren for measuring the expansion rate of the local Universe \cite{LVC2017Nat}.

Inspired by the extensive benefits from the multi-messenger observations of BNS merger event GW170817, the GW astronomy community has been searching for
other possible multi-messenger observation targets, e.g. BBHs and BHNSs. For example, BBH mergers are not likely to produce electromagnetic (EM) emissions except
in a gas rich environment. Graham \textit{et al}. \cite{Graham2020} proposed that the BBH merger GW190521 \cite{GW190521} happened in the accretion disk of an active
galactic nucleus and produced a luminous EM counterpart in the  optical band which was detected by the Zwicky Transient Facility.
But the association between the merger and the optical flare has not been firmly
established \cite{Ashton2021, Nitz2021, Palmese2021, Pan2021}.

Similar to BNS mergers, BHNS mergers are expected to produce short GRBs and kilonovae if the NS is tidally disrupted by the BH. However,
no EM counterpart (short GRBs or kilonovae) associated with these mergers  has been found by detailed follow-up observations of the  BHNS mergers reported by LVKC \cite{Goldstein2019,Hosseinzadeh2019,Coughlin2020,Thakur2020,Alexander2021,Anand2021,Kilpatrick2021}.
The absence of EM counterparts to BHNS mergers is explained by the low probability of NS tidal disruption and faint kilonova signals from the rare NS disruptions \cite{Zhu2021,Fragione2021}.
For the non-disruption systems, EM emissions, as a result of BH interacting with the magnetic field carried by the NS, have also been investigated \cite{Levin2018,Dai2019,Pan2019,Zhong2019,Will2021,DOrazio2022,Neill2022}. These studies show that the resulting EM signals are usually dim for normal NSs
with the surface magnetic field strength $B_\star \lesssim 10^{12}$ Gauss, but the high-energy gamma-ray emission would be much more luminous and potentially observable
for NSs carrying stronger magnetic fields, e.g., magnetars with $B_\star = 10^{14-15}$ Gauss, where the energy budget
is $10^{4-6}$ times higher than in the normal NS case. Notice that such gamma-ray emission is purely from magnetosphere processes \cite{Pan2019}, independent of the post-merger disk formation and possible jet launch for binary NSs.

Following the observation of GW190425 \cite{LIGOScientific:2020aai}, it has been suggested that a fast-merging channel may exist which accounts for a subclass of binary NSs distinct from those found in our galaxy through electromagnetic observations. It is unclear whether  such a fast-merging channel may produce magnetars and merge within their
commonly believed short lifetime ($\sim 10^4$ yr).  However, two recent discoveries PSR J0901-4046 \cite{Caleb2022} and GLEAM-X J162759.5-523504.3 \cite{Hurley2022} provide evidence for an abundant old population of galactic ultra long period magnetars and their age ($\sim 10^{6-7}$ yr) turns out to be much older than previously confirmed galactic magnetars \cite{Beniamini2022}.
On the other hand, recent detection of an unusual gamma-ray burst (GRB 211211A) strongly indicates the common existence of magnetars in compact binary mergers. GRB 211211A is a low-redshift ($z=0.076$)   long gamma-ray burst (GRB) associated with a kilonova and with quasi-periodic oscillations (QPOs) in the X-ray precursor \cite{Rastinejad2022,Xiao2022}. It was accompanied by a long ($>10^4$ sec) high-energy ($>0.1$ GeV) afterglow with excess with respect to
the standard synchrotron and synchrotron self-Compton model of the afterglow \cite{Mei2022,ZhangHM2022}.
This event  suggests   that a magnetar
is most likely involved in the merger, where the QPOs in the precursor are a result of a catastrophic giant flare accompanied
by torsional or crustal oscillations of the magnetar. The long duration of the  GRB produced by a merger is supported
by the prolonged lifetime of the accretion process by the magnetic barrier effect \cite{Xiao2022,Suvorov2022,Gao2022,Zhang2022}.
The long GeV emission is a result of a long deceleration time of the GRB jet in a low-density circumburst medium,
and the GeV excess can be explained with the inverse-Compton scattering of soft photons from the kilonova by hot electrons in the relativistic jet, both of which are also
consistent with the compact binary merger scenario
 \cite{Mei2022,ZhangHM2022}.
As it is difficult to assess the selection effects in finding such a system, determining the rate of similar mergers from this event is highly nontrivial. However, the fact that it happens at such a low redshift implies merging compact binaries with magnetar(s) may be a common class of sources for ground-based gravitational wave detectors.
\footnote{ Nearly one year after the original version of this paper was posted online, a second member
(GRB 230307A) of this new class of long GRBs of merger origin and with a bright X-ray precursor was detected from a low redshift ($z\approx 0.065$) host galaxy \cite{Burns2023,Bulla2023,Yang:2023mqt,Dichiara:2023goh}. Again it is interpreted as a result of magnetar-NS merger \cite{Dichiara:2023goh}.
In a recent work \cite{Yang:2022qmy},
NS-white dwarf mergers as an alternative model
have been invoked for explaining the long duration and the kilonova association of this new class of GRBs .
A possible issue of this alternative model is lack of natural explanation to the bright X-ray precursors.}

Though two members  (GRB 211211A and GRB 230307A) of this new class of long GRBs have been detected,
people still believe magnetars in mergers do not exist or are too rare to detect,
and a popular argument for this belief is based on the non-detection of
magnetars in Galactic BNSs. We now show this argument is not as strong as it seems to be.

Let's first take a step back and see how to reconcile the two observations: no magetar detection in galactic NS binaries and a new class of long gamma bursts of merger origin and with a bright X-ray precursor. Solution a) is that this new class of GRBs have nothing to do with magnetars, and similar to short GRBs, they are simply sourced by norm BNS mergers, and solution b) is that this new class of GRBs are sourced by magnetar-NS mergers in a fast-merging channel.

Many people may take solution a) for granted following the argument that magnetars in BNS mergers must be
too rare to detect
because no magnetars have been detected in the Galactic BNSs.
This solution faces two major issues. One  is that there is no natural explanation to  the long duration and the bright precursor of this new class of GRBs. The other is that the above argument is not reliable and GW 190425 clearly is a counter-example: applying the same argument to BNS mergers, people may incorrectly deny the existence of GW 190425, because no such heavy BNSs have been detected in Galactic BNSs. With solution b) both the long duration and the bright X-ray precursor are naturally explained by the strong magnetic fields carried by the magnetar. Though the fast-merging channel assumed has NOT been confirmed, it is well motivated by another two observation facts:
GW190425 \cite{LIGOScientific:2020aai} and the r-process material in the
Universe
\cite{Komiya2014,Matteucci2014,Safarzadeh:2018ent,Safarzadeh:2018fdy,Zevin:2019obe}.

Comparing the above two solutions, we conclude that solution b) is more plausible, where all the four observation facts
(no magnetars detected in Galactic BNSs, a new class of long GRBs, GW 190425 and the r-process material in the Universe ) are naturally explained.
Therefore we assume that magnetar-NS mergers are the origin of this new class of GRBs. This is also the motivation of this paper.

With this motivation in mind, we now consider another mysterious and intriguing phenomenon closely relating to magnetars,
fast radio bursts (FRBs), which are bright, millisecond-duration radio transients of cosmic origin \cite{Cordes2019}.
Many theoretical models have been proposed for FRBs, among which the most well-studied
models are those that postulate that FRBs arise from the flaring activity of magnetars \cite[see][for  reviews]{Platts2019,Zhang:2020qgp}.
In particular, the recent simultaneous detection of FRBs and X-ray flares originating from the same
galactic magnetar \cite{CHIME2020Nat,Bochenek2020,Mereghetti2020,Li2021Nat,Ridnaia2021,Tavani2021}
clearly show that at least some of FRBs are sourced by magnetar activities.

Based on these observational facts, it is reasonable to speculate that some NS binaries  may  emit FRBs during the inspiral phase as well. Such a scenario has been discussed in \cite{Yang:2020qxt} as a possible explanation for the periodic bursts observed in FRB 180916.J0158+65. Although there are alternative models for the origin of these periodic bursts, e.g. freely precessing magnetars \cite{Levin:2020rhj,Zanazzi:2020vyp} and X-ray binaries with precessing disks \cite{Sridhar:2021zly,Katz:2017ube}, we can use the detection of FRB 180916.J0158+65 to infer the rate of observing similar FRB emitters in merging NS binaries, for the case that the binary interpretation for FRB 180916.J0158+65 is true. The corresponding rate favors a possible detection of such a system in the LIGO A+ era.
The joint observation of FRB bursts and gravitational waves from the same system opens up many opportunities for studying fundamental physics and nuclear astrophysics, as we shall discuss in detail in the main text.

This paper is organized as follows. In Section~\ref{sec:spin}, we introduce the spin dynamics of compact binaries and the FRB timing if an FRB emitter resides in a binary. In Section~\ref{sec:forecast}, we carry out Fisher forecasts on
the precision of constraining model parameters binary NSs with an FRB emitter from a mock observation of FRB arriving times. In Section~\ref{sec:test}, we show that the measurement of NS masses and quadrupole moments from FRB timing can tightly constrain equations of state (EoSs) of NSs. In addition, we also show that multi-band and multi-messenger observations of such a binary will be a sensitive probe to alternative theories of gravity and beyond standard models. In Section~\ref{sec:observ}, we briefly analyze the rate prospect of observing BNSs with an
FRB emitter. Summary and discussion are given in Section~\ref{sec:discussion}.
In the Appendix, we verify the validity of Fisher forecasts by cross-checking the Fisher results with full Bayesian analyses.
In this paper, we use the natural units $\hbar=G=c=1$.

\section{\bf Binary Spin Dynamics and FRB Timing}\label{sec:spin}
FRB emitters currently observed can be classified as repeaters and non-repeaters. In the CHIME/FRB catalog \cite{CHIME-c1},
18 repeaters and 474 non-repeaters have been reported. If the emitter only emits one FRB burst during the observation period, one major question is how to associate it with the gravitational wave observation. The valid association requires the false alarm rate to be much smaller than the expected signal rate. Such a question has been discussed in \cite{LVK-CHIME2022,Wang2022} for current gravitational wave events and also in Sec.~\ref{sec:observ}. In this work, we focus primarily on the repeater scenario, assuming a series of FRB bursts are observed during the year-long observation period.

We assume a merging NS binary with an  FRB repeater that is detectable by radio telescopes on the earth. The FRB emitter may be a magnetar as motivated by the observation by CHIME \cite{CHIME2020Nat}, but a normal NS is also suitable for this analysis if they are capable of generating repeating FRBs, e.g., models of repeating FRBs
from interacting BNSs during the early inspiral phase \cite{Lipunov1996,Lyutikov2018,ZhangB2020}. The emission of FRBs is assumed to be beamed so that FRBs are only observable if the line-of-sight is within the cone of emission. As a result, the arrival time of FRB bursts is modulated by the NS spin precession, which is
governed by the spin-orbit, spin-spin and quadrupole-monopole interactions \cite{Apostolatos1994,Kidder1995,Poisson1998,Racine2008}.
We are to investigate to what precision the binary model parameters can be measured from the FRB timing measurement.

\begin{figure*}
\includegraphics[scale=0.52]{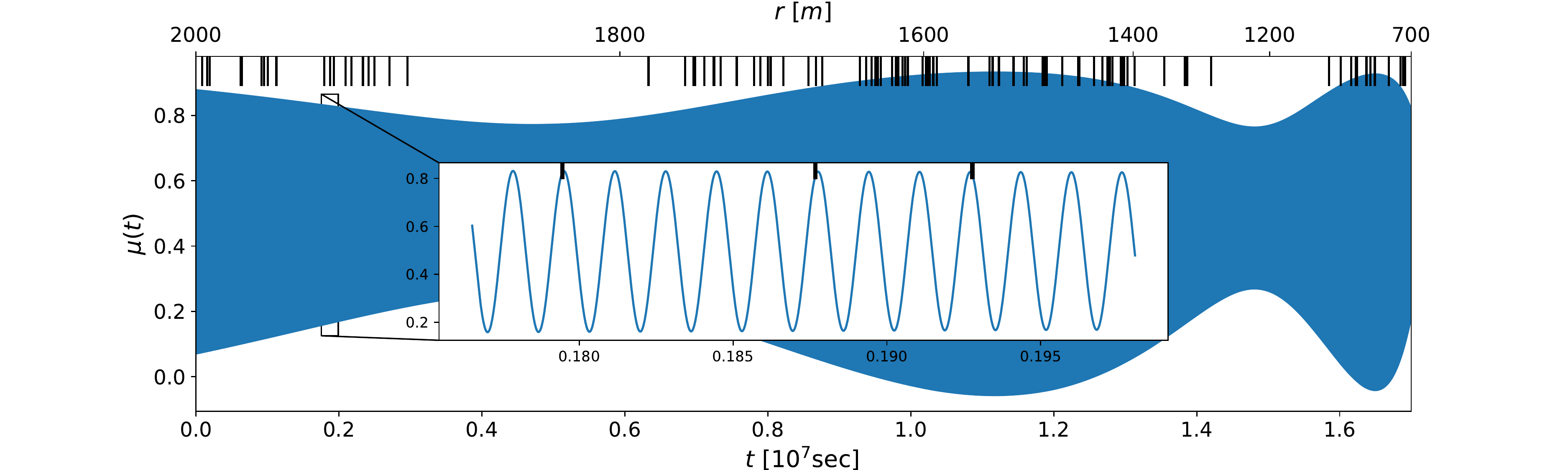}
\caption{\label{fig:mock}
The evolution trajectory of $\mu := \mathbf n_{\rm los}\cdot\hat{\mathbf S}_1$ in the fiducial model,
where the bottom $x$-axis is the time $t\in (0, t_{\rm obs})$, the upper $x$-axis is the binary separation $r$
measured in the total mass $m$ $(:=Gm/c^2)$
and the vertical black ticks mark the observed FRBs. The inset is a zoom-in version of a short time interval.}
\end{figure*}

Consider two stars of masses $m_{1,2}$, dimensionless spins $\chi_{1,2}$, normalized quadrupole moments $\kappa_{1,2}$
(where $\kappa_i:=-Q_i m_i/S_i^2$ with $Q_i$ and $S_i$ being the stellar quadrupole moment and the stellar spin angular momentum, respectively \cite{Laarakkers1999}), and with initial separation $r_0$, binary orbital angular momentum $\mathbf L_{\rm N}$ and spins
$\mathbf S_{1,2}$.
For convenience, we choose to work in a coordinate frame $\hat{x}-\hat{y}-\hat{z}$ with the $\hat{z}$ axis aligned with the initial $\mathbf L_{\rm N}$,
the $\hat{y}$ axis aligned with initial $\mathbf L_{\rm N}\times \mathbf S_1$,
and $\hat{x}$ axis aligned with initial $(\mathbf L_{\rm N}\times \mathbf S_1)\times \mathbf S_1$,
i.e., the initial binary  configuration is parametrized as
\be\label{eq:ini}
\begin{aligned}
  \mathbf L_{\rm N}(t=0) &= \lambda r_0^2\omega_0(0,0,1)\ ,\\
  \mathbf S_1(t=0) &= m_1^2\chi_1(\sin\theta_1, 0, \cos\theta_1)\ ,\\
  \mathbf S_2(t=0) &= m_2^2\chi_2(\sin\theta_2\cos\phi_{12}, \sin\theta_2\sin\phi_{12}, \cos\theta_2)\ ,
\end{aligned}
\ee
where the reduced mass $\lambda=m_1m_2/(m_1+m_2)=m_1m_2/m$ with $m$ representing the total mass and the initial orbital angular frequency $\omega_0$ is determined by the Kepler's law
\be
r^2\omega^2=\frac{m}{r} \left[1+F_\omega(r; m_{1,2}, \mathbf{S}_{1,2})\right]\ ,
\ee
where $F_\omega$ may be approximated by the Post-Newtonian (PN) corrections
(see Eq.~(4.5) in Ref.~\cite{Kidder1995} for the explicit expression).

The evolution equations of the binary spins, the orbital angular momentum and the separation accurate to 2PN order are as follows \cite{Kidder1995, Racine2008},
\be\label{eq:dyn}
\begin{aligned}
  \frac{d\mathbf S_1}{dt} &= \mathbf{\Omega}_1\times \mathbf S_1\ ,\\
  \frac{d\mathbf S_2}{dt} &= \mathbf{\Omega}_2\times \mathbf S_2\ , \\
  \frac{d\mathbf L_{\rm N}}{dt} &= -\mathbf{\Omega}_1\times \mathbf S_1 -\mathbf{\Omega}_2\times \mathbf S_2
  +\left(\frac{d\mathbf L_{\rm N}}{dt}\right)_{\rm rr}\ , \\
  \frac{dr}{dt} &= -\frac{64}{5}\eta \frac{m^3}{r^3} \left[1 + F_r(r; m_{1,2}, \mathbf{S}_{1,2}) \right]\ ,
\end{aligned}
\ee
where the mass ratio $\eta=m_1m_2/m^2$,
\be\label{eq:Omega}
\begin{aligned}
  &\mathbf{\Omega}_1
  = \frac{1}{2r^3} \left[ \left(4+3\frac{m_2}{m_1}\right)\mathbf L_{\rm N} + \mathbf S_2
  -3 \hat{\mathbf L}_{\rm N}\cdot \left(\frac{m_2}{m_1}\kappa_1\mathbf S_1 + \mathbf S_2 \right)\hat{\mathbf L}_{\rm N}  \right]\ ,\\
  &\mathbf{\Omega}_2
  = \frac{1}{2r^3} \left[ \left(4+3\frac{m_1}{m_2}\right)\mathbf L_{\rm N} + \mathbf S_1
  -3 \hat{\mathbf L}_{\rm N}\cdot \left(\frac{m_1}{m_2}\kappa_2\mathbf S_2 + \mathbf S_1 \right)\hat{\mathbf L}_{\rm N}  \right]\ ,\\
  &\left(\frac{d\mathbf L_{\rm N}}{dt}\right)_{\rm rr}
  = \frac{d\mathbf L_{\rm N}}{dr}\frac{dr}{dt}\ ,
\end{aligned}
\ee
with $\hat{\mathbf L}_{\rm N}$ representing the unit vector in the $\mathbf L_{\rm N}$ direction and see Eq.~(12) in Ref.~\cite{Kidder1995} for the PN correction $F_r$.

In general, FRB emission is not isotropic. We adopt a simple model assuming
the probability of observing a burst from a NS depends on the angle between the NS spin axis $\hat{\mathbf S}_1$ and the line of sight (l.o.s.)
\be
\mathbf n_{\rm los} = (\sin\theta_{\rm los}\cos\phi_{\rm los}, \sin\theta_{\rm los}\sin\phi_{\rm los}, \cos\theta_{\rm los})\ .
\ee
As an example, we choose the phenomenological dependence as
\be\label{eq:pdf}
p_{\rm geom}(\mu; \sigma_\mu, \mu_c, \mu_{\rm m}) \propto \exp\left\{-\frac{(\Delta\mu)^2}{2\sigma_\mu^2}\right\}\mathcal{H}(\mu_c-\Delta\mu)\ ,
\ee
 where  $\Delta\mu:=|\mu-\mu_{\rm m}|$,
$\mu=\mu(t):=\mathbf n_{\rm los}\cdot \hat{\mathbf S}_1(t)$, with $\mu_{\rm m}$ the direction where emission probability maximizes (say, magnetic poles), and the absolute value $|\mu|$ takes both north and south pole directions into account.
An implicit assumption underlying this phenomenological dependence
is that the star rotation period is much shorter than the spin precession period, thus the emission anisotropy
only depends on the polar angle $\arccos(\mu)$ w.r.t. the spin direction after averaging over the rotation period.
Note that we did not assume FRB emissions have well defined directions. We only adopt
a rather conservative assumption that FRB emissions contain some anisotropy in the polar direction w.r.t.
the star spin axis after averaging
over the star rotation period. Therefore no matter FRBs are random in the toroidal direction \cite{Zhu:2023spq}
or have a preferred toroidal direction \cite{CHIMEFRB:2021fvq}, our assumption is not affected.
As we will show later that a mild emission anisotropy is sufficient for extracting the
information of binary spin dynamics from the FRB timing.

 In addition to the geometry dependence $p_{\rm geom}$ of the FRB emission, there could be a long timescale modulation of the FRB emission activity
(say, due to some environmental change during the observation time $t_{\rm obs}$), which is hard to model modulation from first principle.
In this work, we again adopt a phenomenological approach by considering the following activity modulations $p_{\rm act}(t)$:
\be \label{eq:pdf2}
\begin{aligned}
    &p_{\rm act}(t) \propto 1 + A_1\times(t/t_{\rm obs}) + A_2\times(t/t_{\rm obs})^2\ ,
\end{aligned}
\ee
and the probability density function
\be \label{eq:pdf0}
p(t;{\bf\Theta}):= p_{\rm geom}(\mu(t); \mu_{\rm m}, \sigma_\mu, \mu_c) p_{\rm act}(t; A_1, A_2)\ ,
\ee is normalized as
$1=\int_0^{t_{\rm obs}} p(t;{\bf\Theta})\ dt$.

As a fiducial model, we consider an FRB emitter-NS binary with $p_{\rm act}={\rm const.}$ and other
model parameters summarized in the 2nd row of Table~\ref{table}.
We initialize and then evolve the binary  according to Eqs.~(\ref{eq:ini}) and (\ref{eq:dyn}), respectively.
In Fig.~\ref{fig:mock},  we show $\mu(t)$ for $t\in(0, t_{\rm obs})$ with the observation time $t_{\rm obs}=1.7\times 10^7\ {\rm sec}$ and the arrival times of $100$ mock FRB pulses, which are randomly sampled according to the probability density function
$p(t;\mathbf{\Theta})$ defined in Eq.~(\ref{eq:pdf0}).
The FRBs arrive when $\mu(t)$ is close to peaks, i.e. when the magnetar spin axis is nearly aligned with the l.o.s., and
the  probability $p(t;\mathbf{\Theta})$ is large as assumed in the fiducial model.

\section{\bf Bayesian Analyses and Fisher Forecasts}\label{sec:forecast}

In order to give forecasts to what precision the binary parameters can be constrained from the FRB timing,
we can either do full Bayesian analyses with mock data or  the Fisher information matrix.
For convenience, we first check the validity of Fisher forecasts with full Bayesian analyses for a fiducial model
(see Appendix for details) and explore more models in a larger parameter space using the more efficient Fisher matrix method.
In this section, we start with `clean' Fisher forecasts assuming vanishing measurement uncertainties of FRB pulse arriving times and
and the independence between FRB emission pulses, then deal with these complications in  subsection B.

\subsection{Clean Fisher forecasts}
Let $f(X;{\bf \Theta})$ be the probability density function for an observable random variable $X$ given model parameters ${\bf \Theta}$,
the Fisher information is defined as \cite{Fisher1922}
\be
\begin{aligned}
   F_{\alpha\beta}
   &= -\braket{\frac{\partial^2 \ln f}{\partial \Theta_\alpha\partial \Theta_\beta} }\ ,\\
   &:= \int \frac{\partial^2 \ln f}{\partial \Theta_\alpha\partial \Theta_\beta} f(x;{\bf \Theta}) dx\ , \\
   &=-\int \left(\frac{f_{,\alpha\beta}}{f}-\frac{f_{,\alpha}f_{,\beta}}{f^2} \right) f dx \ ,
\end{aligned}
\ee
where $f_{,\alpha}:=\partial f/\partial \Theta_{\alpha}$ and $f_{,\alpha\beta}:=\partial^2 f/\partial \Theta_{\alpha}\partial \Theta_{\beta}$.
If $\bf{\Theta}$ is the true model parameters, one finds
\be
\int f_{,\alpha\beta}\ dx = \frac{\partial}{\partial\Theta_\alpha\partial\Theta_\beta} \int f(x;{\bf\Theta}) dx =
 \frac{\partial}{\partial\Theta_\alpha\partial\Theta_\beta} 1 = 0\ ,
\ee
and therefore
\be
F_{\alpha\beta} = \int f(x;{\bf \Theta}) (\ln f)_{,\alpha}(\ln f)_{,\beta} \ dx \ .
\ee

In our case, the observable is the pulse arriving times, i.e., $X=\{t_k\}$ with $k\in(1,..., N_{\rm puls})$.
\emph{If} the pulse arriving times are independent with each other, the probability density function can be written
as $f(X;{\bf\Theta}) = \prod_{k=1}^{N_{\rm puls}} p(t_k;\mathbf{\Theta}) $, and consequently
\be\label{eq:Fisher}
  F_{\alpha\beta}
   =N_{\rm puls} \int_0^{t_{\rm obs}}  p(t; \mathbf{\Theta}) \frac{\partial \ln p}{\partial \Theta_\alpha}
  \frac{\partial \ln p}{\partial \Theta_\beta}\ dt\ ,
\ee
where $p(t|{\bf\Theta})$ is the probability density function defined in Eq.~(\ref{eq:pdf0}).
With the Fisher matrix,
the 1-$\sigma$ uncertainty of parameter $\Theta_\alpha$ is given by $\sigma({\Theta}_\alpha) = \sqrt{(F^{-1})_{\alpha\alpha}}$,
where $\alpha, \beta\in (1,...,17)$.

\begin{figure}
\includegraphics[scale=0.22]{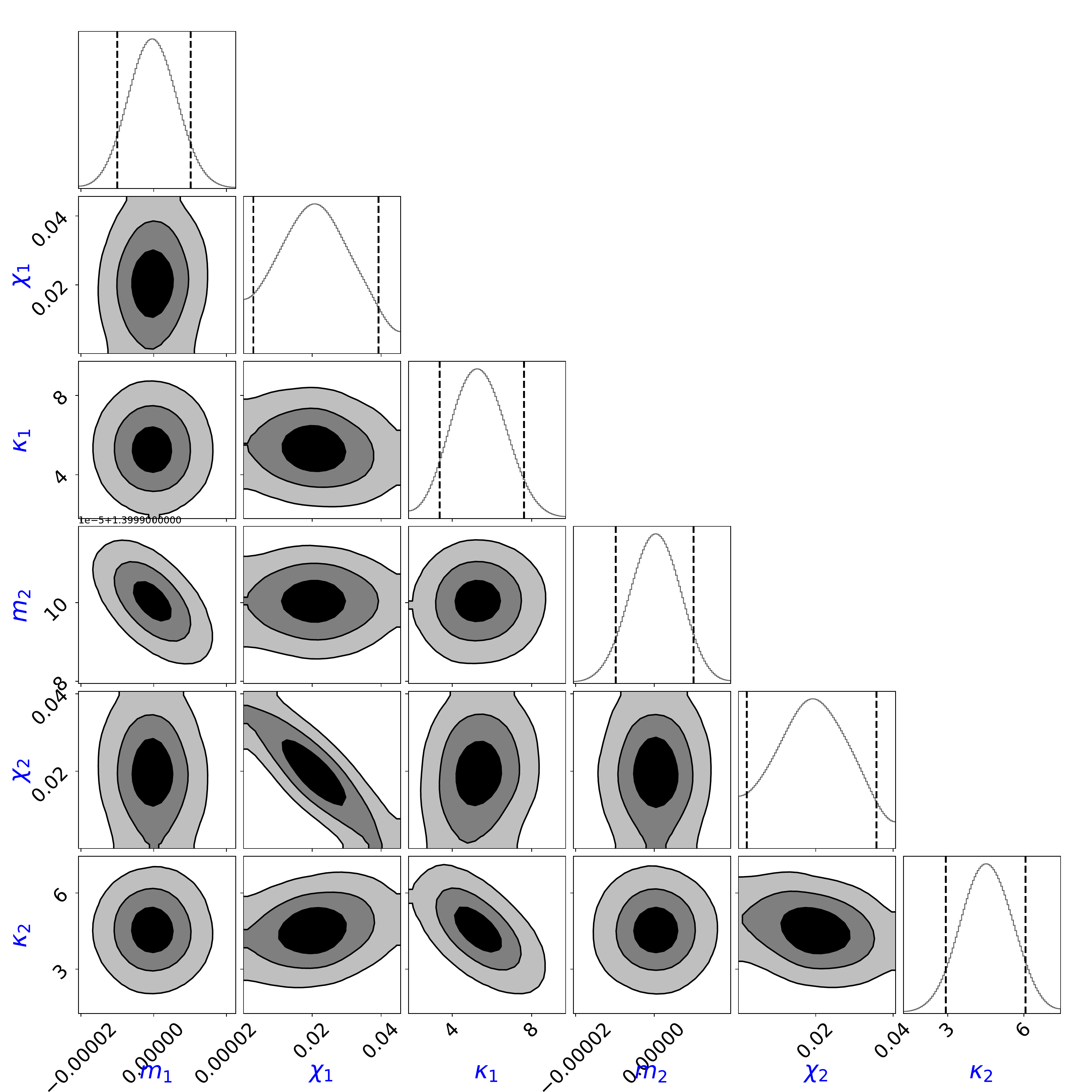}
\caption{\label{fig:corner} The forecast result using the FRB timing of the fidicial model (Fig.~\ref{fig:mock} and Table~\ref{table}).
Posterior contours of part of the model parameters reconstructed from the Fisher forecast,
where each pair of the vertical dashed lines marks the 2-$\sigma$ confidence level.}
\end{figure}

\begin{figure}
\includegraphics[scale=0.55]{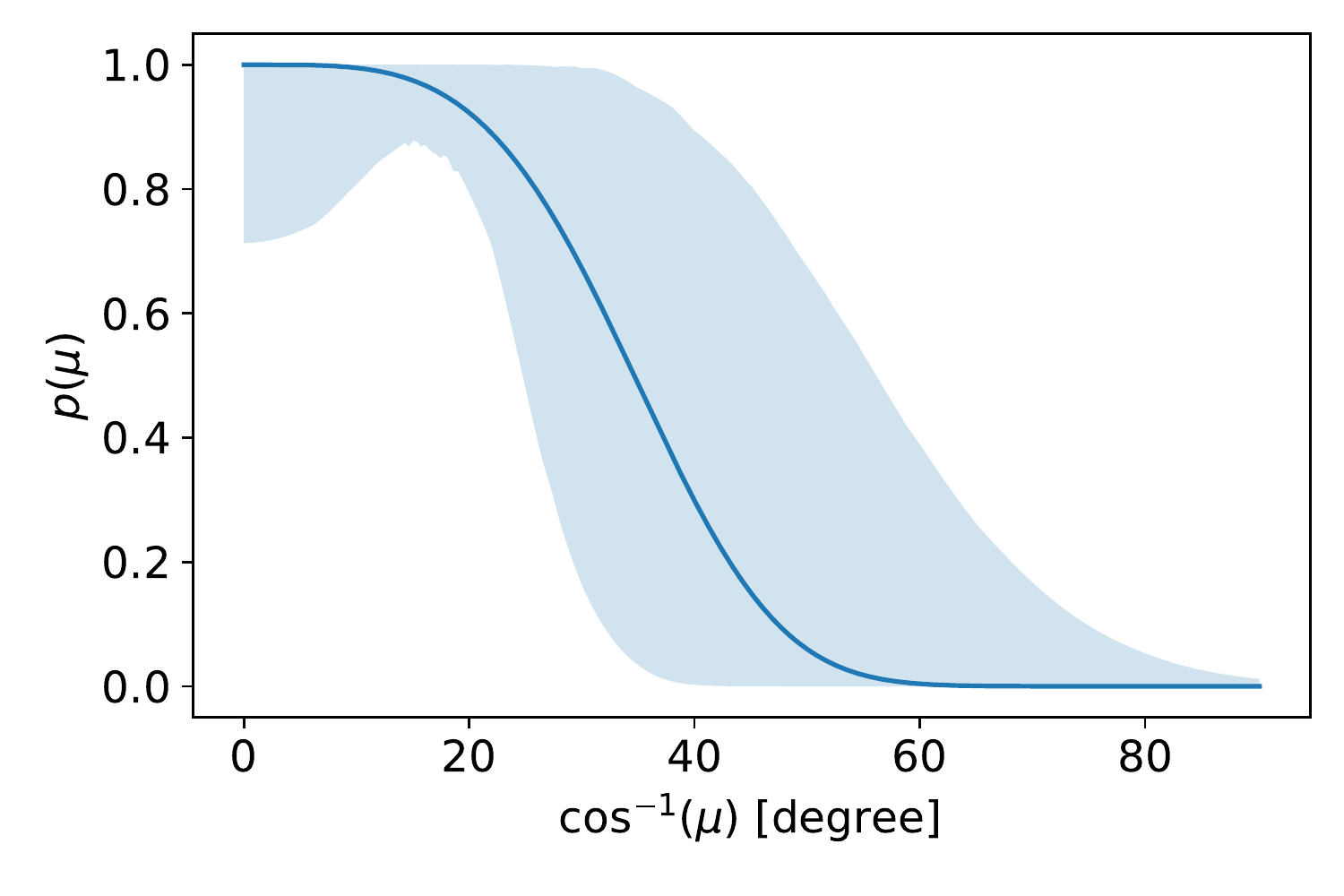}
\caption{\label{fig:pdf} The forecast reconstruction of probability density function $p_{\rm geom}(\mu;\mu_{\rm m},\sigma_\mu, \mu_c)$ in the fiducial model (Fig.~\ref{fig:mock} and Table~\ref{table}),
where the solid line is the injected function and the shaded region is the 2-$\sigma$ uncertainty region of the reconstruction.
}
\end{figure}

The injection values of model parameters $\mathbf{\Theta}$ are listed in Table~\ref{table} and
and we take the expected total number of FRB pulses as $N_{\rm puls}=100$.
In order to accurately calculate the Fisher matrix, we uniformly sample $2^{20}$ points in the range of
$(0, t_{\rm obs})$ in computing the integral in Eq.~(\ref{eq:Fisher}).
The forecasted 1-$\sigma$ uncertainties of all the model parameters
and part of the parameter correlations are summarized in Table~\ref{table} and Fig.~\ref{fig:corner}, respectively.
In the fiducial model, we find both NS masses are expected to be measured to high precision
with $\sigma(m_i/M_\odot) \approx 10^{-5}$ ($i=1,2$), and the total mass ($m$)
and the chirp mass $ m_c=\eta^{3/5}m$ can be measured to even higher precision
with $\sigma(m/M_\odot, m_c/M_\odot) \approx 2 \times 10^{-6}$;
the dimensionless spins are also well constrained with uncertainty $\sigma(\chi_i)< 0.01$;
the normalized quadrupole moments are constrained to good precision with $\sigma(\kappa_i)/\kappa_i\approx 15\%$.
The intrinsic binary orientations ($\theta_1, \theta_2$) and
extrinsic orientations ($\theta_{\rm los}, \phi_{\rm los}$)
can also be measured with low uncertainties $\lesssim 15$ degrees.
In addition, the angular dependence of the FRB emission pattern
$p_{\rm geom}(\mu)$ is also well constrained
(see Fig.~\ref{fig:pdf} for the forecast uncertainty of the reconstructed function $p_{\rm geom}(\mu)$ ).  In the fiducial model (and some of the sample models), the injection values of stellar masses, spins and quadrupole moments are the same for the two stars, but the parameter uncertainties of the two stars
  slightly differ simply because the FRB timing depends only on $\hat{\bf S}_1(t)$,
  which has an asymmetric dependence on the two stars.

\begin{table*}[t]
  \centering
  \resizebox{2.1\columnwidth}{!}{%
  \begin{tabular}{ c | cccc ccccc ccc ccc cc}
    \hline
      & $m_1/M_\odot$ & $\chi_1$ & $\theta_1$ & $\kappa_1$ & $m_2/M_\odot$ & $\chi_2$ & $\theta_2$ & $\kappa_2$ &
    $\phi_{12}$ & $r_0/m$ & $\theta_{\rm los}$ & $\phi_{\rm los}$ & $\mu_{\rm m}$ &$\sigma_\mu$ & $\mu_c$ & $A_1$ & $A_2$\\ \hline
    $\mathbf{\Theta}_{\rm inj}|_{\rm fid}$ & $1.4$ & 0.02 & 0.5 & 4.83
    & $1.4$ & 0.02 & 0.5 & 4.83 & $0.75\pi$
    & 2000 & 0.8 & $-0.5\pi$ & 0.8 & 0.15 & 0.2 & 0 & 0\\
    $\sigma(\mathbf{\Theta})|_{\rm fid}$ & $7.0\times10^{-6}$ & $0.01$& $0.15$& $1.0$
    &$6.4\times10^{-6}$ & $9.0\times10^{-3}$& $0.12$& $0.72$ & $0.40$ & $1.4\times10^{-3}$
    & $0.25$& $0.049$
    & 0.036 & $0.065$& $0.089$ & 0.67 & 0.67\\
    \hline
    $\mathbf{\Theta}_{\rm inj}|_{\rm M1}$ &  & 0.1 & &  &   &  0.1 &   &  &   &  &  &   &   &  \\
    $\sigma(\mathbf{\Theta})|_{\rm M1}$ & $3.4\times10^{-5}$ & $7.0\times10^{-3}$& $0.018$& $0.14$
    &$3.4\times10^{-5}$ & $6.6\times10^{-3}$& $0.019$& $0.12$ & $0.071$
    & $1.2\times10^{-3}$& $0.094$& $0.046$
    & 0.038 & $0.037$& $0.056$ & 0.68 & 0.66 \\
    \hline
    $\mathbf{\Theta}_{\rm inj}|_{\rm M2}$ &  &  & &  &   &  0.1 &   &  &   &  &  &   &   &  \\
    $\sigma(\mathbf{\Theta})|_{\rm M2}$ & $2.7\times10^{-6}$ & $7.0\times10^{-3}$& $0.033$& $0.23$
    &$2.0\times10^{-6}$ & $5.1\times10^{-3}$& $0.016$& $0.12$ & $0.035$
    & $1.2\times10^{-3}$ & $0.062$& $0.05$
    & $0.043$& $0.038$ & 0.041 & 0.69 & 0.65 \\
    \hline
    $\mathbf{\Theta}_{\rm inj}|_{\rm M3}$ &  & &  & 10 &  &  &  & 10 &  &  & &   &  & \\
    $\sigma(\mathbf{\Theta})|_{\rm M3}$ & $1.1\times10^{-5}$ & $4.4\times10^{-3}$& $0.052$& $1.6$
        &$1.1\times10^{-5}$ & $3.6\times10^{-3}$& $0.045$& $1.3$ & $0.14$
        & $1.3\times10^{-3}$ & $0.13$& $0.060$
        & 0.049 & $0.052$& $0.065$ &  0.67 & 0.66 \\
        \hline
            $\mathbf{\Theta}_{\rm inj}|_{\rm M4}$ &  & &  &  &  &  &  &  &  &  & & & $\sqrt{3}/2$  & 0.3 & 0.3 \\
            $\sigma(\mathbf{\Theta})|_{\rm M4}$ & $7.5\times10^{-6}$ & $9.2\times10^{-3}$& $0.16$& $1.1$
                    &$7.3\times10^{-6}$ & $8.1\times10^{-3}$& $0.12$& $0.72$ & $0.38$
                    & $1.1\times10^{-3}$
                    & $0.19$& $0.070$
                    &0.068 & 0.080 & 0.10 & 0.67 & 0.66  \\
                    \hline
    $\mathbf{\Theta}_{\rm inj}|_{\rm M5}$ &  & &  & 6.48 &  &  &  & 6.48 &  &  & &   &  &  \\
    $\sigma(\mathbf{\Theta})|_{\rm M5}$ &$8.4\times10^{-6}$ & $6.0\times10^{-3}$& $0.087$& $1.1$
                                        &$8.2\times10^{-6}$ & $5.1\times10^{-3}$& $0.061$& $0.73$ & $0.24$
                                        &$1.1\times10^{-3}$ & $0.15$& $0.069$
                                        & 0.063 & $0.062$& $0.071$& 0.67 & 0.66 \\
                                            \hline
    $\mathbf{\Theta}_{\rm inj}|_{\rm M6}$ &  & 0.1 &  & 6.48 &  &  0.1 &  & 6.48 &  &  & &   &  &  \\
$\sigma(\mathbf{\Theta})|_{\rm M6}$ &$4.0\times10^{-5}$ & $5.1\times10^{-3}$& $0.021$& $0.20$
                                    &$3.9\times10^{-5}$ & $4.7\times10^{-3}$& $0.020$& $0.17$ & $0.065$
                                    &$1.1\times10^{-3}$ & $0.12$& $0.053$
                                    & 0.053 &$0.049$& $0.047$& 0.66 & 0.66 \\
    \hline
  \end{tabular}
  }
  \caption{Injection values and expected 1-$\sigma$ uncertainties of the 17 model parameters,
  where $m=m_1+m_2$ is the total mass  and we have used the natural units with
  $r/m := r/(Gm/c^2)\approx r/(4.2\ {\rm km})$.
  The 2nd/3rd rows are for the fiducial model,
  where the injection value $\kappa_i=4.83$ for $m_i=1.4 M_\odot$ is expected for NSs with equation of state AP4.
  The injection values in other models considered (M1, M2, M3, M4, M5, M6) are the same as in the fiducial model except those
  explicitly specified.  \label{table}}
\end{table*}

Now let us  briefly comment on the parameter correlations and the reason that model parameters are well constrained by the FRB timing. Starting with stellar masses $m_1$ and $m_2$, they are negatively correlated because of their similar roles in the evolution of the binary
orbital angular momentum $\mathbf{L}_{\rm N}$, which is the main drive of stellar precession [Eq.~(\ref{eq:Omega})].
Denoting $\delta m_1|_{\delta p/p\approx1}$ as the amount of mass change required to shift the probability density function by $\mathcal O(1)$, i.e., the $\mu(t; m_1+\delta m_1|_{\delta p/p\approx1})$ peaks are shifted by $\sim \sigma_\mu$ from
the $\mu(t; m_1)$ peaks.  If there was no parameter degeneracy,  the 1-$\sigma$ uncertainty of $m_1$ should be $\sigma(m_1)\approx F_{m_1, m_1}^{-1/2} \approx  N_{\rm puls}^{-1/2}\delta m_1|_{\delta p/p\approx1}$ [Eq.~(\ref{eq:Fisher})].
From Eqs.~(\ref{eq:dyn},\ref{eq:Omega}), the stellar precession is dominated by the spin-orbit coupling with the precession
rate $\mathbf{\Omega}_1\propto \mathbf{L}_{\rm N}$, and therefore $N_{\rm prec} \propto \Omega_1 t_{\rm obs}$,
where $N_{\rm prec}$ is the number of peaks in $\mu(t)$,
i.e., the number of precession cycles that star 1 has gone through. Consequently, we have $\delta N_{\rm prec}/N_{\rm prec}\approx \delta \Omega_1/\Omega_1$, i.e.,
\be
 \frac{\sigma_\mu}{2\pi N_{\rm prec}}\approx \frac{\delta |\mathbf{\Omega}_1|}{|\mathbf{\Omega}_1|}
 \approx\frac{\delta |\mathbf{L}_{\rm N}|}{|\mathbf{L}_{\rm N}|}\approx
 \frac{\delta m_1|_{\delta p/p\approx1}}{m}\ ,
\ee
where $N_{\rm prec}\approx 2500$
in the fiducial model (Fig.~\ref{fig:mock}).
As a result, we obtain $\sigma(m_1/M_\odot) \approx 3\times 10^{-6} $,
which is consistent with the rigorous Fisher result considering that
the $m_1-m_2$ degeneracy degrades the constraint by a factor of a few.
Following the same argument, we expect that $m_2$ and $r_0$ should be constrained with a similar fractional uncertainty $\sim 10^{-5}$. The constraints on the spins $\chi_{1,2}$ can be understood via its role in the precession frequency of the magnetar spin $\mathbf S_1$ in the same way.
In addition,
it is clear the $\chi_1-\chi_2$ correlation comes from the spin-spin interaction
and the $\chi_i-\kappa_i$ correlations come from the quadrupole-monopole interaction (Fig.~\ref{fig:corner}).

We also perform  forecasts for sample models (M1, M2, M3, M4, M5, M6) with slightly different values of model parameters from those in the fiducial model (see Table~\ref{table}). In M1, we consider a binary with 5 times
higher spins ($\chi_1=\chi_2=0.1$), and we find the constraints on $\kappa_i$ become $\approx 6$ times tighter with $\sigma(\kappa_i)/\kappa_i\approx 3 \%$ as expected. For comparison, we consider a binary with a low-spin magnetar
and a normal NS ($\chi_1=0.02, \chi_2=0.1$) in M2, and we find the constraint on $\kappa_2$ becomes $\approx 6$ times tighter and the constraint on $\kappa_1$ also improves by a factor $\approx 4$ (compared with in the fiducial model). For an ultra long rotation magnetar and a normal NS ($\chi_1 \leq 0.001, \chi_2=0.1$), we find that $\kappa_1$ is not well constrained while $\kappa_2$ can still be measured with percent level fractional uncertainty.
Therefore, a measurement of $\kappa_i$ with a percent level
uncertainty is possible as long as one of the stars is moderately fast spinning ($\chi=0.1$).
In M3, we consider a binary with 2 times larger quadrupole moments
($\kappa_1=\kappa_2=10$), and we find the constraints on $\chi_i$ becomes $(2-3)$ times tighter with $\sigma(\chi_i) < 4\times 10^{-3}$ as expected.
In M4, we consider the same binary in which the FRB emission peak is $30$ degrees off the spin axis ($\mu_{\rm m}=\sqrt{3}/2$) and is allowed in a larger solid angle
($\sigma_\mu=0.3, \mu_c=0.3$), and we find little change in the constraints on the binary model parameters.
This result is easy to understand, because the FRB pulses signal the spin precession evolution
when the spin direction and the l.o.s. subtend a specific angle (which is $\sim 0$ in the fiducial model
and is $\sim 30$ degrees in M4), no matter the pulses come from the spin direction or the magnetic pole direction.  M5 and M6 are the same as the fiducial model and M1, respectively, except with slightly different
quadrupole moments $\kappa_i=6.48$, which is reserved for the purpose of later discussion in Sec.~\ref{sec:univ_rel}.
 For all the models considered, the FRB emission activity parameters $A_{1,2}$ are of little correlation with other model parameters, the constraints of $A_{1,2}$ varies little from model to model. This result is also easy to understand:
$\partial\ln p/\partial A_i = \partial\ln p_{\rm act}/\partial A_i $ and
$\partial\ln p/\partial \Theta_\alpha = \partial\ln p_{\rm geom}/\partial \Theta_\alpha$ for other parameters,
where the latter derivatives vary on the precession timescale, while the former vary on a much longer timescale.

We have been considering monitoring
the binary system with an initial binary separation $r_0=2000 m$ for half a year before the final coalescence, which is a rather conservative scenario. With a longer observation time $t_{\rm obs}$ and a larger initial separation $r_0\propto t_{\rm obs}^{1/4}$, the binary parameters can be measured with even higher precision. For example, the measurement uncertainty of stellar masses scales as
$\sigma(m_i) \propto N_{\rm puls}^{-1/2} (N_{\rm prec}^{\rm spin-orbit})^{-1}$,
where the number of FRB pulses detected $N_{\rm puls}\propto t_{\rm obs}$,
and the number of precession cycles driven by the spin-orbit coupling
$N_{\rm prec}^{\rm spin-orbit}\propto \Omega^{\rm spin-orbit} t_{\rm obs} \propto t_{\rm obs}^{3/8}$.
In a similary way, $\sigma(\chi_i) \propto N_{\rm puls}^{-1/2} (N_{\rm prec}^{\rm spin-spin})^{-1}$
and $\sigma(\kappa_i) \propto N_{\rm puls}^{-1/2} (N_{\rm prec}^{\rm mono-quad})^{-1}$,
with $N_{\rm prec}^{\rm spin-spin}\propto \Omega^{\rm spin-spin} t_{\rm obs} \propto t_{\rm obs}^{1/4}$
and $N_{\rm prec}^{\rm mono-quad}\propto t_{\rm obs}^{1/4}$.

\subsection{Realistic complications}
In the previous subsection, we have assumed vanishing uncertainty of pulse arriving times, otherwise the Fisher matrix
should be  formulated as
\be \label{eq:Fisher_full}
  F_{\alpha\beta}
   =N_{\rm puls} \int_0^{t_{\rm obs}}  p(t; \mathbf{\Theta}) \frac{\partial \ln \braket{p}}{\partial \Theta_\alpha}
  \frac{\partial \ln \braket{p}}{\partial \Theta_\beta}\ dt\ ,
\ee
with
\be
\braket{p(t)} := \int \frac{1}{\sqrt{2\pi}\sigma_t} \exp\left\{-\frac{(t-t')^2}{2\sigma_t^2}\right\} p(t') \ dt'\ ,
\ee
where $\sigma_t$ is the uncertainty of FRB pulse arriving times.

The uncertainty $\sigma_t$ is in general sourced by
the changing dispersion measure (DM) between pulses ($\sigma_t^{\rm DM}$ ) due to
the solar wind and other local factors within our galaxy and the FRB
host galaxy \cite{Yang2017},  the intrinsic pulse width ($\sigma_t^{\rm int}$),
and the pulse width broaden by scattering $\sigma_t^{\rm sct}$.
Taking FRB 180916 as an example, the DM changes over 1 year is found to be
$\Delta{\rm DM}\approx 0.5 {\rm pc} \ {\rm cm}^{-3}$  \cite{CHIME16d,Chawla2020}, which translates to an FRB arriving time uncertainty as
\be
\sigma_t^{\rm DM} = 4.15 \ {\rm ms} \ \left( \frac{\Delta{\rm DM}}{{\rm pc} \ {\rm cm}^{-3}}\right)
\left(\frac{\nu_{\rm FRB}}{\rm GHz} \right)^{-2}\ .
\ee
As measured by the Green Bank Telescope, the intrinsic pulse width is $\sigma_t^{\rm int} < 6$ ms, and
the scattering timescale is $\sigma_t^{\rm sct} < 1.7$ ms at 350 MHz \cite{Chawla2020},
and should be much lower at higher frequency (say GHz) considering the scaling relation $\sigma_t^{\rm sct}\propto \nu_{\rm FRB}^{-4}$. Putting all the three contributions together, we obtain a conservative upper bound $\sigma_t < 10$ ms.

In the binary model we are considering, the stellar spin precession timescale is $t_{\rm obs}/N_{\rm prec}$,
and the timescale on which the probability density function $p(t)$ varies is
$t_{\rm obs}/N_{\rm prec}\times \sigma_\mu/(2\pi) \approx  162$ sec which is at least 4 orders of magnitude higher than the uncertainty of pulse arriving times $\sigma_t$. Therefore the vanishing $\sigma_t$ assumption should be valid and our numerical results with Eq.~(\ref{eq:Fisher_full}) also confirmed this expectation.

~\\

The second assumption we have used in the previous subsection is the independence of each FRB pulse.
As a result, we can simplify the probability density function as $f(X;{\bf\Theta}) = \prod_{k=1}^{N} p(t_k;\mathbf{\Theta}) $. In fact, the bursts are clustered in time as pointed in Refs.~ \cite{Connor2016,Oppermann2018}.
As an example investigated in Ref.~\cite{Oppermann2018}, the mean burst rate of FRB 121102 is $r\approx 6\ {\rm day}^{-1}$, while
the inter-burst intervals cluster around a much shorter time $\delta t\approx 10$ mins ($\ll r^{-1}$), i.e., the pulse arriving times
are non-Poisson. The inter-burst intervals $t_{k+1}-t_k$ in FRB 121102 can be described by the Weibull distribution instead \cite{Oppermann2018}
\be
\mathcal{W}(x; k_w, r) = k_wx^{-1} \left[x r\Gamma(1+1/k_w)\right]^{k_w} e^{-\left[xr \Gamma(1+1/k_w)\right]^{k_w}}\ ,
\ee
with the shape parameter $k_w\approx 0.3$.

If the clustered-in-time behavior in FRB 121102 is typical, the FRB pulses in a same cluster should not be independent, and
the effective number of independent pulses should be less than the total number
$N^{\rm eff}_{\rm puls} < N_{\rm puls}$, and the forecast parameter uncertainties in the previous subsection should be increased by a factor
$\sqrt{N_{\rm puls}/N^{\rm eff}_{\rm puls}}$.
As a rough estimate of the effective number $N^{\rm eff}_{\rm puls}$, we may use the condition $|t_i-t_j|\gg \delta t$ as the criteria
of the independence between two pulses $i$ and $j$. In the fiducial model, the typical interval of detected bursts is $t_{\rm obs}/N_{\rm puls}\approx 2$ days $\approx300\ \delta t$ if the pulses are not strongly clustered.
If the pulses are strongly clustered as in FRB 121102, the typical interval between clusters is even larger,
and the bursts in different clusters should be independent, and $N^{\rm eff}_{\rm puls}$ is approximately the number of pulse clusters.

In practice, the cluster property of FRB bursts can be incorporated in the probability density function $f(X;{\bf\Theta}, k_w, r)$. Following Ref.~\cite{Oppermann2018}, the probability $f(X=\{t_1,...,t_N\};{\bf\Theta}, k_w, r)$ can be calculated by marginalizing the arriving time $t_{ N+1}$ of next
pulse,
\be
\begin{aligned}
&f(t_1, ..., t_{N};{\bf\Theta}, k_w, r)  \\
&=\int_{ t_{N}}^\infty \mathcal{H}(t_{\rm obs}-t_{N})\mathcal{H}(t_{N+1}-t_{\rm obs})
f(t_1, ..., t_{N}, t_{N+1};{\bf\Theta}, k_w, r) \ d t_{N+1}\ ,
\end{aligned}
\ee
where the product of the two Heaviside functions
imposes the condition that $t_N < t_{\rm obs} < t_{N+1}$, and
\be
\begin{aligned}
&f(t_1, ..., t_{N}, t_{N+1};{\bf\Theta}, k_w, r) \\
&= \mathcal{P}(t_1;{\bf\Theta}, k_w, r)
\prod_{i=1}^{N} \mathcal{P}(t_{i+1}|t_i, {\bf\Theta}, k_w, r) \ .
\end{aligned}
\ee
with $\mathcal{P}(t_1;{\bf\Theta}, k_w, r)$ the probability density function of first pulse arrival at $t_1$,
and  $\mathcal{P}(t_{i+1}|t_i, {\bf\Theta}, k_w, r)$ is the conditional probability density function of next pulse arrival at $t_{i+1}$
given the previous one $t_i$.
The explicit expressions for both $\mathcal{P}(t_1) $ and $\mathcal{P}(t_{i+1}|t_i)$ have been derived in Ref.~\cite{Oppermann2018}.
The implementation of the full probability density function $f(X;{\bf\Theta}, k_w, r)$ in Bayesian analyses is beyond the scope of the current paper.

~\\

These simple forecasts  show that the FRB timing from the FRB  binary is a promising   observable for measuring the binary parameters (the NS masses, spins, quadrupole moments, intrinsic/extrinsic orientations, and the FRB emission pattern).
Even with 3rd-generation ground-based gravitational wave detectors, measurement accuracy of the same parameters from merging binaries is not as high as that from FRB observations \cite{Smith2021,Samajdar:2020xrd}. However, the FRB timing mainly probes the binary dynamics at large separations
where the inspiral timescale is long so that we expect to observe the majority of FRB pulses,
while the GW signal is sensitive to dynamics at small separations ($r/m < 170$ for $f_{\rm gw} > 10$ Hz).
As a result, other observables including the binary coalescence time ($t_c$) and the star tidal deformabilities ($\Lambda_{1,2}$) that are sensitive to high-frequency dynamics, can be measured from the GWs with much better precision (as we will show in Sec.~\ref{sec:tidal}).
In the next section, we will investigate a few applications of FRB timing alone and
the benefit of multi-band and multi-messenger observations with both FRB and GW observations.

\section{\bf Applications}\label{sec:test}

\subsection{Nuclear Astrophysics and Radius Measurement
}
\label{sec:nuclear}

The first application we consider is nuclear astrophysics. One of the most well-studied relations between NS observables is the mass-radius relation that depends strongly on the underlying EoSs of nuclear matter. This means that independent measurements of these quantities allow us to probe nuclear physics. A similar test can be performed with independent measurements of the mass and quadrupole moment. This is because the relation between these quantities also depends sensitively on the EoSs.

\begin{figure}
\includegraphics[width=\linewidth]{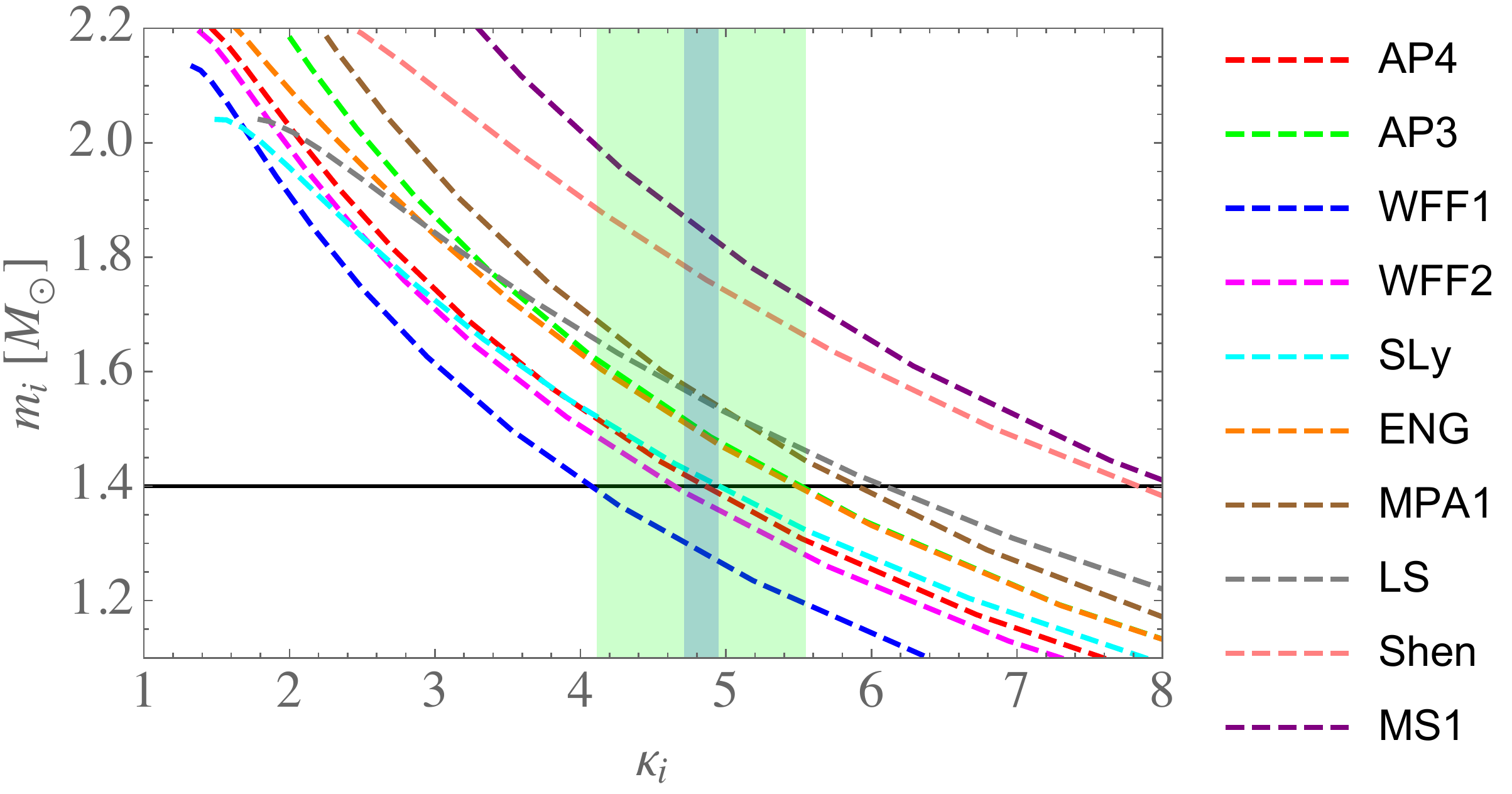}
\caption{\label{fig:M-kappa}
Relation between the mass and quadrupole moment of NSs with various EoSs. Shaded regions show the projected measurement errors on $\kappa_i$ from FRB observations, found from the Fisher analysis in Table~\ref{table}. We assumed the fiducial values of $m_i =1.4M_\odot$ (shown by the horizontal line) and $\kappa_i = 4.83$. The spins are assumed as $\chi_i=0.02$ (green) and 0.1 (blue). The measurement error on $m_i$ is not visible as it is too small.
}
\end{figure}

Figure~\ref{fig:M-kappa} presents such relation with 10 EoSs considered e.g. in~\cite{Yagi:2016qmr}. Observe how the relations depend on a different choice of EoSs. To put the test into context, we consider a fiducial model of $m_i=1.4M_\odot$, $\kappa_i = 4.83$ (corresponding to the AP4 EoS) and $\chi_i=0.02$ or 0.1. The measurement errors of $\kappa_i$ for these models are given in Table~\ref{table} (we use the one for $\kappa_2$). In Fig.~\ref{fig:M-kappa}, we show such errors as green and blue shaded regions respectively. The allowed region for the mass is shown by a horizontal \emph{line} at $1.4M_\odot$ as the measurement error of the mass is too small and not visible. If such observations are realized, the allowed EoSs are those that cross the $m_i = 1.4M_\odot$ line in the shaded regions. For the $\chi_i=0.02$ case, there are several EoSs consistent with the green shaded region at $1.4M_\odot$, while for the $\chi_i=0.1$ case, AP4 is consistent while WFF2 and SLy are only marginally consistent, and all the other EoSs are inconsistent.

Let us now consider converting the measurement errors on $\kappa_i$ to those on the radius, $R_i$. This can be done by taking the advantage of the universal relation between $\kappa_i$ and the stellar compactness  $C_i(:=m_i/R_i)$~\cite{Urbanec:2013fs} that does not depend sensitively on the EoSs, whose fit can be found e.g. in~\cite{Yagi:2016bkt}\footnote{For magnetars, $\kappa_i$ originates not only from spin but also from magnetic field, though the latter contribution matters only for neutron stars with slow rotation~\cite{Haskell:2013vha}.}. Using this relation $\kappa_i(m_i/R_i)$ and the fact that the errors on the masses are small and negligible, the measurement errors on $\kappa_i$ can be converted to those on $R_i$ from the relation
\begin{equation}
\label{eq:radius_error}
\sigma(\kappa_i) \approx \sigma(R_i)  \frac{\partial \kappa_i(m_i/R_i)}{\partial R_i}\,.
\end{equation}
We note that the universal relation between $\kappa_i$ and $R_i$ holds among hadronic EoSs only, and it is not applicable to quark stars~\cite{Yagi:2016bkt}.

\begin{figure}
\includegraphics[width=0.9\linewidth]{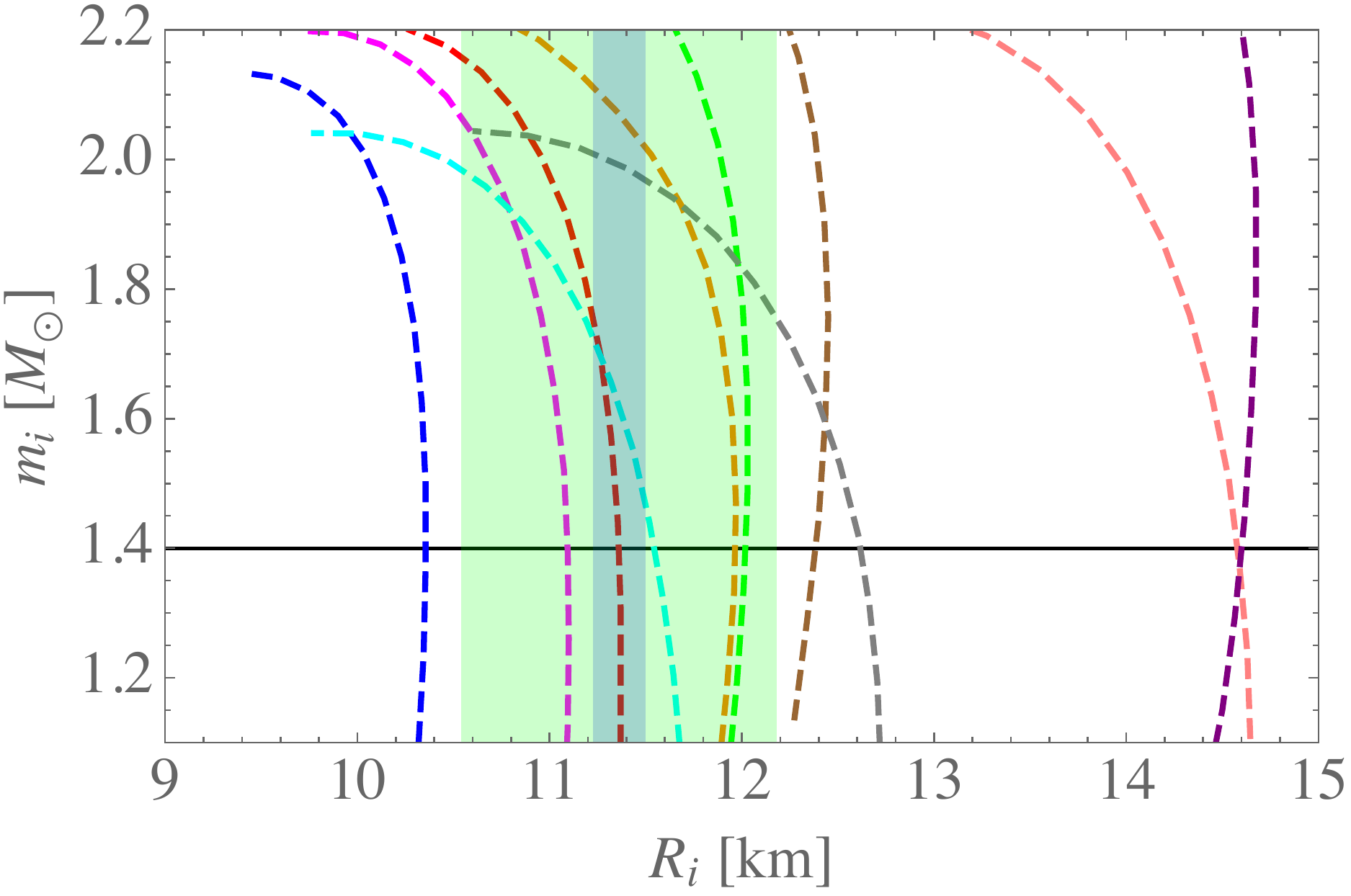}
\caption{\label{fig:M-R}
The relation between the mass and radius of NSs with various EoSs. The meaning of different curves and shaded regions are similar to those in Fig.~\ref{fig:M-kappa}. The measurement errors on the radius are estimated through Eq.~\eqref{eq:radius_error}.
}
\end{figure}

Figure~\ref{fig:M-R} presents the mass-radius relation of NSs with the radius measurement errors converted from those on $\kappa_i$. To be precise, $\sigma(R_i)$ are given by $\sigma(R_i) = {0.82}$ km for $\chi_i=0.02$ and $\sigma(R_i) =  0.14$ km for $\chi_i = 0.1$. The fiducial radius of our model corresponds to $R_i = 11.4$ km. This means that for a NS with $\chi_i = 0.1$, the radius can be inferred down to $\mathcal{O}(1\%)$ with the FRB observations!
The fractional error on the radius is  smaller than that of the quadrupole moment. Indeed, from the figure, we see that for $\chi_i = 0.1$ (blue shade), the only consistent equation of state out of all 10 considered here is AP4, and even WFF2 and SLy are now inconsistent with the projected measurement errors. This shows that the FRB observations can be very powerful in probing nuclear physics, though we stress that the above measurement errors of $R_i$ need to be taken with care as we assumed that nuclear matter obeys hadronic EoSs.

\subsection{Tidal Deformability Measurement}
\label{sec:tidal}

In principle the FRB observations will help the GW observations in better determining the tidal deformability. This is because the prior information on the masses from the FRB observations may break the degeneracy between the tidal deformability and masses in GW observations. We will investigate this in this subsection by carrying out two Fisher analyses, one with GW alone and another using the mass prior information from the FRB observations.

The gravitational waveform in the frequency domain is given by
\be
h(f)=\mathcal{A} f^{-7/6} e^{i \left[\Psi_{\rm pp}(f)+ \Psi_{\rm td}(f) \right]}\ ,
\ee
where $\mathcal{A}$ is the amplitude,
\be
\begin{aligned}
\label{eq:phase_pp}
  \Psi_{\rm pp}(f)
  &= 2\pi ft_c -\phi_c -\frac{\pi}{4} \\
  &+ \frac{3}{128\eta v^5}\left[1+\frac{20}{9}\left(\frac{743}{336}+\frac{11}{4}\eta \right)v^2 + ... \right]
\end{aligned}
\ee
is the phase with $t_c, \phi_c$ being the coalescence time and the coalescence phase \cite[see][for the full expression accurate to 3.5 PN order]{Buonanno2009}, $v=(\pi m f)^{1/3}$ and
\be
\Psi_{\rm td}(f) = -\frac{9v^5}{16\eta} \left[ \frac{m_2+12m_1}{m_2}\Lambda_1\left(\frac{m_1}{m}\right)^5
+  \frac{m_1+12m_2}{m_1}\Lambda_2\left(\frac{m_2}{m}\right)^5 \right]
\ee
is the tidal contribution to the phase at the leading PN order induced by the (dimensionless) stellar tidal deformabilities $\Lambda_i$ \cite{Flanagan2008,Hinderer2010}.

We consider the same compact binary as in the previous section and forecast the constraints of GW model parameters $\Theta_\alpha=\{\mathcal{A}, t_c, \phi_c, m, m_c, \Lambda\}$, where
\be
\Lambda := \frac{16}{13}\left[\frac{m_2+12m_1}{m_2}\Lambda_1\left(\frac{m_1}{m}\right)^5
+  \frac{m_1+12m_2}{m_1}\Lambda_2\left(\frac{m_2}{m}\right)^5 \right]\ . \nonumber
\ee
The fiducial model parameters are chosen as follows: $t_c=\phi_c= 0$,  $m=2.8 M_\odot$,  $m_c=m/4^{3/5}$,  $\Lambda=260$
and $\mathcal{A}$ is chosen such that the signal-to-noise ratio (SNR) of GWs from the binary with LIGO A+ equals 20,
where
\be
{\rm SNR^2} = 4 \int_{f_l}^{f_h} \frac{h(f)h^*(f)}{S_{\rm n}(f)} df\ ,
\ee
with $S_{\rm n}(f)$ being the noise spectral density (the one for LIGO A+ can be found in \cite{ligoA}), $f_l=10$ Hz and $\pi f_h=\omega|_{r=10 m}$.
The corresponding Fisher matrix is defined as
\be
F_{\alpha\beta}^{\rm GW} = 4 \mathcal{R} \int_{f_l}^{f_h} \frac{h_{,\alpha}(f) h_{,\beta}^*(f)}{S_{\rm n}(f)} df\ ,
\ee
where $h_{,\alpha} \equiv \partial h / \partial \Theta_\alpha$, $\mathcal{R}$ denotes the real part.
We find the measurability of the tidal deformability parameter as $\sigma(\Lambda)|_{\rm GW}=260$.
For comparison, we consider multi-messenger/multi-band observations by imposing priors $\sigma(m/M_\odot)=1.8\times 10^{-6}$ and $\sigma(m_c/M_\odot)=2.3\times 10^{-6}$
in the above Fisher forecast from the FRB observations in Table~\ref{table}. We obtain an improved constraint $\sigma(\Lambda)|_{\rm FRB+GW}=200$ due to the breakage of the degeneracy between the masses and tidal deformability parameter.

We also perform similar forecasts assuming GWs from the same binary are detected by a 3G detector, Cosmic Explorer (CE)
(with the baseline 40 km design \cite{Evans2021}). We find that the GW SNR$=367$,
and $\sigma(\Lambda)|_{\rm GW}=\sigma(\Lambda)|_{\rm FRB+GW}=25$. This is expected as 3G detectors can already measure the masses with high precisions.

\subsection{Testing Gravity with Universal Relations}
\label{sec:univ_rel}

Let us now study the application of multimessenger observations on tests of gravity. As discussed in Sec.~\ref{sec:nuclear}, relations between two NS observables typically depend on the underlying EoSs. Therefore, although such relations should also depend on the gravitational theory and independent measurement of the two quantities can be used, in principle, to probe gravity, there is some degeneracy between uncertainties in nuclear physics and gravitational physics. One can avoid this problem by using universal relations mentioned earlier, as such relations do not depend sensitively on EoSs, and thus, one can probe gravity without being contaminated by uncertainties in nuclear physics~\cite{Yagi:2013bca,Yagi:2013awa,Yagi:2016bkt,Gupta:2017vsl,Silva:2020acr}.

\subsubsection{Q-Love Relation}
\label{sec:Q-Love}

Examples of such universal relations include the I-Love-Q relations~\cite{Yagi:2013bca,Yagi:2013awa} between the moment of inertia, tidal Love number (or tidal deformability), and quadrupole moment. The one that is relevant for our analysis is the Q-Love relation, as one can combine the measurements of the quadrupole moment from FRBs and tidal deformability from GWs\footnote{The universal I-Love relation for magnetars has recently been studied in~\cite{Zhu:2020imp} and the effect of magnetic field was shown to be negligible even for the magnetic field strengh of $10^{15}$G.}

Let us first work in a model-independent framework. Following~\cite{Silva:2020acr} for the parameterized I-Love relation, we can construct a parameterized Q-Love relation as follows:
\begin{equation}
\label{eq:Q-Love-par}
\kappa_i (\Lambda_i) = \kappa_i^{\mathrm(GR)}(\Lambda_i) + \beta_\Lambda \Lambda_i^{-b_\Lambda/5}\,,
\end{equation}
where $\kappa_i^{\mathrm(GR)}(\Lambda_i) $ is the universal relation between $\kappa_i$ and $\Lambda_i$ in GR, while $(\beta_\Lambda,b_\Lambda)$ are generic non-GR parameters describing a deviation in the relation from GR. The non-GR term is parameterized to be proportional to $\Lambda_i^{-b_\Lambda/5}$ based on the fact that the stellar compactness is proportional to $\Lambda_i^{-1/5}$ in the Newtonian limit.

\begin{figure*}
\includegraphics[width=8.5cm]{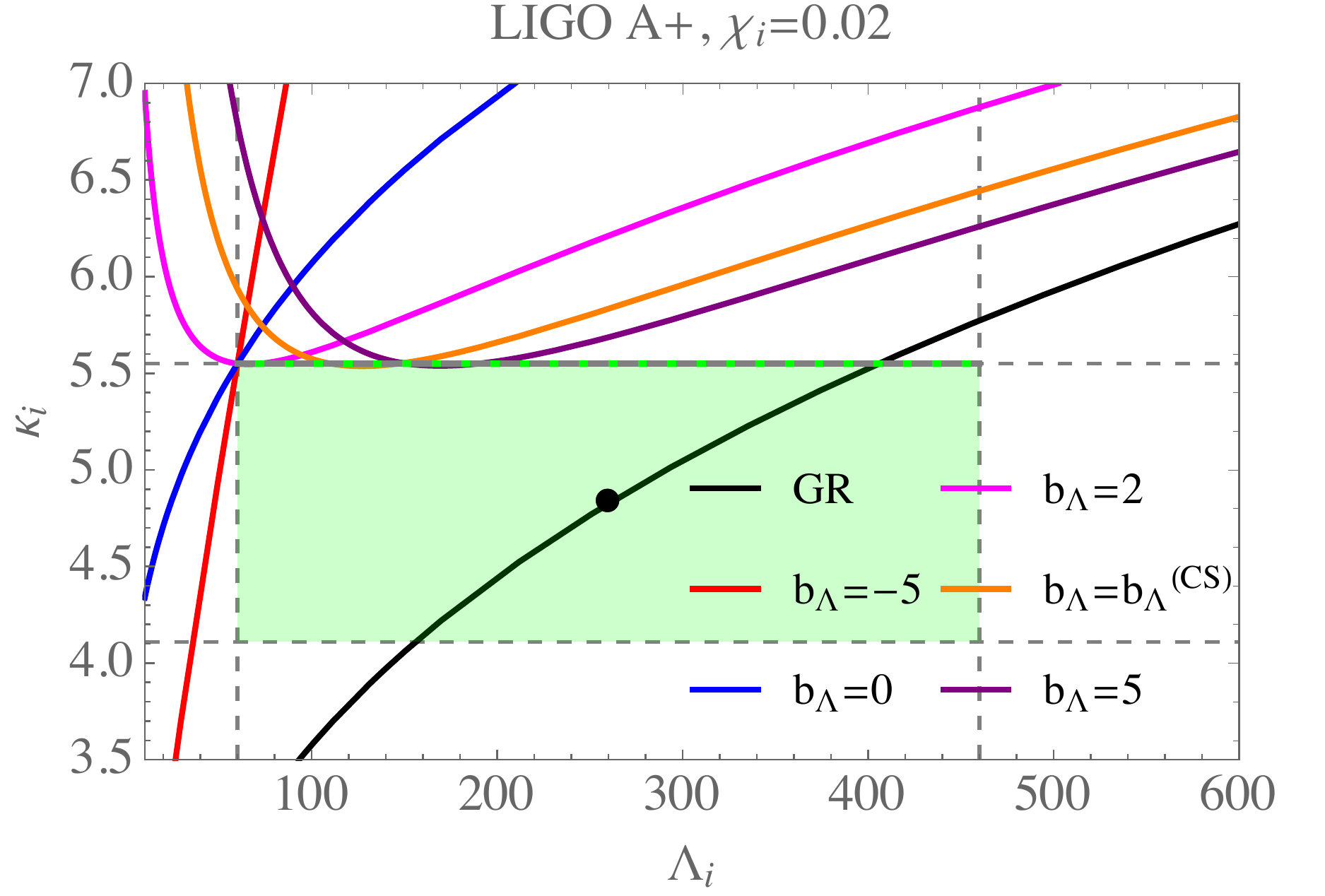}
\includegraphics[width=8.5cm]{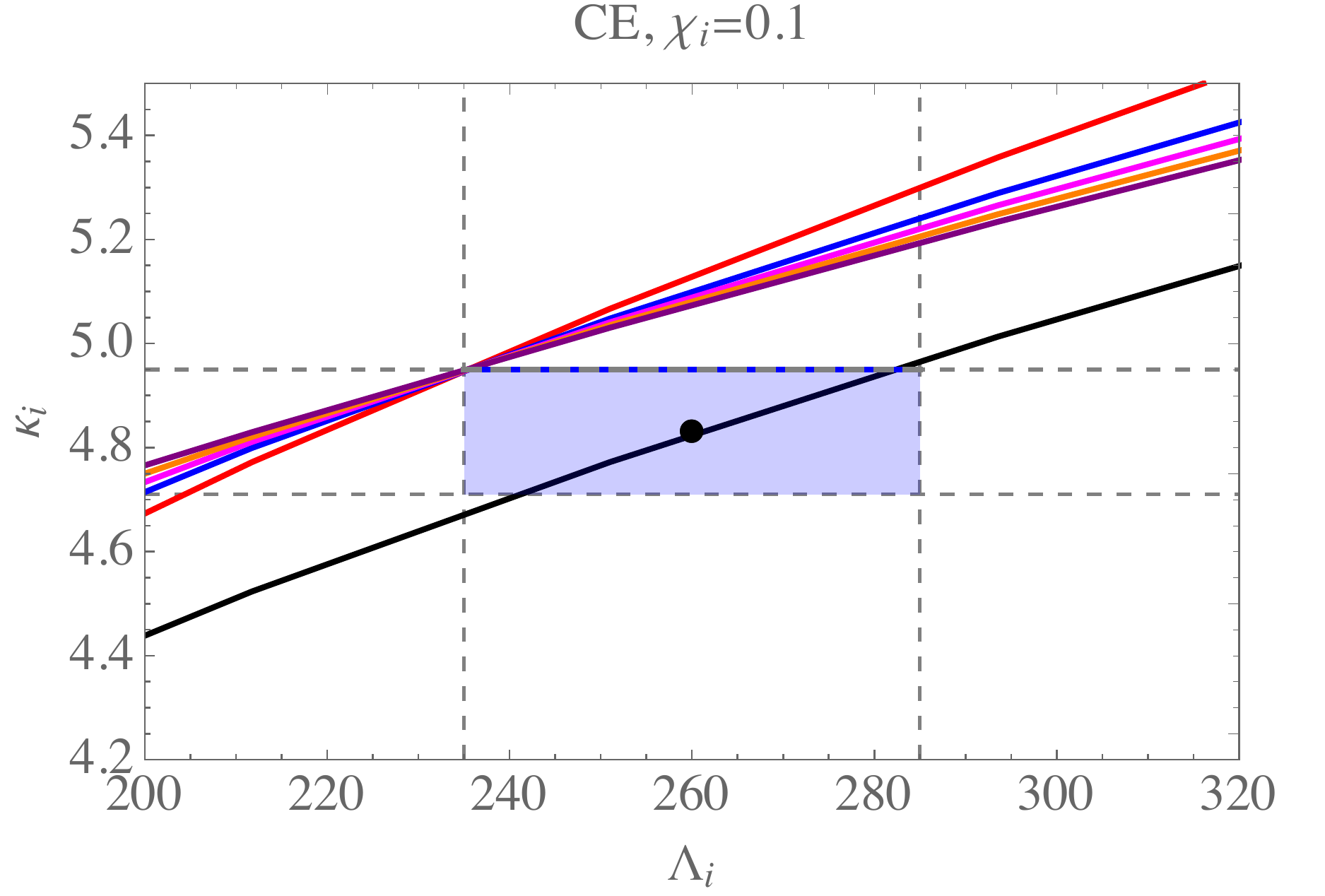}
\caption{\label{fig:Q-Love} Multimessenger tests of gravity using the parameterized Q-Love relation. The black curves are the GR relations between the quadrupole moment and tidal deformability, while colored curves are for parameterized non-GR relations. $b_\Lambda^\mathrm{(CS)}$ is the index $b_\Lambda$ for dCS gravity given in Eq.~\eqref{eq:beta_b_mapping_QLove}. For each $b_\Lambda$, we choose $\beta$ such that the parameterized relation is marginally consistent with the projected measurement errors on $\kappa_i$ and $\Lambda_i$ shown by the green or blue shaded region. We choose the fiducial values as $(\Lambda_i,\kappa_i) = (260,4.83)$ that are shown by the black dots. Similar to Figs.~\ref{fig:M-kappa} and~\ref{fig:M-R}, we consider two cases, (i) LIGO A+ and the NS spin of $\chi_i = 0.02$ (left panel), and (ii) CE and $\chi_i=0.1$ (right panel).
}
\end{figure*}

\begin{figure}
\includegraphics[width=8.5cm]{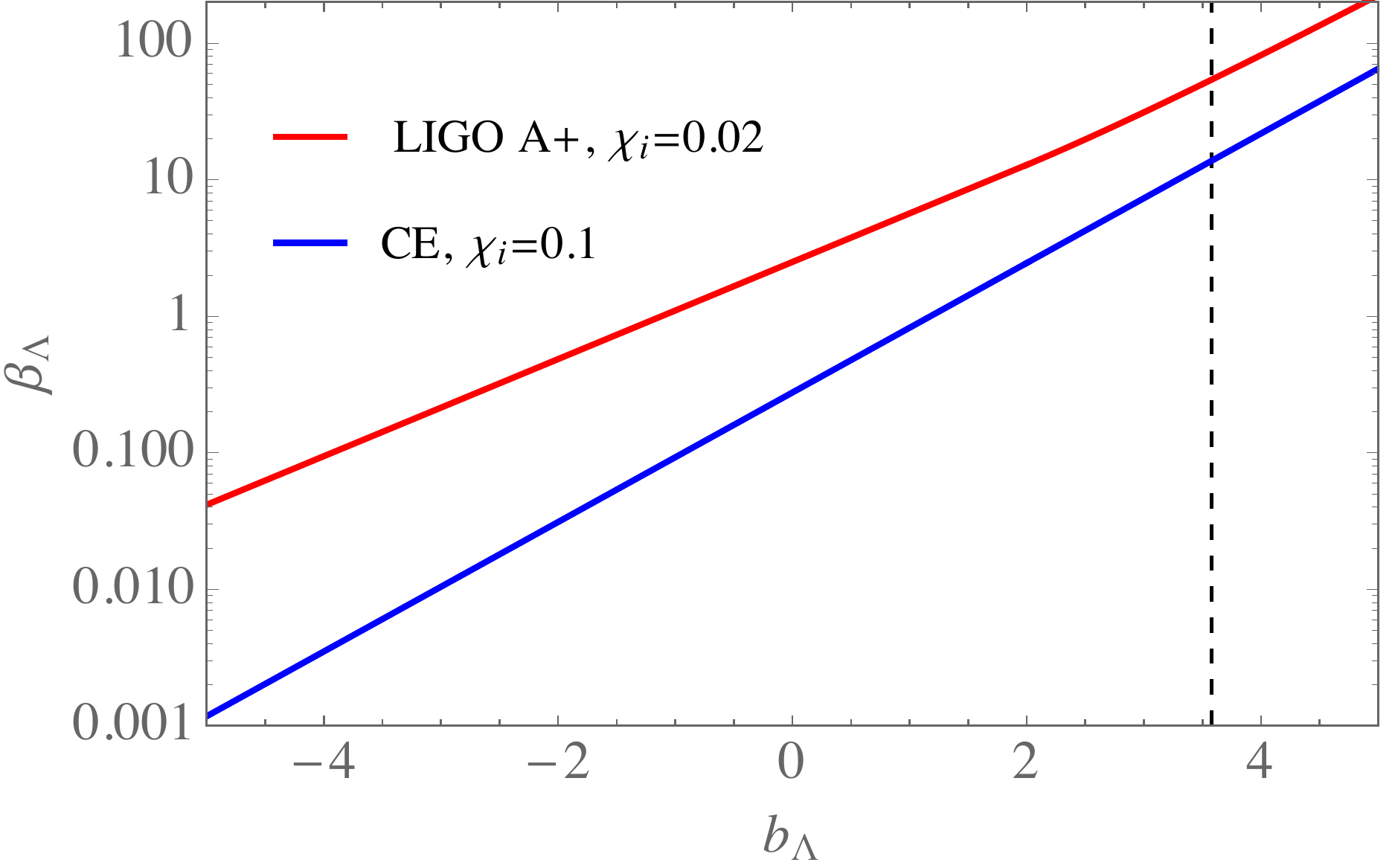}
\caption{\label{fig:beta-bound}
Upper bound on $\beta_\Lambda$ as a function of $b_\Lambda$ from the multimessenger observations of FRBs and GWs. The vertical dashed line corresponds to $b_\Lambda=b_\Lambda^\mathrm{(CS)}$.
}
\end{figure}

Figure~\ref{fig:Q-Love} presents the Q-Love relation in both GR and non-GR cases, with the latter given by several example combinations of $(\beta_\Lambda,b_\Lambda)$. For each value of $b_\Lambda$, we choose $\beta_\Lambda$ so that the relation is marginally consistent with the measurement errors of $\kappa_i$ and $\Lambda_i$ that are found in Table~\ref{table} (we use the one for $\kappa_2$) and Sec.~\ref{sec:tidal}. We choose the fiducial values of $(\Lambda_i,\kappa_i) = (260,4.83)$ and consider two cases: (i) using LIGO A+ for the GW measurement and assuming the spin of $\chi_i=0.02$ for FRB observations, and (ii) similar but for CE and $\chi_i=0.1$. Strictly speaking, the tidal deformability error estimated in Sec.~\ref{sec:tidal} is on $\Lambda$ and not on $\Lambda_i$. It would be difficult to measure individual tidal deformability from GW observations because there is a large degeneracy between $\Lambda_1$ and $\Lambda_2$. Here, we simply use the error on $\Lambda$ for that on $\Lambda_i$ as $\Lambda = \Lambda_i$ for an equal-mass binary. For an unequal mass binary, one can Taylor expand $\Lambda_i$ about a fiducial mass (like 1.4$M_\odot$) and search for the first Taylor coefficient ($\Lambda_i$ at the fiducial mass) to measure the tidal deformability at a given mass~\cite{Messenger:2011gi,Yagi:2015pkc,Yagi:2016qmr}. Although Fig.~\ref{fig:Q-Love} shows only several examples of $b_\Lambda$, one can repeat this for other values of $b_\Lambda$ and derive maximum values of $\beta_\Lambda$ that are consistent with the projected measurements on $\kappa_i$ and $\Lambda_i$. Figure~\ref{fig:beta-bound} shows such an upper bound on $\beta_\Lambda$ as a function of $b_\Lambda$ for the two cases.

Let us now put the parameterized test into context by considering a specific modified theory of gravity. We choose dynamical Chern-Simons (dCS) gravity~\cite{Jackiw:2003pm,Alexander:2009tp}. In this theory, a pseudo-scalar field is coupled to a Pontryagin density with the coupling constant $\alpha$ (that has a dimension of length squared) in the action. This coupling constant has been constrained to $\sqrt{\alpha} < \mathcal{O}(10^8)$ km from solar system and table top experiments~\cite{Ali-Haimoud:2011zme,Yagi:2012ya}. Recently, Silva \textit{et al}.~\cite{Silva:2022srr} derived a new bound from rindown observations of gravitational wave events as $\sqrt{\alpha} < 38.7$ km. Silva \textit{et al}.~\cite{Silva:2020acr} applied the tidal deformability measurement from LIGO/Virgo and the compactness measurement from NICER to the I-Love relation and found $\sqrt{\alpha} < 8.5$ km. In this last analysis, GR was assumed to be the correct theory when converting the measurement of compactness to that of the moment of inertia, which can be a strong assumption.

\begin{figure}
\includegraphics[width=8.5cm]{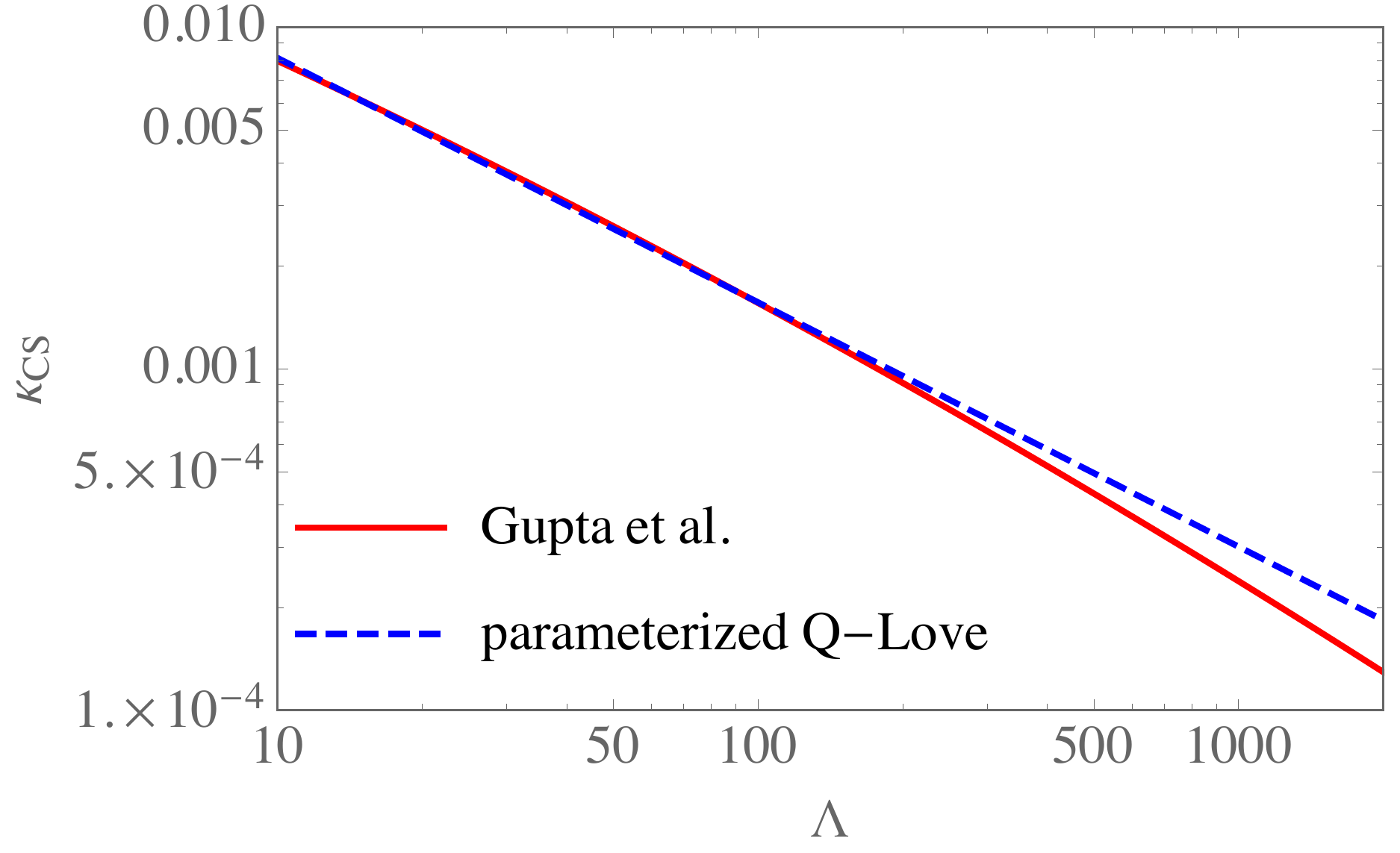}
\caption{\label{fig:dCS-fit}
DCS correction to the quadrupole moment as a function of the tidal deformability. We show the relation found in Gupta \textit{et al}.~\cite{Gupta:2017vsl} and the fit in Eq.~\eqref{eq:Q-Love-par} for the parameterized model.
}
\end{figure}

The I-Love-Q relations in dCS gravity have been studied in~\cite{Gupta:2017vsl}. In particular, the Q-Love relation is given by
\begin{equation}
\kappa_i (\Lambda_i) = \kappa_i^{\mathrm{(GR)}}(\Lambda_i) + \bar \xi_i \,\kappa_\CS(\Lambda_i)\,,
\end{equation}
where $\bar \xi_i = 16\pi \alpha^2/m_i^4$ and the fit to $\kappa_\CS(\Lambda_i)$ can be found in~\cite{Gupta:2017vsl} that is also shown in Fig.~\ref{fig:dCS-fit}. To find the mapping between $(\beta_\Lambda,b_\Lambda)$ and $\alpha$, we can fit the Q-Love relation in dCS to the parameterized form in Eq.~\eqref{eq:Q-Love-par} (a similar mapping for the I-Love relation can be found in~\cite{Silva:2020acr}). We find
\begin{equation}
\label{eq:beta_b_mapping_QLove}
\beta_\Lambda^\mathrm{(CS)} = 4.22 \times 10^{-2} \bar \xi\,, \quad b_\Lambda^\mathrm{(CS)} = 3.58\,.
\end{equation}
$\kappa_\CS$ for this new fit is also shown in Fig.~\ref{fig:dCS-fit}. Notice that the fit is accurate at low $\Lambda$ while it deviates from the correct values at high $\Lambda$. Practically, this is not an issue as the CS correction is much smaller at such high $\Lambda$ than at low $\Lambda$.
Using this mapping, we can convert the upper bound on $\beta_\Lambda$ in Fig.~\ref{fig:beta-bound} at $b_\Lambda=b_\Lambda^\mathrm{(CS)}$ to that on $\sqrt{\alpha}$. The results are summarized in Table~\ref{table:dCS}. These constraints are much stronger than the existing bounds mentioned earlier.

\begin{table}[t]
  \centering
%  \resizebox{2\columnwidth}{!}{%
  \begin{tabular}{ c | cc}
    \hline
      & $\chi_i = 0.02$ & $\chi_i=0.1$\\ \hline
Q-Love & \textbf{4.65} km & \textbf{3.30} km \\
Q-C & \textbf{5.92} km & \textbf{5.54} km\\
    \hline
  \end{tabular}
 % }
  \caption{Projected bounds on the coupling constant $\sqrt{\alpha}$ in dCS gravity using Q-Love or Q-C universal relation. These bounds are stronger than the existing bounds from solar system experiments~\cite{Ali-Haimoud:2011zme}, table-top experiments~\cite{Yagi:2012ya}, GWs~\cite{Silva:2022srr}, and multi-messenger observations~\cite{Silva:2020acr}.}\label{table:dCS}
\end{table}

\subsubsection{Q-C Relation}

\begin{figure}
\includegraphics[width=8.5cm]{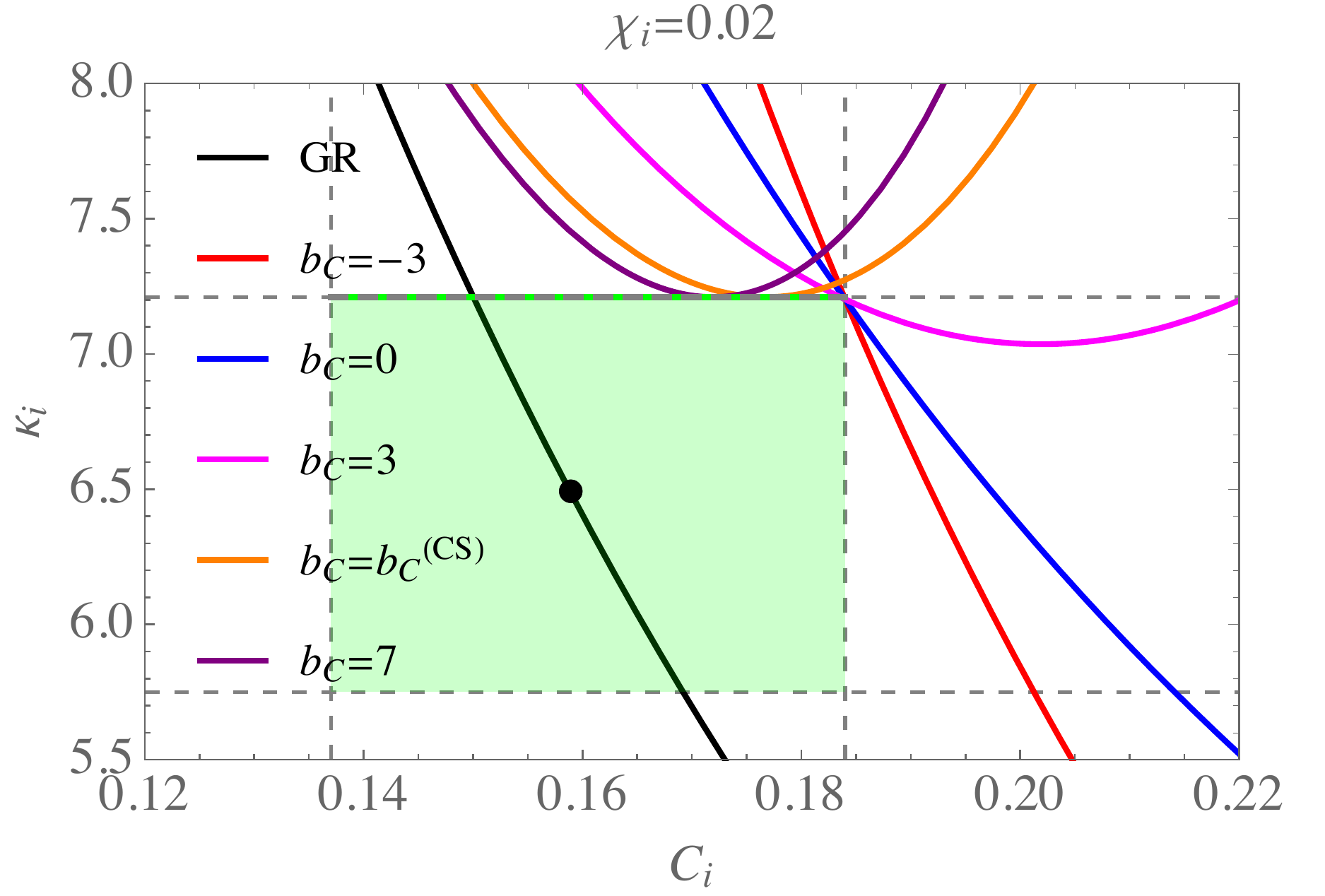}\\
\includegraphics[width=8.5cm]{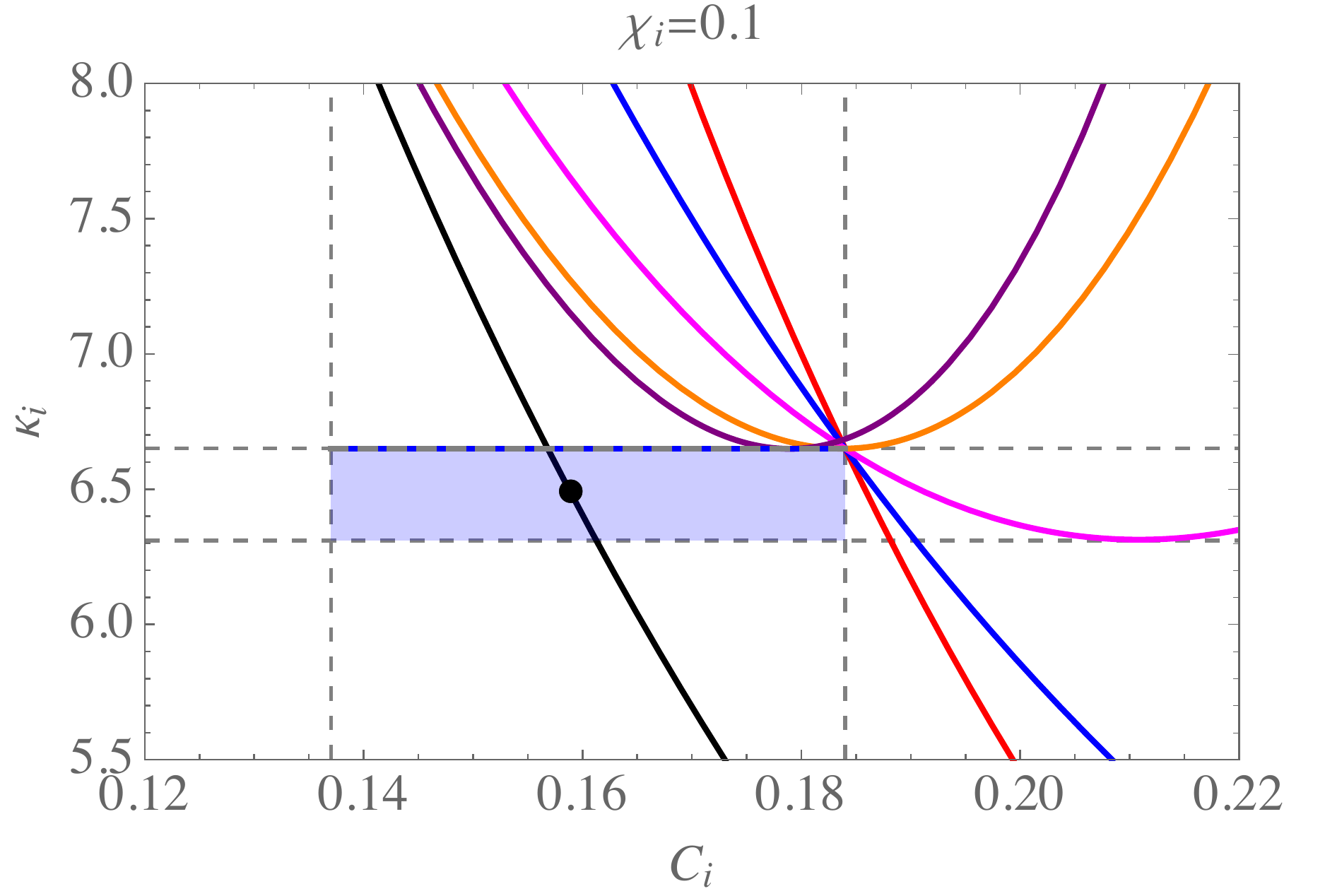}
\caption{\label{fig:Q-C} Similar to Fig.~\ref{fig:Q-Love} but for the relation between the quadrupole moment and compactness. The fiducial values are chosen as $(C_i,\kappa_i) = (0.159,6.48)$.
}
\end{figure}

\begin{figure}
\includegraphics[width=8.5cm]{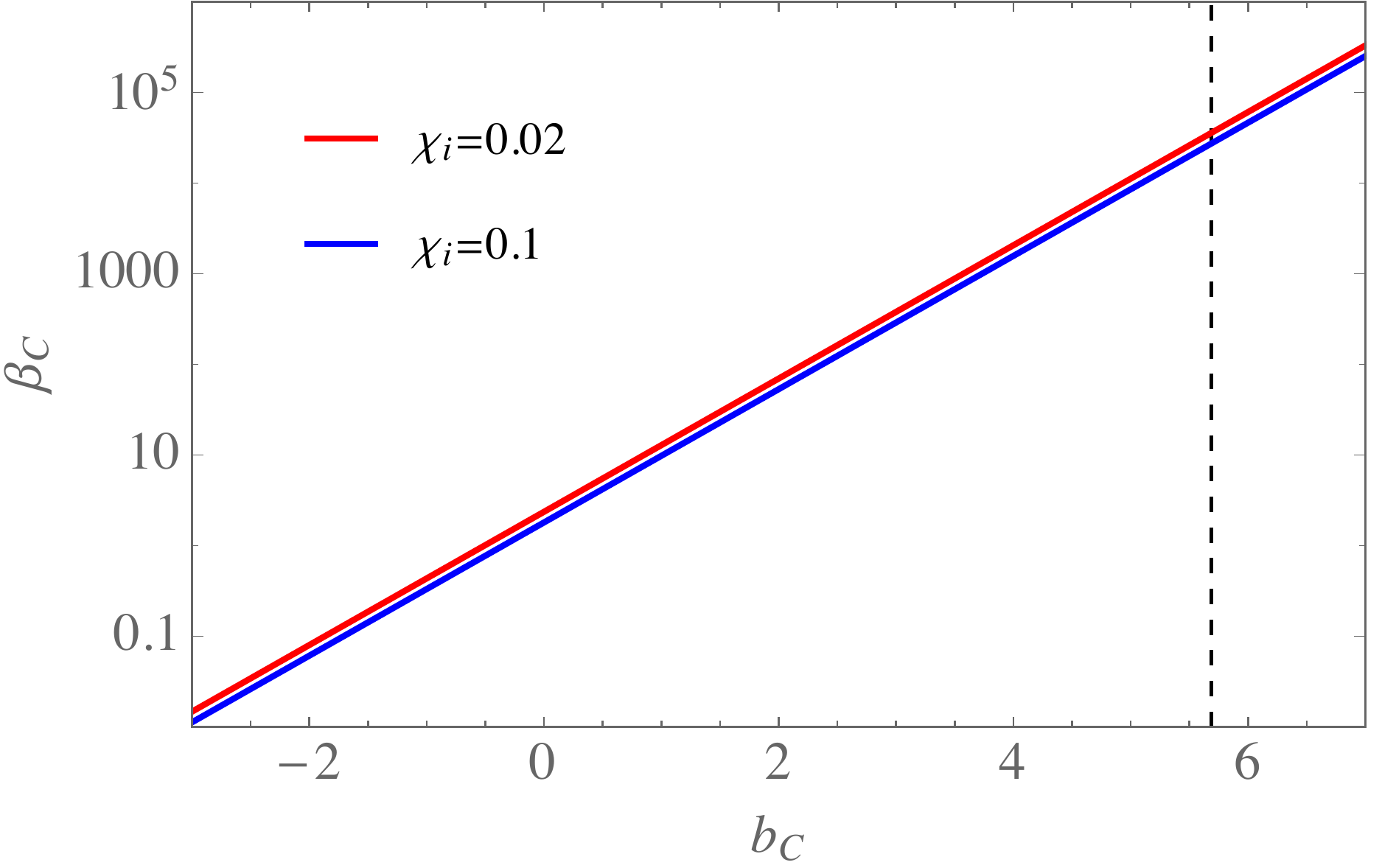}
\caption{\label{fig:beta-bound_QC}
Upper bound on $\beta_C$ as a function of $b_C$ from the multimessenger observations of FRBs and an X-ray pulsar (with NICER). The vertical dashed line corresponds to $b_C=b_C^\mathrm{(CS)}$.
}
\end{figure}

In Sec.~\ref{sec:nuclear}, we explained another universal relation, the one between the quadrupole moment and compactness (the so-called Q-C relation). The compactness of a NS has been measured with x-ray observations with NICER, so one can combine FRB and NICER observations to test gravity through the Q-C relation. Although FRB and NICER sources are different NSs, one can still apply the universal relations derived from a single NS provided we use measurements of NS observables with the \emph{same} mass. To be more specific, the posterior samples on the NS mass and radius from NICER observations have been used to infer the compactness of a NS with a mass of $1.4 M_\odot$, which can be combined with the measurement error on $\kappa_i$ for a NS with $1.4M_\odot$ from FRB. If FRBs are detected for a NS with a mass other than $1.4M_\odot$, say $m_\FRB$, we can derive the inferred value for the compactness of a NS with mass $m_\FRB$ from the NICER observation provided the mass for the NICER source is similar to $m_\FRB$~\cite{Silva:2020acr}\footnote{There are other ways to combine information from two NSs. For example, if we have a measurement of $\kappa_a$ from a NS with mass $m_a$ and $C_b$ from another NS with mass $m_b$, one can construct a new universal relation between $\kappa_a$ and $C_b$ at these specific masses. Instead of having one curve for each EoS, we have one point for each EoS in the $\kappa_a$--$C_b$ plane but repeating this for multiple EoSs provides multiple points which can be approximately connected by a single curve, and thus giving a universal relation~\cite{Saffer:2021gak}.}.

Similar to the Q-Love relation, we begin by constructing a parameterized relation of the form
\begin{equation}
\kappa_i (C_i) = \kappa_i^{\mathrm{(GR)}}(C_i) + \beta_C C_i^{b_C}\,,
\end{equation}
where $\kappa_i^{\mathrm{(GR)}}(C_i)$ is the GR relation while $(\beta_C,b_C)$ are generic non-GR parameters. Figure~\ref{fig:Q-C} presents the Q-C relations with various $(\beta_C,b_C)$. We also present the measurement errors on $C_i$ and $\kappa_i$. We choose the fiducial values of $(C_i,\kappa_i) = (0.159,6.48)$. The measurement error on $\kappa_i$ is given in Table~\ref{table} while the inferred value for $C_i$ with 1.4$M_\odot$ is given by $C_i = 0.159^{+0.025}_{-0.022}$~\cite{Silva:2020acr}.
 For each $b_C$, we choose $\beta_C$ that gives a marginally consistent Q-C relation with the above measurement errors. Figure~\ref{fig:beta-bound_QC} shows the bound on $\beta_C$ as a function of $b_C$ for $\chi_i = 0.02$ and 0.1. Notice that the bounds are not very sensitive to the spin of the NS.

Let us now map the bound on $\beta_C$ to that on dCS parameter $\alpha$. In dCS gravity, the compactness is only modified from GR at second order in spin~\cite{Yagi:2013mbt}. For the fiducial values of spins considered in this paper, such an effect is negligible. Therefore, we can easily turn the Q-Love relation in dCS discussed in Sec.~\ref{sec:Q-Love} to the Q-C relation by using the universal relation between the tidal deformability and compactness in GR~\cite{Maselli:2013mva,Yagi:2016bkt}. We can then derive the mapping between $(\beta_C,b_C)$ and $\alpha$. We find
\begin{equation}
\label{eq:beta_b_mapping_QC}
\beta_C^\mathrm{(CS)} = 10.7 \bar \xi\,, \quad b_C^\mathrm{(CS)} = 5.69\,.
\end{equation}
Finally, we can convert the bound on $\beta_C$ in Fig.~\ref{fig:beta-bound_QC} to that on $\sqrt{\alpha}$. The results are summarized in Table~\ref{table:dCS}. The bounds are similar but slightly weaker than those from the Q-Love relation. Yet these bounds are stronger than all the existing bounds.

\subsection{Testing Beyond Standard Models}\label{sec:bsm}
As an example of beyond standard models, we consider axions which are hypothetical light scalar particles predicted by string theory \cite{Svrcek2006,Mina2010}.
If coupled to matter, axions can mediate long range interactions between compact objects in a binary and thus modify the
binary dynamics. Searching for the axions from gravitational waves of compact binary mergers has been investigated previously, e.g., \cite{Huang2019,Zhang2021}. We will show that the FRB timing is also a sensitive probe to the axions coupled to NSs.

Considering an axion field of mass $m_a$ and a binary NS carrying axion charge $Q_{1,2}$,
which are related to the NS radius $R_{1,2}$ by \cite{Hook2018,Huang2019}
\be\label{eq:Q12}
Q_{1,2} = \pm 4\pi^2 f_a R_{1,2} \ , \ee
where $f_a$ is the axion decay constant.
Accurate to the leading order, the scalar charges induce an effective potential
\be\label{eq:V12}
V_a(r) = -\frac{1}{4\pi}\frac{Q_1Q_2}{r} e^{-m_ar}= -\gamma_a \frac{m_1m_2}{r}e^{-m_ar}\ .
\ee
 with $\gamma_a:=Q_1Q_2/m_1m_2$. As the binary orbits around each other, the axion field also emits scalar waves with power (accurate to the leading order)
\be
\begin{aligned}
  P_a(r,\omega)
  &= \left(\frac{Q_1}{m_1}-\frac{Q_2}{m_2} \right)^2 \left(\frac{m_1m_2}{m_1+m_2} \right)^2
  \frac{r^2\omega^4}{12\pi}\left(1-\frac{m_a^2}{\omega^2} \right)^{3/2}\mathcal{H}(\omega-m_a)\\
  &:= \delta q^2 \left(\frac{m_1m_2}{m_1+m_2} \right)^2
  \frac{r^2\omega^4}{12\pi}\left(1-\frac{m_a^2}{\omega^2} \right)^{3/2}\mathcal{H}(\omega-m_a)\ ,
\end{aligned}
\ee
In the presence of the axion field, the binding energy and the energy dissipation rate of the binary are modified as $E(r) = E_{\rm GR}(r,\omega) + V_a(r)$, and $dE/dt = (dE/dt)_{\rm GR} - P_a(r,\omega)$, respectively, where
\be
\begin{aligned}
  E_{\rm GR} &= \frac{\eta m}{2}r^2\omega^2 \\
  +& \eta m \left\{-\frac{m}{r} +\frac{m}{2r}\left[(3+\eta)r^2\omega^2
  +\frac{m}{r}+ \frac{3}{(1-3\eta)}{8}r^4\omega^4\right] \right\}\ ,\\
\frac{dE}{dt}\Big|_{\rm GR} &=-\frac{32}{5}\eta^2m^2 r^4\omega^6\\
&\times \left\{ \left[1-(1-2\eta)\frac{m}{r} \right]^2 + \left(\frac{769-2772\eta}{336}r^2\omega^2 \right) \right\}
\end{aligned}
\ee
with the  modified Kepler's law due to the axion field given by
\be
r^2\omega^2 = \frac{m}{r} \left[1-(3-\eta)\frac{m}{r} + \gamma_a e^{-m_ar}(1+m_ar)\right]\ ,
\ee
accurate to 1PN order.
As a result, the dynamical equations of the binary are the same as Eqs.~(\ref{eq:dyn}),
except with
\be
\frac{dr}{dt} = \frac{dE/dt}{dE/dr}= \frac{dE/dt|_{\rm GR}-P_a(r,\omega)}{dE/dr|_{\rm GR} + dV_a/dr}\ .
\ee

Let us first study how well one can probe axions with the FRB timing alone.
As an example, we consider the massless axion limit $m_a\rightarrow 0$  and  the same fiducial binary as in Sec. \ref{sec:forecast} (with fiducial $\gamma_a=\delta q^2=0$), and forecast the constraints on axion charge parameters
$\gamma_a$ and $\delta q_a$ (in addition to the 14 base model parameters) using the same FRB timing. We find little change in the constraints on the base model parameters (Table~\ref{table}),
and
\be
\sigma(\gamma_a)=2.4\times 10^{-6},\ \sigma(\delta q^2)=5.4\times 10^{-8}\ .
\ee
The axion decay constant $f_a$ are related to $\gamma_a$ and $\delta q^2$ via Eqs.~(\ref{eq:Q12},\ref{eq:V12}).
In the case of $Q_1Q_2<0$ and $m_1 \approx m_2$, we have $R_1 \approx R_2$ and $Q_1 \approx - Q_2$. Then, $\delta q^2\approx (2Q_1/m_1)^2$ and $\gamma_a \approx(Q_1/m_1)^2 $, and it is clear that
$f_a$ is mainly constrained by the dipole radiation whose power is proportional to $\delta q^2$.
On the other hand, in the case of $Q_1Q_2 > 0$, $\delta q^2\approx (Q_1/m_1)^2(1-R_1/R_2)^2$ which is consistent with zero due to the uncertainty in the NS radii. Therefore $f_a$ is mainly constrained by the axion potential which is proportional to $\gamma_a$.
As a result, we find that ${\rm GeV}/f_a \lesssim 1.2\times 10^{-13}(R/10\ {\rm km})^{1/2}$ and ${\rm GeV}/f_a \lesssim 10^{-14}(R/10\ {\rm km})^{1/2}$ are expected to be excluded at 3$\sigma$ confidence level in the former and latter case, respectively. These projected constraints are orders of magnitude tighter than the existing constraints from GW170817 \cite{Zhang2021}.

\begin{figure}
\includegraphics[scale=0.6]{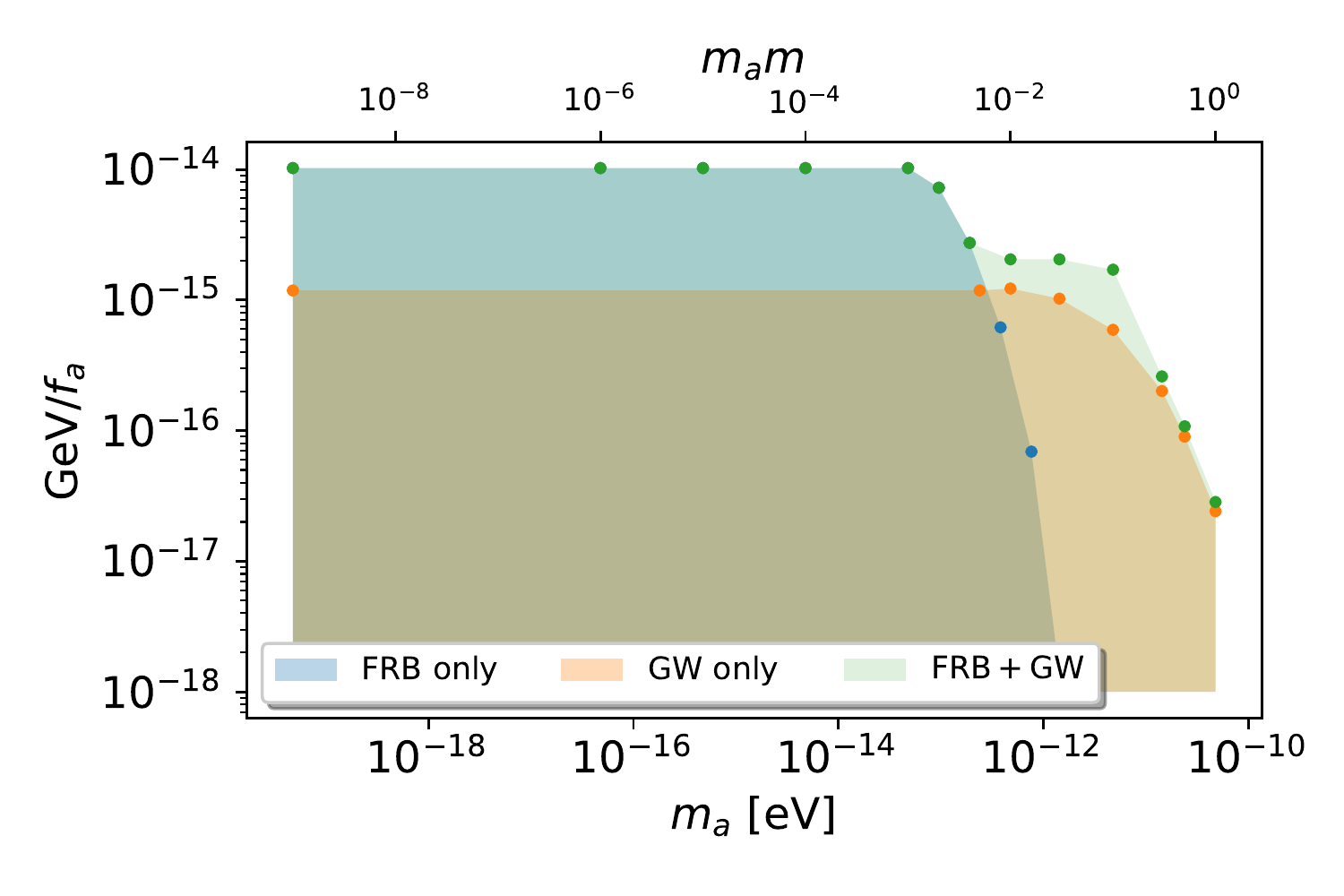}
\caption{\label{fig:bsm} Forecasted exclusion parameter space at 3-$\sigma$ confidence level
from FRB timing alone, GW alone and the combination assuming a $1.4 M_\odot+1.4 M_\odot$ magnetar+NS binary
(more binary model parameters are detailed in Table~\ref{table} and Section~\ref{sec:bsm}
).
}
\end{figure}

In the following discussion, we will make the conservative assumption $Q_1Q_2 > 0$ and $P_a=0$ in forecasting the $f_a$ constraints for a general axion mass $m_a$. The forecast result is shown in Fig.~\ref{fig:bsm}, where the constraint converges to the zero-mass limit result for $m_a\lesssim 10^{-13}$ eV.
However, the FRB timing is not expected to be so constraining in the case of a more massive axion field (say $m_a=\frac{1}{50 m}$)
because the axion charge potential $V_a(r)$ is strongly suppressed at large binary separations where we expect to detect the majority of the burst pulses (Fig.~\ref{fig:mock}).
In that case, multi-band and multi-messenger observations (FRBs and GWs) will be valuable.

Thus, let us now consider adding GW observations.
In the presence of an axion field, the gravitational waveforms of an inspiraling binary
gains an extra phase shift, which (accurate to 1 PN order) is formulated as
\be
h(f)=\mathcal{A} f^{-7/6} e^{i \left[\Psi_{\rm pp}(f)+\delta \Psi_a(f) \right]}\ ,
\ee
where $\mathcal A$ is the amplitude, $\Psi_{\rm pp}$ is the phase in the GR case
\be
\begin{aligned}
\delta\Psi_a(f)=
& \frac{5\gamma_a}{64\eta v^5} e^{-\alpha/v^2}\Big[-4-\frac{32v^2}{\alpha}-\frac{138v^4}{\alpha^2}
  -\frac{360v^6}{\alpha^3}\\
  +\frac{360v^8}{\alpha^4}&(e^{\alpha/v^2}-1)-21\sqrt{\pi}\frac{v^5}{\alpha^{5/2}}e^{\alpha/v^2}
  {\rm Erf}(\sqrt{\alpha/v^2}) \Big]\ ,
\end{aligned}
\ee
with $v=(\pi m f)^{1/3}, \eta=(m_c/m)^{5/3}, \alpha=m_a m$ and Erf being the error function.

We consider the same compact binary as in the case of FRB observations alone and forecast the constraints of GW model parameters $\{\mathcal{A}, t_c, \phi_c, m, m_c, \gamma_a\}$ for different axion masses $m_a$ assuming the binary is observed by LIGO A+. The fiducial model parameters are chosen as follows: $t_c=\phi_c=\gamma_a = 0, m=2.8 M_\odot, m_c=m/4^{3/5}$
and $\mathcal{A}$ is chosen such that the SNR equals 20.
The forecast result with GW observations alone is also shown in Fig.~\ref{fig:bsm}, where we see the heavier axion field can be constrained by the GW signal, simply because LIGO A+ is sensitive to the binary dynamics at smaller separations [Eq.~(\ref{eq:V12})].

Let us finally study the case with multi-messenger/multi-band observations. We give forecasts on
the constraints using the GW signal with the prior information of NS masses from the FRB timing.
As a result, we find the constraint on $\gamma_a$ (or equivalently on $f_a$)
is dominated by the FRB timing for $m_a\lesssim 10^{-13}$ eV,  and for a more massive axion field, the uncertainty $\sigma(\gamma_a)|_{\rm FRB+GW }$  is further lowered by a factor
of few than $\sigma(\gamma_a)|_{\rm GW}$ because
the extra constraints on the NS masses from the FRB timing break the degeneracy between $\gamma_a$ and the chirp mass \cite{Zhang2021}.

\section{Rate Prospect}\label{sec:observ}
In the final part of this paper, we will discuss the rate prospect of observing FRB emitter in merging compact binaries. We will first present a general argument for the rate of forming FRB emitters in binaries, given a number of assumptions. In the case that FRB 180916 is indeed in a binary, we are able to quantitatively estimate the rate of observing FRB repeaters in merging binaries. We will also comment on the detection criteria for non-repeaters in a merging NS binary.

 We  use the volumetric birth rate $R_{\rm FRB}$ of all FRB emitters (either in isolation or in binaries) to infer the number of FRB emitters in close binaries that are going to merge while the FRB emitter is still active
assuming a constant fraction
$f_{\rm bin}:=R_{\rm FRB}^{\rm bin}/R_{\rm FRB}$. Here $R_{\rm FRB}^{\rm bin}$ is the volumetric birth rate of FRB emitters in binaries.
Following Ref.~\cite{Nicholl2017}, the detection rate of FRBs  is formulated as
\be\label{eq:Rd}
\mathcal R(>S_\nu^{\rm lim})  = \int \frac{dz}{1+z} \frac{dV}{dz} \mathcal{B}\mathcal{D}
r_{\rm FRB} (>L_{\nu,{\rm lim}}^{\rm iso}) R_{\rm FRB}(z) \tau_{\rm FRB}\ ,
\ee
where $dV/dz$ is the differential comoving volume as a function of the redshift $z$, $S_\nu^{\rm lim}$ is the flux density threshold for detection,
$L_{\nu,{\rm lim}}^{\rm iso}=4\pi D_{\rm L}^2 S_{\nu}^{\rm lim}$ is
the corresponding isotropic luminosity density limit with  the luminosity distance $D_{\rm L}$,
$R_{\rm FRB}(z)=R_0(1+z)^{3.38}$ is the volumetric birth rate of FRB emitters assuming it follows the cosmic star formation rate,
$\tau_{\rm FRB}$ is the typical lifetime of magnetars as active FRB emitters, $\mathcal B=(\delta\theta)^2/4$ is the FRB emission beaming factor with $\delta\theta$ the half opening angle of FRB emissions,  $\mathcal D$ is the FRB emission duty cycle, and
$r_{\rm FRB} (>L_{\nu,{\rm lim}}^{\rm iso})$ is the average burst rate per FRB emitter in the active phase.
Using existing constraints $r_{\rm FRB} (>L_{\nu,{\rm lim}}^{\rm iso}) \approx 3.5(L_{\nu,{\rm lim}}^{\rm iso}/10^{32} {\rm erg/s/Hz})^{-0.6}\ {\rm day}^{-1}$, $\mathcal R(> S_\nu^{\rm lim}=1\ {\rm Jy})\approx 2870 \ {\rm day}^{-1}$ and Eq.~(\ref{eq:Rd}), the authors of \cite{Nicholl2017} obtained $\mathcal{B}\mathcal{D} R_0\tau_{\rm FRB} \approx 130 \ {\rm Gpc}^{-3}$ .
Then the total birth rate of FRB emitters  is
\be
\int_0^{z_+} dz \frac{dV}{dz}\frac{R_{\rm FRB}(z)}{1+z}   \approx 600 \left(\frac{\tau_{\rm FRB}}{10^3\ {\rm yr}}\right)^{-1} \mathcal{B}^{-1}\mathcal{D}^{-1} \ {\rm yr}^{-1}\ ,
\ee
where we have used $z_+=2$, the redshift where the cosmic star formation rate peaks.
At this stage, let us assume that the birth of FRB emitters is independent of their environment, i.e., whether they  already reside in  binaries or they are isolated NSs. This assumption is at most an approximation, as physically one expects that the emission of FRBs should depend on the evolution stage of NSs. Nevertheless, we shall carry through the analysis to see the corresponding rate under such an assumption.
 The resulting total number  of compact binaries possessing FRB emitters $N(<t_{\rm gw})$
with a merger time shorter than $t_{\rm gw}$ and part of the FRB pulses arriving on the earth, as
\be\label{eq:N2}
\begin{aligned}
N
&=f_{\rm bin} \mathcal{B}_{\rm bin} \int_0^{z_+} dz \frac{dV}{dz}\frac{R_{\rm FRB}(z)}{1+z}   t_{\rm gw}\ ,\\
&\approx   10^3 \left(\frac{f_{\rm bin}}{10^{-2}}\right)\left(\frac{\tau_{\rm FRB}}{10^3\ {\rm yr}}\right)^{-1} \left(\frac{\delta\theta}{0.3}\right)^{-1}\left(\frac{\sin\theta }{0.5}\right)
\left(\frac{\mathcal{D}}{0.4}\right)^{-1} \left(\frac{t_{\rm gw}}{10\ {\rm yr}}\right)\ .
\end{aligned}
\ee
where we have taken $t_{\rm gw}=10$ yr as a typical timescale that multi-band/multi-messenger observation may cover,
and we have used the fact that the emitter spin precession expands the effective FRB emission opening angle with
the beaming factor
$\mathcal{B}_{\rm bin}=\sin\theta\delta\theta$, with $\theta$ the angle between the magnetar spin axis $\mathbf S_1$ and the binary orbital angular momentum direction $\mathbf L_{\rm N}$. As the binary fraction $f_{\rm bin}$ is unconstrained, the above expression should be viewed as a relation between the rate and various physical quantities, instead of being a rate forecast.

 Among the over 500 FRBs reported, a low redshift ($z=0.0337$) repeating
 FRB 180916 showing $\sim 16$ day periodicity is of particular interest  \cite{CHIME16d}. For explaining the periodicity,
 many models have been proposed, e.g., the orbital period of a magnetar-companion star binary, the free precession period of
 a magnetically deformed NS, the spin period of an extremely slow-rotation magnetar, the geodesic precession period of a magnetar in a compact binary, etc.
To date, it is still unclear what is the underlying mechanism for the periodicity, but the predictions of many of these models are falsifiable with further observations \cite{Katz:2020jpv,Wei:2021vco,Li:2021rno}. If the geodetic precession interpretation is true \cite{Yang:2020qxt}, given the low redshift of the source, it means that there are many active FRB repeaters in the universe within binaries $\mathcal{O}(10^2)-\mathcal{O}(10^3)$ years away from merger. This allows us to estimate the rate of observing merging NS binaries with FRB repeaters in this scenario.

Assuming that FRB 180916 is an FRB emitter residing in a compact binary and
the occurrence of FRB 180916-like FRBs follows the Poisson distribution,
it is straightforward to show that the local number density of FRB 180916-like emitters is  given by
\be
n_0(<t_{\rm gw}^{\rm FRB 180916}) V(z\leq 0.0337) \approx 1_{-0.3}^{+2.3}
\ee
at 1-$\sigma$ confidence level, where $t_{\rm gw}^{\rm FRB 180916} \in \mathcal{O}(10^2)-\mathcal{O}(10^3)$ years is the time till merger, as a function of the binary mass ratio \cite{Yang:2020qxt}, and the local volumetric birth rate of FRB emitters residing in binaries
is $R_0^{\rm bin} = n_0(<t_{\rm gw}^{\rm FRB 180916})/t_{\rm gw}^{\rm FRB 180916}$.
Again assuming the birth rate of FRB 180916-like emitters follows the cosmic star formation rate,
$R^{\rm bin}(z)=R_0^{\rm bin}(1+z)^{3.28}$ for a redshift $z\leq z_+\approx 2$, we obtain
the total
number $N(< t_{\rm gw})$ of merging compact binaries with an FRB repeater and
with a merger time shorter than $t_{\rm gw}$ as
\be\label{eq:N1}
\begin{aligned}
  N(<t_{\rm gw})
  &= \int_0^{z_+} dz\frac{dV}{dz} \frac{R^{\rm bin}(z)}{1+z} t_{\rm gw} \\
  &\approx 3000^{+6900}_{-900}\left(\frac{t_{\rm gw}}{10 \ {\rm yr}}\right)
  \left(\frac{t_{\rm gw}^{\rm FRB 180916}}{10^3\ {\rm yr}}\right)^{-1}\ ,
\end{aligned}
\ee
The merger rate of such binaries is simply $\mathcal{R}_{\rm FRB-Bin} =   N(<t_{\rm gw})/t_{\rm gw}$,
therefore we obtain the expected detection rates by LIGO A+ and by 3rd-generation GW detectors as
\be\label{eq:Rmerger}
\begin{aligned}
  \mathcal{R}_{\rm FRB-Bin}^{\rm A+}(z<0.2)
  &\approx 0.3^{+0.69}_{-0.09} \ {\rm yr}^{-1}
  \left(\frac{t_{\rm gw}^{\rm FRB 180916}}{10^3\ {\rm yr}}\right)^{-1}\ , \\
  \mathcal{R}_{\rm FRB-Bin}^{\rm 3G}(z<z_+)
  &\approx 300^{+690}_{-90} \ {\rm yr}^{-1}
  \left(\frac{t_{\rm gw}^{\rm FRB 180916}}{10^3\ {\rm yr}}\right)^{-1}\ ,
\end{aligned}
\ee
where $z=0.2$ is approximately the LIGO A+ horizon for BNS detection
and the horizon of 3G detectors is well beyond $z=z_+$ \cite{Reitze2019}.
Note that the estimate in Eq.~(\ref{eq:Rmerger}) depends on the lifetime of FRB 180916, on which there is
no consensus yet: a kyr or even shorter lifetime is preferred from the birth rate argument \cite{Lu2022},
while the offset of FRB 180916 from a nearby star formation region suggests
a much longer liftelime ($\sim$ Myr) \cite{Tendulkar2021}.

We can also perform a rate estimation from GRB 211211A, assuming a magnetar in the inspiral stage is responsible for the observed QPOs in the precursor and the long duration of the GRB.
The merger rate of compact binaries with a repeating FRB emitter is related to the rate of GRB 211211A like events by
$R^{\rm bin}(z)\approx R^{\rm GRB\ 211211A}(z)\times(1/20)\times(18/474)$,
 considering that 1 FRB has been confirmed from
one of the 20 known galactic magnetars \cite{CHIME2020Nat,Bochenek2020,Mereghetti2020,Li2021Nat,Ridnaia2021,Tavani2021},
and the number  ratio of repeating FRBs over non-repeating FRBs
in the CHIME/FRB catalog   is $18/474$ \cite{CHIME-c1}.
The local rate $R_0^{\rm GRB\ 211211A}$ is approximately
\be
R_0^{\rm GRB\ 211211A} V(z < 0.076) \mathcal{B}_{\rm GRB} T_{\rm obs}\approx 1_{-0.3}^{+2.3}
\ee
with the GRB beaming factor $\mathcal{B}_{\rm GRB}$, and the average time interval to see a  GRB 211211A like event $ T_{\rm obs}$. This is a conservative estimate in which we have not included other selection effects,
e.g., the fraction and efficiency of kilonova follow-up after a long GRB.
Assuming the same redshift dependence $R^{\rm GRB\ 211211A}(z)=R_0^{\rm GRB\ 211211A}(1+z)^{3.28}$, we obtain
\be\label{eq:Rmerger_GRB}
\begin{aligned}
  \mathcal{R}_{\rm FRB-Bin}^{\rm A+}(z<0.2)
  &\approx 0.67^{+1.52}_{-0.20} \ {\rm yr}^{-1}
  \left(\frac{T_{\rm obs}}{10\ {\rm yr}} \right)^{-1}
  \left(\frac{\mathcal{B}_{\rm GRB}}{10^{-2}} \right)^{-1}
  \ , \\
  \mathcal{R}_{\rm FRB-Bin}^{\rm 3G}(z<z_+)
  &\approx 670^{+1500}_{-200} \ {\rm yr}^{-1}
  \left(\frac{T_{\rm obs}}{10\ {\rm yr}} \right)^{-1}
  \left(\frac{\mathcal{B}_{\rm GRB}}{10^{-2}} \right)^{-1}\ .
\end{aligned}
\ee
From the above estimates [Eqs.(\ref{eq:N2},\ref{eq:N1},\ref{eq:Rmerger},\ref{eq:Rmerger_GRB})], FRB repeaters in compact binaries may be a promising multi-messenger source of FRBs and GWs.\footnote{The recent detection of GRB 230307A further
strengthens this conclusion.}

As a comparison, we can estimate the lower limit of short GRB (sGRB)
rate from the BNS merger rate measured by LVK.
With the local merger rate density $R_{\rm BNS}\approx 160 \ {\rm Gpc}^{-3}\ {\rm yr}^{-1}$ \cite{LVK-GWTC3} and assuming the same redshift dependence, it is straightforward to find $\mathcal{R}_{\rm sGRB} \geq \mathcal{R}_{\rm BNS} (z\leq z_+)
\approx 6.3\times 10^4 {\rm yr}^{-1}$, which is two orders of magnitude higher than the estimated merger rate of
compact binaries with an FRB repeater [Eqs.~\eqref{eq:Rmerger},\eqref{eq:Rmerger_GRB}].

In a recent study \cite{Curtin2022}, the FRB-GRB association was searched and no
 GRBs were found within their sample that are coincident with any of the FRBs from the first CHIME/FRB
catalog when applying the joint temporal (up to 7 days)
and spatial criteria (overlapping $3\sigma$ localization regions). This null result is consistent with our
rate estimates above: among all the FRBs in the CHIME/FRB catalog, only one of them (FRB 180916) was found to be periodic, which is possibly a FRB emitter residing in a compact binary that is still long before the
final coalescence producing a sGRB. There might be a few more FRBs residing in compact binaries that are in
even earlier inspiral stage where the spin precession rate of the FRB emitter $\Omega_1\propto r^{-5/2}$ [Eq.~(\ref{eq:Omega})] is too long
to show any period in the bursts. Of course, we do not expect there is any GRB that is associated to
the FRBs in the CHIME/FRB catelog. In the second part of \cite{Curtin2022},
they also determined upper limits on possible radio emission for 39 GRBs (including 5 sGRBs)
6 hours before the GRB high-energy emission. This null result is again consistent with our rate estimates
with $\mathcal{R}_{\rm FRB-Bin}/\mathcal{R}_{\rm sGRB}\leq \mathcal O(10^{-2})$, i.e., $\leq \mathcal O(1\%)$ sGRBs
are produced from compact binaries with a FRB emitter.

We have been focusing on the scenario of a repeating FRB emitter residing in a compact binary in this work.
There are also some proposals invoking interactions between BNS magnetospheres as the FRB emission mechanism
\cite{Lipunov1996,Lyutikov2018,ZhangB2020}
or relating BNS mergers to non-repeating FRBs, e.g., magnetic interactions of a BNS
producing FRBs right before the merger \cite[see e.g.][]{Piro2012,Totani2013,Wang2016,Sridhar2021,Most2022,Most2022b}.
The null result in searching FRB-GRB association \cite{Curtin2022} set an upper limit of the
FRB emission luminosity in the former scenario.
In the latter scenario, associating the FRB to the GW event will be difficult
due to the high false alarm rate. Consider a merger event with its sky localisation constrained within $\delta\Omega_{\rm GW}$
from its GW signal. The average number of  FRBs that coincidentally lie in the same sky area and the same time interval $\Delta t$ is
$
N_{\rm false}=(\delta\Omega/4\pi) \mathcal{R}  \Delta t
$
with $\delta\Omega \approx {\rm max}(\delta\Omega_{\rm GW}, \delta\Omega_{\rm FRB})$,
where  the solid angle uncertainty  in the CHIME/FRB catalog  \cite{CHIME-c1} is at $\delta\Omega_{\rm FRB}\lesssim \mathcal{O}(1) \ {\rm deg}^2$ level.
The detection probability of a true FRB-GW association depends on the FRB production details and the emission beaming factor
$\mathcal{B}$, and in general, should be $N_{\rm true}\approx \mathcal{B}$. Consequently, the false alarm rate is comparable to the true association detection rate with
\be
\frac{N_{\rm false}}{N_{\rm true}}\approx
\frac{\delta\Omega}{10^2 {\rm deg}^2} \frac{\mathcal R}{ 2870\ {\rm day}^{-1}}\frac{\Delta t}{10^2 \ {\rm sec}}\frac{10^{-2}}{\mathcal{B}}\ ,
\ee
and therefore the FRB-GW association is difficult to identify.
On the other hand, if the host galaxy of the BNS is identified (via the EM counterpart),
the identification of the FRB-GW association is possible
because the false alarm rate should be much lower $N_{\rm false}/N_{\rm true}\ll 1$ with $\delta \Omega \approx \delta\Omega_{\rm FRB}$.

\section{\bf Summary and Discussion}\label{sec:discussion}

Recent observations suggest that magnetars commonly reside in merging compact binaries and
magnetars are the sources of at least part of FRBs. It is natural to speculate the  existence of FRB emitters in a class of merging compact binaries, which are ideal
multi-band and multi-messenger observation targets of radio telescopes and ground based GW detectors. In this work, we have performed a preliminary study of what can we learn from observations of such binaries. The key physical process is that FRB arrival times are modulated by the magnetar
spin precession, which encodes the information of the binary dynamics.
As an example, we consider monitoring a fiducial merging magnetar-NS binary for a half year before the final coalescence, and we find the arriving times of 100 FRB pulses from the precessing magnetar enable high-precision measurements of the NS masses, spins, quadrupole moments, binary intrinsic/extrinsic orientations and the angular dependence of an FRB emission pattern (Table~\ref{table}, Figs.~\ref{fig:corner} and \ref{fig:pdf}).
To our knowledge, there is no other way of measuring these quantities to similar precision, even with 3rd-generation GW detectors.

We found that the accurate measurement of stellar masses $m_i$ and quadrupole moments $\kappa_i$ will be an
extremely sensitive probe to the NS EoS (Figs.~\ref{fig:M-kappa},\ref{fig:M-R}), and may also be converted into a
tight (up to $\mathcal{O}(1\%)$) measurement of the NS radius $R_i$ with the aid of $Q-C$ universal relation (Fig.~\ref{fig:M-R}), depending on the spin of NSs.
In combination with measurements of the stellar tidal deformability parameter $\Lambda$ from GW signals and
of stellar compactness $C_i$ by NICER, we found that the $Q-$Love and $Q-C$ universal relations can tightly constrain
alternative theories of gravity, e.g., dCS gravity.
We also found that multi-band and multi-messenger observations of such binaries are useful in testing
beyond standard models, e.g., an axion field that is coupled to matter.

The rate of observing FRB binaries is rather uncertain at this stage, as the physical mechanism of FRB emission and the source classification are still unknown. We are able to obtain a quantitative estimation of the rate in the case that FRB 180916 is of binary origin, which favours detection with future-generation GW detectors. In the future, alternative rate calculations may be performed based on GRB 211211A-like events, once the observational selection effects and the characteristics of magnetar FRB emission are better understood.

\acknowledgements Z. P. and H. Y. are supported by the Natural Sciences and
Engineering Research Council of Canada and in part by
Perimeter Institute for Theoretical Physics. Research at
Perimeter Institute is supported in part by the Government
of Canada through the Department of Innovation, Science
and Economic Development Canada and by the Province of
Ontario through the Ministry of Colleges and Universities.
K.Y. acknowledges support from NSF Grant PHY-1806776, PHY-2207349, a Sloan Foundation Research Fellowship and the Owens Family Foundation.
K.Y. would like to also acknowledge support by the COST Action GWverse CA16104 and JSPS KAKENHI Grants No. JP17H06358.

~\\
\appendix*\label{app}
\section{Validity of Fisher Forecasts}
\begin{table*}[t]
  \centering
  \resizebox{2.0\columnwidth}{!}{%
  \begin{tabular}{ c | cccc ccccc ccc ccc cc}
    \hline
      & $m_1/M_\odot$ & $\chi_1$ & $\theta_1$ & $\kappa_1$ & $m_2/M_\odot$ & $\chi_2$ & $\theta_2$ & $\kappa_2$ &
    $\phi_{12}$ & $r_0/m$ & $\theta_{\rm los}$ & $\phi_{\rm los}$ & $\mu_0$ & $\sigma_\mu$ & $\mu_c$ & $A_1$ & $A_2$ \\ \hline
    $\mathbf{\Theta}_{\rm inj}$ & $1.4$ & 0.1 & 0.5 & 4.83 & $1.4$ & 0.1 & 0.5 & 4.83 & $0.75\pi$ & 160 & 0.8 & $-0.5\pi$
    & 1.0 & 0.15 & 0.2 & 0 & 0 \\
    $\sigma(\mathbf{\Theta})|_{\rm Fisher}$ & $4.6\times10^{-3}$ & $0.14$& $0.23$& $3.4$
    &$3.9\times10^{-3}$ & $0.12$& $0.39$& $2.9$ & $1.1$
    & $2.3\times10^{-3}$ & $0.30$& $0.05$
    & $0.04$& $0.08$ & $0.11$ &0.73 & 0.72\\
    $\sigma(\mathbf{\Theta})|_{\rm MCMC}$ & $^{+5.3}_{-5.7}\times10^{-3}$ & $^{+0.18}_{-0.12}$& $^{+0.15}_{-0.23}$& $^{+1.5}_{-1.0}$
    &$^{+5.6}_{-5.9}\times10^{-3}$ & $^{+0.17}_{-0.16}$& $^{+0.27}_{-0.26}$& $^{+8.8}_{-9.2}$ & $^{+2.1}_{-2.0}$
    & $0.11$ & $^{+0.26}_{-0.20}$& $0.069$
    & $^{+0.15}_{-0.74}$& $^{+0.035}_{-0.026}$ &$^{+0.30}_{-0.38}$ & $^{+0.39}_{-0.56}$ & $^{+0.55}_{-0.40}$ \\
    \hline
  \end{tabular}
  }
  \caption{Injection values and 1-$\sigma$ uncertainties of the model parameters assuming
  that $N_{\rm puls}=100$ FRBs are detected in $t_{\rm obs}=720$ sec.}\label{table:app}
\end{table*}

% \begin{table*}[t]
%   \centering
%   \resizebox{1.8\columnwidth}{!}{%
%   \begin{tabular}{ c | cccc ccccc ccccc c}
%     \hline
%       & $m_1/M_\odot$ & $\chi_1$ & $\theta_1$ & $\kappa_1$ & $m_2/M_\odot$ & $\chi_2$ & $\theta_2$ & $\kappa_2$ &
%     $\phi_{12}$ & $r_0/m$ & $\theta_{\rm los}$ & $\phi_{\rm los}$ & $\mu_0$ & $\sigma_\mu$ & $\mu_c$ \\ \hline
%     $\mathbf{\Theta}_{\rm inj}$ & $1.4$ & 0.1 & 0.5 & 4.83 & $1.4$ & 0.1 & 0.5 & 4.83 & $0.75\pi$ & 160 & 1.0 & $-0.5\pi$ & 0.15 & 0.8  \\
%     $\sigma(\mathbf{\Theta})|_{\rm Fisher}$ & $3.8\times10^{-3}$ & $0.12$& $0.17$& $2.4$
%     &$3.3\times10^{-3}$ & $0.094$& $0.33$& $2.2$ & $0.83$ & $2.3\times10^{-3}$
%     & $0.22$& $0.032$& $0.059$& $0.083$\\
%     $\sigma(\mathbf{\Theta})|_{\rm MCMC}$ & $^{+3.8}_{-4.5}\times10^{-3}$ & $^{+0.063}_{-0.053}$& $^{+0.43}_{-0.18}$& $^{+2.0}_{-1.3}$
%     &$^{+5.9}_{-5.1}\times10^{-3}$ & $^{+0.43}_{-0.22}$& $^{+0.43}_{-0.51}$& $^{+20}_{-7.7}$ & $^{+0.56}_{-1.5}$ & $0.11$
%     & $^{+0.16}_{-0.29}$& $0.074$& $^{+0.049}_{-0.042}$& $^{+0.09}_{-0.15}$ \\
%     \hline
%   \end{tabular}
%   }
%   \caption{Injection values and 1-$\sigma$ uncertainties of the model parameters assuming
%   that $N_{\rm puls}=100$ FRBs are detected in $t_{\rm obs}=720$ sec.}\label{table:app}
% \end{table*}

For verifying the validity of Fisher information in forecasting the model parameter constraints, we consider a mock observation
of FRB pulses, and forecast the model parameter constraints using both the Fisher matrix and full Bayesian analyses.
The binary model parameters are listed in Table~\ref{table:app},
where we choose a small initial separation $r_0=160 m$ and a short observation time $t_{\rm obs}= 720$ sec
for the sake of computational efficiency in MCMC simulations. Same as the models considered in the main text, we choose the expected number of detectable FRB pulses as $N_{\rm puls}=100$ in the Fisher forecast.
Before running MCMC simulations, we also need to generate mock data, i.e., a list of mock FRB pulse arriving times $d=\{t_k\}, (k=1,..., N_{\rm puls})$, in the same way as in Section~\ref{sec:spin}.

\begin{figure*}
\includegraphics[scale=0.32]{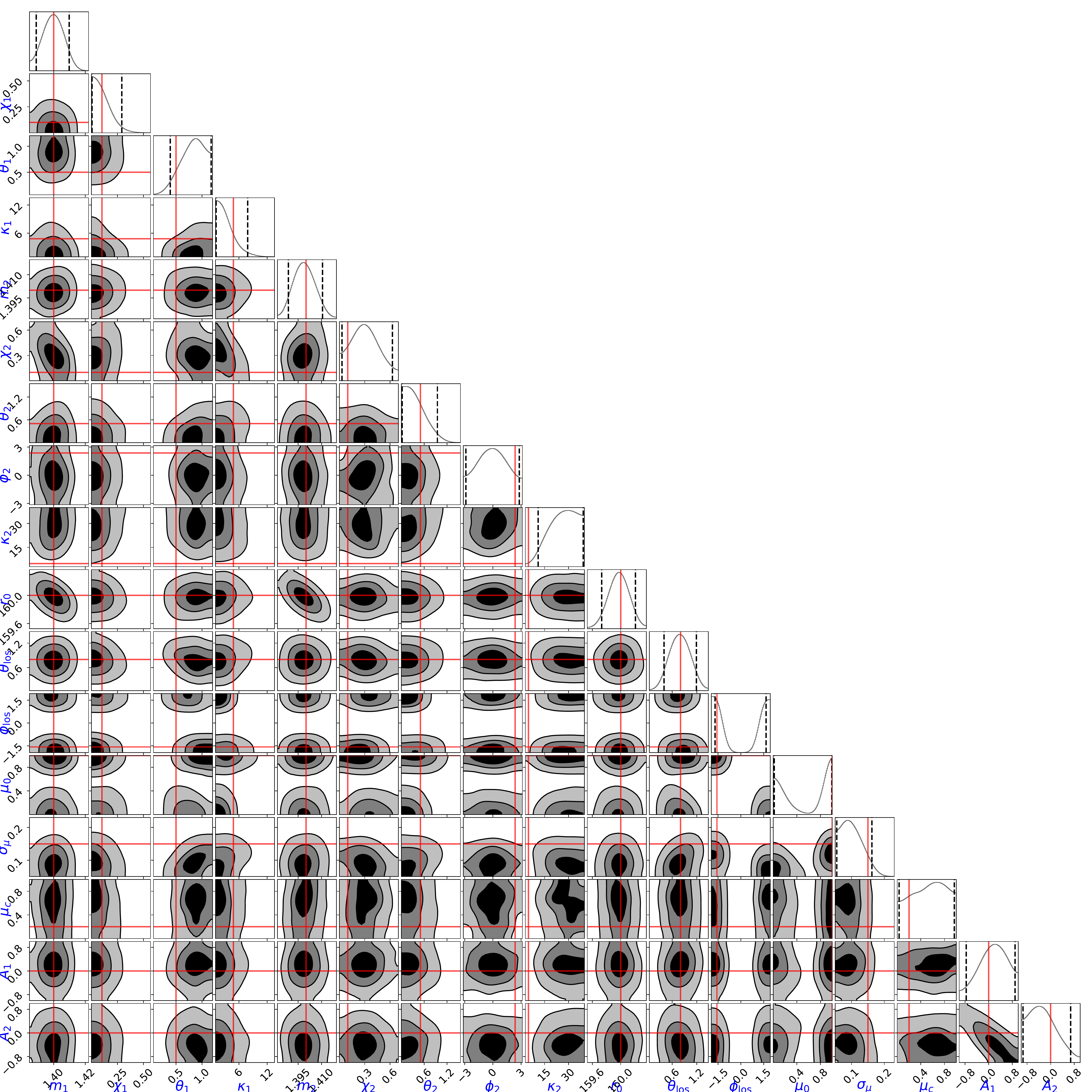}
\caption{\label{fig:app}
Posterior contours of all the model parameters assuming a mock observation detailed in Table~\ref{table:app}, where each pair of the vertical dashed lines marks the 2-$\sigma$ confidence level, and the red/solid lines denote the injection values.
}
\end{figure*}

According to the Bayes theorem, the posterior of parameters given data is
\be \mathcal P(\mathbf{\Theta}|d) \propto \mathcal{L}(d|\mathbf{\Theta}) \pi(\mathbf{\Theta})\ , \ee
where $\mathcal{L}(d|\mathbf{\Theta})$ is the likelihood of detecting data $d$ given a model with parameters $\mathbf{\Theta}$ and $\pi(\mathbf{\Theta})$ is the parameter prior assumed.
In our case, the likelihood is defined as
\be
\mathcal{L}(d|\mathbf{\Theta}) = \prod_{k=1}^{N_{\rm puls}} p(t_k|\mathbf{\Theta})\ ,
\ee
with $p(t|\mathbf{\Theta})$ being the probability density function defined in Eq.~(\ref{eq:pdf0}),
and we impose a sufficiently wide uniform prior for each model parameter. We sample the model parameters using
the dynamical nested sampling method \cite{Higson2019} implemented in the package $\emph{Bilby}$ \cite{Speagle2020}.
The 2-d posteriors and 1-$\sigma$ uncertainties
of all the model parameters are shown in Fig.~\ref{fig:app} and Table~\ref{table:app}, respectively.

From Table~\ref{table:app}, we see that the Fisher forecast uncertainties of all the model parameters
are in agreement with those of full MCMC simulations within a factor of a few, except for $r_0$.
The reason is simply that the $r_0-m_i$ degeneracy largely degrades the $r_0$ constraint in the full MCMC simulations, while the degeneracy is accidentally broken in the Fisher matrix, because only the local curvature of the likelihood is taken into account, which has a small deviation from the curvature in a finite $r_0$ region.  For a longer observation time as considered in the main text, all the model parameters are expected to be constrained with much better precision, and the Fisher matrix should perform much better
where the local curvature is a better approximation (this is well known for Fisher forecasts of GW signals, where Fisher forecasts perform better for GW signals of higher SNRs for the same reason). Therefore, we believe the Fisher forecast results in the main text should be valid.

\bibliography{ms}

\begin{thebibliography}{148}%
\makeatletter
\providecommand \@ifxundefined [1]{%
 \@ifx{#1\undefined}
}%
\providecommand \@ifnum [1]{%
 \ifnum #1\expandafter \@firstoftwo
 \else \expandafter \@secondoftwo
 \fi
}%
\providecommand \@ifx [1]{%
 \ifx #1\expandafter \@firstoftwo
 \else \expandafter \@secondoftwo
 \fi
}%
\providecommand \natexlab [1]{#1}%
\providecommand \enquote  [1]{``#1''}%
\providecommand \bibnamefont  [1]{#1}%
\providecommand \bibfnamefont [1]{#1}%
\providecommand \citenamefont [1]{#1}%
\providecommand \href@noop [0]{\@secondoftwo}%
\providecommand \href [0]{\begingroup \@sanitize@url \@href}%
\providecommand \@href[1]{\@@startlink{#1}\@@href}%
\providecommand \@@href[1]{\endgroup#1\@@endlink}%
\providecommand \@sanitize@url [0]{\catcode `\\12\catcode `\$12\catcode
  `\&12\catcode `\#12\catcode `\^12\catcode `\_12\catcode `\%12\relax}%
\providecommand \@@startlink[1]{}%
\providecommand \@@endlink[0]{}%
\providecommand \url  [0]{\begingroup\@sanitize@url \@url }%
\providecommand \@url [1]{\endgroup\@href {#1}{\urlprefix }}%
\providecommand \urlprefix  [0]{URL }%
\providecommand \Eprint [0]{\href }%
\providecommand \doibase [0]{http://dx.doi.org/}%
\providecommand \selectlanguage [0]{\@gobble}%
\providecommand \bibinfo  [0]{\@secondoftwo}%
\providecommand \bibfield  [0]{\@secondoftwo}%
\providecommand \translation [1]{[#1]}%
\providecommand \BibitemOpen [0]{}%
\providecommand \bibitemStop [0]{}%
\providecommand \bibitemNoStop [0]{.\EOS\space}%
\providecommand \EOS [0]{\spacefactor3000\relax}%
\providecommand \BibitemShut  [1]{\csname bibitem#1\endcsname}%
\let\auto@bib@innerbib\@empty
%</preamble>
\bibitem [{\citenamefont {{LIGO Scientific Collaboration}}\ and\ \citenamefont
  {{Virgo Collaboration}}(2019)}]{LVC-GWTC1}%
  \BibitemOpen
  \bibfield  {author} {\bibinfo {author} {\bibnamefont {{LIGO Scientific
  Collaboration}}}\ and\ \bibinfo {author} {\bibnamefont {{Virgo
  Collaboration}}},\ }\bibfield  {title} {\enquote {\bibinfo {title} {{GWTC-1:
  A Gravitational-Wave Transient Catalog of Compact Binary Mergers Observed by
  LIGO and Virgo during the First and Second Observing Runs}},}\ }\href
  {\doibase 10.1103/PhysRevX.9.031040} {\bibfield  {journal} {\bibinfo
  {journal} {Physical Review X}\ }\textbf {\bibinfo {volume} {9}},\ \bibinfo
  {eid} {031040} (\bibinfo {year} {2019})},\ \Eprint
  {http://arxiv.org/abs/1811.12907} {arXiv:1811.12907 [astro-ph.HE]}
  \BibitemShut {NoStop}%
\bibitem [{\citenamefont {{LIGO Scientific Collaboration}}\ and\ \citenamefont
  {{Virgo Collaboration}}(2021)}]{LVC-GWTC2}%
  \BibitemOpen
  \bibfield  {author} {\bibinfo {author} {\bibnamefont {{LIGO Scientific
  Collaboration}}}\ and\ \bibinfo {author} {\bibnamefont {{Virgo
  Collaboration}}},\ }\bibfield  {title} {\enquote {\bibinfo {title} {{GWTC-2:
  Compact Binary Coalescences Observed by LIGO and Virgo during the First Half
  of the Third Observing Run}},}\ }\href {\doibase 10.1103/PhysRevX.11.021053}
  {\bibfield  {journal} {\bibinfo  {journal} {Physical Review X}\ }\textbf
  {\bibinfo {volume} {11}},\ \bibinfo {eid} {021053} (\bibinfo {year}
  {2021})},\ \Eprint {http://arxiv.org/abs/2010.14527} {arXiv:2010.14527
  [gr-qc]} \BibitemShut {NoStop}%
\bibitem [{\citenamefont {{The LIGO Scientific Collaboration}}\ and\
  \citenamefont {{the Virgo Collaboration}}(2021)}]{LVC-GWTC2.1}%
  \BibitemOpen
  \bibfield  {author} {\bibinfo {author} {\bibnamefont {{The LIGO Scientific
  Collaboration}}}\ and\ \bibinfo {author} {\bibnamefont {{the Virgo
  Collaboration}}},\ }\bibfield  {title} {\enquote {\bibinfo {title}
  {{GWTC-2.1: Deep Extended Catalog of Compact Binary Coalescences Observed by
  LIGO and Virgo During the First Half of the Third Observing Run}},}\
  }\href@noop {} {\bibfield  {journal} {\bibinfo  {journal} {arXiv e-prints}\
  ,\ \bibinfo {eid} {arXiv:2108.01045}} (\bibinfo {year} {2021})},\ \Eprint
  {http://arxiv.org/abs/2108.01045} {arXiv:2108.01045 [gr-qc]} \BibitemShut
  {NoStop}%
\bibitem [{\citenamefont {{The LIGO Scientific Collaboration}}\ \emph
  {et~al.}(2021{\natexlab{a}})\citenamefont {{The LIGO Scientific
  Collaboration}}, \citenamefont {{the Virgo Collaboration}},\ and\
  \citenamefont {{the KAGRA Collaboration}}}]{LVK-GWTC3}%
  \BibitemOpen
  \bibfield  {author} {\bibinfo {author} {\bibnamefont {{The LIGO Scientific
  Collaboration}}}, \bibinfo {author} {\bibnamefont {{the Virgo
  Collaboration}}}, \ and\ \bibinfo {author} {\bibnamefont {{the KAGRA
  Collaboration}}},\ }\bibfield  {title} {\enquote {\bibinfo {title} {{GWTC-3:
  Compact Binary Coalescences Observed by LIGO and Virgo During the Second Part
  of the Third Observing Run}},}\ }\href@noop {} {\bibfield  {journal}
  {\bibinfo  {journal} {arXiv e-prints}\ ,\ \bibinfo {eid} {arXiv:2111.03606}}
  (\bibinfo {year} {2021}{\natexlab{a}})},\ \Eprint
  {http://arxiv.org/abs/2111.03606} {arXiv:2111.03606 [gr-qc]} \BibitemShut
  {NoStop}%
\bibitem [{\citenamefont {{The LIGO Scientific Collaboration}}\ \emph
  {et~al.}(2021{\natexlab{b}})\citenamefont {{The LIGO Scientific
  Collaboration}}, \citenamefont {{the Virgo Collaboration}},\ and\
  \citenamefont {{the KAGRA Collaboration}}}]{LVK2021pop}%
  \BibitemOpen
  \bibfield  {author} {\bibinfo {author} {\bibnamefont {{The LIGO Scientific
  Collaboration}}}, \bibinfo {author} {\bibnamefont {{the Virgo
  Collaboration}}}, \ and\ \bibinfo {author} {\bibnamefont {{the KAGRA
  Collaboration}}},\ }\bibfield  {title} {\enquote {\bibinfo {title} {{The
  population of merging compact binaries inferred using gravitational waves
  through GWTC-3}},}\ }\href@noop {} {\bibfield  {journal} {\bibinfo  {journal}
  {arXiv e-prints}\ ,\ \bibinfo {eid} {arXiv:2111.03634}} (\bibinfo {year}
  {2021}{\natexlab{b}})},\ \Eprint {http://arxiv.org/abs/2111.03634}
  {arXiv:2111.03634 [astro-ph.HE]} \BibitemShut {NoStop}%
\bibitem [{\citenamefont {{The LIGO Scientific Collaboration}}\ \emph
  {et~al.}(2021{\natexlab{c}})\citenamefont {{The LIGO Scientific
  Collaboration}}, \citenamefont {{the Virgo Collaboration}},\ and\
  \citenamefont {{the KAGRA Collaboration}}}]{LVK2021gr}%
  \BibitemOpen
  \bibfield  {author} {\bibinfo {author} {\bibnamefont {{The LIGO Scientific
  Collaboration}}}, \bibinfo {author} {\bibnamefont {{the Virgo
  Collaboration}}}, \ and\ \bibinfo {author} {\bibnamefont {{the KAGRA
  Collaboration}}},\ }\bibfield  {title} {\enquote {\bibinfo {title} {{Tests of
  General Relativity with GWTC-3}},}\ }\href@noop {} {\bibfield  {journal}
  {\bibinfo  {journal} {arXiv e-prints}\ ,\ \bibinfo {eid} {arXiv:2112.06861}}
  (\bibinfo {year} {2021}{\natexlab{c}})},\ \Eprint
  {http://arxiv.org/abs/2112.06861} {arXiv:2112.06861 [gr-qc]} \BibitemShut
  {NoStop}%
\bibitem [{\citenamefont {{The LIGO Scientific Collaboration}}\ \emph
  {et~al.}(2021{\natexlab{d}})\citenamefont {{The LIGO Scientific
  Collaboration}}, \citenamefont {{the Virgo Collaboration}},\ and\
  \citenamefont {{the KAGRA Collaboration}}}]{LVK2021cos}%
  \BibitemOpen
  \bibfield  {author} {\bibinfo {author} {\bibnamefont {{The LIGO Scientific
  Collaboration}}}, \bibinfo {author} {\bibnamefont {{the Virgo
  Collaboration}}}, \ and\ \bibinfo {author} {\bibnamefont {{the KAGRA
  Collaboration}}},\ }\bibfield  {title} {\enquote {\bibinfo {title}
  {{Constraints on the cosmic expansion history from GWTC-3}},}\ }\href@noop {}
  {\bibfield  {journal} {\bibinfo  {journal} {arXiv e-prints}\ ,\ \bibinfo
  {eid} {arXiv:2111.03604}} (\bibinfo {year} {2021}{\natexlab{d}})},\ \Eprint
  {http://arxiv.org/abs/2111.03604} {arXiv:2111.03604 [astro-ph.CO]}
  \BibitemShut {NoStop}%
\bibitem [{\citenamefont {{LIGO Scientific Collaboration}}\ and\ \citenamefont
  {{Virgo Collaboration}}(2016)}]{LVC2016}%
  \BibitemOpen
  \bibfield  {author} {\bibinfo {author} {\bibnamefont {{LIGO Scientific
  Collaboration}}}\ and\ \bibinfo {author} {\bibnamefont {{Virgo
  Collaboration}}},\ }\bibfield  {title} {\enquote {\bibinfo {title}
  {{Observation of Gravitational Waves from a Binary Black Hole Merger}},}\
  }\href {\doibase 10.1103/PhysRevLett.116.061102} {\bibfield  {journal}
  {\bibinfo  {journal} {\prl}\ }\textbf {\bibinfo {volume} {116}},\ \bibinfo
  {eid} {061102} (\bibinfo {year} {2016})},\ \Eprint
  {http://arxiv.org/abs/1602.03837} {arXiv:1602.03837 [gr-qc]} \BibitemShut
  {NoStop}%
\bibitem [{\citenamefont {Sesana}(2016)}]{Sesana2016}%
  \BibitemOpen
  \bibfield  {author} {\bibinfo {author} {\bibfnamefont {Alberto}\ \bibnamefont
  {Sesana}},\ }\bibfield  {title} {\enquote {\bibinfo {title} {Prospects for
  multiband gravitational-wave astronomy after gw150914},}\ }\href {\doibase
  10.1103/PhysRevLett.116.231102} {\bibfield  {journal} {\bibinfo  {journal}
  {Phys. Rev. Lett.}\ }\textbf {\bibinfo {volume} {116}},\ \bibinfo {pages}
  {231102} (\bibinfo {year} {2016})}\BibitemShut {NoStop}%
\bibitem [{\citenamefont {Barausse}\ \emph {et~al.}(2016)\citenamefont
  {Barausse}, \citenamefont {Yunes},\ and\ \citenamefont
  {Chamberlain}}]{Barausse:2016eii}%
  \BibitemOpen
  \bibfield  {author} {\bibinfo {author} {\bibfnamefont {Enrico}\ \bibnamefont
  {Barausse}}, \bibinfo {author} {\bibfnamefont {Nicol\'as}\ \bibnamefont
  {Yunes}}, \ and\ \bibinfo {author} {\bibfnamefont {Katie}\ \bibnamefont
  {Chamberlain}},\ }\bibfield  {title} {\enquote {\bibinfo {title}
  {{Theory-Agnostic Constraints on Black-Hole Dipole Radiation with Multiband
  Gravitational-Wave Astrophysics}},}\ }\href {\doibase
  10.1103/PhysRevLett.116.241104} {\bibfield  {journal} {\bibinfo  {journal}
  {Phys. Rev. Lett.}\ }\textbf {\bibinfo {volume} {116}},\ \bibinfo {pages}
  {241104} (\bibinfo {year} {2016})},\ \Eprint
  {http://arxiv.org/abs/1603.04075} {arXiv:1603.04075 [gr-qc]} \BibitemShut
  {NoStop}%
\bibitem [{\citenamefont {Wong}\ \emph {et~al.}(2018)\citenamefont {Wong},
  \citenamefont {Kovetz}, \citenamefont {Cutler},\ and\ \citenamefont
  {Berti}}]{Wong:2018uwb}%
  \BibitemOpen
  \bibfield  {author} {\bibinfo {author} {\bibfnamefont {Kaze W.~K.}\
  \bibnamefont {Wong}}, \bibinfo {author} {\bibfnamefont {Ely~D.}\ \bibnamefont
  {Kovetz}}, \bibinfo {author} {\bibfnamefont {Curt}\ \bibnamefont {Cutler}}, \
  and\ \bibinfo {author} {\bibfnamefont {Emanuele}\ \bibnamefont {Berti}},\
  }\bibfield  {title} {\enquote {\bibinfo {title} {{Expanding the LISA Horizon
  from the Ground}},}\ }\href {\doibase 10.1103/PhysRevLett.121.251102}
  {\bibfield  {journal} {\bibinfo  {journal} {Phys. Rev. Lett.}\ }\textbf
  {\bibinfo {volume} {121}},\ \bibinfo {pages} {251102} (\bibinfo {year}
  {2018})},\ \Eprint {http://arxiv.org/abs/1808.08247} {arXiv:1808.08247
  [astro-ph.HE]} \BibitemShut {NoStop}%
\bibitem [{\citenamefont {Carson}\ and\ \citenamefont
  {Yagi}(2020{\natexlab{a}})}]{Carson:2019rda}%
  \BibitemOpen
  \bibfield  {author} {\bibinfo {author} {\bibfnamefont {Zack}\ \bibnamefont
  {Carson}}\ and\ \bibinfo {author} {\bibfnamefont {Kent}\ \bibnamefont
  {Yagi}},\ }\bibfield  {title} {\enquote {\bibinfo {title} {{Multi-band
  gravitational wave tests of general relativity}},}\ }\href {\doibase
  10.1088/1361-6382/ab5c9a} {\bibfield  {journal} {\bibinfo  {journal} {Class.
  Quant. Grav.}\ }\textbf {\bibinfo {volume} {37}},\ \bibinfo {pages} {02LT01}
  (\bibinfo {year} {2020}{\natexlab{a}})},\ \Eprint
  {http://arxiv.org/abs/1905.13155} {arXiv:1905.13155 [gr-qc]} \BibitemShut
  {NoStop}%
\bibitem [{\citenamefont {Carson}\ and\ \citenamefont
  {Yagi}(2020{\natexlab{b}})}]{Carson:2019kkh}%
  \BibitemOpen
  \bibfield  {author} {\bibinfo {author} {\bibfnamefont {Zack}\ \bibnamefont
  {Carson}}\ and\ \bibinfo {author} {\bibfnamefont {Kent}\ \bibnamefont
  {Yagi}},\ }\bibfield  {title} {\enquote {\bibinfo {title} {{Parametrized and
  inspiral-merger-ringdown consistency tests of gravity with multiband
  gravitational wave observations}},}\ }\href {\doibase
  10.1103/PhysRevD.101.044047} {\bibfield  {journal} {\bibinfo  {journal}
  {Phys. Rev. D}\ }\textbf {\bibinfo {volume} {101}},\ \bibinfo {pages}
  {044047} (\bibinfo {year} {2020}{\natexlab{b}})},\ \Eprint
  {http://arxiv.org/abs/1911.05258} {arXiv:1911.05258 [gr-qc]} \BibitemShut
  {NoStop}%
\bibitem [{\citenamefont {Gerosa}\ \emph {et~al.}(2019)\citenamefont {Gerosa},
  \citenamefont {Ma}, \citenamefont {Wong}, \citenamefont {Berti},
  \citenamefont {O'Shaughnessy}, \citenamefont {Chen},\ and\ \citenamefont
  {Belczynski}}]{Gerosa:2019dbe}%
  \BibitemOpen
  \bibfield  {author} {\bibinfo {author} {\bibfnamefont {Davide}\ \bibnamefont
  {Gerosa}}, \bibinfo {author} {\bibfnamefont {Sizheng}\ \bibnamefont {Ma}},
  \bibinfo {author} {\bibfnamefont {Kaze W.~K.}\ \bibnamefont {Wong}}, \bibinfo
  {author} {\bibfnamefont {Emanuele}\ \bibnamefont {Berti}}, \bibinfo {author}
  {\bibfnamefont {Richard}\ \bibnamefont {O'Shaughnessy}}, \bibinfo {author}
  {\bibfnamefont {Yanbei}\ \bibnamefont {Chen}}, \ and\ \bibinfo {author}
  {\bibfnamefont {Krzysztof}\ \bibnamefont {Belczynski}},\ }\bibfield  {title}
  {\enquote {\bibinfo {title} {{Multiband gravitational-wave event rates and
  stellar physics}},}\ }\href {\doibase 10.1103/PhysRevD.99.103004} {\bibfield
  {journal} {\bibinfo  {journal} {Phys. Rev. D}\ }\textbf {\bibinfo {volume}
  {99}},\ \bibinfo {pages} {103004} (\bibinfo {year} {2019})},\ \Eprint
  {http://arxiv.org/abs/1902.00021} {arXiv:1902.00021 [astro-ph.HE]}
  \BibitemShut {NoStop}%
\bibitem [{\citenamefont {Jani}\ \emph {et~al.}(2019)\citenamefont {Jani},
  \citenamefont {Shoemaker},\ and\ \citenamefont {Cutler}}]{Jani:2019ffg}%
  \BibitemOpen
  \bibfield  {author} {\bibinfo {author} {\bibfnamefont {Karan}\ \bibnamefont
  {Jani}}, \bibinfo {author} {\bibfnamefont {Deirdre}\ \bibnamefont
  {Shoemaker}}, \ and\ \bibinfo {author} {\bibfnamefont {Curt}\ \bibnamefont
  {Cutler}},\ }\bibfield  {title} {\enquote {\bibinfo {title} {{Detectability
  of Intermediate-Mass Black Holes in Multiband Gravitational Wave
  Astronomy}},}\ }\href {\doibase 10.1038/s41550-019-0932-7} {\bibfield
  {journal} {\bibinfo  {journal} {Nature Astron.}\ }\textbf {\bibinfo {volume}
  {4}},\ \bibinfo {pages} {260--265} (\bibinfo {year} {2019})},\ \Eprint
  {http://arxiv.org/abs/1908.04985} {arXiv:1908.04985 [gr-qc]} \BibitemShut
  {NoStop}%
\bibitem [{\citenamefont {Perkins}\ \emph {et~al.}(2021)\citenamefont
  {Perkins}, \citenamefont {Yunes},\ and\ \citenamefont
  {Berti}}]{Perkins:2020tra}%
  \BibitemOpen
  \bibfield  {author} {\bibinfo {author} {\bibfnamefont {Scott~E.}\
  \bibnamefont {Perkins}}, \bibinfo {author} {\bibfnamefont {Nicol\'as}\
  \bibnamefont {Yunes}}, \ and\ \bibinfo {author} {\bibfnamefont {Emanuele}\
  \bibnamefont {Berti}},\ }\bibfield  {title} {\enquote {\bibinfo {title}
  {{Probing Fundamental Physics with Gravitational Waves: The Next
  Generation}},}\ }\href {\doibase 10.1103/PhysRevD.103.044024} {\bibfield
  {journal} {\bibinfo  {journal} {Phys. Rev. D}\ }\textbf {\bibinfo {volume}
  {103}},\ \bibinfo {pages} {044024} (\bibinfo {year} {2021})},\ \Eprint
  {http://arxiv.org/abs/2010.09010} {arXiv:2010.09010 [gr-qc]} \BibitemShut
  {NoStop}%
\bibitem [{\citenamefont {Gupta}\ \emph {et~al.}(2020)\citenamefont {Gupta},
  \citenamefont {Datta}, \citenamefont {Kastha}, \citenamefont {Borhanian},
  \citenamefont {Arun},\ and\ \citenamefont {Sathyaprakash}}]{Gupta:2020lxa}%
  \BibitemOpen
  \bibfield  {author} {\bibinfo {author} {\bibfnamefont {Anuradha}\
  \bibnamefont {Gupta}}, \bibinfo {author} {\bibfnamefont {Sayantani}\
  \bibnamefont {Datta}}, \bibinfo {author} {\bibfnamefont {Shilpa}\
  \bibnamefont {Kastha}}, \bibinfo {author} {\bibfnamefont {Ssohrab}\
  \bibnamefont {Borhanian}}, \bibinfo {author} {\bibfnamefont {K.~G.}\
  \bibnamefont {Arun}}, \ and\ \bibinfo {author} {\bibfnamefont {B.~S.}\
  \bibnamefont {Sathyaprakash}},\ }\bibfield  {title} {\enquote {\bibinfo
  {title} {{Multiparameter tests of general relativity using multiband
  gravitational-wave observations}},}\ }\href {\doibase
  10.1103/PhysRevLett.125.201101} {\bibfield  {journal} {\bibinfo  {journal}
  {Phys. Rev. Lett.}\ }\textbf {\bibinfo {volume} {125}},\ \bibinfo {pages}
  {201101} (\bibinfo {year} {2020})},\ \Eprint
  {http://arxiv.org/abs/2005.09607} {arXiv:2005.09607 [gr-qc]} \BibitemShut
  {NoStop}%
\bibitem [{\citenamefont {Datta}\ \emph {et~al.}(2021)\citenamefont {Datta},
  \citenamefont {Gupta}, \citenamefont {Kastha}, \citenamefont {Arun},\ and\
  \citenamefont {Sathyaprakash}}]{Datta:2020vcj}%
  \BibitemOpen
  \bibfield  {author} {\bibinfo {author} {\bibfnamefont {Sayantani}\
  \bibnamefont {Datta}}, \bibinfo {author} {\bibfnamefont {Anuradha}\
  \bibnamefont {Gupta}}, \bibinfo {author} {\bibfnamefont {Shilpa}\
  \bibnamefont {Kastha}}, \bibinfo {author} {\bibfnamefont {K.~G.}\
  \bibnamefont {Arun}}, \ and\ \bibinfo {author} {\bibfnamefont {B.~S.}\
  \bibnamefont {Sathyaprakash}},\ }\bibfield  {title} {\enquote {\bibinfo
  {title} {{Tests of general relativity using multiband observations of
  intermediate mass binary black hole mergers}},}\ }\href {\doibase
  10.1103/PhysRevD.103.024036} {\bibfield  {journal} {\bibinfo  {journal}
  {Phys. Rev. D}\ }\textbf {\bibinfo {volume} {103}},\ \bibinfo {pages}
  {024036} (\bibinfo {year} {2021})},\ \Eprint
  {http://arxiv.org/abs/2006.12137} {arXiv:2006.12137 [gr-qc]} \BibitemShut
  {NoStop}%
\bibitem [{\citenamefont {{Abbott}}\ and\ \citenamefont {{et.
  al.}}(2017{\natexlab{a}})}]{GW170817}%
  \BibitemOpen
  \bibfield  {author} {\bibinfo {author} {\bibfnamefont {B.~P.}\ \bibnamefont
  {{Abbott}}}\ and\ \bibinfo {author} {\bibnamefont {{et. al.}}},\ }\bibfield
  {title} {\enquote {\bibinfo {title} {{Multi-messenger Observations of a
  Binary Neutron Star Merger}},}\ }\href {\doibase 10.3847/2041-8213/aa91c9}
  {\bibfield  {journal} {\bibinfo  {journal} {\apjl}\ }\textbf {\bibinfo
  {volume} {848}},\ \bibinfo {eid} {L12} (\bibinfo {year}
  {2017}{\natexlab{a}})},\ \Eprint {http://arxiv.org/abs/1710.05833}
  {arXiv:1710.05833 [astro-ph.HE]} \BibitemShut {NoStop}%
\bibitem [{\citenamefont {{LIGO Scientific Collaboration}}\ and\ \citenamefont
  {{Virgo Collaboration}}(2017{\natexlab{a}})}]{LVC170817}%
  \BibitemOpen
  \bibfield  {author} {\bibinfo {author} {\bibnamefont {{LIGO Scientific
  Collaboration}}}\ and\ \bibinfo {author} {\bibnamefont {{Virgo
  Collaboration}}},\ }\bibfield  {title} {\enquote {\bibinfo {title}
  {{GW170817: Observation of Gravitational Waves from a Binary Neutron Star
  Inspiral}},}\ }\href {\doibase 10.1103/PhysRevLett.119.161101} {\bibfield
  {journal} {\bibinfo  {journal} {\prl}\ }\textbf {\bibinfo {volume} {119}},\
  \bibinfo {eid} {161101} (\bibinfo {year} {2017}{\natexlab{a}})},\ \Eprint
  {http://arxiv.org/abs/1710.05832} {arXiv:1710.05832 [gr-qc]} \BibitemShut
  {NoStop}%
\bibitem [{\citenamefont {{Abbott}}\ and\ \citenamefont {{et.
  al.}}(2017{\natexlab{b}})}]{2017ApJ...848L..13A}%
  \BibitemOpen
  \bibfield  {author} {\bibinfo {author} {\bibfnamefont {B.~P.}\ \bibnamefont
  {{Abbott}}}\ and\ \bibinfo {author} {\bibnamefont {{et. al.}}},\ }\bibfield
  {title} {\enquote {\bibinfo {title} {{Gravitational Waves and Gamma-Rays from
  a Binary Neutron Star Merger: GW170817 and GRB 170817A}},}\ }\href {\doibase
  10.3847/2041-8213/aa920c} {\bibfield  {journal} {\bibinfo  {journal} {\apjl}\
  }\textbf {\bibinfo {volume} {848}},\ \bibinfo {eid} {L13} (\bibinfo {year}
  {2017}{\natexlab{b}})},\ \Eprint {http://arxiv.org/abs/1710.05834}
  {arXiv:1710.05834 [astro-ph.HE]} \BibitemShut {NoStop}%
\bibitem [{\citenamefont {{Margalit}}\ and\ \citenamefont
  {{Metzger}}(2017)}]{Margalit2017}%
  \BibitemOpen
  \bibfield  {author} {\bibinfo {author} {\bibfnamefont {Ben}\ \bibnamefont
  {{Margalit}}}\ and\ \bibinfo {author} {\bibfnamefont {Brian~D.}\ \bibnamefont
  {{Metzger}}},\ }\bibfield  {title} {\enquote {\bibinfo {title} {{Constraining
  the Maximum Mass of Neutron Stars from Multi-messenger Observations of
  GW170817}},}\ }\href {\doibase 10.3847/2041-8213/aa991c} {\bibfield
  {journal} {\bibinfo  {journal} {\apjl}\ }\textbf {\bibinfo {volume} {850}},\
  \bibinfo {eid} {L19} (\bibinfo {year} {2017})},\ \Eprint
  {http://arxiv.org/abs/1710.05938} {arXiv:1710.05938 [astro-ph.HE]}
  \BibitemShut {NoStop}%
\bibitem [{\citenamefont {Essick}\ \emph {et~al.}(2020)\citenamefont {Essick},
  \citenamefont {Landry},\ and\ \citenamefont {Holz}}]{Essick2020}%
  \BibitemOpen
  \bibfield  {author} {\bibinfo {author} {\bibfnamefont {Reed}\ \bibnamefont
  {Essick}}, \bibinfo {author} {\bibfnamefont {Philippe}\ \bibnamefont
  {Landry}}, \ and\ \bibinfo {author} {\bibfnamefont {Daniel~E.}\ \bibnamefont
  {Holz}},\ }\bibfield  {title} {\enquote {\bibinfo {title} {Nonparametric
  inference of neutron star composition, equation of state, and maximum mass
  with gw170817},}\ }\href {\doibase 10.1103/PhysRevD.101.063007} {\bibfield
  {journal} {\bibinfo  {journal} {Phys. Rev. D}\ }\textbf {\bibinfo {volume}
  {101}},\ \bibinfo {pages} {063007} (\bibinfo {year} {2020})}\BibitemShut
  {NoStop}%
\bibitem [{\citenamefont {{Pan}}\ \emph {et~al.}(2020)\citenamefont {{Pan}},
  \citenamefont {{Lyu}}, \citenamefont {{Bonga}}, \citenamefont {{Ortiz}},\
  and\ \citenamefont {{Yang}}}]{Pan2020}%
  \BibitemOpen
  \bibfield  {author} {\bibinfo {author} {\bibfnamefont {Zhen}\ \bibnamefont
  {{Pan}}}, \bibinfo {author} {\bibfnamefont {Zhenwei}\ \bibnamefont {{Lyu}}},
  \bibinfo {author} {\bibfnamefont {B{\'e}atrice}\ \bibnamefont {{Bonga}}},
  \bibinfo {author} {\bibfnamefont {N{\'e}stor}\ \bibnamefont {{Ortiz}}}, \
  and\ \bibinfo {author} {\bibfnamefont {Huan}\ \bibnamefont {{Yang}}},\
  }\bibfield  {title} {\enquote {\bibinfo {title} {{Probing Crust Meltdown in
  Inspiraling Binary Neutron Stars}},}\ }\href {\doibase
  10.1103/PhysRevLett.125.201102} {\bibfield  {journal} {\bibinfo  {journal}
  {\prl}\ }\textbf {\bibinfo {volume} {125}},\ \bibinfo {eid} {201102}
  (\bibinfo {year} {2020})},\ \Eprint {http://arxiv.org/abs/2003.03330}
  {arXiv:2003.03330 [astro-ph.HE]} \BibitemShut {NoStop}%
\bibitem [{\citenamefont {{Hook}}\ and\ \citenamefont
  {{Huang}}(2018)}]{Hook2018}%
  \BibitemOpen
  \bibfield  {author} {\bibinfo {author} {\bibfnamefont {Anson}\ \bibnamefont
  {{Hook}}}\ and\ \bibinfo {author} {\bibfnamefont {Junwu}\ \bibnamefont
  {{Huang}}},\ }\bibfield  {title} {\enquote {\bibinfo {title} {{Probing axions
  with neutron star inspirals and other stellar processes}},}\ }\href {\doibase
  10.1007/JHEP06(2018)036} {\bibfield  {journal} {\bibinfo  {journal} {Journal
  of High Energy Physics}\ }\textbf {\bibinfo {volume} {2018}},\ \bibinfo {eid}
  {36} (\bibinfo {year} {2018})},\ \Eprint {http://arxiv.org/abs/1708.08464}
  {arXiv:1708.08464 [hep-ph]} \BibitemShut {NoStop}%
\bibitem [{\citenamefont {{Huang}}\ \emph {et~al.}(2019)\citenamefont
  {{Huang}}, \citenamefont {{Johnson}}, \citenamefont {{Sagunski}},
  \citenamefont {{Sakellariadou}},\ and\ \citenamefont {{Zhang}}}]{Huang2019}%
  \BibitemOpen
  \bibfield  {author} {\bibinfo {author} {\bibfnamefont {Junwu}\ \bibnamefont
  {{Huang}}}, \bibinfo {author} {\bibfnamefont {Matthew~C.}\ \bibnamefont
  {{Johnson}}}, \bibinfo {author} {\bibfnamefont {Laura}\ \bibnamefont
  {{Sagunski}}}, \bibinfo {author} {\bibfnamefont {Mairi}\ \bibnamefont
  {{Sakellariadou}}}, \ and\ \bibinfo {author} {\bibfnamefont {Jun}\
  \bibnamefont {{Zhang}}},\ }\bibfield  {title} {\enquote {\bibinfo {title}
  {{Prospects for axion searches with Advanced LIGO through binary mergers}},}\
  }\href {\doibase 10.1103/PhysRevD.99.063013} {\bibfield  {journal} {\bibinfo
  {journal} {\prd}\ }\textbf {\bibinfo {volume} {99}},\ \bibinfo {eid} {063013}
  (\bibinfo {year} {2019})},\ \Eprint {http://arxiv.org/abs/1807.02133}
  {arXiv:1807.02133 [hep-ph]} \BibitemShut {NoStop}%
\bibitem [{\citenamefont {{Zhang}}\ \emph {et~al.}(2021)\citenamefont
  {{Zhang}}, \citenamefont {{Lyu}}, \citenamefont {{Huang}}, \citenamefont
  {{Johnson}}, \citenamefont {{Sagunski}}, \citenamefont {{Sakellariadou}},\
  and\ \citenamefont {{Yang}}}]{Zhang2021}%
  \BibitemOpen
  \bibfield  {author} {\bibinfo {author} {\bibfnamefont {Jun}\ \bibnamefont
  {{Zhang}}}, \bibinfo {author} {\bibfnamefont {Zhenwei}\ \bibnamefont
  {{Lyu}}}, \bibinfo {author} {\bibfnamefont {Junwu}\ \bibnamefont {{Huang}}},
  \bibinfo {author} {\bibfnamefont {Matthew~C.}\ \bibnamefont {{Johnson}}},
  \bibinfo {author} {\bibfnamefont {Laura}\ \bibnamefont {{Sagunski}}},
  \bibinfo {author} {\bibfnamefont {Mairi}\ \bibnamefont {{Sakellariadou}}}, \
  and\ \bibinfo {author} {\bibfnamefont {Huan}\ \bibnamefont {{Yang}}},\
  }\bibfield  {title} {\enquote {\bibinfo {title} {{First Constraints on
  Nuclear Coupling of Axionlike Particles from the Binary Neutron Star
  Gravitational Wave Event GW170817}},}\ }\href {\doibase
  10.1103/PhysRevLett.127.161101} {\bibfield  {journal} {\bibinfo  {journal}
  {\prl}\ }\textbf {\bibinfo {volume} {127}},\ \bibinfo {eid} {161101}
  (\bibinfo {year} {2021})},\ \Eprint {http://arxiv.org/abs/2105.13963}
  {arXiv:2105.13963 [hep-ph]} \BibitemShut {NoStop}%
\bibitem [{\citenamefont {{LIGO Scientific Collaboration}}\ and\ \citenamefont
  {{Virgo Collaboration}}(2017{\natexlab{b}})}]{LVC2017Nat}%
  \BibitemOpen
  \bibfield  {author} {\bibinfo {author} {\bibnamefont {{LIGO Scientific
  Collaboration}}}\ and\ \bibinfo {author} {\bibnamefont {{Virgo
  Collaboration}}},\ }\bibfield  {title} {\enquote {\bibinfo {title} {{A
  gravitational-wave standard siren measurement of the Hubble constant}},}\
  }\href {\doibase 10.1038/nature24471} {\bibfield  {journal} {\bibinfo
  {journal} {\nat}\ }\textbf {\bibinfo {volume} {551}},\ \bibinfo {pages}
  {85--88} (\bibinfo {year} {2017}{\natexlab{b}})},\ \Eprint
  {http://arxiv.org/abs/1710.05835} {arXiv:1710.05835 [astro-ph.CO]}
  \BibitemShut {NoStop}%
\bibitem [{\citenamefont {{Graham}}\ \emph {et~al.}(2020)\citenamefont
  {{Graham}}, \citenamefont {{Ford}}, \citenamefont {{McKernan}}, \citenamefont
  {{Ross}}, \citenamefont {{Stern}}, \citenamefont {{Burdge}}, \citenamefont
  {{Coughlin}}, \citenamefont {{Djorgovski}}, \citenamefont {{Drake}},
  \citenamefont {{Duev}}, \citenamefont {{Kasliwal}}, \citenamefont
  {{Mahabal}}, \citenamefont {{van Velzen}}, \citenamefont {{Belecki}},
  \citenamefont {{Bellm}}, \citenamefont {{Burruss}}, \citenamefont {{Cenko}},
  \citenamefont {{Cunningham}}, \citenamefont {{Helou}}, \citenamefont
  {{Kulkarni}}, \citenamefont {{Masci}}, \citenamefont {{Prince}},
  \citenamefont {{Reiley}}, \citenamefont {{Rodriguez}}, \citenamefont
  {{Rusholme}}, \citenamefont {{Smith}},\ and\ \citenamefont
  {{Soumagnac}}}]{Graham2020}%
  \BibitemOpen
  \bibfield  {author} {\bibinfo {author} {\bibfnamefont {M.~J.}\ \bibnamefont
  {{Graham}}}, \bibinfo {author} {\bibfnamefont {K.~E.~S.}\ \bibnamefont
  {{Ford}}}, \bibinfo {author} {\bibfnamefont {B.}~\bibnamefont {{McKernan}}},
  \bibinfo {author} {\bibfnamefont {N.~P.}\ \bibnamefont {{Ross}}}, \bibinfo
  {author} {\bibfnamefont {D.}~\bibnamefont {{Stern}}}, \bibinfo {author}
  {\bibfnamefont {K.}~\bibnamefont {{Burdge}}}, \bibinfo {author}
  {\bibfnamefont {M.}~\bibnamefont {{Coughlin}}}, \bibinfo {author}
  {\bibfnamefont {S.~G.}\ \bibnamefont {{Djorgovski}}}, \bibinfo {author}
  {\bibfnamefont {A.~J.}\ \bibnamefont {{Drake}}}, \bibinfo {author}
  {\bibfnamefont {D.}~\bibnamefont {{Duev}}}, \bibinfo {author} {\bibfnamefont
  {M.}~\bibnamefont {{Kasliwal}}}, \bibinfo {author} {\bibfnamefont {A.~A.}\
  \bibnamefont {{Mahabal}}}, \bibinfo {author} {\bibfnamefont {S.}~\bibnamefont
  {{van Velzen}}}, \bibinfo {author} {\bibfnamefont {J.}~\bibnamefont
  {{Belecki}}}, \bibinfo {author} {\bibfnamefont {E.~C.}\ \bibnamefont
  {{Bellm}}}, \bibinfo {author} {\bibfnamefont {R.}~\bibnamefont {{Burruss}}},
  \bibinfo {author} {\bibfnamefont {S.~B.}\ \bibnamefont {{Cenko}}}, \bibinfo
  {author} {\bibfnamefont {V.}~\bibnamefont {{Cunningham}}}, \bibinfo {author}
  {\bibfnamefont {G.}~\bibnamefont {{Helou}}}, \bibinfo {author} {\bibfnamefont
  {S.~R.}\ \bibnamefont {{Kulkarni}}}, \bibinfo {author} {\bibfnamefont
  {F.~J.}\ \bibnamefont {{Masci}}}, \bibinfo {author} {\bibfnamefont
  {T.}~\bibnamefont {{Prince}}}, \bibinfo {author} {\bibfnamefont
  {D.}~\bibnamefont {{Reiley}}}, \bibinfo {author} {\bibfnamefont
  {H.}~\bibnamefont {{Rodriguez}}}, \bibinfo {author} {\bibfnamefont
  {B.}~\bibnamefont {{Rusholme}}}, \bibinfo {author} {\bibfnamefont {R.~M.}\
  \bibnamefont {{Smith}}}, \ and\ \bibinfo {author} {\bibfnamefont {M.~T.}\
  \bibnamefont {{Soumagnac}}},\ }\bibfield  {title} {\enquote {\bibinfo {title}
  {{Candidate Electromagnetic Counterpart to the Binary Black Hole Merger
  Gravitational-Wave Event S190521g$^{*}$}},}\ }\href {\doibase
  10.1103/PhysRevLett.124.251102} {\bibfield  {journal} {\bibinfo  {journal}
  {\prl}\ }\textbf {\bibinfo {volume} {124}},\ \bibinfo {eid} {251102}
  (\bibinfo {year} {2020})},\ \Eprint {http://arxiv.org/abs/2006.14122}
  {arXiv:2006.14122 [astro-ph.HE]} \BibitemShut {NoStop}%
\bibitem [{\citenamefont {{LIGO Scientific Collaboration}}\ and\ \citenamefont
  {{Virgo Collaboration}}(2020)}]{GW190521}%
  \BibitemOpen
  \bibfield  {author} {\bibinfo {author} {\bibnamefont {{LIGO Scientific
  Collaboration}}}\ and\ \bibinfo {author} {\bibnamefont {{Virgo
  Collaboration}}},\ }\bibfield  {title} {\enquote {\bibinfo {title}
  {{GW190521: A Binary Black Hole Merger with a Total Mass of 150
  M$_{\odot}$}},}\ }\href {\doibase 10.1103/PhysRevLett.125.101102} {\bibfield
  {journal} {\bibinfo  {journal} {\prl}\ }\textbf {\bibinfo {volume} {125}},\
  \bibinfo {eid} {101102} (\bibinfo {year} {2020})},\ \Eprint
  {http://arxiv.org/abs/2009.01075} {arXiv:2009.01075 [gr-qc]} \BibitemShut
  {NoStop}%
\bibitem [{\citenamefont {{Ashton}}\ \emph {et~al.}(2021)\citenamefont
  {{Ashton}}, \citenamefont {{Ackley}}, \citenamefont {{Hernandez}},\ and\
  \citenamefont {{Piotrzkowski}}}]{Ashton2021}%
  \BibitemOpen
  \bibfield  {author} {\bibinfo {author} {\bibfnamefont {Gregory}\ \bibnamefont
  {{Ashton}}}, \bibinfo {author} {\bibfnamefont {Kendall}\ \bibnamefont
  {{Ackley}}}, \bibinfo {author} {\bibfnamefont {Ignacio~Maga{\~n}a}\
  \bibnamefont {{Hernandez}}}, \ and\ \bibinfo {author} {\bibfnamefont
  {Brandon}\ \bibnamefont {{Piotrzkowski}}},\ }\bibfield  {title} {\enquote
  {\bibinfo {title} {{Current observations are insufficient to confidently
  associate the binary black hole merger GW190521 with AGN J124942.3 +
  344929}},}\ }\href {\doibase 10.1088/1361-6382/ac33bb} {\bibfield  {journal}
  {\bibinfo  {journal} {Classical and Quantum Gravity}\ }\textbf {\bibinfo
  {volume} {38}},\ \bibinfo {eid} {235004} (\bibinfo {year} {2021})},\ \Eprint
  {http://arxiv.org/abs/2009.12346} {arXiv:2009.12346 [astro-ph.HE]}
  \BibitemShut {NoStop}%
\bibitem [{\citenamefont {{Nitz}}\ and\ \citenamefont
  {{Capano}}(2021)}]{Nitz2021}%
  \BibitemOpen
  \bibfield  {author} {\bibinfo {author} {\bibfnamefont {Alexander~H.}\
  \bibnamefont {{Nitz}}}\ and\ \bibinfo {author} {\bibfnamefont {Collin~D.}\
  \bibnamefont {{Capano}}},\ }\bibfield  {title} {\enquote {\bibinfo {title}
  {{GW190521 May Be an Intermediate-mass Ratio Inspiral}},}\ }\href {\doibase
  10.3847/2041-8213/abccc5} {\bibfield  {journal} {\bibinfo  {journal} {\apjl}\
  }\textbf {\bibinfo {volume} {907}},\ \bibinfo {eid} {L9} (\bibinfo {year}
  {2021})},\ \Eprint {http://arxiv.org/abs/2010.12558} {arXiv:2010.12558
  [astro-ph.HE]} \BibitemShut {NoStop}%
\bibitem [{\citenamefont {{Palmese}}\ \emph {et~al.}(2021)\citenamefont
  {{Palmese}}, \citenamefont {{Fishbach}}, \citenamefont {{Burke}},
  \citenamefont {{Annis}},\ and\ \citenamefont {{Liu}}}]{Palmese2021}%
  \BibitemOpen
  \bibfield  {author} {\bibinfo {author} {\bibfnamefont {A.}~\bibnamefont
  {{Palmese}}}, \bibinfo {author} {\bibfnamefont {M.}~\bibnamefont
  {{Fishbach}}}, \bibinfo {author} {\bibfnamefont {C.~J.}\ \bibnamefont
  {{Burke}}}, \bibinfo {author} {\bibfnamefont {J.}~\bibnamefont {{Annis}}}, \
  and\ \bibinfo {author} {\bibfnamefont {X.}~\bibnamefont {{Liu}}},\ }\bibfield
   {title} {\enquote {\bibinfo {title} {{Do LIGO/Virgo Black Hole Mergers
  Produce AGN Flares? The Case of GW190521 and Prospects for Reaching a
  Confident Association}},}\ }\href {\doibase 10.3847/2041-8213/ac0883}
  {\bibfield  {journal} {\bibinfo  {journal} {\apjl}\ }\textbf {\bibinfo
  {volume} {914}},\ \bibinfo {eid} {L34} (\bibinfo {year} {2021})},\ \Eprint
  {http://arxiv.org/abs/2103.16069} {arXiv:2103.16069 [astro-ph.HE]}
  \BibitemShut {NoStop}%
\bibitem [{\citenamefont {{Pan}}\ and\ \citenamefont {{Yang}}(2021)}]{Pan2021}%
  \BibitemOpen
  \bibfield  {author} {\bibinfo {author} {\bibfnamefont {Zhen}\ \bibnamefont
  {{Pan}}}\ and\ \bibinfo {author} {\bibfnamefont {Huan}\ \bibnamefont
  {{Yang}}},\ }\bibfield  {title} {\enquote {\bibinfo {title} {{Supercritical
  Accretion of Stellar-mass Compact Objects in Active Galactic Nuclei}},}\
  }\href {\doibase 10.3847/1538-4357/ac249c} {\bibfield  {journal} {\bibinfo
  {journal} {\apj}\ }\textbf {\bibinfo {volume} {923}},\ \bibinfo {eid} {173}
  (\bibinfo {year} {2021})},\ \Eprint {http://arxiv.org/abs/2108.00267}
  {arXiv:2108.00267 [astro-ph.HE]} \BibitemShut {NoStop}%
\bibitem [{\citenamefont {{Goldstein}}\ \emph {et~al.}(2019)\citenamefont
  {{Goldstein}}, \citenamefont {{Andreoni}}, \citenamefont {{Nugent}},
  \citenamefont {{Kasliwal}}, \citenamefont {{Coughlin}}, \citenamefont
  {{Anand}}, \citenamefont {{Bloom}}, \citenamefont {{Mart{\'\i}nez-Palomera}},
  \citenamefont {{Zhang}}, \citenamefont {{Ahumada}}, \citenamefont
  {{Bagdasaryan}}, \citenamefont {{Cooke}}, \citenamefont {{De}}, \citenamefont
  {{Duev}}, \citenamefont {{Fremling}}, \citenamefont {{Gatkine}},
  \citenamefont {{Graham}}, \citenamefont {{Ofek}}, \citenamefont {{Singer}},\
  and\ \citenamefont {{Yan}}}]{Goldstein2019}%
  \BibitemOpen
  \bibfield  {author} {\bibinfo {author} {\bibfnamefont {Daniel~A.}\
  \bibnamefont {{Goldstein}}}, \bibinfo {author} {\bibfnamefont {Igor}\
  \bibnamefont {{Andreoni}}}, \bibinfo {author} {\bibfnamefont {Peter~E.}\
  \bibnamefont {{Nugent}}}, \bibinfo {author} {\bibfnamefont {Mansi~M.}\
  \bibnamefont {{Kasliwal}}}, \bibinfo {author} {\bibfnamefont {Michael~W.}\
  \bibnamefont {{Coughlin}}}, \bibinfo {author} {\bibfnamefont {Shreya}\
  \bibnamefont {{Anand}}}, \bibinfo {author} {\bibfnamefont {Joshua~S.}\
  \bibnamefont {{Bloom}}}, \bibinfo {author} {\bibfnamefont {Jorge}\
  \bibnamefont {{Mart{\'\i}nez-Palomera}}}, \bibinfo {author} {\bibfnamefont
  {Keming}\ \bibnamefont {{Zhang}}}, \bibinfo {author} {\bibfnamefont
  {Tom{\'a}s}\ \bibnamefont {{Ahumada}}}, \bibinfo {author} {\bibfnamefont
  {Ashot}\ \bibnamefont {{Bagdasaryan}}}, \bibinfo {author} {\bibfnamefont
  {Jeff}\ \bibnamefont {{Cooke}}}, \bibinfo {author} {\bibfnamefont {Kishalay}\
  \bibnamefont {{De}}}, \bibinfo {author} {\bibfnamefont {Dmitry~A.}\
  \bibnamefont {{Duev}}}, \bibinfo {author} {\bibfnamefont {U.~Christoffer}\
  \bibnamefont {{Fremling}}}, \bibinfo {author} {\bibfnamefont {Pradip}\
  \bibnamefont {{Gatkine}}}, \bibinfo {author} {\bibfnamefont {Matthew}\
  \bibnamefont {{Graham}}}, \bibinfo {author} {\bibfnamefont {Eran~O.}\
  \bibnamefont {{Ofek}}}, \bibinfo {author} {\bibfnamefont {Leo~P.}\
  \bibnamefont {{Singer}}}, \ and\ \bibinfo {author} {\bibfnamefont {Lin}\
  \bibnamefont {{Yan}}},\ }\bibfield  {title} {\enquote {\bibinfo {title}
  {{GROWTH on S190426c: Real-time Search for a Counterpart to the Probable
  Neutron Star-Black Hole Merger using an Automated Difference Imaging Pipeline
  for DECam}},}\ }\href {\doibase 10.3847/2041-8213/ab3046} {\bibfield
  {journal} {\bibinfo  {journal} {\apjl}\ }\textbf {\bibinfo {volume} {881}},\
  \bibinfo {eid} {L7} (\bibinfo {year} {2019})},\ \Eprint
  {http://arxiv.org/abs/1905.06980} {arXiv:1905.06980 [astro-ph.HE]}
  \BibitemShut {NoStop}%
\bibitem [{\citenamefont {{Hosseinzadeh}}\ \emph {et~al.}(2019)\citenamefont
  {{Hosseinzadeh}}, \citenamefont {{Cowperthwaite}}, \citenamefont {{Gomez}},
  \citenamefont {{Villar}}, \citenamefont {{Nicholl}}, \citenamefont
  {{Margutti}}, \citenamefont {{Berger}}, \citenamefont {{Chornock}},
  \citenamefont {{Paterson}}, \citenamefont {{Fong}}, \citenamefont
  {{Savchenko}}, \citenamefont {{Short}}, \citenamefont {{Alexander}},
  \citenamefont {{Blanchard}}, \citenamefont {{Braga}}, \citenamefont
  {{Calkins}}, \citenamefont {{Cartier}}, \citenamefont {{Coppejans}},
  \citenamefont {{Eftekhari}}, \citenamefont {{Laskar}}, \citenamefont {{Ly}},
  \citenamefont {{Patton}}, \citenamefont {{Pelisoli}}, \citenamefont
  {{Reichart}}, \citenamefont {{Terreran}},\ and\ \citenamefont
  {{Williams}}}]{Hosseinzadeh2019}%
  \BibitemOpen
  \bibfield  {author} {\bibinfo {author} {\bibfnamefont {G.}~\bibnamefont
  {{Hosseinzadeh}}}, \bibinfo {author} {\bibfnamefont {P.~S.}\ \bibnamefont
  {{Cowperthwaite}}}, \bibinfo {author} {\bibfnamefont {S.}~\bibnamefont
  {{Gomez}}}, \bibinfo {author} {\bibfnamefont {V.~A.}\ \bibnamefont
  {{Villar}}}, \bibinfo {author} {\bibfnamefont {M.}~\bibnamefont {{Nicholl}}},
  \bibinfo {author} {\bibfnamefont {R.}~\bibnamefont {{Margutti}}}, \bibinfo
  {author} {\bibfnamefont {E.}~\bibnamefont {{Berger}}}, \bibinfo {author}
  {\bibfnamefont {R.}~\bibnamefont {{Chornock}}}, \bibinfo {author}
  {\bibfnamefont {K.}~\bibnamefont {{Paterson}}}, \bibinfo {author}
  {\bibfnamefont {W.}~\bibnamefont {{Fong}}}, \bibinfo {author} {\bibfnamefont
  {V.}~\bibnamefont {{Savchenko}}}, \bibinfo {author} {\bibfnamefont
  {P.}~\bibnamefont {{Short}}}, \bibinfo {author} {\bibfnamefont {K.~D.}\
  \bibnamefont {{Alexander}}}, \bibinfo {author} {\bibfnamefont {P.~K.}\
  \bibnamefont {{Blanchard}}}, \bibinfo {author} {\bibfnamefont
  {J.}~\bibnamefont {{Braga}}}, \bibinfo {author} {\bibfnamefont {M.~L.}\
  \bibnamefont {{Calkins}}}, \bibinfo {author} {\bibfnamefont {R.}~\bibnamefont
  {{Cartier}}}, \bibinfo {author} {\bibfnamefont {D.~L.}\ \bibnamefont
  {{Coppejans}}}, \bibinfo {author} {\bibfnamefont {T.}~\bibnamefont
  {{Eftekhari}}}, \bibinfo {author} {\bibfnamefont {T.}~\bibnamefont
  {{Laskar}}}, \bibinfo {author} {\bibfnamefont {C.}~\bibnamefont {{Ly}}},
  \bibinfo {author} {\bibfnamefont {L.}~\bibnamefont {{Patton}}}, \bibinfo
  {author} {\bibfnamefont {I.}~\bibnamefont {{Pelisoli}}}, \bibinfo {author}
  {\bibfnamefont {D.~E.}\ \bibnamefont {{Reichart}}}, \bibinfo {author}
  {\bibfnamefont {G.}~\bibnamefont {{Terreran}}}, \ and\ \bibinfo {author}
  {\bibfnamefont {P.~K.~G.}\ \bibnamefont {{Williams}}},\ }\bibfield  {title}
  {\enquote {\bibinfo {title} {{Follow-up of the Neutron Star Bearing
  Gravitational-wave Candidate Events S190425z and S190426c with MMT and
  SOAR}},}\ }\href {\doibase 10.3847/2041-8213/ab271c} {\bibfield  {journal}
  {\bibinfo  {journal} {\apjl}\ }\textbf {\bibinfo {volume} {880}},\ \bibinfo
  {eid} {L4} (\bibinfo {year} {2019})},\ \Eprint
  {http://arxiv.org/abs/1905.02186} {arXiv:1905.02186 [astro-ph.HE]}
  \BibitemShut {NoStop}%
\bibitem [{\citenamefont {{Coughlin}}\ \emph {et~al.}(2020)\citenamefont
  {{Coughlin}}, \citenamefont {{Dietrich}}, \citenamefont {{Antier}},
  \citenamefont {{Bulla}}, \citenamefont {{Foucart}}, \citenamefont
  {{Hotokezaka}}, \citenamefont {{Raaijmakers}}, \citenamefont {{Hinderer}},\
  and\ \citenamefont {{Nissanke}}}]{Coughlin2020}%
  \BibitemOpen
  \bibfield  {author} {\bibinfo {author} {\bibfnamefont {Michael~W.}\
  \bibnamefont {{Coughlin}}}, \bibinfo {author} {\bibfnamefont {Tim}\
  \bibnamefont {{Dietrich}}}, \bibinfo {author} {\bibfnamefont {Sarah}\
  \bibnamefont {{Antier}}}, \bibinfo {author} {\bibfnamefont {Mattia}\
  \bibnamefont {{Bulla}}}, \bibinfo {author} {\bibfnamefont {Francois}\
  \bibnamefont {{Foucart}}}, \bibinfo {author} {\bibfnamefont {Kenta}\
  \bibnamefont {{Hotokezaka}}}, \bibinfo {author} {\bibfnamefont {Geert}\
  \bibnamefont {{Raaijmakers}}}, \bibinfo {author} {\bibfnamefont {Tanja}\
  \bibnamefont {{Hinderer}}}, \ and\ \bibinfo {author} {\bibfnamefont {Samaya}\
  \bibnamefont {{Nissanke}}},\ }\bibfield  {title} {\enquote {\bibinfo {title}
  {{Implications of the search for optical counterparts during the first six
  months of the Advanced LIGO's and Advanced Virgo's third observing run:
  possible limits on the ejecta mass and binary properties}},}\ }\href
  {\doibase 10.1093/mnras/stz3457} {\bibfield  {journal} {\bibinfo  {journal}
  {\mnras}\ }\textbf {\bibinfo {volume} {492}},\ \bibinfo {pages} {863--876}
  (\bibinfo {year} {2020})},\ \Eprint {http://arxiv.org/abs/1910.11246}
  {arXiv:1910.11246 [astro-ph.HE]} \BibitemShut {NoStop}%
\bibitem [{\citenamefont {{Thakur}}\ \emph {et~al.}(2020)\citenamefont
  {{Thakur}}, \citenamefont {{Dichiara}}, \citenamefont {{Troja}},
  \citenamefont {{Chase}}, \citenamefont {{S{\'a}nchez-Ram{\'\i}rez}},
  \citenamefont {{Piro}}, \citenamefont {{Fryer}}, \citenamefont {{Butler}},
  \citenamefont {{Watson}}, \citenamefont {{Wollaeger}}, \citenamefont
  {{Ambrosi}}, \citenamefont {{Becerra Gonz{\'a}lez}}, \citenamefont
  {{Becerra}}, \citenamefont {{Bruni}}, \citenamefont {{Cenko}}, \citenamefont
  {{Cusumano}}, \citenamefont {{D'A{\`\i}}}, \citenamefont {{Durbak}},
  \citenamefont {{Fontes}}, \citenamefont {{Gatkine}}, \citenamefont
  {{Hungerford}}, \citenamefont {{Korobkin}}, \citenamefont {{Kutyrev}},
  \citenamefont {{Lee}}, \citenamefont {{Lotti}}, \citenamefont {{Minervini}},
  \citenamefont {{Novara}}, \citenamefont {{Parola}}, \citenamefont
  {{Pereyra}}, \citenamefont {{Ricci}}, \citenamefont {{Tiengo}},\ and\
  \citenamefont {{Veilleux}}}]{Thakur2020}%
  \BibitemOpen
  \bibfield  {author} {\bibinfo {author} {\bibfnamefont {A.~L.}\ \bibnamefont
  {{Thakur}}}, \bibinfo {author} {\bibfnamefont {S.}~\bibnamefont
  {{Dichiara}}}, \bibinfo {author} {\bibfnamefont {E.}~\bibnamefont {{Troja}}},
  \bibinfo {author} {\bibfnamefont {E.~A.}\ \bibnamefont {{Chase}}}, \bibinfo
  {author} {\bibfnamefont {R.}~\bibnamefont {{S{\'a}nchez-Ram{\'\i}rez}}},
  \bibinfo {author} {\bibfnamefont {L.}~\bibnamefont {{Piro}}}, \bibinfo
  {author} {\bibfnamefont {C.~L.}\ \bibnamefont {{Fryer}}}, \bibinfo {author}
  {\bibfnamefont {N.~R.}\ \bibnamefont {{Butler}}}, \bibinfo {author}
  {\bibfnamefont {A.~M.}\ \bibnamefont {{Watson}}}, \bibinfo {author}
  {\bibfnamefont {R.~T.}\ \bibnamefont {{Wollaeger}}}, \bibinfo {author}
  {\bibfnamefont {E.}~\bibnamefont {{Ambrosi}}}, \bibinfo {author}
  {\bibfnamefont {J.}~\bibnamefont {{Becerra Gonz{\'a}lez}}}, \bibinfo {author}
  {\bibfnamefont {R.~L.}\ \bibnamefont {{Becerra}}}, \bibinfo {author}
  {\bibfnamefont {G.}~\bibnamefont {{Bruni}}}, \bibinfo {author} {\bibfnamefont
  {S.~B.}\ \bibnamefont {{Cenko}}}, \bibinfo {author} {\bibfnamefont
  {G.}~\bibnamefont {{Cusumano}}}, \bibinfo {author} {\bibfnamefont
  {A.}~\bibnamefont {{D'A{\`\i}}}}, \bibinfo {author} {\bibfnamefont
  {J.}~\bibnamefont {{Durbak}}}, \bibinfo {author} {\bibfnamefont {C.~J.}\
  \bibnamefont {{Fontes}}}, \bibinfo {author} {\bibfnamefont {P.}~\bibnamefont
  {{Gatkine}}}, \bibinfo {author} {\bibfnamefont {A.~L.}\ \bibnamefont
  {{Hungerford}}}, \bibinfo {author} {\bibfnamefont {O.}~\bibnamefont
  {{Korobkin}}}, \bibinfo {author} {\bibfnamefont {A.~S.}\ \bibnamefont
  {{Kutyrev}}}, \bibinfo {author} {\bibfnamefont {W.~H.}\ \bibnamefont
  {{Lee}}}, \bibinfo {author} {\bibfnamefont {S.}~\bibnamefont {{Lotti}}},
  \bibinfo {author} {\bibfnamefont {G.}~\bibnamefont {{Minervini}}}, \bibinfo
  {author} {\bibfnamefont {G.}~\bibnamefont {{Novara}}}, \bibinfo {author}
  {\bibfnamefont {V.~La}\ \bibnamefont {{Parola}}}, \bibinfo {author}
  {\bibfnamefont {M.}~\bibnamefont {{Pereyra}}}, \bibinfo {author}
  {\bibfnamefont {R.}~\bibnamefont {{Ricci}}}, \bibinfo {author} {\bibfnamefont
  {A.}~\bibnamefont {{Tiengo}}}, \ and\ \bibinfo {author} {\bibfnamefont
  {S.}~\bibnamefont {{Veilleux}}},\ }\bibfield  {title} {\enquote {\bibinfo
  {title} {{A search for optical and near-infrared counterparts of the compact
  binary merger GW190814}},}\ }\href {\doibase 10.1093/mnras/staa2798}
  {\bibfield  {journal} {\bibinfo  {journal} {\mnras}\ }\textbf {\bibinfo
  {volume} {499}},\ \bibinfo {pages} {3868--3883} (\bibinfo {year} {2020})},\
  \Eprint {http://arxiv.org/abs/2007.04998} {arXiv:2007.04998 [astro-ph.HE]}
  \BibitemShut {NoStop}%
\bibitem [{\citenamefont {{Alexander}}\ \emph {et~al.}(2021)\citenamefont
  {{Alexander}}, \citenamefont {{Schroeder}}, \citenamefont {{Paterson}},
  \citenamefont {{Fong}}, \citenamefont {{Cowperthwaite}}, \citenamefont
  {{Gomez}}, \citenamefont {{Margalit}}, \citenamefont {{Margutti}},
  \citenamefont {{Berger}}, \citenamefont {{Blanchard}}, \citenamefont
  {{Chornock}}, \citenamefont {{Eftekhari}}, \citenamefont {{Laskar}},
  \citenamefont {{Metzger}}, \citenamefont {{Nicholl}}, \citenamefont
  {{Villar}},\ and\ \citenamefont {{Williams}}}]{Alexander2021}%
  \BibitemOpen
  \bibfield  {author} {\bibinfo {author} {\bibfnamefont {K.~D.}\ \bibnamefont
  {{Alexander}}}, \bibinfo {author} {\bibfnamefont {G.}~\bibnamefont
  {{Schroeder}}}, \bibinfo {author} {\bibfnamefont {K.}~\bibnamefont
  {{Paterson}}}, \bibinfo {author} {\bibfnamefont {W.}~\bibnamefont {{Fong}}},
  \bibinfo {author} {\bibfnamefont {P.}~\bibnamefont {{Cowperthwaite}}},
  \bibinfo {author} {\bibfnamefont {S.}~\bibnamefont {{Gomez}}}, \bibinfo
  {author} {\bibfnamefont {B.}~\bibnamefont {{Margalit}}}, \bibinfo {author}
  {\bibfnamefont {R.}~\bibnamefont {{Margutti}}}, \bibinfo {author}
  {\bibfnamefont {E.}~\bibnamefont {{Berger}}}, \bibinfo {author}
  {\bibfnamefont {P.}~\bibnamefont {{Blanchard}}}, \bibinfo {author}
  {\bibfnamefont {R.}~\bibnamefont {{Chornock}}}, \bibinfo {author}
  {\bibfnamefont {T.}~\bibnamefont {{Eftekhari}}}, \bibinfo {author}
  {\bibfnamefont {T.}~\bibnamefont {{Laskar}}}, \bibinfo {author}
  {\bibfnamefont {B.~D.}\ \bibnamefont {{Metzger}}}, \bibinfo {author}
  {\bibfnamefont {M.}~\bibnamefont {{Nicholl}}}, \bibinfo {author}
  {\bibfnamefont {V.~A.}\ \bibnamefont {{Villar}}}, \ and\ \bibinfo {author}
  {\bibfnamefont {P.~K.~G.}\ \bibnamefont {{Williams}}},\ }\bibfield  {title}
  {\enquote {\bibinfo {title} {{A Late-time Galaxy-targeted Search for the
  Radio Counterpart of GW190814}},}\ }\href {\doibase 10.3847/1538-4357/ac281a}
  {\bibfield  {journal} {\bibinfo  {journal} {\apj}\ }\textbf {\bibinfo
  {volume} {923}},\ \bibinfo {eid} {66} (\bibinfo {year} {2021})},\ \Eprint
  {http://arxiv.org/abs/2102.08957} {arXiv:2102.08957 [astro-ph.HE]}
  \BibitemShut {NoStop}%
\bibitem [{\citenamefont {{Anand}}\ \emph {et~al.}(2021)\citenamefont
  {{Anand}}, \citenamefont {{Coughlin}}, \citenamefont {{Kasliwal}},
  \citenamefont {{Bulla}}, \citenamefont {{Ahumada}}, \citenamefont
  {{Sagu{\'e}s Carracedo}}, \citenamefont {{Almualla}}, \citenamefont
  {{Andreoni}}, \citenamefont {{Stein}}, \citenamefont {{Foucart}},
  \citenamefont {{Singer}}, \citenamefont {{Sollerman}}, \citenamefont
  {{Bellm}}, \citenamefont {{Bolin}}, \citenamefont {{Caballero-Garc{\'\i}a}},
  \citenamefont {{Castro-Tirado}}, \citenamefont {{Cenko}}, \citenamefont
  {{De}}, \citenamefont {{Dekany}}, \citenamefont {{Duev}}, \citenamefont
  {{Feeney}}, \citenamefont {{Fremling}}, \citenamefont {{Goldstein}},
  \citenamefont {{Golkhou}}, \citenamefont {{Graham}}, \citenamefont
  {{Guessoum}}, \citenamefont {{Hankins}}, \citenamefont {{Hu}}, \citenamefont
  {{Kong}}, \citenamefont {{Kool}}, \citenamefont {{Kulkarni}}, \citenamefont
  {{Kumar}}, \citenamefont {{Laher}}, \citenamefont {{Masci}}, \citenamefont
  {{Mr{\'o}z}}, \citenamefont {{Nissanke}}, \citenamefont {{Porter}},
  \citenamefont {{Reusch}}, \citenamefont {{Riddle}}, \citenamefont {{Rosnet}},
  \citenamefont {{Rusholme}}, \citenamefont {{Serabyn}}, \citenamefont
  {{S{\'a}nchez-Ram{\'\i}rez}}, \citenamefont {{Rigault}}, \citenamefont
  {{Shupe}}, \citenamefont {{Smith}}, \citenamefont {{Soumagnac}},
  \citenamefont {{Walters}},\ and\ \citenamefont {{Valeev}}}]{Anand2021}%
  \BibitemOpen
  \bibfield  {author} {\bibinfo {author} {\bibfnamefont {Shreya}\ \bibnamefont
  {{Anand}}}, \bibinfo {author} {\bibfnamefont {Michael~W.}\ \bibnamefont
  {{Coughlin}}}, \bibinfo {author} {\bibfnamefont {Mansi~M.}\ \bibnamefont
  {{Kasliwal}}}, \bibinfo {author} {\bibfnamefont {Mattia}\ \bibnamefont
  {{Bulla}}}, \bibinfo {author} {\bibfnamefont {Tom{\'a}s}\ \bibnamefont
  {{Ahumada}}}, \bibinfo {author} {\bibfnamefont {Ana}\ \bibnamefont
  {{Sagu{\'e}s Carracedo}}}, \bibinfo {author} {\bibfnamefont {Mouza}\
  \bibnamefont {{Almualla}}}, \bibinfo {author} {\bibfnamefont {Igor}\
  \bibnamefont {{Andreoni}}}, \bibinfo {author} {\bibfnamefont {Robert}\
  \bibnamefont {{Stein}}}, \bibinfo {author} {\bibfnamefont {Francois}\
  \bibnamefont {{Foucart}}}, \bibinfo {author} {\bibfnamefont {Leo~P.}\
  \bibnamefont {{Singer}}}, \bibinfo {author} {\bibfnamefont {Jesper}\
  \bibnamefont {{Sollerman}}}, \bibinfo {author} {\bibfnamefont {Eric~C.}\
  \bibnamefont {{Bellm}}}, \bibinfo {author} {\bibfnamefont {Bryce}\
  \bibnamefont {{Bolin}}}, \bibinfo {author} {\bibfnamefont {M.~D.}\
  \bibnamefont {{Caballero-Garc{\'\i}a}}}, \bibinfo {author} {\bibfnamefont
  {Alberto~J.}\ \bibnamefont {{Castro-Tirado}}}, \bibinfo {author}
  {\bibfnamefont {S.~Bradley}\ \bibnamefont {{Cenko}}}, \bibinfo {author}
  {\bibfnamefont {Kishalay}\ \bibnamefont {{De}}}, \bibinfo {author}
  {\bibfnamefont {Richard~G.}\ \bibnamefont {{Dekany}}}, \bibinfo {author}
  {\bibfnamefont {Dmitry~A.}\ \bibnamefont {{Duev}}}, \bibinfo {author}
  {\bibfnamefont {Michael}\ \bibnamefont {{Feeney}}}, \bibinfo {author}
  {\bibfnamefont {Christoffer}\ \bibnamefont {{Fremling}}}, \bibinfo {author}
  {\bibfnamefont {Daniel~A.}\ \bibnamefont {{Goldstein}}}, \bibinfo {author}
  {\bibfnamefont {V.~Zach}\ \bibnamefont {{Golkhou}}}, \bibinfo {author}
  {\bibfnamefont {Matthew~J.}\ \bibnamefont {{Graham}}}, \bibinfo {author}
  {\bibfnamefont {Nidhal}\ \bibnamefont {{Guessoum}}}, \bibinfo {author}
  {\bibfnamefont {Matthew~J.}\ \bibnamefont {{Hankins}}}, \bibinfo {author}
  {\bibfnamefont {Youdong}\ \bibnamefont {{Hu}}}, \bibinfo {author}
  {\bibfnamefont {Albert K.~H.}\ \bibnamefont {{Kong}}}, \bibinfo {author}
  {\bibfnamefont {Erik~C.}\ \bibnamefont {{Kool}}}, \bibinfo {author}
  {\bibfnamefont {S.~R.}\ \bibnamefont {{Kulkarni}}}, \bibinfo {author}
  {\bibfnamefont {Harsh}\ \bibnamefont {{Kumar}}}, \bibinfo {author}
  {\bibfnamefont {Russ~R.}\ \bibnamefont {{Laher}}}, \bibinfo {author}
  {\bibfnamefont {Frank~J.}\ \bibnamefont {{Masci}}}, \bibinfo {author}
  {\bibfnamefont {Przemek}\ \bibnamefont {{Mr{\'o}z}}}, \bibinfo {author}
  {\bibfnamefont {Samaya}\ \bibnamefont {{Nissanke}}}, \bibinfo {author}
  {\bibfnamefont {Michael}\ \bibnamefont {{Porter}}}, \bibinfo {author}
  {\bibfnamefont {Simeon}\ \bibnamefont {{Reusch}}}, \bibinfo {author}
  {\bibfnamefont {Reed}\ \bibnamefont {{Riddle}}}, \bibinfo {author}
  {\bibfnamefont {Philippe}\ \bibnamefont {{Rosnet}}}, \bibinfo {author}
  {\bibfnamefont {Ben}\ \bibnamefont {{Rusholme}}}, \bibinfo {author}
  {\bibfnamefont {Eugene}\ \bibnamefont {{Serabyn}}}, \bibinfo {author}
  {\bibfnamefont {R.}~\bibnamefont {{S{\'a}nchez-Ram{\'\i}rez}}}, \bibinfo
  {author} {\bibfnamefont {Mickael}\ \bibnamefont {{Rigault}}}, \bibinfo
  {author} {\bibfnamefont {David~L.}\ \bibnamefont {{Shupe}}}, \bibinfo
  {author} {\bibfnamefont {Roger}\ \bibnamefont {{Smith}}}, \bibinfo {author}
  {\bibfnamefont {Maayane~T.}\ \bibnamefont {{Soumagnac}}}, \bibinfo {author}
  {\bibfnamefont {Richard}\ \bibnamefont {{Walters}}}, \ and\ \bibinfo {author}
  {\bibfnamefont {Azamat~F.}\ \bibnamefont {{Valeev}}},\ }\bibfield  {title}
  {\enquote {\bibinfo {title} {{Optical follow-up of the neutron star-black
  hole mergers S200105ae and S200115j}},}\ }\href {\doibase
  10.1038/s41550-020-1183-3} {\bibfield  {journal} {\bibinfo  {journal} {Nature
  Astronomy}\ }\textbf {\bibinfo {volume} {5}},\ \bibinfo {pages} {46--53}
  (\bibinfo {year} {2021})},\ \Eprint {http://arxiv.org/abs/2009.07210}
  {arXiv:2009.07210 [astro-ph.HE]} \BibitemShut {NoStop}%
\bibitem [{\citenamefont {{Kilpatrick}}\ \emph {et~al.}(2021)\citenamefont
  {{Kilpatrick}}, \citenamefont {{Coulter}}, \citenamefont {{Arcavi}},
  \citenamefont {{Brink}}, \citenamefont {{Dimitriadis}}, \citenamefont
  {{Filippenko}}, \citenamefont {{Foley}}, \citenamefont {{Howell}},
  \citenamefont {{Jones}}, \citenamefont {{Kasen}}, \citenamefont {{Makler}},
  \citenamefont {{Piro}}, \citenamefont {{Rojas-Bravo}}, \citenamefont
  {{Sand}}, \citenamefont {{Swift}}, \citenamefont {{Tucker}}, \citenamefont
  {{Zheng}}, \citenamefont {{Allam}}, \citenamefont {{Annis}}, \citenamefont
  {{Antilen}}, \citenamefont {{Bachmann}}, \citenamefont {{Bloom}},
  \citenamefont {{Bom}}, \citenamefont {{Bostroem}}, \citenamefont {{Brout}},
  \citenamefont {{Burke}}, \citenamefont {{Butler}}, \citenamefont {{Butner}},
  \citenamefont {{Campillay}}, \citenamefont {{Clever}}, \citenamefont
  {{Conselice}}, \citenamefont {{Cooke}}, \citenamefont {{Dage}}, \citenamefont
  {{de Carvalho}}, \citenamefont {{de Jaeger}}, \citenamefont {{Desai}},
  \citenamefont {{Garcia}}, \citenamefont {{Garcia-Bellido}}, \citenamefont
  {{Gill}}, \citenamefont {{Girish}}, \citenamefont {{Hallakoun}},
  \citenamefont {{Herner}}, \citenamefont {{Hiramatsu}}, \citenamefont
  {{Holz}}, \citenamefont {{Huber}}, \citenamefont {{Kawash}}, \citenamefont
  {{McCully}}, \citenamefont {{Medallon}}, \citenamefont {{Metzger}},
  \citenamefont {{Modak}}, \citenamefont {{Morgan}}, \citenamefont
  {{Mu{\~n}oz}}, \citenamefont {{Mu{\~n}oz-Elgueta}}, \citenamefont
  {{Murakami}}, \citenamefont {{Felipe Olivares}}, \citenamefont {{Palmese}},
  \citenamefont {{Patra}}, \citenamefont {{Pereira}}, \citenamefont {{Pessi}},
  \citenamefont {{Pineda-Garcia}}, \citenamefont {{Quirola-V{\'a}squez}},
  \citenamefont {{Ramirez-Ruiz}}, \citenamefont {{Rembold}}, \citenamefont
  {{Rest}}, \citenamefont {{Rodr{\'\i}guez}}, \citenamefont {{Santana-Silva}},
  \citenamefont {{Sherman}}, \citenamefont {{Siebert}}, \citenamefont
  {{Smith}}, \citenamefont {{Smith}}, \citenamefont {{Soares-Santos}},
  \citenamefont {{Stacey}}, \citenamefont {{Stahl}}, \citenamefont {{Strader}},
  \citenamefont {{Strasburger}}, \citenamefont {{Sunseri}}, \citenamefont
  {{Tinyanont}}, \citenamefont {{Tucker}}, \citenamefont {{Ulloa}},
  \citenamefont {{Valenti}}, \citenamefont {{Vasylyev}}, \citenamefont
  {{Wiesner}},\ and\ \citenamefont {{Zhang}}}]{Kilpatrick2021}%
  \BibitemOpen
  \bibfield  {author} {\bibinfo {author} {\bibfnamefont {Charles~D.}\
  \bibnamefont {{Kilpatrick}}}, \bibinfo {author} {\bibfnamefont {David~A.}\
  \bibnamefont {{Coulter}}}, \bibinfo {author} {\bibfnamefont {Iair}\
  \bibnamefont {{Arcavi}}}, \bibinfo {author} {\bibfnamefont {Thomas~G.}\
  \bibnamefont {{Brink}}}, \bibinfo {author} {\bibfnamefont {Georgios}\
  \bibnamefont {{Dimitriadis}}}, \bibinfo {author} {\bibfnamefont {Alexei~V.}\
  \bibnamefont {{Filippenko}}}, \bibinfo {author} {\bibfnamefont {Ryan~J.}\
  \bibnamefont {{Foley}}}, \bibinfo {author} {\bibfnamefont {D.~Andrew}\
  \bibnamefont {{Howell}}}, \bibinfo {author} {\bibfnamefont {David~O.}\
  \bibnamefont {{Jones}}}, \bibinfo {author} {\bibfnamefont {Daniel}\
  \bibnamefont {{Kasen}}}, \bibinfo {author} {\bibfnamefont {Martin}\
  \bibnamefont {{Makler}}}, \bibinfo {author} {\bibfnamefont {Anthony~L.}\
  \bibnamefont {{Piro}}}, \bibinfo {author} {\bibfnamefont {C{\'e}sar}\
  \bibnamefont {{Rojas-Bravo}}}, \bibinfo {author} {\bibfnamefont {David~J.}\
  \bibnamefont {{Sand}}}, \bibinfo {author} {\bibfnamefont {Jonathan~J.}\
  \bibnamefont {{Swift}}}, \bibinfo {author} {\bibfnamefont {Douglas}\
  \bibnamefont {{Tucker}}}, \bibinfo {author} {\bibfnamefont {WeiKang}\
  \bibnamefont {{Zheng}}}, \bibinfo {author} {\bibfnamefont {Sahar~S.}\
  \bibnamefont {{Allam}}}, \bibinfo {author} {\bibfnamefont {James~T.}\
  \bibnamefont {{Annis}}}, \bibinfo {author} {\bibfnamefont {Juanita}\
  \bibnamefont {{Antilen}}}, \bibinfo {author} {\bibfnamefont {Tristan~G.}\
  \bibnamefont {{Bachmann}}}, \bibinfo {author} {\bibfnamefont {Joshua~S.}\
  \bibnamefont {{Bloom}}}, \bibinfo {author} {\bibfnamefont {Clecio~R.}\
  \bibnamefont {{Bom}}}, \bibinfo {author} {\bibfnamefont {K.~Azalee}\
  \bibnamefont {{Bostroem}}}, \bibinfo {author} {\bibfnamefont {Dillon}\
  \bibnamefont {{Brout}}}, \bibinfo {author} {\bibfnamefont {Jamison}\
  \bibnamefont {{Burke}}}, \bibinfo {author} {\bibfnamefont {Robert~E.}\
  \bibnamefont {{Butler}}}, \bibinfo {author} {\bibfnamefont {Melissa}\
  \bibnamefont {{Butner}}}, \bibinfo {author} {\bibfnamefont {Abdo}\
  \bibnamefont {{Campillay}}}, \bibinfo {author} {\bibfnamefont {Karoli~E.}\
  \bibnamefont {{Clever}}}, \bibinfo {author} {\bibfnamefont {Christopher~J.}\
  \bibnamefont {{Conselice}}}, \bibinfo {author} {\bibfnamefont {Jeff}\
  \bibnamefont {{Cooke}}}, \bibinfo {author} {\bibfnamefont {Kristen~C.}\
  \bibnamefont {{Dage}}}, \bibinfo {author} {\bibfnamefont {Reinaldo~R.}\
  \bibnamefont {{de Carvalho}}}, \bibinfo {author} {\bibfnamefont {Thomas}\
  \bibnamefont {{de Jaeger}}}, \bibinfo {author} {\bibfnamefont {Shantanu}\
  \bibnamefont {{Desai}}}, \bibinfo {author} {\bibfnamefont {Alyssa}\
  \bibnamefont {{Garcia}}}, \bibinfo {author} {\bibfnamefont {Juan}\
  \bibnamefont {{Garcia-Bellido}}}, \bibinfo {author} {\bibfnamefont {Mandeep
  S.~S.}\ \bibnamefont {{Gill}}}, \bibinfo {author} {\bibfnamefont {Nachiket}\
  \bibnamefont {{Girish}}}, \bibinfo {author} {\bibfnamefont {Na'ama}\
  \bibnamefont {{Hallakoun}}}, \bibinfo {author} {\bibfnamefont {Kenneth}\
  \bibnamefont {{Herner}}}, \bibinfo {author} {\bibfnamefont {Daichi}\
  \bibnamefont {{Hiramatsu}}}, \bibinfo {author} {\bibfnamefont {Daniel~E.}\
  \bibnamefont {{Holz}}}, \bibinfo {author} {\bibfnamefont {Grace}\
  \bibnamefont {{Huber}}}, \bibinfo {author} {\bibfnamefont {Adam~M.}\
  \bibnamefont {{Kawash}}}, \bibinfo {author} {\bibfnamefont {Curtis}\
  \bibnamefont {{McCully}}}, \bibinfo {author} {\bibfnamefont {Sophia~A.}\
  \bibnamefont {{Medallon}}}, \bibinfo {author} {\bibfnamefont {Brian~D.}\
  \bibnamefont {{Metzger}}}, \bibinfo {author} {\bibfnamefont {Shaunak}\
  \bibnamefont {{Modak}}}, \bibinfo {author} {\bibfnamefont {Robert}\
  \bibnamefont {{Morgan}}}, \bibinfo {author} {\bibfnamefont {Ricardo~R.}\
  \bibnamefont {{Mu{\~n}oz}}}, \bibinfo {author} {\bibfnamefont {Nahir}\
  \bibnamefont {{Mu{\~n}oz-Elgueta}}}, \bibinfo {author} {\bibfnamefont
  {Yukei~S.}\ \bibnamefont {{Murakami}}}, \bibinfo {author} {\bibfnamefont
  {E.}~\bibnamefont {{Felipe Olivares}}}, \bibinfo {author} {\bibfnamefont
  {Antonella}\ \bibnamefont {{Palmese}}}, \bibinfo {author} {\bibfnamefont
  {Kishore~C.}\ \bibnamefont {{Patra}}}, \bibinfo {author} {\bibfnamefont
  {Maria E.~S.}\ \bibnamefont {{Pereira}}}, \bibinfo {author} {\bibfnamefont
  {Thallis~L.}\ \bibnamefont {{Pessi}}}, \bibinfo {author} {\bibfnamefont
  {J.}~\bibnamefont {{Pineda-Garcia}}}, \bibinfo {author} {\bibfnamefont
  {Jonathan}\ \bibnamefont {{Quirola-V{\'a}squez}}}, \bibinfo {author}
  {\bibfnamefont {Enrico}\ \bibnamefont {{Ramirez-Ruiz}}}, \bibinfo {author}
  {\bibfnamefont {Sandro~Barboza}\ \bibnamefont {{Rembold}}}, \bibinfo {author}
  {\bibfnamefont {Armin}\ \bibnamefont {{Rest}}}, \bibinfo {author}
  {\bibfnamefont {{\'O}smar}\ \bibnamefont {{Rodr{\'\i}guez}}}, \bibinfo
  {author} {\bibfnamefont {Luidhy}\ \bibnamefont {{Santana-Silva}}}, \bibinfo
  {author} {\bibfnamefont {Nora~F.}\ \bibnamefont {{Sherman}}}, \bibinfo
  {author} {\bibfnamefont {Matthew~R.}\ \bibnamefont {{Siebert}}}, \bibinfo
  {author} {\bibfnamefont {Carli}\ \bibnamefont {{Smith}}}, \bibinfo {author}
  {\bibfnamefont {J.~Allyn}\ \bibnamefont {{Smith}}}, \bibinfo {author}
  {\bibfnamefont {Marcelle}\ \bibnamefont {{Soares-Santos}}}, \bibinfo {author}
  {\bibfnamefont {Holland}\ \bibnamefont {{Stacey}}}, \bibinfo {author}
  {\bibfnamefont {Benjamin~E.}\ \bibnamefont {{Stahl}}}, \bibinfo {author}
  {\bibfnamefont {Jay}\ \bibnamefont {{Strader}}}, \bibinfo {author}
  {\bibfnamefont {Erika}\ \bibnamefont {{Strasburger}}}, \bibinfo {author}
  {\bibfnamefont {James}\ \bibnamefont {{Sunseri}}}, \bibinfo {author}
  {\bibfnamefont {Samaporn}\ \bibnamefont {{Tinyanont}}}, \bibinfo {author}
  {\bibfnamefont {Brad~E.}\ \bibnamefont {{Tucker}}}, \bibinfo {author}
  {\bibfnamefont {Natalie}\ \bibnamefont {{Ulloa}}}, \bibinfo {author}
  {\bibfnamefont {Stefano}\ \bibnamefont {{Valenti}}}, \bibinfo {author}
  {\bibfnamefont {Sergiy~S.}\ \bibnamefont {{Vasylyev}}}, \bibinfo {author}
  {\bibfnamefont {Matthew~P.}\ \bibnamefont {{Wiesner}}}, \ and\ \bibinfo
  {author} {\bibfnamefont {Keto~D.}\ \bibnamefont {{Zhang}}},\ }\bibfield
  {title} {\enquote {\bibinfo {title} {{The Gravity Collective: A Search for
  the Electromagnetic Counterpart to the Neutron Star-Black Hole Merger
  GW190814}},}\ }\href {\doibase 10.3847/1538-4357/ac23c6} {\bibfield
  {journal} {\bibinfo  {journal} {\apj}\ }\textbf {\bibinfo {volume} {923}},\
  \bibinfo {eid} {258} (\bibinfo {year} {2021})},\ \Eprint
  {http://arxiv.org/abs/2106.06897} {arXiv:2106.06897 [astro-ph.HE]}
  \BibitemShut {NoStop}%
\bibitem [{\citenamefont {{Zhu}}\ \emph {et~al.}(2021)\citenamefont {{Zhu}},
  \citenamefont {{Wu}}, \citenamefont {{Yang}}, \citenamefont {{Zhang}},
  \citenamefont {{Yu}}, \citenamefont {{Gao}}, \citenamefont {{Cao}},\ and\
  \citenamefont {{Liu}}}]{Zhu2021}%
  \BibitemOpen
  \bibfield  {author} {\bibinfo {author} {\bibfnamefont {Jin-Ping}\
  \bibnamefont {{Zhu}}}, \bibinfo {author} {\bibfnamefont {Shichao}\
  \bibnamefont {{Wu}}}, \bibinfo {author} {\bibfnamefont {Yuan-Pei}\
  \bibnamefont {{Yang}}}, \bibinfo {author} {\bibfnamefont {Bing}\ \bibnamefont
  {{Zhang}}}, \bibinfo {author} {\bibfnamefont {Yun-Wei}\ \bibnamefont {{Yu}}},
  \bibinfo {author} {\bibfnamefont {He}~\bibnamefont {{Gao}}}, \bibinfo
  {author} {\bibfnamefont {Zhoujian}\ \bibnamefont {{Cao}}}, \ and\ \bibinfo
  {author} {\bibfnamefont {Liang-Duan}\ \bibnamefont {{Liu}}},\ }\bibfield
  {title} {\enquote {\bibinfo {title} {{No Detectable Kilonova Counterpart is
  Expected for O3 Neutron Star-Black Hole Candidates}},}\ }\href {\doibase
  10.3847/1538-4357/ac19a7} {\bibfield  {journal} {\bibinfo  {journal} {\apj}\
  }\textbf {\bibinfo {volume} {921}},\ \bibinfo {eid} {156} (\bibinfo {year}
  {2021})},\ \Eprint {http://arxiv.org/abs/2106.15781} {arXiv:2106.15781
  [astro-ph.HE]} \BibitemShut {NoStop}%
\bibitem [{\citenamefont {{Fragione}}(2021)}]{Fragione2021}%
  \BibitemOpen
  \bibfield  {author} {\bibinfo {author} {\bibfnamefont {Giacomo}\ \bibnamefont
  {{Fragione}}},\ }\bibfield  {title} {\enquote {\bibinfo {title}
  {{Black-hole-Neutron-star Mergers Are Unlikely Multimessenger Sources}},}\
  }\href {\doibase 10.3847/2041-8213/ac3bcd} {\bibfield  {journal} {\bibinfo
  {journal} {\apjl}\ }\textbf {\bibinfo {volume} {923}},\ \bibinfo {eid} {L2}
  (\bibinfo {year} {2021})},\ \Eprint {http://arxiv.org/abs/2110.09604}
  {arXiv:2110.09604 [astro-ph.HE]} \BibitemShut {NoStop}%
\bibitem [{\citenamefont {{Levin}}\ \emph {et~al.}(2018)\citenamefont
  {{Levin}}, \citenamefont {{D'Orazio}},\ and\ \citenamefont
  {{Garcia-Saenz}}}]{Levin2018}%
  \BibitemOpen
  \bibfield  {author} {\bibinfo {author} {\bibfnamefont {Janna}\ \bibnamefont
  {{Levin}}}, \bibinfo {author} {\bibfnamefont {Daniel~J.}\ \bibnamefont
  {{D'Orazio}}}, \ and\ \bibinfo {author} {\bibfnamefont {Sebastian}\
  \bibnamefont {{Garcia-Saenz}}},\ }\bibfield  {title} {\enquote {\bibinfo
  {title} {{Black hole pulsar}},}\ }\href {\doibase 10.1103/PhysRevD.98.123002}
  {\bibfield  {journal} {\bibinfo  {journal} {\prd}\ }\textbf {\bibinfo
  {volume} {98}},\ \bibinfo {eid} {123002} (\bibinfo {year} {2018})},\ \Eprint
  {http://arxiv.org/abs/1808.07887} {arXiv:1808.07887 [astro-ph.HE]}
  \BibitemShut {NoStop}%
\bibitem [{\citenamefont {{Dai}}(2019)}]{Dai2019}%
  \BibitemOpen
  \bibfield  {author} {\bibinfo {author} {\bibfnamefont {Z.~G.}\ \bibnamefont
  {{Dai}}},\ }\bibfield  {title} {\enquote {\bibinfo {title} {{Inspiral of a
  Spinning Black Hole-Magnetized Neutron Star Binary: Increasing Charge and
  Electromagnetic Emission}},}\ }\href {\doibase 10.3847/2041-8213/ab0b45}
  {\bibfield  {journal} {\bibinfo  {journal} {\apjl}\ }\textbf {\bibinfo
  {volume} {873}},\ \bibinfo {eid} {L13} (\bibinfo {year} {2019})},\ \Eprint
  {http://arxiv.org/abs/1902.07939} {arXiv:1902.07939 [astro-ph.HE]}
  \BibitemShut {NoStop}%
\bibitem [{\citenamefont {{Pan}}\ and\ \citenamefont {{Yang}}(2019)}]{Pan2019}%
  \BibitemOpen
  \bibfield  {author} {\bibinfo {author} {\bibfnamefont {Zhen}\ \bibnamefont
  {{Pan}}}\ and\ \bibinfo {author} {\bibfnamefont {Huan}\ \bibnamefont
  {{Yang}}},\ }\bibfield  {title} {\enquote {\bibinfo {title} {{Black hole
  discharge: Very-high-energy gamma rays from black hole-neutron star
  mergers}},}\ }\href {\doibase 10.1103/PhysRevD.100.043025} {\bibfield
  {journal} {\bibinfo  {journal} {\prd}\ }\textbf {\bibinfo {volume} {100}},\
  \bibinfo {eid} {043025} (\bibinfo {year} {2019})},\ \Eprint
  {http://arxiv.org/abs/1905.04775} {arXiv:1905.04775 [astro-ph.HE]}
  \BibitemShut {NoStop}%
\bibitem [{\citenamefont {{Zhong}}\ \emph {et~al.}(2019)\citenamefont
  {{Zhong}}, \citenamefont {{Dai}},\ and\ \citenamefont {{Deng}}}]{Zhong2019}%
  \BibitemOpen
  \bibfield  {author} {\bibinfo {author} {\bibfnamefont {Shu-Qing}\
  \bibnamefont {{Zhong}}}, \bibinfo {author} {\bibfnamefont {Zi-Gao}\
  \bibnamefont {{Dai}}}, \ and\ \bibinfo {author} {\bibfnamefont {Can-Min}\
  \bibnamefont {{Deng}}},\ }\bibfield  {title} {\enquote {\bibinfo {title}
  {{Electromagnetic Emission Post Spinning Black Hole Magnetized Neutron Star
  Mergers}},}\ }\href {\doibase 10.3847/2041-8213/ab40c5} {\bibfield  {journal}
  {\bibinfo  {journal} {\apjl}\ }\textbf {\bibinfo {volume} {883}},\ \bibinfo
  {eid} {L19} (\bibinfo {year} {2019})},\ \Eprint
  {http://arxiv.org/abs/1909.00494} {arXiv:1909.00494 [astro-ph.HE]}
  \BibitemShut {NoStop}%
\bibitem [{\citenamefont {{East}}\ \emph {et~al.}(2021)\citenamefont {{East}},
  \citenamefont {{Lehner}}, \citenamefont {{Liebling}},\ and\ \citenamefont
  {{Palenzuela}}}]{Will2021}%
  \BibitemOpen
  \bibfield  {author} {\bibinfo {author} {\bibfnamefont {William~E.}\
  \bibnamefont {{East}}}, \bibinfo {author} {\bibfnamefont {Luis}\ \bibnamefont
  {{Lehner}}}, \bibinfo {author} {\bibfnamefont {Steven~L.}\ \bibnamefont
  {{Liebling}}}, \ and\ \bibinfo {author} {\bibfnamefont {Carlos}\ \bibnamefont
  {{Palenzuela}}},\ }\bibfield  {title} {\enquote {\bibinfo {title}
  {{Multimessenger Signals from Black Hole-Neutron Star Mergers without
  Significant Tidal Disruption}},}\ }\href {\doibase 10.3847/2041-8213/abf566}
  {\bibfield  {journal} {\bibinfo  {journal} {\apjl}\ }\textbf {\bibinfo
  {volume} {912}},\ \bibinfo {eid} {L18} (\bibinfo {year} {2021})},\ \Eprint
  {http://arxiv.org/abs/2101.12214} {arXiv:2101.12214 [astro-ph.HE]}
  \BibitemShut {NoStop}%
\bibitem [{\citenamefont {{D'Orazio}}\ \emph {et~al.}(2022)\citenamefont
  {{D'Orazio}}, \citenamefont {{Haiman}}, \citenamefont {{Levin}},
  \citenamefont {{Samsing}},\ and\ \citenamefont
  {{Vigna-G{\'o}mez}}}]{DOrazio2022}%
  \BibitemOpen
  \bibfield  {author} {\bibinfo {author} {\bibfnamefont {Daniel~J.}\
  \bibnamefont {{D'Orazio}}}, \bibinfo {author} {\bibfnamefont {Zolt{\'a}n}\
  \bibnamefont {{Haiman}}}, \bibinfo {author} {\bibfnamefont {Janna}\
  \bibnamefont {{Levin}}}, \bibinfo {author} {\bibfnamefont {Johan}\
  \bibnamefont {{Samsing}}}, \ and\ \bibinfo {author} {\bibfnamefont
  {Alejandro}\ \bibnamefont {{Vigna-G{\'o}mez}}},\ }\bibfield  {title}
  {\enquote {\bibinfo {title} {{Multimessenger Constraints on Magnetic Fields
  in Merging Black Hole-Neutron Star Binaries}},}\ }\href {\doibase
  10.3847/1538-4357/ac4bdb} {\bibfield  {journal} {\bibinfo  {journal} {\apj}\
  }\textbf {\bibinfo {volume} {927}},\ \bibinfo {eid} {56} (\bibinfo {year}
  {2022})},\ \Eprint {http://arxiv.org/abs/2112.01979} {arXiv:2112.01979
  [astro-ph.HE]} \BibitemShut {NoStop}%
\bibitem [{\citenamefont {{Neill}}\ \emph {et~al.}(2022)\citenamefont
  {{Neill}}, \citenamefont {{Tsang}}, \citenamefont {{van Eerten}},
  \citenamefont {{Ryan}},\ and\ \citenamefont {{Newton}}}]{Neill2022}%
  \BibitemOpen
  \bibfield  {author} {\bibinfo {author} {\bibfnamefont {Duncan}\ \bibnamefont
  {{Neill}}}, \bibinfo {author} {\bibfnamefont {David}\ \bibnamefont
  {{Tsang}}}, \bibinfo {author} {\bibfnamefont {Hendrik}\ \bibnamefont {{van
  Eerten}}}, \bibinfo {author} {\bibfnamefont {Geoffrey}\ \bibnamefont
  {{Ryan}}}, \ and\ \bibinfo {author} {\bibfnamefont {William~G.}\ \bibnamefont
  {{Newton}}},\ }\bibfield  {title} {\enquote {\bibinfo {title} {{Resonant
  shattering flares in black hole-neutron star and binary neutron star
  mergers}},}\ }\href {\doibase 10.1093/mnras/stac1645} {\bibfield  {journal}
  {\bibinfo  {journal} {\mnras}\ }\textbf {\bibinfo {volume} {514}},\ \bibinfo
  {pages} {5385--5402} (\bibinfo {year} {2022})},\ \Eprint
  {http://arxiv.org/abs/2111.03686} {arXiv:2111.03686 [astro-ph.HE]}
  \BibitemShut {NoStop}%
\bibitem [{\citenamefont {Abbott}\ \emph {et~al.}(2020)\citenamefont {Abbott}
  \emph {et~al.}}]{LIGOScientific:2020aai}%
  \BibitemOpen
  \bibfield  {author} {\bibinfo {author} {\bibfnamefont {B.~P.}\ \bibnamefont
  {Abbott}} \emph {et~al.} (\bibinfo {collaboration} {LIGO Scientific,
  Virgo}),\ }\bibfield  {title} {\enquote {\bibinfo {title} {{GW190425:
  Observation of a Compact Binary Coalescence with Total Mass $\sim 3.4
  M_{\odot}$}},}\ }\href {\doibase 10.3847/2041-8213/ab75f5} {\bibfield
  {journal} {\bibinfo  {journal} {Astrophys. J. Lett.}\ }\textbf {\bibinfo
  {volume} {892}},\ \bibinfo {pages} {L3} (\bibinfo {year} {2020})},\ \Eprint
  {http://arxiv.org/abs/2001.01761} {arXiv:2001.01761 [astro-ph.HE]}
  \BibitemShut {NoStop}%
\bibitem [{\citenamefont {{Caleb}}\ \emph {et~al.}(2022)\citenamefont
  {{Caleb}}, \citenamefont {{Heywood}}, \citenamefont {{Rajwade}},
  \citenamefont {{Malenta}}, \citenamefont {{Stappers}}, \citenamefont
  {{Barr}}, \citenamefont {{Chen}}, \citenamefont {{Morello}}, \citenamefont
  {{Sanidas}}, \citenamefont {{van den Eijnden}}, \citenamefont {{Kramer}},
  \citenamefont {{Buckley}}, \citenamefont {{Brink}}, \citenamefont {{Motta}},
  \citenamefont {{Woudt}}, \citenamefont {{Weltevrede}}, \citenamefont
  {{Jankowski}}, \citenamefont {{Surnis}}, \citenamefont {{Buchner}},
  \citenamefont {{Bezuidenhout}}, \citenamefont {{Driessen}},\ and\
  \citenamefont {{Fender}}}]{Caleb2022}%
  \BibitemOpen
  \bibfield  {author} {\bibinfo {author} {\bibfnamefont {Manisha}\ \bibnamefont
  {{Caleb}}}, \bibinfo {author} {\bibfnamefont {Ian}\ \bibnamefont
  {{Heywood}}}, \bibinfo {author} {\bibfnamefont {Kaustubh}\ \bibnamefont
  {{Rajwade}}}, \bibinfo {author} {\bibfnamefont {Mateusz}\ \bibnamefont
  {{Malenta}}}, \bibinfo {author} {\bibfnamefont {Benjamin~Willem}\
  \bibnamefont {{Stappers}}}, \bibinfo {author} {\bibfnamefont {Ewan}\
  \bibnamefont {{Barr}}}, \bibinfo {author} {\bibfnamefont {Weiwei}\
  \bibnamefont {{Chen}}}, \bibinfo {author} {\bibfnamefont {Vincent}\
  \bibnamefont {{Morello}}}, \bibinfo {author} {\bibfnamefont {Sotiris}\
  \bibnamefont {{Sanidas}}}, \bibinfo {author} {\bibfnamefont {Jakob}\
  \bibnamefont {{van den Eijnden}}}, \bibinfo {author} {\bibfnamefont
  {Michael}\ \bibnamefont {{Kramer}}}, \bibinfo {author} {\bibfnamefont
  {David}\ \bibnamefont {{Buckley}}}, \bibinfo {author} {\bibfnamefont {Jaco}\
  \bibnamefont {{Brink}}}, \bibinfo {author} {\bibfnamefont {Sara~Elisa}\
  \bibnamefont {{Motta}}}, \bibinfo {author} {\bibfnamefont {Patrick}\
  \bibnamefont {{Woudt}}}, \bibinfo {author} {\bibfnamefont {Patrick}\
  \bibnamefont {{Weltevrede}}}, \bibinfo {author} {\bibfnamefont {Fabian}\
  \bibnamefont {{Jankowski}}}, \bibinfo {author} {\bibfnamefont {Mayuresh}\
  \bibnamefont {{Surnis}}}, \bibinfo {author} {\bibfnamefont {Sarah}\
  \bibnamefont {{Buchner}}}, \bibinfo {author} {\bibfnamefont
  {Mechiel~Christiaan}\ \bibnamefont {{Bezuidenhout}}}, \bibinfo {author}
  {\bibfnamefont {Laura~Nicole}\ \bibnamefont {{Driessen}}}, \ and\ \bibinfo
  {author} {\bibfnamefont {Rob}\ \bibnamefont {{Fender}}},\ }\bibfield  {title}
  {\enquote {\bibinfo {title} {{Discovery of a radio-emitting neutron star with
  an ultra-long spin period of 76 s}},}\ }\href {\doibase
  10.1038/s41550-022-01688-x} {\bibfield  {journal} {\bibinfo  {journal}
  {Nature Astronomy}\ }\textbf {\bibinfo {volume} {6}},\ \bibinfo {pages}
  {828--836} (\bibinfo {year} {2022})},\ \Eprint
  {http://arxiv.org/abs/2206.01346} {arXiv:2206.01346 [astro-ph.HE]}
  \BibitemShut {NoStop}%
\bibitem [{\citenamefont {{Hurley-Walker}}\ \emph {et~al.}(2022)\citenamefont
  {{Hurley-Walker}}, \citenamefont {{Zhang}}, \citenamefont {{Bahramian}},
  \citenamefont {{McSweeney}}, \citenamefont {{O'Doherty}}, \citenamefont
  {{Hancock}}, \citenamefont {{Morgan}}, \citenamefont {{Anderson}},
  \citenamefont {{Heald}},\ and\ \citenamefont {{Galvin}}}]{Hurley2022}%
  \BibitemOpen
  \bibfield  {author} {\bibinfo {author} {\bibfnamefont {N.}~\bibnamefont
  {{Hurley-Walker}}}, \bibinfo {author} {\bibfnamefont {X.}~\bibnamefont
  {{Zhang}}}, \bibinfo {author} {\bibfnamefont {A.}~\bibnamefont
  {{Bahramian}}}, \bibinfo {author} {\bibfnamefont {S.~J.}\ \bibnamefont
  {{McSweeney}}}, \bibinfo {author} {\bibfnamefont {T.~N.}\ \bibnamefont
  {{O'Doherty}}}, \bibinfo {author} {\bibfnamefont {P.~J.}\ \bibnamefont
  {{Hancock}}}, \bibinfo {author} {\bibfnamefont {J.~S.}\ \bibnamefont
  {{Morgan}}}, \bibinfo {author} {\bibfnamefont {G.~E.}\ \bibnamefont
  {{Anderson}}}, \bibinfo {author} {\bibfnamefont {G.~H.}\ \bibnamefont
  {{Heald}}}, \ and\ \bibinfo {author} {\bibfnamefont {T.~J.}\ \bibnamefont
  {{Galvin}}},\ }\bibfield  {title} {\enquote {\bibinfo {title} {{A radio
  transient with unusually slow periodic emission}},}\ }\href {\doibase
  10.1038/s41586-021-04272-x} {\bibfield  {journal} {\bibinfo  {journal}
  {\nat}\ }\textbf {\bibinfo {volume} {601}},\ \bibinfo {pages} {526--530}
  (\bibinfo {year} {2022})}\BibitemShut {NoStop}%
\bibitem [{\citenamefont {{Beniamini}}\ \emph {et~al.}(2022)\citenamefont
  {{Beniamini}}, \citenamefont {{Wadiasingh}}, \citenamefont {{Hare}},
  \citenamefont {{Rajwade}}, \citenamefont {{Younes}},\ and\ \citenamefont
  {{van der Horst}}}]{Beniamini2022}%
  \BibitemOpen
  \bibfield  {author} {\bibinfo {author} {\bibfnamefont {P.}~\bibnamefont
  {{Beniamini}}}, \bibinfo {author} {\bibfnamefont {Z.}~\bibnamefont
  {{Wadiasingh}}}, \bibinfo {author} {\bibfnamefont {J.}~\bibnamefont
  {{Hare}}}, \bibinfo {author} {\bibfnamefont {K.}~\bibnamefont {{Rajwade}}},
  \bibinfo {author} {\bibfnamefont {G.}~\bibnamefont {{Younes}}}, \ and\
  \bibinfo {author} {\bibfnamefont {A.~J.}\ \bibnamefont {{van der Horst}}},\
  }\bibfield  {title} {\enquote {\bibinfo {title} {{Evidence for an abundant
  old population of Galactic Ultra long period magnetars and implications for
  fast radio bursts}},}\ }\href@noop {} {\bibfield  {journal} {\bibinfo
  {journal} {arXiv e-prints}\ ,\ \bibinfo {eid} {arXiv:2210.09323}} (\bibinfo
  {year} {2022})},\ \Eprint {http://arxiv.org/abs/2210.09323} {arXiv:2210.09323
  [astro-ph.HE]} \BibitemShut {NoStop}%
\bibitem [{\citenamefont {{Rastinejad}}\ \emph {et~al.}(2022)\citenamefont
  {{Rastinejad}}, \citenamefont {{Gompertz}}, \citenamefont {{Levan}},
  \citenamefont {{Fong}}, \citenamefont {{Nicholl}}, \citenamefont {{Lamb}},
  \citenamefont {{Malesani}}, \citenamefont {{Nugent}}, \citenamefont
  {{Oates}}, \citenamefont {{Tanvir}}, \citenamefont {{de Ugarte Postigo}},
  \citenamefont {{Kilpatrick}}, \citenamefont {{Moore}}, \citenamefont
  {{Metzger}}, \citenamefont {{Ravasio}}, \citenamefont {{Rossi}},
  \citenamefont {{Schroeder}}, \citenamefont {{Jencson}}, \citenamefont
  {{Sand}}, \citenamefont {{Smith}}, \citenamefont {{Ag{\"u}{\'\i}
  Fern{\'a}ndez}}, \citenamefont {{Berger}}, \citenamefont {{Blanchard}},
  \citenamefont {{Chornock}}, \citenamefont {{Cobb}}, \citenamefont {{De
  Pasquale}}, \citenamefont {{Fynbo}}, \citenamefont {{Izzo}}, \citenamefont
  {{Kann}}, \citenamefont {{Laskar}}, \citenamefont {{Marini}}, \citenamefont
  {{Paterson}}, \citenamefont {{Rouco Escorial}}, \citenamefont {{Sears}},\
  and\ \citenamefont {{Th{\"o}ne}}}]{Rastinejad2022}%
  \BibitemOpen
  \bibfield  {author} {\bibinfo {author} {\bibfnamefont {J.~C.}\ \bibnamefont
  {{Rastinejad}}}, \bibinfo {author} {\bibfnamefont {B.~P.}\ \bibnamefont
  {{Gompertz}}}, \bibinfo {author} {\bibfnamefont {A.~J.}\ \bibnamefont
  {{Levan}}}, \bibinfo {author} {\bibfnamefont {W.}~\bibnamefont {{Fong}}},
  \bibinfo {author} {\bibfnamefont {M.}~\bibnamefont {{Nicholl}}}, \bibinfo
  {author} {\bibfnamefont {G.~P.}\ \bibnamefont {{Lamb}}}, \bibinfo {author}
  {\bibfnamefont {D.~B.}\ \bibnamefont {{Malesani}}}, \bibinfo {author}
  {\bibfnamefont {A.~E.}\ \bibnamefont {{Nugent}}}, \bibinfo {author}
  {\bibfnamefont {S.~R.}\ \bibnamefont {{Oates}}}, \bibinfo {author}
  {\bibfnamefont {N.~R.}\ \bibnamefont {{Tanvir}}}, \bibinfo {author}
  {\bibfnamefont {A.}~\bibnamefont {{de Ugarte Postigo}}}, \bibinfo {author}
  {\bibfnamefont {C.~D.}\ \bibnamefont {{Kilpatrick}}}, \bibinfo {author}
  {\bibfnamefont {C.~J.}\ \bibnamefont {{Moore}}}, \bibinfo {author}
  {\bibfnamefont {B.~D.}\ \bibnamefont {{Metzger}}}, \bibinfo {author}
  {\bibfnamefont {M.~E.}\ \bibnamefont {{Ravasio}}}, \bibinfo {author}
  {\bibfnamefont {A.}~\bibnamefont {{Rossi}}}, \bibinfo {author} {\bibfnamefont
  {G.}~\bibnamefont {{Schroeder}}}, \bibinfo {author} {\bibfnamefont
  {J.}~\bibnamefont {{Jencson}}}, \bibinfo {author} {\bibfnamefont {D.~J.}\
  \bibnamefont {{Sand}}}, \bibinfo {author} {\bibfnamefont {N.}~\bibnamefont
  {{Smith}}}, \bibinfo {author} {\bibfnamefont {J.~F.}\ \bibnamefont
  {{Ag{\"u}{\'\i} Fern{\'a}ndez}}}, \bibinfo {author} {\bibfnamefont
  {E.}~\bibnamefont {{Berger}}}, \bibinfo {author} {\bibfnamefont {P.~K.}\
  \bibnamefont {{Blanchard}}}, \bibinfo {author} {\bibfnamefont
  {R.}~\bibnamefont {{Chornock}}}, \bibinfo {author} {\bibfnamefont {B.~E.}\
  \bibnamefont {{Cobb}}}, \bibinfo {author} {\bibfnamefont {M.}~\bibnamefont
  {{De Pasquale}}}, \bibinfo {author} {\bibfnamefont {J.~P.~U.}\ \bibnamefont
  {{Fynbo}}}, \bibinfo {author} {\bibfnamefont {L.}~\bibnamefont {{Izzo}}},
  \bibinfo {author} {\bibfnamefont {D.~A.}\ \bibnamefont {{Kann}}}, \bibinfo
  {author} {\bibfnamefont {T.}~\bibnamefont {{Laskar}}}, \bibinfo {author}
  {\bibfnamefont {E.}~\bibnamefont {{Marini}}}, \bibinfo {author}
  {\bibfnamefont {K.}~\bibnamefont {{Paterson}}}, \bibinfo {author}
  {\bibfnamefont {A.}~\bibnamefont {{Rouco Escorial}}}, \bibinfo {author}
  {\bibfnamefont {H.~M.}\ \bibnamefont {{Sears}}}, \ and\ \bibinfo {author}
  {\bibfnamefont {C.~C.}\ \bibnamefont {{Th{\"o}ne}}},\ }\bibfield  {title}
  {\enquote {\bibinfo {title} {{A Kilonova Following a Long-Duration Gamma-Ray
  Burst at 350 Mpc}},}\ }\href@noop {} {\bibfield  {journal} {\bibinfo
  {journal} {arXiv e-prints}\ ,\ \bibinfo {eid} {arXiv:2204.10864}} (\bibinfo
  {year} {2022})},\ \Eprint {http://arxiv.org/abs/2204.10864} {arXiv:2204.10864
  [astro-ph.HE]} \BibitemShut {NoStop}%
\bibitem [{\citenamefont {{Xiao}}\ \emph {et~al.}(2022)\citenamefont {{Xiao}},
  \citenamefont {{Zhang}}, \citenamefont {{Zhu}}, \citenamefont {{Xiong}},
  \citenamefont {{Gao}}, \citenamefont {{Xu}}, \citenamefont {{Zhang}},
  \citenamefont {{Peng}}, \citenamefont {{Li}}, \citenamefont {{Zhang}},
  \citenamefont {{Lu}}, \citenamefont {{Lin}}, \citenamefont {{Liu}},
  \citenamefont {{Zhang}}, \citenamefont {{Ge}}, \citenamefont {{Tuo}},
  \citenamefont {{Xue}}, \citenamefont {{Fu}}, \citenamefont {{Liu}},
  \citenamefont {{Li}}, \citenamefont {{Wang}}, \citenamefont {{Zheng}},
  \citenamefont {{Wang}}, \citenamefont {{Jiang}}, \citenamefont {{Li}},
  \citenamefont {{Liu}}, \citenamefont {{Cao}}, \citenamefont {{Cai}},
  \citenamefont {{Yi}}, \citenamefont {{Zhao}}, \citenamefont {{Xie}},
  \citenamefont {{Li}}, \citenamefont {{Luo}}, \citenamefont {{Liao}},
  \citenamefont {{Song}}, \citenamefont {{Zhang}}, \citenamefont {{Qu}},
  \citenamefont {{Liu}}, \citenamefont {{Li}}, \citenamefont {{Xu}},\ and\
  \citenamefont {{Li}}}]{Xiao2022}%
  \BibitemOpen
  \bibfield  {author} {\bibinfo {author} {\bibfnamefont {Shuo}\ \bibnamefont
  {{Xiao}}}, \bibinfo {author} {\bibfnamefont {Yan-Qiu}\ \bibnamefont
  {{Zhang}}}, \bibinfo {author} {\bibfnamefont {Zi-Pei}\ \bibnamefont {{Zhu}}},
  \bibinfo {author} {\bibfnamefont {Shao-Lin}\ \bibnamefont {{Xiong}}},
  \bibinfo {author} {\bibfnamefont {He}~\bibnamefont {{Gao}}}, \bibinfo
  {author} {\bibfnamefont {Dong}\ \bibnamefont {{Xu}}}, \bibinfo {author}
  {\bibfnamefont {Shuang-Nan}\ \bibnamefont {{Zhang}}}, \bibinfo {author}
  {\bibfnamefont {Wen-Xi}\ \bibnamefont {{Peng}}}, \bibinfo {author}
  {\bibfnamefont {Xiao-Bo}\ \bibnamefont {{Li}}}, \bibinfo {author}
  {\bibfnamefont {Peng}\ \bibnamefont {{Zhang}}}, \bibinfo {author}
  {\bibfnamefont {Fang-Jun}\ \bibnamefont {{Lu}}}, \bibinfo {author}
  {\bibfnamefont {Lin}\ \bibnamefont {{Lin}}}, \bibinfo {author} {\bibfnamefont
  {Liang-Duan}\ \bibnamefont {{Liu}}}, \bibinfo {author} {\bibfnamefont {Zhen}\
  \bibnamefont {{Zhang}}}, \bibinfo {author} {\bibfnamefont {Ming-Yu}\
  \bibnamefont {{Ge}}}, \bibinfo {author} {\bibfnamefont {You-Li}\ \bibnamefont
  {{Tuo}}}, \bibinfo {author} {\bibfnamefont {Wang-Chen}\ \bibnamefont
  {{Xue}}}, \bibinfo {author} {\bibfnamefont {Shao-Yu}\ \bibnamefont {{Fu}}},
  \bibinfo {author} {\bibfnamefont {Xing}\ \bibnamefont {{Liu}}}, \bibinfo
  {author} {\bibfnamefont {An}~\bibnamefont {{Li}}}, \bibinfo {author}
  {\bibfnamefont {Tian-Cong}\ \bibnamefont {{Wang}}}, \bibinfo {author}
  {\bibfnamefont {Chao}\ \bibnamefont {{Zheng}}}, \bibinfo {author}
  {\bibfnamefont {Yue}\ \bibnamefont {{Wang}}}, \bibinfo {author}
  {\bibfnamefont {Shuai-Qing}\ \bibnamefont {{Jiang}}}, \bibinfo {author}
  {\bibfnamefont {Jin-Da}\ \bibnamefont {{Li}}}, \bibinfo {author}
  {\bibfnamefont {Jia-Cong}\ \bibnamefont {{Liu}}}, \bibinfo {author}
  {\bibfnamefont {Zhou-Jian}\ \bibnamefont {{Cao}}}, \bibinfo {author}
  {\bibfnamefont {Ce}~\bibnamefont {{Cai}}}, \bibinfo {author} {\bibfnamefont
  {Qi-Bin}\ \bibnamefont {{Yi}}}, \bibinfo {author} {\bibfnamefont
  {Yi}~\bibnamefont {{Zhao}}}, \bibinfo {author} {\bibfnamefont {Sheng-Lun}\
  \bibnamefont {{Xie}}}, \bibinfo {author} {\bibfnamefont {Cheng-Kui}\
  \bibnamefont {{Li}}}, \bibinfo {author} {\bibfnamefont {Qi}~\bibnamefont
  {{Luo}}}, \bibinfo {author} {\bibfnamefont {Jin-Yuan}\ \bibnamefont
  {{Liao}}}, \bibinfo {author} {\bibfnamefont {Li-Ming}\ \bibnamefont
  {{Song}}}, \bibinfo {author} {\bibfnamefont {Shu}\ \bibnamefont {{Zhang}}},
  \bibinfo {author} {\bibfnamefont {Jin-Lu}\ \bibnamefont {{Qu}}}, \bibinfo
  {author} {\bibfnamefont {Cong-Zhan}\ \bibnamefont {{Liu}}}, \bibinfo {author}
  {\bibfnamefont {Xu-Fang}\ \bibnamefont {{Li}}}, \bibinfo {author}
  {\bibfnamefont {Yu-Peng}\ \bibnamefont {{Xu}}}, \ and\ \bibinfo {author}
  {\bibfnamefont {Ti-Pei}\ \bibnamefont {{Li}}},\ }\bibfield  {title} {\enquote
  {\bibinfo {title} {{The quasi-periodically oscillating precursor of a long
  gamma-ray burst from a binary neutron star merger}},}\ }\href@noop {}
  {\bibfield  {journal} {\bibinfo  {journal} {arXiv e-prints}\ ,\ \bibinfo
  {eid} {arXiv:2205.02186}} (\bibinfo {year} {2022})},\ \Eprint
  {http://arxiv.org/abs/2205.02186} {arXiv:2205.02186 [astro-ph.HE]}
  \BibitemShut {NoStop}%
\bibitem [{\citenamefont {{Mei}}\ \emph {et~al.}(2022)\citenamefont {{Mei}},
  \citenamefont {{Banerjee}}, \citenamefont {{Oganesyan}}, \citenamefont
  {{Sharan Salafia}}, \citenamefont {{Giarratana}}, \citenamefont
  {{Branchesi}}, \citenamefont {{D'Avanzo}}, \citenamefont {{Campana}},
  \citenamefont {{Ghirlanda}}, \citenamefont {{Ronchini}}, \citenamefont
  {{Shukla}},\ and\ \citenamefont {{Tiwari}}}]{Mei2022}%
  \BibitemOpen
  \bibfield  {author} {\bibinfo {author} {\bibfnamefont {Alessio}\ \bibnamefont
  {{Mei}}}, \bibinfo {author} {\bibfnamefont {Biswajit}\ \bibnamefont
  {{Banerjee}}}, \bibinfo {author} {\bibfnamefont {Gor}\ \bibnamefont
  {{Oganesyan}}}, \bibinfo {author} {\bibfnamefont {Om}~\bibnamefont {{Sharan
  Salafia}}}, \bibinfo {author} {\bibfnamefont {Stefano}\ \bibnamefont
  {{Giarratana}}}, \bibinfo {author} {\bibfnamefont {Marica}\ \bibnamefont
  {{Branchesi}}}, \bibinfo {author} {\bibfnamefont {Paolo}\ \bibnamefont
  {{D'Avanzo}}}, \bibinfo {author} {\bibfnamefont {Sergio}\ \bibnamefont
  {{Campana}}}, \bibinfo {author} {\bibfnamefont {Giancarlo}\ \bibnamefont
  {{Ghirlanda}}}, \bibinfo {author} {\bibfnamefont {Samuele}\ \bibnamefont
  {{Ronchini}}}, \bibinfo {author} {\bibfnamefont {Amit}\ \bibnamefont
  {{Shukla}}}, \ and\ \bibinfo {author} {\bibfnamefont {Pawan}\ \bibnamefont
  {{Tiwari}}},\ }\bibfield  {title} {\enquote {\bibinfo {title} {{GeV emission
  from a compact binary merger}},}\ }\href@noop {} {\bibfield  {journal}
  {\bibinfo  {journal} {arXiv e-prints}\ ,\ \bibinfo {eid} {arXiv:2205.08566}}
  (\bibinfo {year} {2022})},\ \Eprint {http://arxiv.org/abs/2205.08566}
  {arXiv:2205.08566 [astro-ph.HE]} \BibitemShut {NoStop}%
\bibitem [{\citenamefont {{Zhang}}\ \emph
  {et~al.}(2022{\natexlab{a}})\citenamefont {{Zhang}}, \citenamefont {{Huang}},
  \citenamefont {{Zheng}}, \citenamefont {{Liu}},\ and\ \citenamefont
  {{Wang}}}]{ZhangHM2022}%
  \BibitemOpen
  \bibfield  {author} {\bibinfo {author} {\bibfnamefont {Hai-Ming}\
  \bibnamefont {{Zhang}}}, \bibinfo {author} {\bibfnamefont {Yi-Yun}\
  \bibnamefont {{Huang}}}, \bibinfo {author} {\bibfnamefont {Jian-He}\
  \bibnamefont {{Zheng}}}, \bibinfo {author} {\bibfnamefont {Ruo-Yu}\
  \bibnamefont {{Liu}}}, \ and\ \bibinfo {author} {\bibfnamefont {Xiang-Yu}\
  \bibnamefont {{Wang}}},\ }\bibfield  {title} {\enquote {\bibinfo {title}
  {{Fermi-LAT Detection of a GeV Afterglow from a Compact Stellar Merger}},}\
  }\href {\doibase 10.3847/2041-8213/ac7b23} {\bibfield  {journal} {\bibinfo
  {journal} {\apjl}\ }\textbf {\bibinfo {volume} {933}},\ \bibinfo {eid} {L22}
  (\bibinfo {year} {2022}{\natexlab{a}})},\ \Eprint
  {http://arxiv.org/abs/2205.09675} {arXiv:2205.09675 [astro-ph.HE]}
  \BibitemShut {NoStop}%
\bibitem [{\citenamefont {{Suvorov}}\ \emph {et~al.}(2022)\citenamefont
  {{Suvorov}}, \citenamefont {{Kuan}},\ and\ \citenamefont
  {{Kokkotas}}}]{Suvorov2022}%
  \BibitemOpen
  \bibfield  {author} {\bibinfo {author} {\bibfnamefont {Arthur~G.}\
  \bibnamefont {{Suvorov}}}, \bibinfo {author} {\bibfnamefont {Hao-Jui}\
  \bibnamefont {{Kuan}}}, \ and\ \bibinfo {author} {\bibfnamefont {Kostas~D.}\
  \bibnamefont {{Kokkotas}}},\ }\bibfield  {title} {\enquote {\bibinfo {title}
  {{Quasi-periodic oscillations in precursor flares via seismic aftershocks
  from resonant shattering}},}\ }\href@noop {} {\bibfield  {journal} {\bibinfo
  {journal} {arXiv e-prints}\ ,\ \bibinfo {eid} {arXiv:2205.11112}} (\bibinfo
  {year} {2022})},\ \Eprint {http://arxiv.org/abs/2205.11112} {arXiv:2205.11112
  [astro-ph.HE]} \BibitemShut {NoStop}%
\bibitem [{\citenamefont {{Gao}}\ \emph {et~al.}(2022)\citenamefont {{Gao}},
  \citenamefont {{Lei}},\ and\ \citenamefont {{Zhu}}}]{Gao2022}%
  \BibitemOpen
  \bibfield  {author} {\bibinfo {author} {\bibfnamefont {He}~\bibnamefont
  {{Gao}}}, \bibinfo {author} {\bibfnamefont {Wei-Hua}\ \bibnamefont {{Lei}}},
  \ and\ \bibinfo {author} {\bibfnamefont {Zi-Pei}\ \bibnamefont {{Zhu}}},\
  }\bibfield  {title} {\enquote {\bibinfo {title} {{GRB 211211A: a Prolonged
  Central Engine under a Strong Magnetic Field Environment}},}\ }\href
  {\doibase 10.3847/2041-8213/ac80c7} {\bibfield  {journal} {\bibinfo
  {journal} {\apjl}\ }\textbf {\bibinfo {volume} {934}},\ \bibinfo {eid} {L12}
  (\bibinfo {year} {2022})},\ \Eprint {http://arxiv.org/abs/2205.05031}
  {arXiv:2205.05031 [astro-ph.HE]} \BibitemShut {NoStop}%
\bibitem [{\citenamefont {{Zhang}}\ \emph
  {et~al.}(2022{\natexlab{b}})\citenamefont {{Zhang}}, \citenamefont {{Yi}},
  \citenamefont {{Zhang}}, \citenamefont {{Xiong}},\ and\ \citenamefont
  {{Xiao}}}]{Zhang2022}%
  \BibitemOpen
  \bibfield  {author} {\bibinfo {author} {\bibfnamefont {Zhen}\ \bibnamefont
  {{Zhang}}}, \bibinfo {author} {\bibfnamefont {Shu-Xu}\ \bibnamefont {{Yi}}},
  \bibinfo {author} {\bibfnamefont {Shuang-Nan}\ \bibnamefont {{Zhang}}},
  \bibinfo {author} {\bibfnamefont {Shao-Lin}\ \bibnamefont {{Xiong}}}, \ and\
  \bibinfo {author} {\bibfnamefont {Shuo}\ \bibnamefont {{Xiao}}},\ }\bibfield
  {title} {\enquote {\bibinfo {title} {{Tidal-induced magnetar super flare at
  the eve of coalescence with its compact companion}},}\ }\href@noop {}
  {\bibfield  {journal} {\bibinfo  {journal} {arXiv e-prints}\ ,\ \bibinfo
  {eid} {arXiv:2207.12324}} (\bibinfo {year} {2022}{\natexlab{b}})},\ \Eprint
  {http://arxiv.org/abs/2207.12324} {arXiv:2207.12324 [astro-ph.HE]}
  \BibitemShut {NoStop}%
\bibitem [{\citenamefont {{Burns}}\ \emph {et~al.}(2023)\citenamefont
  {{Burns}}, \citenamefont {{Goldstein}}, \citenamefont {{Lesage}},
  \citenamefont {{Dalessi}},\ and\ \citenamefont {{Fermi-GBM
  Team.}}}]{Burns2023}%
  \BibitemOpen
  \bibfield  {author} {\bibinfo {author} {\bibfnamefont {E.}~\bibnamefont
  {{Burns}}}, \bibinfo {author} {\bibfnamefont {A.}~\bibnamefont
  {{Goldstein}}}, \bibinfo {author} {\bibfnamefont {S.}~\bibnamefont
  {{Lesage}}}, \bibinfo {author} {\bibfnamefont {S.}~\bibnamefont {{Dalessi}}},
  \ and\ \bibinfo {author} {\bibnamefont {{Fermi-GBM Team.}}},\ }\bibfield
  {title} {\enquote {\bibinfo {title} {{GRB 230307A: possibly the second
  highest GRB energy fluence ever identified}},}\ }\href@noop {} {\bibfield
  {journal} {\bibinfo  {journal} {GRB Coordinates Network}\ }\textbf {\bibinfo
  {volume} {33414}},\ \bibinfo {pages} {1} (\bibinfo {year}
  {2023})}\BibitemShut {NoStop}%
\bibitem [{\citenamefont {{Bulla}}\ \emph {et~al.}(2023)\citenamefont
  {{Bulla}}, \citenamefont {{Camisasca}}, \citenamefont {{Guidorzi}},
  \citenamefont {{Amati}}, \citenamefont {{Rossi}}, \citenamefont {{Stratta}},\
  and\ \citenamefont {{Singh}}}]{Bulla2023}%
  \BibitemOpen
  \bibfield  {author} {\bibinfo {author} {\bibfnamefont {M.}~\bibnamefont
  {{Bulla}}}, \bibinfo {author} {\bibfnamefont {A.~E.}\ \bibnamefont
  {{Camisasca}}}, \bibinfo {author} {\bibfnamefont {C.}~\bibnamefont
  {{Guidorzi}}}, \bibinfo {author} {\bibfnamefont {L.}~\bibnamefont {{Amati}}},
  \bibinfo {author} {\bibfnamefont {A.}~\bibnamefont {{Rossi}}}, \bibinfo
  {author} {\bibfnamefont {G.}~\bibnamefont {{Stratta}}}, \ and\ \bibinfo
  {author} {\bibfnamefont {P.}~\bibnamefont {{Singh}}},\ }\bibfield  {title}
  {\enquote {\bibinfo {title} {{GRB 230307A: good match with kilonova
  models}},}\ }\href@noop {} {\bibfield  {journal} {\bibinfo  {journal} {GRB
  Coordinates Network}\ }\textbf {\bibinfo {volume} {33578}},\ \bibinfo {pages}
  {1} (\bibinfo {year} {2023})}\BibitemShut {NoStop}%
\bibitem [{\citenamefont {Yang}\ \emph {et~al.}(2023)\citenamefont {Yang} \emph
  {et~al.}}]{Yang:2023mqt}%
  \BibitemOpen
  \bibfield  {author} {\bibinfo {author} {\bibfnamefont {Yu-Han}\ \bibnamefont
  {Yang}} \emph {et~al.},\ }\bibfield  {title} {\enquote {\bibinfo {title} {{A
  lanthanide-rich kilonova in the aftermath of a long gamma-ray burst}},}\
  }\href@noop {} {\  (\bibinfo {year} {2023})},\ \Eprint
  {http://arxiv.org/abs/2308.00638} {arXiv:2308.00638 [astro-ph.HE]}
  \BibitemShut {NoStop}%
\bibitem [{\citenamefont {Dichiara}\ \emph {et~al.}(2023)\citenamefont
  {Dichiara}, \citenamefont {Tsang}, \citenamefont {Troja}, \citenamefont
  {Neill}, \citenamefont {Norris},\ and\ \citenamefont
  {Yang}}]{Dichiara:2023goh}%
  \BibitemOpen
  \bibfield  {author} {\bibinfo {author} {\bibfnamefont {S.}~\bibnamefont
  {Dichiara}}, \bibinfo {author} {\bibfnamefont {D.}~\bibnamefont {Tsang}},
  \bibinfo {author} {\bibfnamefont {E.}~\bibnamefont {Troja}}, \bibinfo
  {author} {\bibfnamefont {D.}~\bibnamefont {Neill}}, \bibinfo {author}
  {\bibfnamefont {J.~P.}\ \bibnamefont {Norris}}, \ and\ \bibinfo {author}
  {\bibfnamefont {Y.~H.}\ \bibnamefont {Yang}},\ }\bibfield  {title} {\enquote
  {\bibinfo {title} {{A luminous precursor in the extremely bright GRB
  230307A}},}\ }\href@noop {} {\  (\bibinfo {year} {2023})},\ \Eprint
  {http://arxiv.org/abs/2307.02996} {arXiv:2307.02996 [astro-ph.HE]}
  \BibitemShut {NoStop}%
\bibitem [{\citenamefont {Yang}\ \emph {et~al.}(2022)\citenamefont {Yang},
  \citenamefont {Ai}, \citenamefont {Zhang}, \citenamefont {Zhang},
  \citenamefont {Liu}, \citenamefont {Wang}, \citenamefont {Yang},
  \citenamefont {Yin}, \citenamefont {Li},\ and\ \citenamefont
  {L\"u}}]{Yang:2022qmy}%
  \BibitemOpen
  \bibfield  {author} {\bibinfo {author} {\bibfnamefont {Jun}\ \bibnamefont
  {Yang}}, \bibinfo {author} {\bibfnamefont {Shunke}\ \bibnamefont {Ai}},
  \bibinfo {author} {\bibfnamefont {Bin-Bin}\ \bibnamefont {Zhang}}, \bibinfo
  {author} {\bibfnamefont {Bing}\ \bibnamefont {Zhang}}, \bibinfo {author}
  {\bibfnamefont {Zi-Ke}\ \bibnamefont {Liu}}, \bibinfo {author} {\bibfnamefont
  {Xiangyu~Ivy}\ \bibnamefont {Wang}}, \bibinfo {author} {\bibfnamefont
  {Yu-Han}\ \bibnamefont {Yang}}, \bibinfo {author} {\bibfnamefont {Yi-Han}\
  \bibnamefont {Yin}}, \bibinfo {author} {\bibfnamefont {Ye}~\bibnamefont
  {Li}}, \ and\ \bibinfo {author} {\bibfnamefont {Hou-Jun}\ \bibnamefont
  {L\"u}},\ }\bibfield  {title} {\enquote {\bibinfo {title} {{A long-duration
  gamma-ray burst with a peculiar origin}},}\ }\href {\doibase
  10.1038/s41586-022-05403-8} {\bibfield  {journal} {\bibinfo  {journal}
  {Nature}\ }\textbf {\bibinfo {volume} {612}},\ \bibinfo {pages} {232--235}
  (\bibinfo {year} {2022})},\ \Eprint {http://arxiv.org/abs/2204.12771}
  {arXiv:2204.12771 [astro-ph.HE]} \BibitemShut {NoStop}%
\bibitem [{\citenamefont {{Komiya}}\ \emph {et~al.}(2014)\citenamefont
  {{Komiya}}, \citenamefont {{Yamada}}, \citenamefont {{Suda}},\ and\
  \citenamefont {{Fujimoto}}}]{Komiya2014}%
  \BibitemOpen
  \bibfield  {author} {\bibinfo {author} {\bibfnamefont {Yutaka}\ \bibnamefont
  {{Komiya}}}, \bibinfo {author} {\bibfnamefont {Shimako}\ \bibnamefont
  {{Yamada}}}, \bibinfo {author} {\bibfnamefont {Takuma}\ \bibnamefont
  {{Suda}}}, \ and\ \bibinfo {author} {\bibfnamefont {Masayuki~Y.}\
  \bibnamefont {{Fujimoto}}},\ }\bibfield  {title} {\enquote {\bibinfo {title}
  {{The New Model of Chemical Evolution of r-process Elements Based on the
  Hierarchical Galaxy Formation. I. Ba and Eu}},}\ }\href {\doibase
  10.1088/0004-637X/783/2/132} {\bibfield  {journal} {\bibinfo  {journal}
  {\apj}\ }\textbf {\bibinfo {volume} {783}},\ \bibinfo {eid} {132} (\bibinfo
  {year} {2014})},\ \Eprint {http://arxiv.org/abs/1401.6261} {arXiv:1401.6261
  [astro-ph.GA]} \BibitemShut {NoStop}%
\bibitem [{\citenamefont {{Matteucci}}\ \emph {et~al.}(2014)\citenamefont
  {{Matteucci}}, \citenamefont {{Romano}}, \citenamefont {{Arcones}},
  \citenamefont {{Korobkin}},\ and\ \citenamefont {{Rosswog}}}]{Matteucci2014}%
  \BibitemOpen
  \bibfield  {author} {\bibinfo {author} {\bibfnamefont {F.}~\bibnamefont
  {{Matteucci}}}, \bibinfo {author} {\bibfnamefont {D.}~\bibnamefont
  {{Romano}}}, \bibinfo {author} {\bibfnamefont {A.}~\bibnamefont {{Arcones}}},
  \bibinfo {author} {\bibfnamefont {O.}~\bibnamefont {{Korobkin}}}, \ and\
  \bibinfo {author} {\bibfnamefont {S.}~\bibnamefont {{Rosswog}}},\ }\bibfield
  {title} {\enquote {\bibinfo {title} {{Europium production: neutron star
  mergers versus core-collapse supernovae}},}\ }\href {\doibase
  10.1093/mnras/stt2350} {\bibfield  {journal} {\bibinfo  {journal} {\mnras}\
  }\textbf {\bibinfo {volume} {438}},\ \bibinfo {pages} {2177--2185} (\bibinfo
  {year} {2014})},\ \Eprint {http://arxiv.org/abs/1311.6980} {arXiv:1311.6980
  [astro-ph.GA]} \BibitemShut {NoStop}%
\bibitem [{\citenamefont {Safarzadeh}\ \emph
  {et~al.}(2019{\natexlab{a}})\citenamefont {Safarzadeh}, \citenamefont
  {Sarmento},\ and\ \citenamefont {Scannapieco}}]{Safarzadeh:2018ent}%
  \BibitemOpen
  \bibfield  {author} {\bibinfo {author} {\bibfnamefont {Mohammadtaher}\
  \bibnamefont {Safarzadeh}}, \bibinfo {author} {\bibfnamefont {Richard}\
  \bibnamefont {Sarmento}}, \ and\ \bibinfo {author} {\bibfnamefont {Evan}\
  \bibnamefont {Scannapieco}},\ }\bibfield  {title} {\enquote {\bibinfo {title}
  {{On Neutron Star Mergers as the Source of r-process Enhanced Metal Poor
  Stars in the Milky Way}},}\ }\href {\doibase 10.3847/1538-4357/ab1341}
  {\bibfield  {journal} {\bibinfo  {journal} {Astrophys. J.}\ }\textbf
  {\bibinfo {volume} {876}},\ \bibinfo {pages} {28} (\bibinfo {year}
  {2019}{\natexlab{a}})},\ \Eprint {http://arxiv.org/abs/1812.02779}
  {arXiv:1812.02779 [astro-ph.GA]} \BibitemShut {NoStop}%
\bibitem [{\citenamefont {Safarzadeh}\ \emph
  {et~al.}(2019{\natexlab{b}})\citenamefont {Safarzadeh}, \citenamefont
  {Ramirez-Ruiz}, \citenamefont {Andrews}, \citenamefont {Fragos},
  \citenamefont {Macias},\ and\ \citenamefont
  {Scannapieco}}]{Safarzadeh:2018fdy}%
  \BibitemOpen
  \bibfield  {author} {\bibinfo {author} {\bibfnamefont {Mohammadtaher}\
  \bibnamefont {Safarzadeh}}, \bibinfo {author} {\bibfnamefont {Enrico}\
  \bibnamefont {Ramirez-Ruiz}}, \bibinfo {author} {\bibfnamefont {Jeff~J.}\
  \bibnamefont {Andrews}}, \bibinfo {author} {\bibfnamefont {Tassos}\
  \bibnamefont {Fragos}}, \bibinfo {author} {\bibfnamefont {Phillip}\
  \bibnamefont {Macias}}, \ and\ \bibinfo {author} {\bibfnamefont {Evan}\
  \bibnamefont {Scannapieco}},\ }\bibfield  {title} {\enquote {\bibinfo {title}
  {{r-process Enrichment of the Ultra-faint Dwarf Galaxies by Fast-merging
  Double-neutron Stars}},}\ }\href {\doibase 10.3847/1538-4357/aafe0e}
  {\bibfield  {journal} {\bibinfo  {journal} {Astrophys. J.}\ }\textbf
  {\bibinfo {volume} {872}},\ \bibinfo {pages} {105} (\bibinfo {year}
  {2019}{\natexlab{b}})},\ \Eprint {http://arxiv.org/abs/1810.04176}
  {arXiv:1810.04176 [astro-ph.HE]} \BibitemShut {NoStop}%
\bibitem [{\citenamefont {Zevin}\ \emph {et~al.}(2019)\citenamefont {Zevin},
  \citenamefont {Kremer}, \citenamefont {Siegel}, \citenamefont {Coughlin},
  \citenamefont {Tsang}, \citenamefont {Berry},\ and\ \citenamefont
  {Kalogera}}]{Zevin:2019obe}%
  \BibitemOpen
  \bibfield  {author} {\bibinfo {author} {\bibfnamefont {Michael}\ \bibnamefont
  {Zevin}}, \bibinfo {author} {\bibfnamefont {Kyle}\ \bibnamefont {Kremer}},
  \bibinfo {author} {\bibfnamefont {Daniel~M.}\ \bibnamefont {Siegel}},
  \bibinfo {author} {\bibfnamefont {Scott}\ \bibnamefont {Coughlin}}, \bibinfo
  {author} {\bibfnamefont {Benny T.~H.}\ \bibnamefont {Tsang}}, \bibinfo
  {author} {\bibfnamefont {Christopher P.~L.}\ \bibnamefont {Berry}}, \ and\
  \bibinfo {author} {\bibfnamefont {Vicky}\ \bibnamefont {Kalogera}},\
  }\bibfield  {title} {\enquote {\bibinfo {title} {{Can Neutron-Star Mergers
  Explain the r-process Enrichment in Globular Clusters?}}}\ }\href {\doibase
  10.3847/1538-4357/ab498b} {\  (\bibinfo {year} {2019}),\
  10.3847/1538-4357/ab498b},\ \Eprint {http://arxiv.org/abs/1906.11299}
  {arXiv:1906.11299 [astro-ph.HE]} \BibitemShut {NoStop}%
\bibitem [{\citenamefont {{Cordes}}\ and\ \citenamefont
  {{Chatterjee}}(2019)}]{Cordes2019}%
  \BibitemOpen
  \bibfield  {author} {\bibinfo {author} {\bibfnamefont {James~M.}\
  \bibnamefont {{Cordes}}}\ and\ \bibinfo {author} {\bibfnamefont {Shami}\
  \bibnamefont {{Chatterjee}}},\ }\bibfield  {title} {\enquote {\bibinfo
  {title} {{Fast Radio Bursts: An Extragalactic Enigma}},}\ }\href {\doibase
  10.1146/annurev-astro-091918-104501} {\bibfield  {journal} {\bibinfo
  {journal} {\araa}\ }\textbf {\bibinfo {volume} {57}},\ \bibinfo {pages}
  {417--465} (\bibinfo {year} {2019})},\ \Eprint
  {http://arxiv.org/abs/1906.05878} {arXiv:1906.05878 [astro-ph.HE]}
  \BibitemShut {NoStop}%
\bibitem [{\citenamefont {{Platts}}\ \emph {et~al.}(2019)\citenamefont
  {{Platts}}, \citenamefont {{Weltman}}, \citenamefont {{Walters}},
  \citenamefont {{Tendulkar}}, \citenamefont {{Gordin}},\ and\ \citenamefont
  {{Kandhai}}}]{Platts2019}%
  \BibitemOpen
  \bibfield  {author} {\bibinfo {author} {\bibfnamefont {E.}~\bibnamefont
  {{Platts}}}, \bibinfo {author} {\bibfnamefont {A.}~\bibnamefont {{Weltman}}},
  \bibinfo {author} {\bibfnamefont {A.}~\bibnamefont {{Walters}}}, \bibinfo
  {author} {\bibfnamefont {S.~P.}\ \bibnamefont {{Tendulkar}}}, \bibinfo
  {author} {\bibfnamefont {J.~E.~B.}\ \bibnamefont {{Gordin}}}, \ and\ \bibinfo
  {author} {\bibfnamefont {S.}~\bibnamefont {{Kandhai}}},\ }\bibfield  {title}
  {\enquote {\bibinfo {title} {{A living theory catalogue for fast radio
  bursts}},}\ }\href {\doibase 10.1016/j.physrep.2019.06.003} {\bibfield
  {journal} {\bibinfo  {journal} {Phys.Rep}\ }\textbf {\bibinfo {volume}
  {821}},\ \bibinfo {pages} {1--27} (\bibinfo {year} {2019})},\ \Eprint
  {http://arxiv.org/abs/1810.05836} {arXiv:1810.05836 [astro-ph.HE]}
  \BibitemShut {NoStop}%
\bibitem [{\citenamefont {Zhang}(2020)}]{Zhang:2020qgp}%
  \BibitemOpen
  \bibfield  {author} {\bibinfo {author} {\bibfnamefont {Bing}\ \bibnamefont
  {Zhang}},\ }\bibfield  {title} {\enquote {\bibinfo {title} {{The Physical
  Mechanisms of Fast Radio Bursts}},}\ }\href {\doibase
  10.1038/s41586-020-2828-1} {\bibfield  {journal} {\bibinfo  {journal}
  {Nature}\ }\textbf {\bibinfo {volume} {587}},\ \bibinfo {pages} {45--53}
  (\bibinfo {year} {2020})},\ \Eprint {http://arxiv.org/abs/2011.03500}
  {arXiv:2011.03500 [astro-ph.HE]} \BibitemShut {NoStop}%
\bibitem [{\citenamefont {{CHIME/FRB Collaboration}}\ \emph
  {et~al.}(2020)\citenamefont {{CHIME/FRB Collaboration}}, \citenamefont
  {{Andersen}}, \citenamefont {{Bandura}}, \citenamefont {{Bhardwaj}},
  \citenamefont {{Bij}}, \citenamefont {{Boyce}}, \citenamefont {{Boyle}},
  \citenamefont {{Brar}}, \citenamefont {{Cassanelli}}, \citenamefont
  {{Chawla}}, \citenamefont {{Chen}}, \citenamefont {{Cliche}}, \citenamefont
  {{Cook}}, \citenamefont {{Cubranic}}, \citenamefont {{Curtin}}, \citenamefont
  {{Denman}}, \citenamefont {{Dobbs}}, \citenamefont {{Dong}}, \citenamefont
  {{Fandino}}, \citenamefont {{Fonseca}}, \citenamefont {{Gaensler}},
  \citenamefont {{Giri}}, \citenamefont {{Good}}, \citenamefont {{Halpern}},
  \citenamefont {{Hill}}, \citenamefont {{Hinshaw}}, \citenamefont
  {{H{\"o}fer}}, \citenamefont {{Josephy}}, \citenamefont {{Kania}},
  \citenamefont {{Kaspi}}, \citenamefont {{Landecker}}, \citenamefont
  {{Leung}}, \citenamefont {{Li}}, \citenamefont {{Lin}}, \citenamefont
  {{Masui}}, \citenamefont {{McKinven}}, \citenamefont {{Mena-Parra}},
  \citenamefont {{Merryfield}}, \citenamefont {{Meyers}}, \citenamefont
  {{Michilli}}, \citenamefont {{Milutinovic}}, \citenamefont {{Mirhosseini}},
  \citenamefont {{M{\"u}nchmeyer}}, \citenamefont {{Naidu}}, \citenamefont
  {{Newburgh}}, \citenamefont {{Ng}}, \citenamefont {{Patel}}, \citenamefont
  {{Pen}}, \citenamefont {{Pinsonneault-Marotte}}, \citenamefont {{Pleunis}},
  \citenamefont {{Quine}}, \citenamefont {{Rafiei-Ravandi}}, \citenamefont
  {{Rahman}}, \citenamefont {{Ransom}}, \citenamefont {{Renard}}, \citenamefont
  {{Sanghavi}}, \citenamefont {{Scholz}}, \citenamefont {{Shaw}}, \citenamefont
  {{Shin}}, \citenamefont {{Siegel}}, \citenamefont {{Singh}}, \citenamefont
  {{Smegal}}, \citenamefont {{Smith}}, \citenamefont {{Stairs}}, \citenamefont
  {{Tan}}, \citenamefont {{Tendulkar}}, \citenamefont {{Tretyakov}},
  \citenamefont {{Vanderlinde}}, \citenamefont {{Wang}}, \citenamefont
  {{Wulf}},\ and\ \citenamefont {{Zwaniga}}}]{CHIME2020Nat}%
  \BibitemOpen
  \bibfield  {author} {\bibinfo {author} {\bibnamefont {{CHIME/FRB
  Collaboration}}}, \bibinfo {author} {\bibfnamefont {B.~C.}\ \bibnamefont
  {{Andersen}}}, \bibinfo {author} {\bibfnamefont {K.~M.}\ \bibnamefont
  {{Bandura}}}, \bibinfo {author} {\bibfnamefont {M.}~\bibnamefont
  {{Bhardwaj}}}, \bibinfo {author} {\bibfnamefont {A.}~\bibnamefont {{Bij}}},
  \bibinfo {author} {\bibfnamefont {M.~M.}\ \bibnamefont {{Boyce}}}, \bibinfo
  {author} {\bibfnamefont {P.~J.}\ \bibnamefont {{Boyle}}}, \bibinfo {author}
  {\bibfnamefont {C.}~\bibnamefont {{Brar}}}, \bibinfo {author} {\bibfnamefont
  {T.}~\bibnamefont {{Cassanelli}}}, \bibinfo {author} {\bibfnamefont
  {P.}~\bibnamefont {{Chawla}}}, \bibinfo {author} {\bibfnamefont
  {T.}~\bibnamefont {{Chen}}}, \bibinfo {author} {\bibfnamefont {J.~F.}\
  \bibnamefont {{Cliche}}}, \bibinfo {author} {\bibfnamefont {A.}~\bibnamefont
  {{Cook}}}, \bibinfo {author} {\bibfnamefont {D.}~\bibnamefont {{Cubranic}}},
  \bibinfo {author} {\bibfnamefont {A.~P.}\ \bibnamefont {{Curtin}}}, \bibinfo
  {author} {\bibfnamefont {N.~T.}\ \bibnamefont {{Denman}}}, \bibinfo {author}
  {\bibfnamefont {M.}~\bibnamefont {{Dobbs}}}, \bibinfo {author} {\bibfnamefont
  {F.~Q.}\ \bibnamefont {{Dong}}}, \bibinfo {author} {\bibfnamefont
  {M.}~\bibnamefont {{Fandino}}}, \bibinfo {author} {\bibfnamefont
  {E.}~\bibnamefont {{Fonseca}}}, \bibinfo {author} {\bibfnamefont {B.~M.}\
  \bibnamefont {{Gaensler}}}, \bibinfo {author} {\bibfnamefont
  {U.}~\bibnamefont {{Giri}}}, \bibinfo {author} {\bibfnamefont {D.~C.}\
  \bibnamefont {{Good}}}, \bibinfo {author} {\bibfnamefont {M.}~\bibnamefont
  {{Halpern}}}, \bibinfo {author} {\bibfnamefont {A.~S.}\ \bibnamefont
  {{Hill}}}, \bibinfo {author} {\bibfnamefont {G.~F.}\ \bibnamefont
  {{Hinshaw}}}, \bibinfo {author} {\bibfnamefont {C.}~\bibnamefont
  {{H{\"o}fer}}}, \bibinfo {author} {\bibfnamefont {A.}~\bibnamefont
  {{Josephy}}}, \bibinfo {author} {\bibfnamefont {J.~W.}\ \bibnamefont
  {{Kania}}}, \bibinfo {author} {\bibfnamefont {V.~M.}\ \bibnamefont
  {{Kaspi}}}, \bibinfo {author} {\bibfnamefont {T.~L.}\ \bibnamefont
  {{Landecker}}}, \bibinfo {author} {\bibfnamefont {C.}~\bibnamefont
  {{Leung}}}, \bibinfo {author} {\bibfnamefont {D.~Z.}\ \bibnamefont {{Li}}},
  \bibinfo {author} {\bibfnamefont {H.~H.}\ \bibnamefont {{Lin}}}, \bibinfo
  {author} {\bibfnamefont {K.~W.}\ \bibnamefont {{Masui}}}, \bibinfo {author}
  {\bibfnamefont {R.}~\bibnamefont {{McKinven}}}, \bibinfo {author}
  {\bibfnamefont {J.}~\bibnamefont {{Mena-Parra}}}, \bibinfo {author}
  {\bibfnamefont {M.}~\bibnamefont {{Merryfield}}}, \bibinfo {author}
  {\bibfnamefont {B.~W.}\ \bibnamefont {{Meyers}}}, \bibinfo {author}
  {\bibfnamefont {D.}~\bibnamefont {{Michilli}}}, \bibinfo {author}
  {\bibfnamefont {N.}~\bibnamefont {{Milutinovic}}}, \bibinfo {author}
  {\bibfnamefont {A.}~\bibnamefont {{Mirhosseini}}}, \bibinfo {author}
  {\bibfnamefont {M.}~\bibnamefont {{M{\"u}nchmeyer}}}, \bibinfo {author}
  {\bibfnamefont {A.}~\bibnamefont {{Naidu}}}, \bibinfo {author} {\bibfnamefont
  {L.~B.}\ \bibnamefont {{Newburgh}}}, \bibinfo {author} {\bibfnamefont
  {C.}~\bibnamefont {{Ng}}}, \bibinfo {author} {\bibfnamefont {C.}~\bibnamefont
  {{Patel}}}, \bibinfo {author} {\bibfnamefont {U.~L.}\ \bibnamefont {{Pen}}},
  \bibinfo {author} {\bibfnamefont {T.}~\bibnamefont {{Pinsonneault-Marotte}}},
  \bibinfo {author} {\bibfnamefont {Z.}~\bibnamefont {{Pleunis}}}, \bibinfo
  {author} {\bibfnamefont {B.~M.}\ \bibnamefont {{Quine}}}, \bibinfo {author}
  {\bibfnamefont {M.}~\bibnamefont {{Rafiei-Ravandi}}}, \bibinfo {author}
  {\bibfnamefont {M.}~\bibnamefont {{Rahman}}}, \bibinfo {author}
  {\bibfnamefont {S.~M.}\ \bibnamefont {{Ransom}}}, \bibinfo {author}
  {\bibfnamefont {A.}~\bibnamefont {{Renard}}}, \bibinfo {author}
  {\bibfnamefont {P.}~\bibnamefont {{Sanghavi}}}, \bibinfo {author}
  {\bibfnamefont {P.}~\bibnamefont {{Scholz}}}, \bibinfo {author}
  {\bibfnamefont {J.~R.}\ \bibnamefont {{Shaw}}}, \bibinfo {author}
  {\bibfnamefont {K.}~\bibnamefont {{Shin}}}, \bibinfo {author} {\bibfnamefont
  {S.~R.}\ \bibnamefont {{Siegel}}}, \bibinfo {author} {\bibfnamefont
  {S.}~\bibnamefont {{Singh}}}, \bibinfo {author} {\bibfnamefont {R.~J.}\
  \bibnamefont {{Smegal}}}, \bibinfo {author} {\bibfnamefont {K.~M.}\
  \bibnamefont {{Smith}}}, \bibinfo {author} {\bibfnamefont {I.~H.}\
  \bibnamefont {{Stairs}}}, \bibinfo {author} {\bibfnamefont {C.~M.}\
  \bibnamefont {{Tan}}}, \bibinfo {author} {\bibfnamefont {S.~P.}\ \bibnamefont
  {{Tendulkar}}}, \bibinfo {author} {\bibfnamefont {I.}~\bibnamefont
  {{Tretyakov}}}, \bibinfo {author} {\bibfnamefont {K.}~\bibnamefont
  {{Vanderlinde}}}, \bibinfo {author} {\bibfnamefont {H.}~\bibnamefont
  {{Wang}}}, \bibinfo {author} {\bibfnamefont {D.}~\bibnamefont {{Wulf}}}, \
  and\ \bibinfo {author} {\bibfnamefont {A.~V.}\ \bibnamefont {{Zwaniga}}},\
  }\bibfield  {title} {\enquote {\bibinfo {title} {{A bright
  millisecond-duration radio burst from a Galactic magnetar}},}\ }\href
  {\doibase 10.1038/s41586-020-2863-y} {\bibfield  {journal} {\bibinfo
  {journal} {\nat}\ }\textbf {\bibinfo {volume} {587}},\ \bibinfo {pages}
  {54--58} (\bibinfo {year} {2020})},\ \Eprint
  {http://arxiv.org/abs/2005.10324} {arXiv:2005.10324 [astro-ph.HE]}
  \BibitemShut {NoStop}%
\bibitem [{\citenamefont {{Bochenek}}\ \emph {et~al.}(2020)\citenamefont
  {{Bochenek}}, \citenamefont {{Ravi}}, \citenamefont {{Belov}}, \citenamefont
  {{Hallinan}}, \citenamefont {{Kocz}}, \citenamefont {{Kulkarni}},\ and\
  \citenamefont {{McKenna}}}]{Bochenek2020}%
  \BibitemOpen
  \bibfield  {author} {\bibinfo {author} {\bibfnamefont {C.~D.}\ \bibnamefont
  {{Bochenek}}}, \bibinfo {author} {\bibfnamefont {V.}~\bibnamefont {{Ravi}}},
  \bibinfo {author} {\bibfnamefont {K.~V.}\ \bibnamefont {{Belov}}}, \bibinfo
  {author} {\bibfnamefont {G.}~\bibnamefont {{Hallinan}}}, \bibinfo {author}
  {\bibfnamefont {J.}~\bibnamefont {{Kocz}}}, \bibinfo {author} {\bibfnamefont
  {S.~R.}\ \bibnamefont {{Kulkarni}}}, \ and\ \bibinfo {author} {\bibfnamefont
  {D.~L.}\ \bibnamefont {{McKenna}}},\ }\bibfield  {title} {\enquote {\bibinfo
  {title} {{A fast radio burst associated with a Galactic magnetar}},}\ }\href
  {\doibase 10.1038/s41586-020-2872-x} {\bibfield  {journal} {\bibinfo
  {journal} {\nat}\ }\textbf {\bibinfo {volume} {587}},\ \bibinfo {pages}
  {59--62} (\bibinfo {year} {2020})},\ \Eprint
  {http://arxiv.org/abs/2005.10828} {arXiv:2005.10828 [astro-ph.HE]}
  \BibitemShut {NoStop}%
\bibitem [{\citenamefont {{Mereghetti}}\ \emph {et~al.}(2020)\citenamefont
  {{Mereghetti}}, \citenamefont {{Savchenko}}, \citenamefont {{Ferrigno}},
  \citenamefont {{G{\"o}tz}}, \citenamefont {{Rigoselli}}, \citenamefont
  {{Tiengo}}, \citenamefont {{Bazzano}}, \citenamefont {{Bozzo}}, \citenamefont
  {{Coleiro}}, \citenamefont {{Courvoisier}}, \citenamefont {{Doyle}},
  \citenamefont {{Goldwurm}}, \citenamefont {{Hanlon}}, \citenamefont
  {{Jourdain}}, \citenamefont {{von Kienlin}}, \citenamefont {{Lutovinov}},
  \citenamefont {{Martin-Carrillo}}, \citenamefont {{Molkov}}, \citenamefont
  {{Natalucci}}, \citenamefont {{Onori}}, \citenamefont {{Panessa}},
  \citenamefont {{Rodi}}, \citenamefont {{Rodriguez}}, \citenamefont
  {{S{\'a}nchez-Fern{\'a}ndez}}, \citenamefont {{Sunyaev}},\ and\ \citenamefont
  {{Ubertini}}}]{Mereghetti2020}%
  \BibitemOpen
  \bibfield  {author} {\bibinfo {author} {\bibfnamefont {S.}~\bibnamefont
  {{Mereghetti}}}, \bibinfo {author} {\bibfnamefont {V.}~\bibnamefont
  {{Savchenko}}}, \bibinfo {author} {\bibfnamefont {C.}~\bibnamefont
  {{Ferrigno}}}, \bibinfo {author} {\bibfnamefont {D.}~\bibnamefont
  {{G{\"o}tz}}}, \bibinfo {author} {\bibfnamefont {M.}~\bibnamefont
  {{Rigoselli}}}, \bibinfo {author} {\bibfnamefont {A.}~\bibnamefont
  {{Tiengo}}}, \bibinfo {author} {\bibfnamefont {A.}~\bibnamefont {{Bazzano}}},
  \bibinfo {author} {\bibfnamefont {E.}~\bibnamefont {{Bozzo}}}, \bibinfo
  {author} {\bibfnamefont {A.}~\bibnamefont {{Coleiro}}}, \bibinfo {author}
  {\bibfnamefont {T.~J.~L.}\ \bibnamefont {{Courvoisier}}}, \bibinfo {author}
  {\bibfnamefont {M.}~\bibnamefont {{Doyle}}}, \bibinfo {author} {\bibfnamefont
  {A.}~\bibnamefont {{Goldwurm}}}, \bibinfo {author} {\bibfnamefont
  {L.}~\bibnamefont {{Hanlon}}}, \bibinfo {author} {\bibfnamefont
  {E.}~\bibnamefont {{Jourdain}}}, \bibinfo {author} {\bibfnamefont
  {A.}~\bibnamefont {{von Kienlin}}}, \bibinfo {author} {\bibfnamefont
  {A.}~\bibnamefont {{Lutovinov}}}, \bibinfo {author} {\bibfnamefont
  {A.}~\bibnamefont {{Martin-Carrillo}}}, \bibinfo {author} {\bibfnamefont
  {S.}~\bibnamefont {{Molkov}}}, \bibinfo {author} {\bibfnamefont
  {L.}~\bibnamefont {{Natalucci}}}, \bibinfo {author} {\bibfnamefont
  {F.}~\bibnamefont {{Onori}}}, \bibinfo {author} {\bibfnamefont
  {F.}~\bibnamefont {{Panessa}}}, \bibinfo {author} {\bibfnamefont
  {J.}~\bibnamefont {{Rodi}}}, \bibinfo {author} {\bibfnamefont
  {J.}~\bibnamefont {{Rodriguez}}}, \bibinfo {author} {\bibfnamefont
  {C.}~\bibnamefont {{S{\'a}nchez-Fern{\'a}ndez}}}, \bibinfo {author}
  {\bibfnamefont {R.}~\bibnamefont {{Sunyaev}}}, \ and\ \bibinfo {author}
  {\bibfnamefont {P.}~\bibnamefont {{Ubertini}}},\ }\bibfield  {title}
  {\enquote {\bibinfo {title} {{INTEGRAL Discovery of a Burst with Associated
  Radio Emission from the Magnetar SGR 1935+2154}},}\ }\href {\doibase
  10.3847/2041-8213/aba2cf} {\bibfield  {journal} {\bibinfo  {journal} {\apjl}\
  }\textbf {\bibinfo {volume} {898}},\ \bibinfo {eid} {L29} (\bibinfo {year}
  {2020})},\ \Eprint {http://arxiv.org/abs/2005.06335} {arXiv:2005.06335
  [astro-ph.HE]} \BibitemShut {NoStop}%
\bibitem [{\citenamefont {{Li}}\ \emph {et~al.}(2021)\citenamefont {{Li}},
  \citenamefont {{Lin}}, \citenamefont {{Xiong}}, \citenamefont {{Ge}},
  \citenamefont {{Li}}, \citenamefont {{Li}}, \citenamefont {{Lu}},
  \citenamefont {{Zhang}}, \citenamefont {{Tuo}}, \citenamefont {{Nang}},
  \citenamefont {{Zhang}}, \citenamefont {{Xiao}}, \citenamefont {{Chen}},
  \citenamefont {{Song}}, \citenamefont {{Xu}}, \citenamefont {{Liu}},
  \citenamefont {{Jia}}, \citenamefont {{Cao}}, \citenamefont {{Qu}},
  \citenamefont {{Zhang}}, \citenamefont {{Gu}}, \citenamefont {{Liao}},
  \citenamefont {{Zhao}}, \citenamefont {{Tan}}, \citenamefont {{Nie}},
  \citenamefont {{Zhao}}, \citenamefont {{Zheng}}, \citenamefont {{Zheng}},
  \citenamefont {{Luo}}, \citenamefont {{Cai}}, \citenamefont {{Li}},
  \citenamefont {{Xue}}, \citenamefont {{Bu}}, \citenamefont {{Chang}},
  \citenamefont {{Chen}}, \citenamefont {{Chen}}, \citenamefont {{Chen}},
  \citenamefont {{Chen}}, \citenamefont {{Chen}}, \citenamefont {{Cui}},
  \citenamefont {{Cui}}, \citenamefont {{Deng}}, \citenamefont {{Dong}},
  \citenamefont {{Du}}, \citenamefont {{Fu}}, \citenamefont {{Gao}},
  \citenamefont {{Gao}}, \citenamefont {{Gao}}, \citenamefont {{Gu}},
  \citenamefont {{Guan}}, \citenamefont {{Guo}}, \citenamefont {{Han}},
  \citenamefont {{Huang}}, \citenamefont {{Huo}}, \citenamefont {{Jiang}},
  \citenamefont {{Jiang}}, \citenamefont {{Jin}}, \citenamefont {{Jin}},
  \citenamefont {{Kong}}, \citenamefont {{Li}}, \citenamefont {{Li}},
  \citenamefont {{Li}}, \citenamefont {{Li}}, \citenamefont {{Li}},
  \citenamefont {{Li}}, \citenamefont {{Li}}, \citenamefont {{Liang}},
  \citenamefont {{Liu}}, \citenamefont {{Liu}}, \citenamefont {{Liu}},
  \citenamefont {{Liu}}, \citenamefont {{Liu}}, \citenamefont {{Lu}},
  \citenamefont {{Lu}}, \citenamefont {{Luo}}, \citenamefont {{Ma}},
  \citenamefont {{Meng}}, \citenamefont {{Ou}}, \citenamefont {{Sai}},
  \citenamefont {{Shang}}, \citenamefont {{Song}}, \citenamefont {{Sun}},
  \citenamefont {{Tao}}, \citenamefont {{Wang}}, \citenamefont {{Wang}},
  \citenamefont {{Wang}}, \citenamefont {{Wang}}, \citenamefont {{Wang}},
  \citenamefont {{Wen}}, \citenamefont {{Wu}}, \citenamefont {{Wu}},
  \citenamefont {{Wu}}, \citenamefont {{Xiao}}, \citenamefont {{Xu}},
  \citenamefont {{Yang}}, \citenamefont {{Yang}}, \citenamefont {{Yang}},
  \citenamefont {{Yang}}, \citenamefont {{Yi}}, \citenamefont {{Yin}},
  \citenamefont {{You}}, \citenamefont {{Zhang}}, \citenamefont {{Zhang}},
  \citenamefont {{Zhang}}, \citenamefont {{Zhang}}, \citenamefont {{Zhang}},
  \citenamefont {{Zhang}}, \citenamefont {{Zhang}}, \citenamefont {{Zhang}},
  \citenamefont {{Zhang}}, \citenamefont {{Zhang}}, \citenamefont {{Zhang}},
  \citenamefont {{Zhang}}, \citenamefont {{Zhang}}, \citenamefont {{Zhang}},
  \citenamefont {{Zhang}}, \citenamefont {{Zhang}}, \citenamefont {{Zhou}},
  \citenamefont {{Zhou}}, \citenamefont {{Zhu}}, \citenamefont {{Zhu}},\ and\
  \citenamefont {{Zhuang}}}]{Li2021Nat}%
  \BibitemOpen
  \bibfield  {author} {\bibinfo {author} {\bibfnamefont {C.~K.}\ \bibnamefont
  {{Li}}}, \bibinfo {author} {\bibfnamefont {L.}~\bibnamefont {{Lin}}},
  \bibinfo {author} {\bibfnamefont {S.~L.}\ \bibnamefont {{Xiong}}}, \bibinfo
  {author} {\bibfnamefont {M.~Y.}\ \bibnamefont {{Ge}}}, \bibinfo {author}
  {\bibfnamefont {X.~B.}\ \bibnamefont {{Li}}}, \bibinfo {author}
  {\bibfnamefont {T.~P.}\ \bibnamefont {{Li}}}, \bibinfo {author}
  {\bibfnamefont {F.~J.}\ \bibnamefont {{Lu}}}, \bibinfo {author}
  {\bibfnamefont {S.~N.}\ \bibnamefont {{Zhang}}}, \bibinfo {author}
  {\bibfnamefont {Y.~L.}\ \bibnamefont {{Tuo}}}, \bibinfo {author}
  {\bibfnamefont {Y.}~\bibnamefont {{Nang}}}, \bibinfo {author} {\bibfnamefont
  {B.}~\bibnamefont {{Zhang}}}, \bibinfo {author} {\bibfnamefont
  {S.}~\bibnamefont {{Xiao}}}, \bibinfo {author} {\bibfnamefont
  {Y.}~\bibnamefont {{Chen}}}, \bibinfo {author} {\bibfnamefont {L.~M.}\
  \bibnamefont {{Song}}}, \bibinfo {author} {\bibfnamefont {Y.~P.}\
  \bibnamefont {{Xu}}}, \bibinfo {author} {\bibfnamefont {C.~Z.}\ \bibnamefont
  {{Liu}}}, \bibinfo {author} {\bibfnamefont {S.~M.}\ \bibnamefont {{Jia}}},
  \bibinfo {author} {\bibfnamefont {X.~L.}\ \bibnamefont {{Cao}}}, \bibinfo
  {author} {\bibfnamefont {J.~L.}\ \bibnamefont {{Qu}}}, \bibinfo {author}
  {\bibfnamefont {S.}~\bibnamefont {{Zhang}}}, \bibinfo {author} {\bibfnamefont
  {Y.~D.}\ \bibnamefont {{Gu}}}, \bibinfo {author} {\bibfnamefont {J.~Y.}\
  \bibnamefont {{Liao}}}, \bibinfo {author} {\bibfnamefont {X.~F.}\
  \bibnamefont {{Zhao}}}, \bibinfo {author} {\bibfnamefont {Y.}~\bibnamefont
  {{Tan}}}, \bibinfo {author} {\bibfnamefont {J.~Y.}\ \bibnamefont {{Nie}}},
  \bibinfo {author} {\bibfnamefont {H.~S.}\ \bibnamefont {{Zhao}}}, \bibinfo
  {author} {\bibfnamefont {S.~J.}\ \bibnamefont {{Zheng}}}, \bibinfo {author}
  {\bibfnamefont {Y.~G.}\ \bibnamefont {{Zheng}}}, \bibinfo {author}
  {\bibfnamefont {Q.}~\bibnamefont {{Luo}}}, \bibinfo {author} {\bibfnamefont
  {C.}~\bibnamefont {{Cai}}}, \bibinfo {author} {\bibfnamefont
  {B.}~\bibnamefont {{Li}}}, \bibinfo {author} {\bibfnamefont {W.~C.}\
  \bibnamefont {{Xue}}}, \bibinfo {author} {\bibfnamefont {Q.~C.}\ \bibnamefont
  {{Bu}}}, \bibinfo {author} {\bibfnamefont {Z.}~\bibnamefont {{Chang}}},
  \bibinfo {author} {\bibfnamefont {G.}~\bibnamefont {{Chen}}}, \bibinfo
  {author} {\bibfnamefont {L.}~\bibnamefont {{Chen}}}, \bibinfo {author}
  {\bibfnamefont {T.~X.}\ \bibnamefont {{Chen}}}, \bibinfo {author}
  {\bibfnamefont {Y.~B.}\ \bibnamefont {{Chen}}}, \bibinfo {author}
  {\bibfnamefont {Y.~P.}\ \bibnamefont {{Chen}}}, \bibinfo {author}
  {\bibfnamefont {W.}~\bibnamefont {{Cui}}}, \bibinfo {author} {\bibfnamefont
  {W.~W.}\ \bibnamefont {{Cui}}}, \bibinfo {author} {\bibfnamefont {J.~K.}\
  \bibnamefont {{Deng}}}, \bibinfo {author} {\bibfnamefont {Y.~W.}\
  \bibnamefont {{Dong}}}, \bibinfo {author} {\bibfnamefont {Y.~Y.}\
  \bibnamefont {{Du}}}, \bibinfo {author} {\bibfnamefont {M.~X.}\ \bibnamefont
  {{Fu}}}, \bibinfo {author} {\bibfnamefont {G.~H.}\ \bibnamefont {{Gao}}},
  \bibinfo {author} {\bibfnamefont {H.}~\bibnamefont {{Gao}}}, \bibinfo
  {author} {\bibfnamefont {M.}~\bibnamefont {{Gao}}}, \bibinfo {author}
  {\bibfnamefont {Y.~D.}\ \bibnamefont {{Gu}}}, \bibinfo {author}
  {\bibfnamefont {J.}~\bibnamefont {{Guan}}}, \bibinfo {author} {\bibfnamefont
  {C.~C.}\ \bibnamefont {{Guo}}}, \bibinfo {author} {\bibfnamefont {D.~W.}\
  \bibnamefont {{Han}}}, \bibinfo {author} {\bibfnamefont {Y.}~\bibnamefont
  {{Huang}}}, \bibinfo {author} {\bibfnamefont {J.}~\bibnamefont {{Huo}}},
  \bibinfo {author} {\bibfnamefont {L.~H.}\ \bibnamefont {{Jiang}}}, \bibinfo
  {author} {\bibfnamefont {W.~C.}\ \bibnamefont {{Jiang}}}, \bibinfo {author}
  {\bibfnamefont {J.}~\bibnamefont {{Jin}}}, \bibinfo {author} {\bibfnamefont
  {Y.~J.}\ \bibnamefont {{Jin}}}, \bibinfo {author} {\bibfnamefont {L.~D.}\
  \bibnamefont {{Kong}}}, \bibinfo {author} {\bibfnamefont {G.}~\bibnamefont
  {{Li}}}, \bibinfo {author} {\bibfnamefont {M.~S.}\ \bibnamefont {{Li}}},
  \bibinfo {author} {\bibfnamefont {W.}~\bibnamefont {{Li}}}, \bibinfo {author}
  {\bibfnamefont {X.}~\bibnamefont {{Li}}}, \bibinfo {author} {\bibfnamefont
  {X.~F.}\ \bibnamefont {{Li}}}, \bibinfo {author} {\bibfnamefont {Y.~G.}\
  \bibnamefont {{Li}}}, \bibinfo {author} {\bibfnamefont {Z.~W.}\ \bibnamefont
  {{Li}}}, \bibinfo {author} {\bibfnamefont {X.~H.}\ \bibnamefont {{Liang}}},
  \bibinfo {author} {\bibfnamefont {B.~S.}\ \bibnamefont {{Liu}}}, \bibinfo
  {author} {\bibfnamefont {G.~Q.}\ \bibnamefont {{Liu}}}, \bibinfo {author}
  {\bibfnamefont {H.~W.}\ \bibnamefont {{Liu}}}, \bibinfo {author}
  {\bibfnamefont {X.~J.}\ \bibnamefont {{Liu}}}, \bibinfo {author}
  {\bibfnamefont {Y.~N.}\ \bibnamefont {{Liu}}}, \bibinfo {author}
  {\bibfnamefont {B.}~\bibnamefont {{Lu}}}, \bibinfo {author} {\bibfnamefont
  {X.~F.}\ \bibnamefont {{Lu}}}, \bibinfo {author} {\bibfnamefont
  {T.}~\bibnamefont {{Luo}}}, \bibinfo {author} {\bibfnamefont
  {X.}~\bibnamefont {{Ma}}}, \bibinfo {author} {\bibfnamefont {B.}~\bibnamefont
  {{Meng}}}, \bibinfo {author} {\bibfnamefont {G.}~\bibnamefont {{Ou}}},
  \bibinfo {author} {\bibfnamefont {N.}~\bibnamefont {{Sai}}}, \bibinfo
  {author} {\bibfnamefont {R.~C.}\ \bibnamefont {{Shang}}}, \bibinfo {author}
  {\bibfnamefont {X.~Y.}\ \bibnamefont {{Song}}}, \bibinfo {author}
  {\bibfnamefont {L.}~\bibnamefont {{Sun}}}, \bibinfo {author} {\bibfnamefont
  {L.}~\bibnamefont {{Tao}}}, \bibinfo {author} {\bibfnamefont
  {C.}~\bibnamefont {{Wang}}}, \bibinfo {author} {\bibfnamefont {G.~F.}\
  \bibnamefont {{Wang}}}, \bibinfo {author} {\bibfnamefont {J.}~\bibnamefont
  {{Wang}}}, \bibinfo {author} {\bibfnamefont {W.~S.}\ \bibnamefont {{Wang}}},
  \bibinfo {author} {\bibfnamefont {Y.~S.}\ \bibnamefont {{Wang}}}, \bibinfo
  {author} {\bibfnamefont {X.~Y.}\ \bibnamefont {{Wen}}}, \bibinfo {author}
  {\bibfnamefont {B.~B.}\ \bibnamefont {{Wu}}}, \bibinfo {author}
  {\bibfnamefont {B.~Y.}\ \bibnamefont {{Wu}}}, \bibinfo {author}
  {\bibfnamefont {M.}~\bibnamefont {{Wu}}}, \bibinfo {author} {\bibfnamefont
  {G.~C.}\ \bibnamefont {{Xiao}}}, \bibinfo {author} {\bibfnamefont
  {H.}~\bibnamefont {{Xu}}}, \bibinfo {author} {\bibfnamefont {J.~W.}\
  \bibnamefont {{Yang}}}, \bibinfo {author} {\bibfnamefont {S.}~\bibnamefont
  {{Yang}}}, \bibinfo {author} {\bibfnamefont {Y.~J.}\ \bibnamefont {{Yang}}},
  \bibinfo {author} {\bibfnamefont {Yi-Jung}\ \bibnamefont {{Yang}}}, \bibinfo
  {author} {\bibfnamefont {Q.~B.}\ \bibnamefont {{Yi}}}, \bibinfo {author}
  {\bibfnamefont {Q.~Q.}\ \bibnamefont {{Yin}}}, \bibinfo {author}
  {\bibfnamefont {Y.}~\bibnamefont {{You}}}, \bibinfo {author} {\bibfnamefont
  {A.~M.}\ \bibnamefont {{Zhang}}}, \bibinfo {author} {\bibfnamefont {C.~M.}\
  \bibnamefont {{Zhang}}}, \bibinfo {author} {\bibfnamefont {F.}~\bibnamefont
  {{Zhang}}}, \bibinfo {author} {\bibfnamefont {H.~M.}\ \bibnamefont
  {{Zhang}}}, \bibinfo {author} {\bibfnamefont {J.}~\bibnamefont {{Zhang}}},
  \bibinfo {author} {\bibfnamefont {T.}~\bibnamefont {{Zhang}}}, \bibinfo
  {author} {\bibfnamefont {W.}~\bibnamefont {{Zhang}}}, \bibinfo {author}
  {\bibfnamefont {W.~C.}\ \bibnamefont {{Zhang}}}, \bibinfo {author}
  {\bibfnamefont {W.~Z.}\ \bibnamefont {{Zhang}}}, \bibinfo {author}
  {\bibfnamefont {Y.}~\bibnamefont {{Zhang}}}, \bibinfo {author} {\bibfnamefont
  {Yue}\ \bibnamefont {{Zhang}}}, \bibinfo {author} {\bibfnamefont {Y.~F.}\
  \bibnamefont {{Zhang}}}, \bibinfo {author} {\bibfnamefont {Y.~J.}\
  \bibnamefont {{Zhang}}}, \bibinfo {author} {\bibfnamefont {Z.}~\bibnamefont
  {{Zhang}}}, \bibinfo {author} {\bibfnamefont {Zhi}\ \bibnamefont {{Zhang}}},
  \bibinfo {author} {\bibfnamefont {Z.~L.}\ \bibnamefont {{Zhang}}}, \bibinfo
  {author} {\bibfnamefont {D.~K.}\ \bibnamefont {{Zhou}}}, \bibinfo {author}
  {\bibfnamefont {J.~F.}\ \bibnamefont {{Zhou}}}, \bibinfo {author}
  {\bibfnamefont {Y.}~\bibnamefont {{Zhu}}}, \bibinfo {author} {\bibfnamefont
  {Y.~X.}\ \bibnamefont {{Zhu}}}, \ and\ \bibinfo {author} {\bibfnamefont
  {R.~L.}\ \bibnamefont {{Zhuang}}},\ }\bibfield  {title} {\enquote {\bibinfo
  {title} {{HXMT identification of a non-thermal X-ray burst from SGR
  J1935+2154 and with FRB 200428}},}\ }\href {\doibase
  10.1038/s41550-021-01302-6} {\bibfield  {journal} {\bibinfo  {journal}
  {Nature Astronomy}\ }\textbf {\bibinfo {volume} {5}},\ \bibinfo {pages}
  {378--384} (\bibinfo {year} {2021})},\ \Eprint
  {http://arxiv.org/abs/2005.11071} {arXiv:2005.11071 [astro-ph.HE]}
  \BibitemShut {NoStop}%
\bibitem [{\citenamefont {{Ridnaia}}\ \emph {et~al.}(2021)\citenamefont
  {{Ridnaia}}, \citenamefont {{Svinkin}}, \citenamefont {{Frederiks}},
  \citenamefont {{Bykov}}, \citenamefont {{Popov}}, \citenamefont {{Aptekar}},
  \citenamefont {{Golenetskii}}, \citenamefont {{Lysenko}}, \citenamefont
  {{Tsvetkova}}, \citenamefont {{Ulanov}},\ and\ \citenamefont
  {{Cline}}}]{Ridnaia2021}%
  \BibitemOpen
  \bibfield  {author} {\bibinfo {author} {\bibfnamefont {A.}~\bibnamefont
  {{Ridnaia}}}, \bibinfo {author} {\bibfnamefont {D.}~\bibnamefont
  {{Svinkin}}}, \bibinfo {author} {\bibfnamefont {D.}~\bibnamefont
  {{Frederiks}}}, \bibinfo {author} {\bibfnamefont {A.}~\bibnamefont
  {{Bykov}}}, \bibinfo {author} {\bibfnamefont {S.}~\bibnamefont {{Popov}}},
  \bibinfo {author} {\bibfnamefont {R.}~\bibnamefont {{Aptekar}}}, \bibinfo
  {author} {\bibfnamefont {S.}~\bibnamefont {{Golenetskii}}}, \bibinfo {author}
  {\bibfnamefont {A.}~\bibnamefont {{Lysenko}}}, \bibinfo {author}
  {\bibfnamefont {A.}~\bibnamefont {{Tsvetkova}}}, \bibinfo {author}
  {\bibfnamefont {M.}~\bibnamefont {{Ulanov}}}, \ and\ \bibinfo {author}
  {\bibfnamefont {T.~L.}\ \bibnamefont {{Cline}}},\ }\bibfield  {title}
  {\enquote {\bibinfo {title} {{A peculiar hard X-ray counterpart of a Galactic
  fast radio burst}},}\ }\href {\doibase 10.1038/s41550-020-01265-0} {\bibfield
   {journal} {\bibinfo  {journal} {Nature Astronomy}\ }\textbf {\bibinfo
  {volume} {5}},\ \bibinfo {pages} {372--377} (\bibinfo {year} {2021})},\
  \Eprint {http://arxiv.org/abs/2005.11178} {arXiv:2005.11178 [astro-ph.HE]}
  \BibitemShut {NoStop}%
\bibitem [{\citenamefont {{Tavani}}\ \emph {et~al.}(2021)\citenamefont
  {{Tavani}}, \citenamefont {{Casentini}}, \citenamefont {{Ursi}},
  \citenamefont {{Verrecchia}}, \citenamefont {{Addis}}, \citenamefont
  {{Antonelli}}, \citenamefont {{Argan}}, \citenamefont {{Barbiellini}},
  \citenamefont {{Baroncelli}}, \citenamefont {{Bernardi}}, \citenamefont
  {{Bianchi}}, \citenamefont {{Bulgarelli}}, \citenamefont {{Caraveo}},
  \citenamefont {{Cardillo}}, \citenamefont {{Cattaneo}}, \citenamefont
  {{Chen}}, \citenamefont {{Costa}}, \citenamefont {{Del Monte}}, \citenamefont
  {{Di Cocco}}, \citenamefont {{Di Persio}}, \citenamefont {{Donnarumma}},
  \citenamefont {{Evangelista}}, \citenamefont {{Feroci}}, \citenamefont
  {{Ferrari}}, \citenamefont {{Fioretti}}, \citenamefont {{Fuschino}},
  \citenamefont {{Galli}}, \citenamefont {{Gianotti}}, \citenamefont
  {{Giuliani}}, \citenamefont {{Labanti}}, \citenamefont {{Lazzarotto}},
  \citenamefont {{Lipari}}, \citenamefont {{Longo}}, \citenamefont
  {{Lucarelli}}, \citenamefont {{Magro}}, \citenamefont {{Marisaldi}},
  \citenamefont {{Mereghetti}}, \citenamefont {{Morelli}}, \citenamefont
  {{Morselli}}, \citenamefont {{Naldi}}, \citenamefont {{Pacciani}},
  \citenamefont {{Parmiggiani}}, \citenamefont {{Paoletti}}, \citenamefont
  {{Pellizzoni}}, \citenamefont {{Perri}}, \citenamefont {{Perotti}},
  \citenamefont {{Piano}}, \citenamefont {{Picozza}}, \citenamefont {{Pilia}},
  \citenamefont {{Pittori}}, \citenamefont {{Puccetti}}, \citenamefont
  {{Pupillo}}, \citenamefont {{Rapisarda}}, \citenamefont {{Rappoldi}},
  \citenamefont {{Rubini}}, \citenamefont {{Setti}}, \citenamefont
  {{Soffitta}}, \citenamefont {{Trifoglio}}, \citenamefont {{Trois}},
  \citenamefont {{Vercellone}}, \citenamefont {{Vittorini}}, \citenamefont
  {{Giommi}},\ and\ \citenamefont {{D'Amico}}}]{Tavani2021}%
  \BibitemOpen
  \bibfield  {author} {\bibinfo {author} {\bibfnamefont {M.}~\bibnamefont
  {{Tavani}}}, \bibinfo {author} {\bibfnamefont {C.}~\bibnamefont
  {{Casentini}}}, \bibinfo {author} {\bibfnamefont {A.}~\bibnamefont {{Ursi}}},
  \bibinfo {author} {\bibfnamefont {F.}~\bibnamefont {{Verrecchia}}}, \bibinfo
  {author} {\bibfnamefont {A.}~\bibnamefont {{Addis}}}, \bibinfo {author}
  {\bibfnamefont {L.~A.}\ \bibnamefont {{Antonelli}}}, \bibinfo {author}
  {\bibfnamefont {A.}~\bibnamefont {{Argan}}}, \bibinfo {author} {\bibfnamefont
  {G.}~\bibnamefont {{Barbiellini}}}, \bibinfo {author} {\bibfnamefont
  {L.}~\bibnamefont {{Baroncelli}}}, \bibinfo {author} {\bibfnamefont
  {G.}~\bibnamefont {{Bernardi}}}, \bibinfo {author} {\bibfnamefont
  {G.}~\bibnamefont {{Bianchi}}}, \bibinfo {author} {\bibfnamefont
  {A.}~\bibnamefont {{Bulgarelli}}}, \bibinfo {author} {\bibfnamefont
  {P.}~\bibnamefont {{Caraveo}}}, \bibinfo {author} {\bibfnamefont
  {M.}~\bibnamefont {{Cardillo}}}, \bibinfo {author} {\bibfnamefont {P.~W.}\
  \bibnamefont {{Cattaneo}}}, \bibinfo {author} {\bibfnamefont {A.~W.}\
  \bibnamefont {{Chen}}}, \bibinfo {author} {\bibfnamefont {E.}~\bibnamefont
  {{Costa}}}, \bibinfo {author} {\bibfnamefont {E.}~\bibnamefont {{Del
  Monte}}}, \bibinfo {author} {\bibfnamefont {G.}~\bibnamefont {{Di Cocco}}},
  \bibinfo {author} {\bibfnamefont {G.}~\bibnamefont {{Di Persio}}}, \bibinfo
  {author} {\bibfnamefont {I.}~\bibnamefont {{Donnarumma}}}, \bibinfo {author}
  {\bibfnamefont {Y.}~\bibnamefont {{Evangelista}}}, \bibinfo {author}
  {\bibfnamefont {M.}~\bibnamefont {{Feroci}}}, \bibinfo {author}
  {\bibfnamefont {A.}~\bibnamefont {{Ferrari}}}, \bibinfo {author}
  {\bibfnamefont {V.}~\bibnamefont {{Fioretti}}}, \bibinfo {author}
  {\bibfnamefont {F.}~\bibnamefont {{Fuschino}}}, \bibinfo {author}
  {\bibfnamefont {M.}~\bibnamefont {{Galli}}}, \bibinfo {author} {\bibfnamefont
  {F.}~\bibnamefont {{Gianotti}}}, \bibinfo {author} {\bibfnamefont
  {A.}~\bibnamefont {{Giuliani}}}, \bibinfo {author} {\bibfnamefont
  {C.}~\bibnamefont {{Labanti}}}, \bibinfo {author} {\bibfnamefont
  {F.}~\bibnamefont {{Lazzarotto}}}, \bibinfo {author} {\bibfnamefont
  {P.}~\bibnamefont {{Lipari}}}, \bibinfo {author} {\bibfnamefont
  {F.}~\bibnamefont {{Longo}}}, \bibinfo {author} {\bibfnamefont
  {F.}~\bibnamefont {{Lucarelli}}}, \bibinfo {author} {\bibfnamefont
  {A.}~\bibnamefont {{Magro}}}, \bibinfo {author} {\bibfnamefont
  {M.}~\bibnamefont {{Marisaldi}}}, \bibinfo {author} {\bibfnamefont
  {S.}~\bibnamefont {{Mereghetti}}}, \bibinfo {author} {\bibfnamefont
  {E.}~\bibnamefont {{Morelli}}}, \bibinfo {author} {\bibfnamefont
  {A.}~\bibnamefont {{Morselli}}}, \bibinfo {author} {\bibfnamefont
  {G.}~\bibnamefont {{Naldi}}}, \bibinfo {author} {\bibfnamefont
  {L.}~\bibnamefont {{Pacciani}}}, \bibinfo {author} {\bibfnamefont
  {N.}~\bibnamefont {{Parmiggiani}}}, \bibinfo {author} {\bibfnamefont
  {F.}~\bibnamefont {{Paoletti}}}, \bibinfo {author} {\bibfnamefont
  {A.}~\bibnamefont {{Pellizzoni}}}, \bibinfo {author} {\bibfnamefont
  {M.}~\bibnamefont {{Perri}}}, \bibinfo {author} {\bibfnamefont
  {F.}~\bibnamefont {{Perotti}}}, \bibinfo {author} {\bibfnamefont
  {G.}~\bibnamefont {{Piano}}}, \bibinfo {author} {\bibfnamefont
  {P.}~\bibnamefont {{Picozza}}}, \bibinfo {author} {\bibfnamefont
  {M.}~\bibnamefont {{Pilia}}}, \bibinfo {author} {\bibfnamefont
  {C.}~\bibnamefont {{Pittori}}}, \bibinfo {author} {\bibfnamefont
  {S.}~\bibnamefont {{Puccetti}}}, \bibinfo {author} {\bibfnamefont
  {G.}~\bibnamefont {{Pupillo}}}, \bibinfo {author} {\bibfnamefont
  {M.}~\bibnamefont {{Rapisarda}}}, \bibinfo {author} {\bibfnamefont
  {A.}~\bibnamefont {{Rappoldi}}}, \bibinfo {author} {\bibfnamefont
  {A.}~\bibnamefont {{Rubini}}}, \bibinfo {author} {\bibfnamefont
  {G.}~\bibnamefont {{Setti}}}, \bibinfo {author} {\bibfnamefont
  {P.}~\bibnamefont {{Soffitta}}}, \bibinfo {author} {\bibfnamefont
  {M.}~\bibnamefont {{Trifoglio}}}, \bibinfo {author} {\bibfnamefont
  {A.}~\bibnamefont {{Trois}}}, \bibinfo {author} {\bibfnamefont
  {S.}~\bibnamefont {{Vercellone}}}, \bibinfo {author} {\bibfnamefont
  {V.}~\bibnamefont {{Vittorini}}}, \bibinfo {author} {\bibfnamefont
  {P.}~\bibnamefont {{Giommi}}}, \ and\ \bibinfo {author} {\bibfnamefont
  {F.}~\bibnamefont {{D'Amico}}},\ }\bibfield  {title} {\enquote {\bibinfo
  {title} {{An X-ray burst from a magnetar enlightening the mechanism of fast
  radio bursts}},}\ }\href {\doibase 10.1038/s41550-020-01276-x} {\bibfield
  {journal} {\bibinfo  {journal} {Nature Astronomy}\ }\textbf {\bibinfo
  {volume} {5}},\ \bibinfo {pages} {401--407} (\bibinfo {year} {2021})},\
  \Eprint {http://arxiv.org/abs/2005.12164} {arXiv:2005.12164 [astro-ph.HE]}
  \BibitemShut {NoStop}%
\bibitem [{\citenamefont {Yang}\ and\ \citenamefont
  {Zou}(2020)}]{Yang:2020qxt}%
  \BibitemOpen
  \bibfield  {author} {\bibinfo {author} {\bibfnamefont {Huan}\ \bibnamefont
  {Yang}}\ and\ \bibinfo {author} {\bibfnamefont {Yuan-Chuan}\ \bibnamefont
  {Zou}},\ }\bibfield  {title} {\enquote {\bibinfo {title} {{Orbit-induced Spin
  Precession as a Possible Origin for Periodicity in Periodically Repeating
  Fast Radio Bursts}},}\ }\href {\doibase 10.3847/2041-8213/ab800f} {\bibfield
  {journal} {\bibinfo  {journal} {Astrophys. J. Lett.}\ }\textbf {\bibinfo
  {volume} {893}},\ \bibinfo {pages} {L31} (\bibinfo {year} {2020})},\ \Eprint
  {http://arxiv.org/abs/2002.02553} {arXiv:2002.02553 [astro-ph.HE]}
  \BibitemShut {NoStop}%
\bibitem [{\citenamefont {Levin}\ \emph {et~al.}(2020)\citenamefont {Levin},
  \citenamefont {Beloborodov},\ and\ \citenamefont
  {Bransgrove}}]{Levin:2020rhj}%
  \BibitemOpen
  \bibfield  {author} {\bibinfo {author} {\bibfnamefont {Yuri}\ \bibnamefont
  {Levin}}, \bibinfo {author} {\bibfnamefont {Andrei~M.}\ \bibnamefont
  {Beloborodov}}, \ and\ \bibinfo {author} {\bibfnamefont {Ashley}\
  \bibnamefont {Bransgrove}},\ }\bibfield  {title} {\enquote {\bibinfo {title}
  {{Precessing flaring magnetar as a source of repeating FRB
  180916.J0158+65}},}\ }\href {\doibase 10.3847/2041-8213/ab8c4c} {\bibfield
  {journal} {\bibinfo  {journal} {Astrophys. J. Lett.}\ }\textbf {\bibinfo
  {volume} {895}},\ \bibinfo {pages} {L30} (\bibinfo {year} {2020})},\ \Eprint
  {http://arxiv.org/abs/2002.04595} {arXiv:2002.04595 [astro-ph.HE]}
  \BibitemShut {NoStop}%
\bibitem [{\citenamefont {Zanazzi}\ and\ \citenamefont
  {Lai}(2020)}]{Zanazzi:2020vyp}%
  \BibitemOpen
  \bibfield  {author} {\bibinfo {author} {\bibfnamefont {J.~J.}\ \bibnamefont
  {Zanazzi}}\ and\ \bibinfo {author} {\bibfnamefont {Dong}\ \bibnamefont
  {Lai}},\ }\bibfield  {title} {\enquote {\bibinfo {title} {{Periodic Fast
  Radio Bursts with Neutron Star Free Precession}},}\ }\href {\doibase
  10.3847/2041-8213/ab7cdd} {\bibfield  {journal} {\bibinfo  {journal}
  {Astrophys. J.}\ }\textbf {\bibinfo {volume} {892}},\ \bibinfo {pages} {L15}
  (\bibinfo {year} {2020})},\ \Eprint {http://arxiv.org/abs/2002.05752}
  {arXiv:2002.05752 [astro-ph.HE]} \BibitemShut {NoStop}%
\bibitem [{\citenamefont {Sridhar}\ \emph {et~al.}(2021)\citenamefont
  {Sridhar}, \citenamefont {Metzger}, \citenamefont {Beniamini}, \citenamefont
  {Margalit}, \citenamefont {Renzo}, \citenamefont {Sironi},\ and\
  \citenamefont {Kovlakas}}]{Sridhar:2021zly}%
  \BibitemOpen
  \bibfield  {author} {\bibinfo {author} {\bibfnamefont {Navin}\ \bibnamefont
  {Sridhar}}, \bibinfo {author} {\bibfnamefont {Brian~D.}\ \bibnamefont
  {Metzger}}, \bibinfo {author} {\bibfnamefont {Paz}\ \bibnamefont
  {Beniamini}}, \bibinfo {author} {\bibfnamefont {Ben}\ \bibnamefont
  {Margalit}}, \bibinfo {author} {\bibfnamefont {Mathieu}\ \bibnamefont
  {Renzo}}, \bibinfo {author} {\bibfnamefont {Lorenzo}\ \bibnamefont {Sironi}},
  \ and\ \bibinfo {author} {\bibfnamefont {Konstantinos}\ \bibnamefont
  {Kovlakas}},\ }\bibfield  {title} {\enquote {\bibinfo {title} {{Periodic Fast
  Radio Bursts from Ultra-luminous X-Ray\textendash{}like Binaries}},}\ }\href
  {\doibase 10.3847/1538-4357/ac0140} {\bibfield  {journal} {\bibinfo
  {journal} {Astrophys. J.}\ }\textbf {\bibinfo {volume} {917}},\ \bibinfo
  {pages} {13} (\bibinfo {year} {2021})},\ \Eprint
  {http://arxiv.org/abs/2102.06138} {arXiv:2102.06138 [astro-ph.HE]}
  \BibitemShut {NoStop}%
\bibitem [{\citenamefont {Katz}(2017)}]{Katz:2017ube}%
  \BibitemOpen
  \bibfield  {author} {\bibinfo {author} {\bibfnamefont {J.~I.}\ \bibnamefont
  {Katz}},\ }\bibfield  {title} {\enquote {\bibinfo {title} {{FRB as Products
  of Accretion Disc Funnels}},}\ }\href {\doibase 10.1093/mnrasl/slx113}
  {\bibfield  {journal} {\bibinfo  {journal} {Mon. Not. Roy. Astron. Soc.}\
  }\textbf {\bibinfo {volume} {471}},\ \bibinfo {pages} {L92--L95} (\bibinfo
  {year} {2017})},\ \Eprint {http://arxiv.org/abs/1704.08301} {arXiv:1704.08301
  [astro-ph.HE]} \BibitemShut {NoStop}%
\bibitem [{\citenamefont {{CHIME/FRB Collaboration}}(2021)}]{CHIME-c1}%
  \BibitemOpen
  \bibfield  {author} {\bibinfo {author} {\bibnamefont {{CHIME/FRB
  Collaboration}}},\ }\bibfield  {title} {\enquote {\bibinfo {title} {{The
  First CHIME/FRB Fast Radio Burst Catalog}},}\ }\href {\doibase
  10.3847/1538-4365/ac33ab} {\bibfield  {journal} {\bibinfo  {journal} {\apjs}\
  }\textbf {\bibinfo {volume} {257}},\ \bibinfo {eid} {59} (\bibinfo {year}
  {2021})},\ \Eprint {http://arxiv.org/abs/2106.04352} {arXiv:2106.04352
  [astro-ph.HE]} \BibitemShut {NoStop}%
\bibitem [{\citenamefont {{The LIGO Scientific Collaboration}}\ \emph
  {et~al.}(2022)\citenamefont {{The LIGO Scientific Collaboration}},
  \citenamefont {{the Virgo Collaboration}}, \citenamefont {{the KAGRA
  Collaboration}},\ and\ \citenamefont {{the CHIME/FRB
  Collaboration}}}]{LVK-CHIME2022}%
  \BibitemOpen
  \bibfield  {author} {\bibinfo {author} {\bibnamefont {{The LIGO Scientific
  Collaboration}}}, \bibinfo {author} {\bibnamefont {{the Virgo
  Collaboration}}}, \bibinfo {author} {\bibnamefont {{the KAGRA
  Collaboration}}}, \ and\ \bibinfo {author} {\bibnamefont {{the CHIME/FRB
  Collaboration}}},\ }\bibfield  {title} {\enquote {\bibinfo {title} {{Search
  for Gravitational Waves Associated with Fast Radio Bursts Detected by
  CHIME/FRB During the LIGO--Virgo Observing Run O3a}},}\ }\href@noop {}
  {\bibfield  {journal} {\bibinfo  {journal} {arXiv e-prints}\ ,\ \bibinfo
  {eid} {arXiv:2203.12038}} (\bibinfo {year} {2022})},\ \Eprint
  {http://arxiv.org/abs/2203.12038} {arXiv:2203.12038 [astro-ph.HE]}
  \BibitemShut {NoStop}%
\bibitem [{\citenamefont {{Wang}}\ and\ \citenamefont
  {{Nitz}}(2022)}]{Wang2022}%
  \BibitemOpen
  \bibfield  {author} {\bibinfo {author} {\bibfnamefont {Yi-Fan}\ \bibnamefont
  {{Wang}}}\ and\ \bibinfo {author} {\bibfnamefont {Alexander~H.}\ \bibnamefont
  {{Nitz}}},\ }\bibfield  {title} {\enquote {\bibinfo {title} {{Search for
  Coincident Gravitational Wave and Fast Radio Burst Events from 4-OGC and the
  First CHIME/FRB Catalog}},}\ }\href@noop {} {\bibfield  {journal} {\bibinfo
  {journal} {arXiv e-prints}\ ,\ \bibinfo {eid} {arXiv:2203.17222}} (\bibinfo
  {year} {2022})},\ \Eprint {http://arxiv.org/abs/2203.17222} {arXiv:2203.17222
  [astro-ph.HE]} \BibitemShut {NoStop}%
\bibitem [{\citenamefont {{Lipunov}}\ and\ \citenamefont
  {{Panchenko}}(1996)}]{Lipunov1996}%
  \BibitemOpen
  \bibfield  {author} {\bibinfo {author} {\bibfnamefont {V.~M.}\ \bibnamefont
  {{Lipunov}}}\ and\ \bibinfo {author} {\bibfnamefont {I.~E.}\ \bibnamefont
  {{Panchenko}}},\ }\bibfield  {title} {\enquote {\bibinfo {title} {{Pulsars
  revived by gravitational waves.}}}\ }\href {\doibase
  10.48550/arXiv.astro-ph/9608155} {\bibfield  {journal} {\bibinfo  {journal}
  {Astron.Astrophys}\ }\textbf {\bibinfo {volume} {312}},\ \bibinfo {pages}
  {937--940} (\bibinfo {year} {1996})},\ \Eprint
  {http://arxiv.org/abs/astro-ph/9608155} {arXiv:astro-ph/9608155 [astro-ph]}
  \BibitemShut {NoStop}%
\bibitem [{\citenamefont {{Lyutikov}}(2018)}]{Lyutikov2018}%
  \BibitemOpen
  \bibfield  {author} {\bibinfo {author} {\bibfnamefont {Maxim}\ \bibnamefont
  {{Lyutikov}}},\ }\bibfield  {title} {\enquote {\bibinfo {title}
  {{Electrodynamics of double neutron star mergers}},}\ }\href {\doibase
  10.48550/arXiv.1809.10478} {\bibfield  {journal} {\bibinfo  {journal} {arXiv
  e-prints}\ ,\ \bibinfo {eid} {arXiv:1809.10478}} (\bibinfo {year} {2018})},\
  \Eprint {http://arxiv.org/abs/1809.10478} {arXiv:1809.10478 [astro-ph.HE]}
  \BibitemShut {NoStop}%
\bibitem [{\citenamefont {{Zhang}}(2020)}]{ZhangB2020}%
  \BibitemOpen
  \bibfield  {author} {\bibinfo {author} {\bibfnamefont {Bing}\ \bibnamefont
  {{Zhang}}},\ }\bibfield  {title} {\enquote {\bibinfo {title} {{Fast Radio
  Bursts from Interacting Binary Neutron Star Systems}},}\ }\href {\doibase
  10.3847/2041-8213/ab7244} {\bibfield  {journal} {\bibinfo  {journal} {\apjl}\
  }\textbf {\bibinfo {volume} {890}},\ \bibinfo {eid} {L24} (\bibinfo {year}
  {2020})},\ \Eprint {http://arxiv.org/abs/2002.00335} {arXiv:2002.00335
  [astro-ph.HE]} \BibitemShut {NoStop}%
\bibitem [{\citenamefont {Apostolatos}\ \emph {et~al.}(1994)\citenamefont
  {Apostolatos}, \citenamefont {Cutler}, \citenamefont {Sussman},\ and\
  \citenamefont {Thorne}}]{Apostolatos1994}%
  \BibitemOpen
  \bibfield  {author} {\bibinfo {author} {\bibfnamefont {Theocharis~A.}\
  \bibnamefont {Apostolatos}}, \bibinfo {author} {\bibfnamefont {Curt}\
  \bibnamefont {Cutler}}, \bibinfo {author} {\bibfnamefont {Gerald~J.}\
  \bibnamefont {Sussman}}, \ and\ \bibinfo {author} {\bibfnamefont {Kip~S.}\
  \bibnamefont {Thorne}},\ }\bibfield  {title} {\enquote {\bibinfo {title}
  {Spin-induced orbital precession and its modulation of the gravitational
  waveforms from merging binaries},}\ }\href {\doibase
  10.1103/PhysRevD.49.6274} {\bibfield  {journal} {\bibinfo  {journal} {Phys.
  Rev. D}\ }\textbf {\bibinfo {volume} {49}},\ \bibinfo {pages} {6274--6297}
  (\bibinfo {year} {1994})}\BibitemShut {NoStop}%
\bibitem [{\citenamefont {{Kidder}}(1995)}]{Kidder1995}%
  \BibitemOpen
  \bibfield  {author} {\bibinfo {author} {\bibfnamefont {Lawrence~E.}\
  \bibnamefont {{Kidder}}},\ }\bibfield  {title} {\enquote {\bibinfo {title}
  {{Coalescing binary systems of compact objects to (post)$^{5/2}$-Newtonian
  order. V. Spin effects}},}\ }\href {\doibase 10.1103/PhysRevD.52.821}
  {\bibfield  {journal} {\bibinfo  {journal} {\prd}\ }\textbf {\bibinfo
  {volume} {52}},\ \bibinfo {pages} {821--847} (\bibinfo {year} {1995})},\
  \Eprint {http://arxiv.org/abs/gr-qc/9506022} {arXiv:gr-qc/9506022 [gr-qc]}
  \BibitemShut {NoStop}%
\bibitem [{\citenamefont {{Poisson}}(1998)}]{Poisson1998}%
  \BibitemOpen
  \bibfield  {author} {\bibinfo {author} {\bibfnamefont {Eric}\ \bibnamefont
  {{Poisson}}},\ }\bibfield  {title} {\enquote {\bibinfo {title}
  {{Gravitational waves from inspiraling compact binaries: The
  quadrupole-moment term}},}\ }\href {\doibase 10.1103/PhysRevD.57.5287}
  {\bibfield  {journal} {\bibinfo  {journal} {\prd}\ }\textbf {\bibinfo
  {volume} {57}},\ \bibinfo {pages} {5287--5290} (\bibinfo {year} {1998})},\
  \Eprint {http://arxiv.org/abs/gr-qc/9709032} {arXiv:gr-qc/9709032 [gr-qc]}
  \BibitemShut {NoStop}%
\bibitem [{\citenamefont {{Racine}}(2008)}]{Racine2008}%
  \BibitemOpen
  \bibfield  {author} {\bibinfo {author} {\bibfnamefont {{\'E}tienne}\
  \bibnamefont {{Racine}}},\ }\bibfield  {title} {\enquote {\bibinfo {title}
  {{Analysis of spin precession in binary black hole systems including
  quadrupole-monopole interaction}},}\ }\href {\doibase
  10.1103/PhysRevD.78.044021} {\bibfield  {journal} {\bibinfo  {journal}
  {\prd}\ }\textbf {\bibinfo {volume} {78}},\ \bibinfo {eid} {044021} (\bibinfo
  {year} {2008})},\ \Eprint {http://arxiv.org/abs/0803.1820} {arXiv:0803.1820
  [gr-qc]} \BibitemShut {NoStop}%
\bibitem [{\citenamefont {{Laarakkers}}\ and\ \citenamefont
  {{Poisson}}(1999)}]{Laarakkers1999}%
  \BibitemOpen
  \bibfield  {author} {\bibinfo {author} {\bibfnamefont {William~G.}\
  \bibnamefont {{Laarakkers}}}\ and\ \bibinfo {author} {\bibfnamefont {Eric}\
  \bibnamefont {{Poisson}}},\ }\bibfield  {title} {\enquote {\bibinfo {title}
  {{Quadrupole Moments of Rotating Neutron Stars}},}\ }\href {\doibase
  10.1086/306732} {\bibfield  {journal} {\bibinfo  {journal} {\apj}\ }\textbf
  {\bibinfo {volume} {512}},\ \bibinfo {pages} {282--287} (\bibinfo {year}
  {1999})},\ \Eprint {http://arxiv.org/abs/gr-qc/9709033} {arXiv:gr-qc/9709033
  [gr-qc]} \BibitemShut {NoStop}%
\bibitem [{\citenamefont {Zhu}\ \emph {et~al.}(2023)\citenamefont {Zhu} \emph
  {et~al.}}]{Zhu:2023spq}%
  \BibitemOpen
  \bibfield  {author} {\bibinfo {author} {\bibfnamefont {Weiwei}\ \bibnamefont
  {Zhu}} \emph {et~al.},\ }\bibfield  {title} {\enquote {\bibinfo {title} {{A
  radio pulsar phase from SGR J1935+2154 provides clues to the magnetar FRB
  mechanism}},}\ }\href {\doibase 10.1126/sciadv.adf6198} {\bibfield  {journal}
  {\bibinfo  {journal} {Sci. Adv.}\ }\textbf {\bibinfo {volume} {9}},\ \bibinfo
  {pages} {adf6198} (\bibinfo {year} {2023})},\ \Eprint
  {http://arxiv.org/abs/2307.16124} {arXiv:2307.16124 [astro-ph.HE]}
  \BibitemShut {NoStop}%
\bibitem [{\citenamefont {Andersen}\ \emph {et~al.}(2022)\citenamefont
  {Andersen} \emph {et~al.}}]{CHIMEFRB:2021fvq}%
  \BibitemOpen
  \bibfield  {author} {\bibinfo {author} {\bibfnamefont {Bridget~C.}\
  \bibnamefont {Andersen}} \emph {et~al.} (\bibinfo {collaboration}
  {CHIME/FRB}),\ }\bibfield  {title} {\enquote {\bibinfo {title} {{Sub-second
  periodicity in a fast radio burst}},}\ }\href {\doibase
  10.1038/s41586-022-04841-8} {\bibfield  {journal} {\bibinfo  {journal}
  {Nature}\ }\textbf {\bibinfo {volume} {607}},\ \bibinfo {pages} {256--259}
  (\bibinfo {year} {2022})},\ \Eprint {http://arxiv.org/abs/2107.08463}
  {arXiv:2107.08463 [astro-ph.HE]} \BibitemShut {NoStop}%
\bibitem [{\citenamefont {{Fisher}}(1922)}]{Fisher1922}%
  \BibitemOpen
  \bibfield  {author} {\bibinfo {author} {\bibfnamefont {R.~A.}\ \bibnamefont
  {{Fisher}}},\ }\bibfield  {title} {\enquote {\bibinfo {title} {{On the
  Mathematical Foundations of Theoretical Statistics}},}\ }\href {\doibase
  10.1098/rsta.1922.0009} {\bibfield  {journal} {\bibinfo  {journal}
  {Philosophical Transactions of the Royal Society of London Series A}\
  }\textbf {\bibinfo {volume} {222}},\ \bibinfo {pages} {309--368} (\bibinfo
  {year} {1922})}\BibitemShut {NoStop}%
\bibitem [{\citenamefont {{Yang}}\ and\ \citenamefont
  {{Zhang}}(2017)}]{Yang2017}%
  \BibitemOpen
  \bibfield  {author} {\bibinfo {author} {\bibfnamefont {Yuan-Pei}\
  \bibnamefont {{Yang}}}\ and\ \bibinfo {author} {\bibfnamefont {Bing}\
  \bibnamefont {{Zhang}}},\ }\bibfield  {title} {\enquote {\bibinfo {title}
  {{Dispersion Measure Variation of Repeating Fast Radio Burst Sources}},}\
  }\href {\doibase 10.3847/1538-4357/aa8721} {\bibfield  {journal} {\bibinfo
  {journal} {\apj}\ }\textbf {\bibinfo {volume} {847}},\ \bibinfo {eid} {22}
  (\bibinfo {year} {2017})},\ \Eprint {http://arxiv.org/abs/1707.02923}
  {arXiv:1707.02923 [astro-ph.HE]} \BibitemShut {NoStop}%
\bibitem [{\citenamefont {{Chime/Frb Collaboration}}\ \emph
  {et~al.}(2020)\citenamefont {{Chime/Frb Collaboration}}, \citenamefont
  {{Amiri}}, \citenamefont {{Andersen}}, \citenamefont {{Bandura}},
  \citenamefont {{Bhardwaj}}, \citenamefont {{Boyle}}, \citenamefont {{Brar}},
  \citenamefont {{Chawla}}, \citenamefont {{Chen}}, \citenamefont {{Cliche}},
  \citenamefont {{Cubranic}}, \citenamefont {{Deng}}, \citenamefont {{Denman}},
  \citenamefont {{Dobbs}}, \citenamefont {{Dong}}, \citenamefont {{Fandino}},
  \citenamefont {{Fonseca}}, \citenamefont {{Gaensler}}, \citenamefont
  {{Giri}}, \citenamefont {{Good}}, \citenamefont {{Halpern}}, \citenamefont
  {{Hessels}}, \citenamefont {{Hill}}, \citenamefont {{H{\"o}fer}},
  \citenamefont {{Josephy}}, \citenamefont {{Kania}}, \citenamefont
  {{Karuppusamy}}, \citenamefont {{Kaspi}}, \citenamefont {{Keimpema}},
  \citenamefont {{Kirsten}}, \citenamefont {{Landecker}}, \citenamefont
  {{Lang}}, \citenamefont {{Leung}}, \citenamefont {{Li}}, \citenamefont
  {{Lin}}, \citenamefont {{Marcote}}, \citenamefont {{Masui}}, \citenamefont
  {{McKinven}}, \citenamefont {{Mena-Parra}}, \citenamefont {{Merryfield}},
  \citenamefont {{Michilli}}, \citenamefont {{Milutinovic}}, \citenamefont
  {{Mirhosseini}}, \citenamefont {{Naidu}}, \citenamefont {{Newburgh}},
  \citenamefont {{Ng}}, \citenamefont {{Nimmo}}, \citenamefont {{Paragi}},
  \citenamefont {{Patel}}, \citenamefont {{Pen}}, \citenamefont
  {{Pinsonneault-Marotte}}, \citenamefont {{Pleunis}}, \citenamefont
  {{Rafiei-Ravandi}}, \citenamefont {{Rahman}}, \citenamefont {{Ransom}},
  \citenamefont {{Renard}}, \citenamefont {{Sanghavi}}, \citenamefont
  {{Scholz}}, \citenamefont {{Shaw}}, \citenamefont {{Shin}}, \citenamefont
  {{Siegel}}, \citenamefont {{Singh}}, \citenamefont {{Smegal}}, \citenamefont
  {{Smith}}, \citenamefont {{Stairs}}, \citenamefont {{Tendulkar}},
  \citenamefont {{Tretyakov}}, \citenamefont {{Vanderlinde}}, \citenamefont
  {{Wang}}, \citenamefont {{Wang}}, \citenamefont {{Wulf}}, \citenamefont
  {{Yadav}},\ and\ \citenamefont {{Zwaniga}}}]{CHIME16d}%
  \BibitemOpen
  \bibfield  {author} {\bibinfo {author} {\bibnamefont {{Chime/Frb
  Collaboration}}}, \bibinfo {author} {\bibfnamefont {M.}~\bibnamefont
  {{Amiri}}}, \bibinfo {author} {\bibfnamefont {B.~C.}\ \bibnamefont
  {{Andersen}}}, \bibinfo {author} {\bibfnamefont {K.~M.}\ \bibnamefont
  {{Bandura}}}, \bibinfo {author} {\bibfnamefont {M.}~\bibnamefont
  {{Bhardwaj}}}, \bibinfo {author} {\bibfnamefont {P.~J.}\ \bibnamefont
  {{Boyle}}}, \bibinfo {author} {\bibfnamefont {C.}~\bibnamefont {{Brar}}},
  \bibinfo {author} {\bibfnamefont {P.}~\bibnamefont {{Chawla}}}, \bibinfo
  {author} {\bibfnamefont {T.}~\bibnamefont {{Chen}}}, \bibinfo {author}
  {\bibfnamefont {J.~F.}\ \bibnamefont {{Cliche}}}, \bibinfo {author}
  {\bibfnamefont {D.}~\bibnamefont {{Cubranic}}}, \bibinfo {author}
  {\bibfnamefont {M.}~\bibnamefont {{Deng}}}, \bibinfo {author} {\bibfnamefont
  {N.~T.}\ \bibnamefont {{Denman}}}, \bibinfo {author} {\bibfnamefont
  {M.}~\bibnamefont {{Dobbs}}}, \bibinfo {author} {\bibfnamefont {F.~Q.}\
  \bibnamefont {{Dong}}}, \bibinfo {author} {\bibfnamefont {M.}~\bibnamefont
  {{Fandino}}}, \bibinfo {author} {\bibfnamefont {E.}~\bibnamefont
  {{Fonseca}}}, \bibinfo {author} {\bibfnamefont {B.~M.}\ \bibnamefont
  {{Gaensler}}}, \bibinfo {author} {\bibfnamefont {U.}~\bibnamefont {{Giri}}},
  \bibinfo {author} {\bibfnamefont {D.~C.}\ \bibnamefont {{Good}}}, \bibinfo
  {author} {\bibfnamefont {M.}~\bibnamefont {{Halpern}}}, \bibinfo {author}
  {\bibfnamefont {J.~W.~T.}\ \bibnamefont {{Hessels}}}, \bibinfo {author}
  {\bibfnamefont {A.~S.}\ \bibnamefont {{Hill}}}, \bibinfo {author}
  {\bibfnamefont {C.}~\bibnamefont {{H{\"o}fer}}}, \bibinfo {author}
  {\bibfnamefont {A.}~\bibnamefont {{Josephy}}}, \bibinfo {author}
  {\bibfnamefont {J.~W.}\ \bibnamefont {{Kania}}}, \bibinfo {author}
  {\bibfnamefont {R.}~\bibnamefont {{Karuppusamy}}}, \bibinfo {author}
  {\bibfnamefont {V.~M.}\ \bibnamefont {{Kaspi}}}, \bibinfo {author}
  {\bibfnamefont {A.}~\bibnamefont {{Keimpema}}}, \bibinfo {author}
  {\bibfnamefont {F.}~\bibnamefont {{Kirsten}}}, \bibinfo {author}
  {\bibfnamefont {T.~L.}\ \bibnamefont {{Landecker}}}, \bibinfo {author}
  {\bibfnamefont {D.~A.}\ \bibnamefont {{Lang}}}, \bibinfo {author}
  {\bibfnamefont {C.}~\bibnamefont {{Leung}}}, \bibinfo {author} {\bibfnamefont
  {D.~Z.}\ \bibnamefont {{Li}}}, \bibinfo {author} {\bibfnamefont {H.~H.}\
  \bibnamefont {{Lin}}}, \bibinfo {author} {\bibfnamefont {B.}~\bibnamefont
  {{Marcote}}}, \bibinfo {author} {\bibfnamefont {K.~W.}\ \bibnamefont
  {{Masui}}}, \bibinfo {author} {\bibfnamefont {R.}~\bibnamefont {{McKinven}}},
  \bibinfo {author} {\bibfnamefont {J.}~\bibnamefont {{Mena-Parra}}}, \bibinfo
  {author} {\bibfnamefont {M.}~\bibnamefont {{Merryfield}}}, \bibinfo {author}
  {\bibfnamefont {D.}~\bibnamefont {{Michilli}}}, \bibinfo {author}
  {\bibfnamefont {N.}~\bibnamefont {{Milutinovic}}}, \bibinfo {author}
  {\bibfnamefont {A.}~\bibnamefont {{Mirhosseini}}}, \bibinfo {author}
  {\bibfnamefont {A.}~\bibnamefont {{Naidu}}}, \bibinfo {author} {\bibfnamefont
  {L.~B.}\ \bibnamefont {{Newburgh}}}, \bibinfo {author} {\bibfnamefont
  {C.}~\bibnamefont {{Ng}}}, \bibinfo {author} {\bibfnamefont {K.}~\bibnamefont
  {{Nimmo}}}, \bibinfo {author} {\bibfnamefont {Z.}~\bibnamefont {{Paragi}}},
  \bibinfo {author} {\bibfnamefont {C.}~\bibnamefont {{Patel}}}, \bibinfo
  {author} {\bibfnamefont {U.~L.}\ \bibnamefont {{Pen}}}, \bibinfo {author}
  {\bibfnamefont {T.}~\bibnamefont {{Pinsonneault-Marotte}}}, \bibinfo {author}
  {\bibfnamefont {Z.}~\bibnamefont {{Pleunis}}}, \bibinfo {author}
  {\bibfnamefont {M.}~\bibnamefont {{Rafiei-Ravandi}}}, \bibinfo {author}
  {\bibfnamefont {M.}~\bibnamefont {{Rahman}}}, \bibinfo {author}
  {\bibfnamefont {S.~M.}\ \bibnamefont {{Ransom}}}, \bibinfo {author}
  {\bibfnamefont {A.}~\bibnamefont {{Renard}}}, \bibinfo {author}
  {\bibfnamefont {P.}~\bibnamefont {{Sanghavi}}}, \bibinfo {author}
  {\bibfnamefont {P.}~\bibnamefont {{Scholz}}}, \bibinfo {author}
  {\bibfnamefont {J.~R.}\ \bibnamefont {{Shaw}}}, \bibinfo {author}
  {\bibfnamefont {K.}~\bibnamefont {{Shin}}}, \bibinfo {author} {\bibfnamefont
  {S.~R.}\ \bibnamefont {{Siegel}}}, \bibinfo {author} {\bibfnamefont
  {S.}~\bibnamefont {{Singh}}}, \bibinfo {author} {\bibfnamefont {R.~J.}\
  \bibnamefont {{Smegal}}}, \bibinfo {author} {\bibfnamefont {K.~M.}\
  \bibnamefont {{Smith}}}, \bibinfo {author} {\bibfnamefont {I.~H.}\
  \bibnamefont {{Stairs}}}, \bibinfo {author} {\bibfnamefont {S.~P.}\
  \bibnamefont {{Tendulkar}}}, \bibinfo {author} {\bibfnamefont
  {I.}~\bibnamefont {{Tretyakov}}}, \bibinfo {author} {\bibfnamefont
  {K.}~\bibnamefont {{Vanderlinde}}}, \bibinfo {author} {\bibfnamefont
  {H.}~\bibnamefont {{Wang}}}, \bibinfo {author} {\bibfnamefont
  {X.}~\bibnamefont {{Wang}}}, \bibinfo {author} {\bibfnamefont
  {D.}~\bibnamefont {{Wulf}}}, \bibinfo {author} {\bibfnamefont
  {P.}~\bibnamefont {{Yadav}}}, \ and\ \bibinfo {author} {\bibfnamefont
  {A.~V.}\ \bibnamefont {{Zwaniga}}},\ }\bibfield  {title} {\enquote {\bibinfo
  {title} {{Periodic activity from a fast radio burst source}},}\ }\href
  {\doibase 10.1038/s41586-020-2398-2} {\bibfield  {journal} {\bibinfo
  {journal} {\nat}\ }\textbf {\bibinfo {volume} {582}},\ \bibinfo {pages}
  {351--355} (\bibinfo {year} {2020})},\ \Eprint
  {http://arxiv.org/abs/2001.10275} {arXiv:2001.10275 [astro-ph.HE]}
  \BibitemShut {NoStop}%
\bibitem [{\citenamefont {{Chawla}}\ \emph {et~al.}(2020)\citenamefont
  {{Chawla}}, \citenamefont {{Andersen}}, \citenamefont {{Bhardwaj}},
  \citenamefont {{Fonseca}}, \citenamefont {{Josephy}}, \citenamefont
  {{Kaspi}}, \citenamefont {{Michilli}}, \citenamefont {{Pleunis}},
  \citenamefont {{Bandura}}, \citenamefont {{Bassa}}, \citenamefont {{Boyle}},
  \citenamefont {{Brar}}, \citenamefont {{Cassanelli}}, \citenamefont
  {{Cubranic}}, \citenamefont {{Dobbs}}, \citenamefont {{Dong}}, \citenamefont
  {{Gaensler}}, \citenamefont {{Good}}, \citenamefont {{Hessels}},
  \citenamefont {{Landecker}}, \citenamefont {{Leung}}, \citenamefont {{Li}},
  \citenamefont {{Lin}}, \citenamefont {{Masui}}, \citenamefont {{Mckinven}},
  \citenamefont {{Mena-Parra}}, \citenamefont {{Merryfield}}, \citenamefont
  {{Meyers}}, \citenamefont {{Naidu}}, \citenamefont {{Ng}}, \citenamefont
  {{Patel}}, \citenamefont {{Rafiei-Ravandi}}, \citenamefont {{Rahman}},
  \citenamefont {{Sanghavi}}, \citenamefont {{Scholz}}, \citenamefont {{Shin}},
  \citenamefont {{Smith}}, \citenamefont {{Stairs}}, \citenamefont
  {{Tendulkar}},\ and\ \citenamefont {{Vanderlinde}}}]{Chawla2020}%
  \BibitemOpen
  \bibfield  {author} {\bibinfo {author} {\bibfnamefont {P.}~\bibnamefont
  {{Chawla}}}, \bibinfo {author} {\bibfnamefont {B.~C.}\ \bibnamefont
  {{Andersen}}}, \bibinfo {author} {\bibfnamefont {M.}~\bibnamefont
  {{Bhardwaj}}}, \bibinfo {author} {\bibfnamefont {E.}~\bibnamefont
  {{Fonseca}}}, \bibinfo {author} {\bibfnamefont {A.}~\bibnamefont
  {{Josephy}}}, \bibinfo {author} {\bibfnamefont {V.~M.}\ \bibnamefont
  {{Kaspi}}}, \bibinfo {author} {\bibfnamefont {D.}~\bibnamefont {{Michilli}}},
  \bibinfo {author} {\bibfnamefont {Z.}~\bibnamefont {{Pleunis}}}, \bibinfo
  {author} {\bibfnamefont {K.~M.}\ \bibnamefont {{Bandura}}}, \bibinfo {author}
  {\bibfnamefont {C.~G.}\ \bibnamefont {{Bassa}}}, \bibinfo {author}
  {\bibfnamefont {P.~J.}\ \bibnamefont {{Boyle}}}, \bibinfo {author}
  {\bibfnamefont {C.}~\bibnamefont {{Brar}}}, \bibinfo {author} {\bibfnamefont
  {T.}~\bibnamefont {{Cassanelli}}}, \bibinfo {author} {\bibfnamefont
  {D.}~\bibnamefont {{Cubranic}}}, \bibinfo {author} {\bibfnamefont
  {M.}~\bibnamefont {{Dobbs}}}, \bibinfo {author} {\bibfnamefont {F.~Q.}\
  \bibnamefont {{Dong}}}, \bibinfo {author} {\bibfnamefont {B.~M.}\
  \bibnamefont {{Gaensler}}}, \bibinfo {author} {\bibfnamefont {D.~C.}\
  \bibnamefont {{Good}}}, \bibinfo {author} {\bibfnamefont {J.~W.~T.}\
  \bibnamefont {{Hessels}}}, \bibinfo {author} {\bibfnamefont {T.~L.}\
  \bibnamefont {{Landecker}}}, \bibinfo {author} {\bibfnamefont
  {C.}~\bibnamefont {{Leung}}}, \bibinfo {author} {\bibfnamefont {D.~Z.}\
  \bibnamefont {{Li}}}, \bibinfo {author} {\bibfnamefont {H.~.~H.}\
  \bibnamefont {{Lin}}}, \bibinfo {author} {\bibfnamefont {K.}~\bibnamefont
  {{Masui}}}, \bibinfo {author} {\bibfnamefont {R.}~\bibnamefont {{Mckinven}}},
  \bibinfo {author} {\bibfnamefont {J.}~\bibnamefont {{Mena-Parra}}}, \bibinfo
  {author} {\bibfnamefont {M.}~\bibnamefont {{Merryfield}}}, \bibinfo {author}
  {\bibfnamefont {B.~W.}\ \bibnamefont {{Meyers}}}, \bibinfo {author}
  {\bibfnamefont {A.}~\bibnamefont {{Naidu}}}, \bibinfo {author} {\bibfnamefont
  {C.}~\bibnamefont {{Ng}}}, \bibinfo {author} {\bibfnamefont {C.}~\bibnamefont
  {{Patel}}}, \bibinfo {author} {\bibfnamefont {M.}~\bibnamefont
  {{Rafiei-Ravandi}}}, \bibinfo {author} {\bibfnamefont {M.}~\bibnamefont
  {{Rahman}}}, \bibinfo {author} {\bibfnamefont {P.}~\bibnamefont
  {{Sanghavi}}}, \bibinfo {author} {\bibfnamefont {P.}~\bibnamefont
  {{Scholz}}}, \bibinfo {author} {\bibfnamefont {K.}~\bibnamefont {{Shin}}},
  \bibinfo {author} {\bibfnamefont {K.~M.}\ \bibnamefont {{Smith}}}, \bibinfo
  {author} {\bibfnamefont {I.~H.}\ \bibnamefont {{Stairs}}}, \bibinfo {author}
  {\bibfnamefont {S.~P.}\ \bibnamefont {{Tendulkar}}}, \ and\ \bibinfo {author}
  {\bibfnamefont {K.}~\bibnamefont {{Vanderlinde}}},\ }\bibfield  {title}
  {\enquote {\bibinfo {title} {{Detection of Repeating FRB 180916.J0158+65 Down
  to Frequencies of 300 MHz}},}\ }\href {\doibase 10.3847/2041-8213/ab96bf}
  {\bibfield  {journal} {\bibinfo  {journal} {\apjl}\ }\textbf {\bibinfo
  {volume} {896}},\ \bibinfo {eid} {L41} (\bibinfo {year} {2020})},\ \Eprint
  {http://arxiv.org/abs/2004.02862} {arXiv:2004.02862 [astro-ph.HE]}
  \BibitemShut {NoStop}%
\bibitem [{\citenamefont {{Connor}}\ \emph {et~al.}(2016)\citenamefont
  {{Connor}}, \citenamefont {{Pen}},\ and\ \citenamefont
  {{Oppermann}}}]{Connor2016}%
  \BibitemOpen
  \bibfield  {author} {\bibinfo {author} {\bibfnamefont {Liam}\ \bibnamefont
  {{Connor}}}, \bibinfo {author} {\bibfnamefont {Ue-Li}\ \bibnamefont {{Pen}}},
  \ and\ \bibinfo {author} {\bibfnamefont {Niels}\ \bibnamefont
  {{Oppermann}}},\ }\bibfield  {title} {\enquote {\bibinfo {title} {{FRB
  repetition and non-Poissonian statistics}},}\ }\href {\doibase
  10.1093/mnrasl/slw026} {\bibfield  {journal} {\bibinfo  {journal} {\mnras}\
  }\textbf {\bibinfo {volume} {458}},\ \bibinfo {pages} {L89--L93} (\bibinfo
  {year} {2016})},\ \Eprint {http://arxiv.org/abs/1601.04051} {arXiv:1601.04051
  [astro-ph.HE]} \BibitemShut {NoStop}%
\bibitem [{\citenamefont {{Oppermann}}\ \emph {et~al.}(2018)\citenamefont
  {{Oppermann}}, \citenamefont {{Yu}},\ and\ \citenamefont
  {{Pen}}}]{Oppermann2018}%
  \BibitemOpen
  \bibfield  {author} {\bibinfo {author} {\bibfnamefont {Niels}\ \bibnamefont
  {{Oppermann}}}, \bibinfo {author} {\bibfnamefont {Hao-Ran}\ \bibnamefont
  {{Yu}}}, \ and\ \bibinfo {author} {\bibfnamefont {Ue-Li}\ \bibnamefont
  {{Pen}}},\ }\bibfield  {title} {\enquote {\bibinfo {title} {{On the
  non-Poissonian repetition pattern of FRB121102}},}\ }\href {\doibase
  10.1093/mnras/sty004} {\bibfield  {journal} {\bibinfo  {journal} {\mnras}\
  }\textbf {\bibinfo {volume} {475}},\ \bibinfo {pages} {5109--5115} (\bibinfo
  {year} {2018})},\ \Eprint {http://arxiv.org/abs/1705.04881} {arXiv:1705.04881
  [astro-ph.HE]} \BibitemShut {NoStop}%
\bibitem [{\citenamefont {{Smith}}\ \emph {et~al.}(2021)\citenamefont
  {{Smith}}, \citenamefont {{Borhanian}}, \citenamefont {{Sathyaprakash}},
  \citenamefont {{Hernandez Vivanco}}, \citenamefont {{Field}}, \citenamefont
  {{Lasky}}, \citenamefont {{Mandel}}, \citenamefont {{Morisaki}},
  \citenamefont {{Ottaway}}, \citenamefont {{Slagmolen}}, \citenamefont
  {{Thrane}}, \citenamefont {{T{\"o}yr{\"a}}},\ and\ \citenamefont
  {{Vitale}}}]{Smith2021}%
  \BibitemOpen
  \bibfield  {author} {\bibinfo {author} {\bibfnamefont {Rory}\ \bibnamefont
  {{Smith}}}, \bibinfo {author} {\bibfnamefont {Ssohrab}\ \bibnamefont
  {{Borhanian}}}, \bibinfo {author} {\bibfnamefont {Bangalore}\ \bibnamefont
  {{Sathyaprakash}}}, \bibinfo {author} {\bibfnamefont {Francisco}\
  \bibnamefont {{Hernandez Vivanco}}}, \bibinfo {author} {\bibfnamefont
  {Scott~E.}\ \bibnamefont {{Field}}}, \bibinfo {author} {\bibfnamefont {Paul}\
  \bibnamefont {{Lasky}}}, \bibinfo {author} {\bibfnamefont {Ilya}\
  \bibnamefont {{Mandel}}}, \bibinfo {author} {\bibfnamefont {Soichiro}\
  \bibnamefont {{Morisaki}}}, \bibinfo {author} {\bibfnamefont {David}\
  \bibnamefont {{Ottaway}}}, \bibinfo {author} {\bibfnamefont {Bram J.~J.}\
  \bibnamefont {{Slagmolen}}}, \bibinfo {author} {\bibfnamefont {Eric}\
  \bibnamefont {{Thrane}}}, \bibinfo {author} {\bibfnamefont {Daniel}\
  \bibnamefont {{T{\"o}yr{\"a}}}}, \ and\ \bibinfo {author} {\bibfnamefont
  {Salvatore}\ \bibnamefont {{Vitale}}},\ }\bibfield  {title} {\enquote
  {\bibinfo {title} {{Bayesian Inference for Gravitational Waves from Binary
  Neutron Star Mergers in Third Generation Observatories}},}\ }\href {\doibase
  10.1103/PhysRevLett.127.081102} {\bibfield  {journal} {\bibinfo  {journal}
  {\prl}\ }\textbf {\bibinfo {volume} {127}},\ \bibinfo {eid} {081102}
  (\bibinfo {year} {2021})},\ \Eprint {http://arxiv.org/abs/2103.12274}
  {arXiv:2103.12274 [gr-qc]} \BibitemShut {NoStop}%
\bibitem [{\citenamefont {Samajdar}\ and\ \citenamefont
  {Dietrich}(2020)}]{Samajdar:2020xrd}%
  \BibitemOpen
  \bibfield  {author} {\bibinfo {author} {\bibfnamefont {Anuradha}\
  \bibnamefont {Samajdar}}\ and\ \bibinfo {author} {\bibfnamefont {Tim}\
  \bibnamefont {Dietrich}},\ }\bibfield  {title} {\enquote {\bibinfo {title}
  {{Constructing Love-Q-Relations with Gravitational Wave Detections}},}\
  }\href {\doibase 10.1103/PhysRevD.101.124014} {\bibfield  {journal} {\bibinfo
   {journal} {Phys. Rev. D}\ }\textbf {\bibinfo {volume} {101}},\ \bibinfo
  {pages} {124014} (\bibinfo {year} {2020})},\ \Eprint
  {http://arxiv.org/abs/2002.07918} {arXiv:2002.07918 [gr-qc]} \BibitemShut
  {NoStop}%
\bibitem [{\citenamefont {Yagi}\ and\ \citenamefont
  {Yunes}(2017{\natexlab{a}})}]{Yagi:2016qmr}%
  \BibitemOpen
  \bibfield  {author} {\bibinfo {author} {\bibfnamefont {Kent}\ \bibnamefont
  {Yagi}}\ and\ \bibinfo {author} {\bibfnamefont {Nicolas}\ \bibnamefont
  {Yunes}},\ }\bibfield  {title} {\enquote {\bibinfo {title} {{Approximate
  Universal Relations among Tidal Parameters for Neutron Star Binaries}},}\
  }\href {\doibase 10.1088/1361-6382/34/1/015006} {\bibfield  {journal}
  {\bibinfo  {journal} {Class. Quant. Grav.}\ }\textbf {\bibinfo {volume}
  {34}},\ \bibinfo {pages} {015006} (\bibinfo {year} {2017}{\natexlab{a}})},\
  \Eprint {http://arxiv.org/abs/1608.06187} {arXiv:1608.06187 [gr-qc]}
  \BibitemShut {NoStop}%
\bibitem [{\citenamefont {Urbanec}\ \emph {et~al.}(2013)\citenamefont
  {Urbanec}, \citenamefont {Miller},\ and\ \citenamefont
  {Stuchlik}}]{Urbanec:2013fs}%
  \BibitemOpen
  \bibfield  {author} {\bibinfo {author} {\bibfnamefont {Martin}\ \bibnamefont
  {Urbanec}}, \bibinfo {author} {\bibfnamefont {John~C.}\ \bibnamefont
  {Miller}}, \ and\ \bibinfo {author} {\bibfnamefont {Zdenek}\ \bibnamefont
  {Stuchlik}},\ }\bibfield  {title} {\enquote {\bibinfo {title} {{Quadrupole
  moments of rotating neutron stars and strange stars}},}\ }\href {\doibase
  10.1093/mnras/stt858} {\bibfield  {journal} {\bibinfo  {journal} {Mon. Not.
  Roy. Astron. Soc.}\ }\textbf {\bibinfo {volume} {433}},\ \bibinfo {pages}
  {1903} (\bibinfo {year} {2013})},\ \Eprint {http://arxiv.org/abs/1301.5925}
  {arXiv:1301.5925 [astro-ph.SR]} \BibitemShut {NoStop}%
\bibitem [{\citenamefont {Yagi}\ and\ \citenamefont
  {Yunes}(2017{\natexlab{b}})}]{Yagi:2016bkt}%
  \BibitemOpen
  \bibfield  {author} {\bibinfo {author} {\bibfnamefont {Kent}\ \bibnamefont
  {Yagi}}\ and\ \bibinfo {author} {\bibfnamefont {Nicol\'as}\ \bibnamefont
  {Yunes}},\ }\bibfield  {title} {\enquote {\bibinfo {title} {{Approximate
  Universal Relations for Neutron Stars and Quark Stars}},}\ }\href {\doibase
  10.1016/j.physrep.2017.03.002} {\bibfield  {journal} {\bibinfo  {journal}
  {Phys. Rept.}\ }\textbf {\bibinfo {volume} {681}},\ \bibinfo {pages} {1--72}
  (\bibinfo {year} {2017}{\natexlab{b}})},\ \Eprint
  {http://arxiv.org/abs/1608.02582} {arXiv:1608.02582 [gr-qc]} \BibitemShut
  {NoStop}%
\bibitem [{\citenamefont {Haskell}\ \emph {et~al.}(2014)\citenamefont
  {Haskell}, \citenamefont {Ciolfi}, \citenamefont {Pannarale},\ and\
  \citenamefont {Rezzolla}}]{Haskell:2013vha}%
  \BibitemOpen
  \bibfield  {author} {\bibinfo {author} {\bibfnamefont {Brynmor}\ \bibnamefont
  {Haskell}}, \bibinfo {author} {\bibfnamefont {Riccardo}\ \bibnamefont
  {Ciolfi}}, \bibinfo {author} {\bibfnamefont {Francesco}\ \bibnamefont
  {Pannarale}}, \ and\ \bibinfo {author} {\bibfnamefont {Luciano}\ \bibnamefont
  {Rezzolla}},\ }\bibfield  {title} {\enquote {\bibinfo {title} {{On the
  universality of I-Love-Q relations in magnetized neutron stars}},}\ }\href
  {\doibase 10.1093/mnrasl/slt161} {\bibfield  {journal} {\bibinfo  {journal}
  {Mon. Not. Roy. Astron. Soc.}\ }\textbf {\bibinfo {volume} {438}},\ \bibinfo
  {pages} {L71--L75} (\bibinfo {year} {2014})},\ \Eprint
  {http://arxiv.org/abs/1309.3885} {arXiv:1309.3885 [astro-ph.SR]} \BibitemShut
  {NoStop}%
\bibitem [{\citenamefont {{Buonanno}}\ \emph {et~al.}(2009)\citenamefont
  {{Buonanno}}, \citenamefont {{Iyer}}, \citenamefont {{Ochsner}},
  \citenamefont {{Pan}},\ and\ \citenamefont {{Sathyaprakash}}}]{Buonanno2009}%
  \BibitemOpen
  \bibfield  {author} {\bibinfo {author} {\bibfnamefont {Alessandra}\
  \bibnamefont {{Buonanno}}}, \bibinfo {author} {\bibfnamefont {Bala~R.}\
  \bibnamefont {{Iyer}}}, \bibinfo {author} {\bibfnamefont {Evan}\ \bibnamefont
  {{Ochsner}}}, \bibinfo {author} {\bibfnamefont {Yi}~\bibnamefont {{Pan}}}, \
  and\ \bibinfo {author} {\bibfnamefont {B.~S.}\ \bibnamefont
  {{Sathyaprakash}}},\ }\bibfield  {title} {\enquote {\bibinfo {title}
  {{Comparison of post-Newtonian templates for compact binary inspiral signals
  in gravitational-wave detectors}},}\ }\href {\doibase
  10.1103/PhysRevD.80.084043} {\bibfield  {journal} {\bibinfo  {journal}
  {\prd}\ }\textbf {\bibinfo {volume} {80}},\ \bibinfo {eid} {084043} (\bibinfo
  {year} {2009})},\ \Eprint {http://arxiv.org/abs/0907.0700} {arXiv:0907.0700
  [gr-qc]} \BibitemShut {NoStop}%
\bibitem [{\citenamefont {{Flanagan}}\ and\ \citenamefont
  {{Hinderer}}(2008)}]{Flanagan2008}%
  \BibitemOpen
  \bibfield  {author} {\bibinfo {author} {\bibfnamefont {{\'E}anna~{\'E}.}\
  \bibnamefont {{Flanagan}}}\ and\ \bibinfo {author} {\bibfnamefont {Tanja}\
  \bibnamefont {{Hinderer}}},\ }\bibfield  {title} {\enquote {\bibinfo {title}
  {{Constraining neutron-star tidal Love numbers with gravitational-wave
  detectors}},}\ }\href {\doibase 10.1103/PhysRevD.77.021502} {\bibfield
  {journal} {\bibinfo  {journal} {\prd}\ }\textbf {\bibinfo {volume} {77}},\
  \bibinfo {eid} {021502} (\bibinfo {year} {2008})},\ \Eprint
  {http://arxiv.org/abs/0709.1915} {arXiv:0709.1915 [astro-ph]} \BibitemShut
  {NoStop}%
\bibitem [{\citenamefont {{Hinderer}}\ \emph {et~al.}(2010)\citenamefont
  {{Hinderer}}, \citenamefont {{Lackey}}, \citenamefont {{Lang}},\ and\
  \citenamefont {{Read}}}]{Hinderer2010}%
  \BibitemOpen
  \bibfield  {author} {\bibinfo {author} {\bibfnamefont {Tanja}\ \bibnamefont
  {{Hinderer}}}, \bibinfo {author} {\bibfnamefont {Benjamin~D.}\ \bibnamefont
  {{Lackey}}}, \bibinfo {author} {\bibfnamefont {Ryan~N.}\ \bibnamefont
  {{Lang}}}, \ and\ \bibinfo {author} {\bibfnamefont {Jocelyn~S.}\ \bibnamefont
  {{Read}}},\ }\bibfield  {title} {\enquote {\bibinfo {title} {{Tidal
  deformability of neutron stars with realistic equations of state and their
  gravitational wave signatures in binary inspiral}},}\ }\href {\doibase
  10.1103/PhysRevD.81.123016} {\bibfield  {journal} {\bibinfo  {journal}
  {\prd}\ }\textbf {\bibinfo {volume} {81}},\ \bibinfo {eid} {123016} (\bibinfo
  {year} {2010})},\ \Eprint {http://arxiv.org/abs/0911.3535} {arXiv:0911.3535
  [astro-ph.HE]} \BibitemShut {NoStop}%
\bibitem [{lig(2018)}]{ligoA}%
  \BibitemOpen
  \bibfield  {title} {\enquote {\bibinfo {title} {{The A plus design curve}},}\
  }\href@noop {} {\bibfield  {journal} {\bibinfo  {journal} {Tech. Report No.
  LIGO-T1800042}\ } (\bibinfo {year} {2018})}\BibitemShut {NoStop}%
\bibitem [{\citenamefont {{Evans}}\ \emph {et~al.}(2021)\citenamefont
  {{Evans}}, \citenamefont {{Adhikari}}, \citenamefont {{Afle}}, \citenamefont
  {{Ballmer}}, \citenamefont {{Biscoveanu}}, \citenamefont {{Borhanian}},
  \citenamefont {{Brown}}, \citenamefont {{Chen}}, \citenamefont
  {{Eisenstein}}, \citenamefont {{Gruson}}, \citenamefont {{Gupta}},
  \citenamefont {{Hall}}, \citenamefont {{Huxford}}, \citenamefont {{Kamai}},
  \citenamefont {{Kashyap}}, \citenamefont {{Kissel}}, \citenamefont {{Kuns}},
  \citenamefont {{Landry}}, \citenamefont {{Lenon}}, \citenamefont
  {{Lovelace}}, \citenamefont {{McCuller}}, \citenamefont {{Ng}}, \citenamefont
  {{Nitz}}, \citenamefont {{Read}}, \citenamefont {{Sathyaprakash}},
  \citenamefont {{Shoemaker}}, \citenamefont {{Slagmolen}}, \citenamefont
  {{Smith}}, \citenamefont {{Srivastava}}, \citenamefont {{Sun}}, \citenamefont
  {{Vitale}},\ and\ \citenamefont {{Weiss}}}]{Evans2021}%
  \BibitemOpen
  \bibfield  {author} {\bibinfo {author} {\bibfnamefont {Matthew}\ \bibnamefont
  {{Evans}}}, \bibinfo {author} {\bibfnamefont {Rana~X}\ \bibnamefont
  {{Adhikari}}}, \bibinfo {author} {\bibfnamefont {Chaitanya}\ \bibnamefont
  {{Afle}}}, \bibinfo {author} {\bibfnamefont {Stefan~W.}\ \bibnamefont
  {{Ballmer}}}, \bibinfo {author} {\bibfnamefont {Sylvia}\ \bibnamefont
  {{Biscoveanu}}}, \bibinfo {author} {\bibfnamefont {Ssohrab}\ \bibnamefont
  {{Borhanian}}}, \bibinfo {author} {\bibfnamefont {Duncan~A.}\ \bibnamefont
  {{Brown}}}, \bibinfo {author} {\bibfnamefont {Yanbei}\ \bibnamefont
  {{Chen}}}, \bibinfo {author} {\bibfnamefont {Robert}\ \bibnamefont
  {{Eisenstein}}}, \bibinfo {author} {\bibfnamefont {Alexandra}\ \bibnamefont
  {{Gruson}}}, \bibinfo {author} {\bibfnamefont {Anuradha}\ \bibnamefont
  {{Gupta}}}, \bibinfo {author} {\bibfnamefont {Evan~D.}\ \bibnamefont
  {{Hall}}}, \bibinfo {author} {\bibfnamefont {Rachael}\ \bibnamefont
  {{Huxford}}}, \bibinfo {author} {\bibfnamefont {Brittany}\ \bibnamefont
  {{Kamai}}}, \bibinfo {author} {\bibfnamefont {Rahul}\ \bibnamefont
  {{Kashyap}}}, \bibinfo {author} {\bibfnamefont {Jeff~S.}\ \bibnamefont
  {{Kissel}}}, \bibinfo {author} {\bibfnamefont {Kevin}\ \bibnamefont
  {{Kuns}}}, \bibinfo {author} {\bibfnamefont {Philippe}\ \bibnamefont
  {{Landry}}}, \bibinfo {author} {\bibfnamefont {Amber}\ \bibnamefont
  {{Lenon}}}, \bibinfo {author} {\bibfnamefont {Geoffrey}\ \bibnamefont
  {{Lovelace}}}, \bibinfo {author} {\bibfnamefont {Lee}\ \bibnamefont
  {{McCuller}}}, \bibinfo {author} {\bibfnamefont {Ken K.~Y.}\ \bibnamefont
  {{Ng}}}, \bibinfo {author} {\bibfnamefont {Alexander~H.}\ \bibnamefont
  {{Nitz}}}, \bibinfo {author} {\bibfnamefont {Jocelyn}\ \bibnamefont
  {{Read}}}, \bibinfo {author} {\bibfnamefont {B.~S.}\ \bibnamefont
  {{Sathyaprakash}}}, \bibinfo {author} {\bibfnamefont {David~H.}\ \bibnamefont
  {{Shoemaker}}}, \bibinfo {author} {\bibfnamefont {Bram J.~J.}\ \bibnamefont
  {{Slagmolen}}}, \bibinfo {author} {\bibfnamefont {Joshua~R.}\ \bibnamefont
  {{Smith}}}, \bibinfo {author} {\bibfnamefont {Varun}\ \bibnamefont
  {{Srivastava}}}, \bibinfo {author} {\bibfnamefont {Ling}\ \bibnamefont
  {{Sun}}}, \bibinfo {author} {\bibfnamefont {Salvatore}\ \bibnamefont
  {{Vitale}}}, \ and\ \bibinfo {author} {\bibfnamefont {Rainer}\ \bibnamefont
  {{Weiss}}},\ }\bibfield  {title} {\enquote {\bibinfo {title} {{A Horizon
  Study for Cosmic Explorer: Science, Observatories, and Community}},}\
  }\href@noop {} {\bibfield  {journal} {\bibinfo  {journal} {arXiv e-prints}\
  ,\ \bibinfo {eid} {arXiv:2109.09882}} (\bibinfo {year} {2021})},\ \Eprint
  {http://arxiv.org/abs/2109.09882} {arXiv:2109.09882 [astro-ph.IM]}
  \BibitemShut {NoStop}%
\bibitem [{\citenamefont {Yagi}\ and\ \citenamefont
  {Yunes}(2013{\natexlab{a}})}]{Yagi:2013bca}%
  \BibitemOpen
  \bibfield  {author} {\bibinfo {author} {\bibfnamefont {Kent}\ \bibnamefont
  {Yagi}}\ and\ \bibinfo {author} {\bibfnamefont {Nicolas}\ \bibnamefont
  {Yunes}},\ }\bibfield  {title} {\enquote {\bibinfo {title} {{I-Love-Q}},}\
  }\href {\doibase 10.1126/science.1236462} {\bibfield  {journal} {\bibinfo
  {journal} {Science}\ }\textbf {\bibinfo {volume} {341}},\ \bibinfo {pages}
  {365--368} (\bibinfo {year} {2013}{\natexlab{a}})},\ \Eprint
  {http://arxiv.org/abs/1302.4499} {arXiv:1302.4499 [gr-qc]} \BibitemShut
  {NoStop}%
\bibitem [{\citenamefont {Yagi}\ and\ \citenamefont
  {Yunes}(2013{\natexlab{b}})}]{Yagi:2013awa}%
  \BibitemOpen
  \bibfield  {author} {\bibinfo {author} {\bibfnamefont {Kent}\ \bibnamefont
  {Yagi}}\ and\ \bibinfo {author} {\bibfnamefont {Nicolas}\ \bibnamefont
  {Yunes}},\ }\bibfield  {title} {\enquote {\bibinfo {title} {{I-Love-Q
  Relations in Neutron Stars and their Applications to Astrophysics,
  Gravitational Waves and Fundamental Physics}},}\ }\href {\doibase
  10.1103/PhysRevD.88.023009} {\bibfield  {journal} {\bibinfo  {journal} {Phys.
  Rev. D}\ }\textbf {\bibinfo {volume} {88}},\ \bibinfo {pages} {023009}
  (\bibinfo {year} {2013}{\natexlab{b}})},\ \Eprint
  {http://arxiv.org/abs/1303.1528} {arXiv:1303.1528 [gr-qc]} \BibitemShut
  {NoStop}%
\bibitem [{\citenamefont {Gupta}\ \emph {et~al.}(2018)\citenamefont {Gupta},
  \citenamefont {Majumder}, \citenamefont {Yagi},\ and\ \citenamefont
  {Yunes}}]{Gupta:2017vsl}%
  \BibitemOpen
  \bibfield  {author} {\bibinfo {author} {\bibfnamefont {Toral}\ \bibnamefont
  {Gupta}}, \bibinfo {author} {\bibfnamefont {Barun}\ \bibnamefont {Majumder}},
  \bibinfo {author} {\bibfnamefont {Kent}\ \bibnamefont {Yagi}}, \ and\
  \bibinfo {author} {\bibfnamefont {Nicol\'as}\ \bibnamefont {Yunes}},\
  }\bibfield  {title} {\enquote {\bibinfo {title} {{I-Love-Q Relations for
  Neutron Stars in dynamical Chern Simons Gravity}},}\ }\href {\doibase
  10.1088/1361-6382/aa9c68} {\bibfield  {journal} {\bibinfo  {journal} {Class.
  Quant. Grav.}\ }\textbf {\bibinfo {volume} {35}},\ \bibinfo {pages} {025009}
  (\bibinfo {year} {2018})},\ \Eprint {http://arxiv.org/abs/1710.07862}
  {arXiv:1710.07862 [gr-qc]} \BibitemShut {NoStop}%
\bibitem [{\citenamefont {Silva}\ \emph {et~al.}(2021)\citenamefont {Silva},
  \citenamefont {Holgado}, \citenamefont {C\'ardenas-Avenda\~no},\ and\
  \citenamefont {Yunes}}]{Silva:2020acr}%
  \BibitemOpen
  \bibfield  {author} {\bibinfo {author} {\bibfnamefont {Hector~O.}\
  \bibnamefont {Silva}}, \bibinfo {author} {\bibfnamefont {A.~Miguel}\
  \bibnamefont {Holgado}}, \bibinfo {author} {\bibfnamefont {Alejandro}\
  \bibnamefont {C\'ardenas-Avenda\~no}}, \ and\ \bibinfo {author}
  {\bibfnamefont {Nicol\'as}\ \bibnamefont {Yunes}},\ }\bibfield  {title}
  {\enquote {\bibinfo {title} {{Astrophysical and theoretical physics
  implications from multimessenger neutron star observations}},}\ }\href
  {\doibase 10.1103/PhysRevLett.126.181101} {\bibfield  {journal} {\bibinfo
  {journal} {Phys. Rev. Lett.}\ }\textbf {\bibinfo {volume} {126}},\ \bibinfo
  {pages} {181101} (\bibinfo {year} {2021})},\ \Eprint
  {http://arxiv.org/abs/2004.01253} {arXiv:2004.01253 [gr-qc]} \BibitemShut
  {NoStop}%
\bibitem [{\citenamefont {Zhu}\ \emph {et~al.}(2020)\citenamefont {Zhu},
  \citenamefont {Li},\ and\ \citenamefont {Rezzolla}}]{Zhu:2020imp}%
  \BibitemOpen
  \bibfield  {author} {\bibinfo {author} {\bibfnamefont {Zhenyu}\ \bibnamefont
  {Zhu}}, \bibinfo {author} {\bibfnamefont {Ang}\ \bibnamefont {Li}}, \ and\
  \bibinfo {author} {\bibfnamefont {Luciano}\ \bibnamefont {Rezzolla}},\
  }\bibfield  {title} {\enquote {\bibinfo {title} {{Tidal deformability and
  gravitational-wave phase evolution of magnetized compact-star binaries}},}\
  }\href {\doibase 10.1103/PhysRevD.102.084058} {\bibfield  {journal} {\bibinfo
   {journal} {Phys. Rev. D}\ }\textbf {\bibinfo {volume} {102}},\ \bibinfo
  {pages} {084058} (\bibinfo {year} {2020})},\ \Eprint
  {http://arxiv.org/abs/2005.02677} {arXiv:2005.02677 [astro-ph.HE]}
  \BibitemShut {NoStop}%
\bibitem [{\citenamefont {Messenger}\ and\ \citenamefont
  {Read}(2012)}]{Messenger:2011gi}%
  \BibitemOpen
  \bibfield  {author} {\bibinfo {author} {\bibfnamefont {C.}~\bibnamefont
  {Messenger}}\ and\ \bibinfo {author} {\bibfnamefont {J.}~\bibnamefont
  {Read}},\ }\bibfield  {title} {\enquote {\bibinfo {title} {{Measuring a
  cosmological distance-redshift relationship using only gravitational wave
  observations of binary neutron star coalescences}},}\ }\href {\doibase
  10.1103/PhysRevLett.108.091101} {\bibfield  {journal} {\bibinfo  {journal}
  {Phys. Rev. Lett.}\ }\textbf {\bibinfo {volume} {108}},\ \bibinfo {pages}
  {091101} (\bibinfo {year} {2012})},\ \Eprint {http://arxiv.org/abs/1107.5725}
  {arXiv:1107.5725 [gr-qc]} \BibitemShut {NoStop}%
\bibitem [{\citenamefont {Yagi}\ and\ \citenamefont
  {Yunes}(2016)}]{Yagi:2015pkc}%
  \BibitemOpen
  \bibfield  {author} {\bibinfo {author} {\bibfnamefont {Kent}\ \bibnamefont
  {Yagi}}\ and\ \bibinfo {author} {\bibfnamefont {Nicolas}\ \bibnamefont
  {Yunes}},\ }\bibfield  {title} {\enquote {\bibinfo {title} {{Binary Love
  Relations}},}\ }\href {\doibase 10.1088/0264-9381/33/13/13LT01} {\bibfield
  {journal} {\bibinfo  {journal} {Class. Quant. Grav.}\ }\textbf {\bibinfo
  {volume} {33}},\ \bibinfo {pages} {13LT01} (\bibinfo {year} {2016})},\
  \Eprint {http://arxiv.org/abs/1512.02639} {arXiv:1512.02639 [gr-qc]}
  \BibitemShut {NoStop}%
\bibitem [{\citenamefont {Jackiw}\ and\ \citenamefont
  {Pi}(2003)}]{Jackiw:2003pm}%
  \BibitemOpen
  \bibfield  {author} {\bibinfo {author} {\bibfnamefont {R.}~\bibnamefont
  {Jackiw}}\ and\ \bibinfo {author} {\bibfnamefont {S.~Y.}\ \bibnamefont
  {Pi}},\ }\bibfield  {title} {\enquote {\bibinfo {title} {{Chern-Simons
  modification of general relativity}},}\ }\href {\doibase
  10.1103/PhysRevD.68.104012} {\bibfield  {journal} {\bibinfo  {journal} {Phys.
  Rev. D}\ }\textbf {\bibinfo {volume} {68}},\ \bibinfo {pages} {104012}
  (\bibinfo {year} {2003})},\ \Eprint {http://arxiv.org/abs/gr-qc/0308071}
  {arXiv:gr-qc/0308071} \BibitemShut {NoStop}%
\bibitem [{\citenamefont {Alexander}\ and\ \citenamefont
  {Yunes}(2009)}]{Alexander:2009tp}%
  \BibitemOpen
  \bibfield  {author} {\bibinfo {author} {\bibfnamefont {Stephon}\ \bibnamefont
  {Alexander}}\ and\ \bibinfo {author} {\bibfnamefont {Nicolas}\ \bibnamefont
  {Yunes}},\ }\bibfield  {title} {\enquote {\bibinfo {title} {{Chern-Simons
  Modified General Relativity}},}\ }\href {\doibase
  10.1016/j.physrep.2009.07.002} {\bibfield  {journal} {\bibinfo  {journal}
  {Phys. Rept.}\ }\textbf {\bibinfo {volume} {480}},\ \bibinfo {pages} {1--55}
  (\bibinfo {year} {2009})},\ \Eprint {http://arxiv.org/abs/0907.2562}
  {arXiv:0907.2562 [hep-th]} \BibitemShut {NoStop}%
\bibitem [{\citenamefont {Ali-Haimoud}\ and\ \citenamefont
  {Chen}(2011)}]{Ali-Haimoud:2011zme}%
  \BibitemOpen
  \bibfield  {author} {\bibinfo {author} {\bibfnamefont {Yacine}\ \bibnamefont
  {Ali-Haimoud}}\ and\ \bibinfo {author} {\bibfnamefont {Yanbei}\ \bibnamefont
  {Chen}},\ }\bibfield  {title} {\enquote {\bibinfo {title} {{Slowly-rotating
  stars and black holes in dynamical Chern-Simons gravity}},}\ }\href {\doibase
  10.1103/PhysRevD.84.124033} {\bibfield  {journal} {\bibinfo  {journal} {Phys.
  Rev. D}\ }\textbf {\bibinfo {volume} {84}},\ \bibinfo {pages} {124033}
  (\bibinfo {year} {2011})},\ \Eprint {http://arxiv.org/abs/1110.5329}
  {arXiv:1110.5329 [astro-ph.HE]} \BibitemShut {NoStop}%
\bibitem [{\citenamefont {Yagi}\ \emph {et~al.}(2012)\citenamefont {Yagi},
  \citenamefont {Yunes},\ and\ \citenamefont {Tanaka}}]{Yagi:2012ya}%
  \BibitemOpen
  \bibfield  {author} {\bibinfo {author} {\bibfnamefont {Kent}\ \bibnamefont
  {Yagi}}, \bibinfo {author} {\bibfnamefont {Nicolas}\ \bibnamefont {Yunes}}, \
  and\ \bibinfo {author} {\bibfnamefont {Takahiro}\ \bibnamefont {Tanaka}},\
  }\bibfield  {title} {\enquote {\bibinfo {title} {{Slowly Rotating Black Holes
  in Dynamical Chern-Simons Gravity: Deformation Quadratic in the Spin}},}\
  }\href {\doibase 10.1103/PhysRevD.86.044037} {\bibfield  {journal} {\bibinfo
  {journal} {Phys. Rev. D}\ }\textbf {\bibinfo {volume} {86}},\ \bibinfo
  {pages} {044037} (\bibinfo {year} {2012})},\ \bibinfo {note} {[Erratum:
  Phys.Rev.D 89, 049902 (2014)]},\ \Eprint {http://arxiv.org/abs/1206.6130}
  {arXiv:1206.6130 [gr-qc]} \BibitemShut {NoStop}%
\bibitem [{\citenamefont {Silva}\ \emph {et~al.}(2022)\citenamefont {Silva},
  \citenamefont {Ghosh},\ and\ \citenamefont {Buonanno}}]{Silva:2022srr}%
  \BibitemOpen
  \bibfield  {author} {\bibinfo {author} {\bibfnamefont {Hector~O.}\
  \bibnamefont {Silva}}, \bibinfo {author} {\bibfnamefont {Abhirup}\
  \bibnamefont {Ghosh}}, \ and\ \bibinfo {author} {\bibfnamefont {Alessandra}\
  \bibnamefont {Buonanno}},\ }\bibfield  {title} {\enquote {\bibinfo {title}
  {{Black-hole ringdown as a probe of higher-curvature gravity theories}},}\
  }\href@noop {} {\  (\bibinfo {year} {2022})},\ \Eprint
  {http://arxiv.org/abs/2205.05132} {arXiv:2205.05132 [gr-qc]} \BibitemShut
  {NoStop}%
\bibitem [{\citenamefont {Saffer}\ and\ \citenamefont
  {Yagi}(2021)}]{Saffer:2021gak}%
  \BibitemOpen
  \bibfield  {author} {\bibinfo {author} {\bibfnamefont {Alexander}\
  \bibnamefont {Saffer}}\ and\ \bibinfo {author} {\bibfnamefont {Kent}\
  \bibnamefont {Yagi}},\ }\bibfield  {title} {\enquote {\bibinfo {title}
  {{Tidal deformabilities of neutron stars in scalar-Gauss-Bonnet gravity and
  their applications to multimessenger tests of gravity}},}\ }\href {\doibase
  10.1103/PhysRevD.104.124052} {\bibfield  {journal} {\bibinfo  {journal}
  {Phys. Rev. D}\ }\textbf {\bibinfo {volume} {104}},\ \bibinfo {pages}
  {124052} (\bibinfo {year} {2021})},\ \Eprint
  {http://arxiv.org/abs/2110.02997} {arXiv:2110.02997 [gr-qc]} \BibitemShut
  {NoStop}%
\bibitem [{\citenamefont {Yagi}\ \emph {et~al.}(2013)\citenamefont {Yagi},
  \citenamefont {Stein}, \citenamefont {Yunes},\ and\ \citenamefont
  {Tanaka}}]{Yagi:2013mbt}%
  \BibitemOpen
  \bibfield  {author} {\bibinfo {author} {\bibfnamefont {Kent}\ \bibnamefont
  {Yagi}}, \bibinfo {author} {\bibfnamefont {Leo~C.}\ \bibnamefont {Stein}},
  \bibinfo {author} {\bibfnamefont {Nicolas}\ \bibnamefont {Yunes}}, \ and\
  \bibinfo {author} {\bibfnamefont {Takahiro}\ \bibnamefont {Tanaka}},\
  }\bibfield  {title} {\enquote {\bibinfo {title} {{Isolated and Binary Neutron
  Stars in Dynamical Chern-Simons Gravity}},}\ }\href {\doibase
  10.1103/PhysRevD.87.084058} {\bibfield  {journal} {\bibinfo  {journal} {Phys.
  Rev. D}\ }\textbf {\bibinfo {volume} {87}},\ \bibinfo {pages} {084058}
  (\bibinfo {year} {2013})},\ \bibinfo {note} {[Erratum: Phys.Rev.D 93, 089909
  (2016)]},\ \Eprint {http://arxiv.org/abs/1302.1918} {arXiv:1302.1918 [gr-qc]}
  \BibitemShut {NoStop}%
\bibitem [{\citenamefont {Maselli}\ \emph {et~al.}(2013)\citenamefont
  {Maselli}, \citenamefont {Cardoso}, \citenamefont {Ferrari}, \citenamefont
  {Gualtieri},\ and\ \citenamefont {Pani}}]{Maselli:2013mva}%
  \BibitemOpen
  \bibfield  {author} {\bibinfo {author} {\bibfnamefont {Andrea}\ \bibnamefont
  {Maselli}}, \bibinfo {author} {\bibfnamefont {Vitor}\ \bibnamefont
  {Cardoso}}, \bibinfo {author} {\bibfnamefont {Valeria}\ \bibnamefont
  {Ferrari}}, \bibinfo {author} {\bibfnamefont {Leonardo}\ \bibnamefont
  {Gualtieri}}, \ and\ \bibinfo {author} {\bibfnamefont {Paolo}\ \bibnamefont
  {Pani}},\ }\bibfield  {title} {\enquote {\bibinfo {title}
  {{Equation-of-state-independent relations in neutron stars}},}\ }\href
  {\doibase 10.1103/PhysRevD.88.023007} {\bibfield  {journal} {\bibinfo
  {journal} {Phys. Rev. D}\ }\textbf {\bibinfo {volume} {88}},\ \bibinfo
  {pages} {023007} (\bibinfo {year} {2013})},\ \Eprint
  {http://arxiv.org/abs/1304.2052} {arXiv:1304.2052 [gr-qc]} \BibitemShut
  {NoStop}%
\bibitem [{\citenamefont {{Svrcek}}\ and\ \citenamefont
  {{Witten}}(2006)}]{Svrcek2006}%
  \BibitemOpen
  \bibfield  {author} {\bibinfo {author} {\bibfnamefont {Peter}\ \bibnamefont
  {{Svrcek}}}\ and\ \bibinfo {author} {\bibfnamefont {Edward}\ \bibnamefont
  {{Witten}}},\ }\bibfield  {title} {\enquote {\bibinfo {title} {{Axions in
  string theory}},}\ }\href {\doibase 10.1088/1126-6708/2006/06/051} {\bibfield
   {journal} {\bibinfo  {journal} {Journal of High Energy Physics}\ }\textbf
  {\bibinfo {volume} {2006}},\ \bibinfo {eid} {051} (\bibinfo {year} {2006})},\
  \Eprint {http://arxiv.org/abs/hep-th/0605206} {arXiv:hep-th/0605206 [hep-th]}
  \BibitemShut {NoStop}%
\bibitem [{\citenamefont {{Arvanitaki}}\ \emph {et~al.}(2010)\citenamefont
  {{Arvanitaki}}, \citenamefont {{Dimopoulos}}, \citenamefont {{Dubovsky}},
  \citenamefont {{Kaloper}},\ and\ \citenamefont {{March-Russell}}}]{Mina2010}%
  \BibitemOpen
  \bibfield  {author} {\bibinfo {author} {\bibfnamefont {Asimina}\ \bibnamefont
  {{Arvanitaki}}}, \bibinfo {author} {\bibfnamefont {Savas}\ \bibnamefont
  {{Dimopoulos}}}, \bibinfo {author} {\bibfnamefont {Sergei}\ \bibnamefont
  {{Dubovsky}}}, \bibinfo {author} {\bibfnamefont {Nemanja}\ \bibnamefont
  {{Kaloper}}}, \ and\ \bibinfo {author} {\bibfnamefont {John}\ \bibnamefont
  {{March-Russell}}},\ }\bibfield  {title} {\enquote {\bibinfo {title} {{String
  axiverse}},}\ }\href {\doibase 10.1103/PhysRevD.81.123530} {\bibfield
  {journal} {\bibinfo  {journal} {\prd}\ }\textbf {\bibinfo {volume} {81}},\
  \bibinfo {eid} {123530} (\bibinfo {year} {2010})},\ \Eprint
  {http://arxiv.org/abs/0905.4720} {arXiv:0905.4720 [hep-th]} \BibitemShut
  {NoStop}%
\bibitem [{\citenamefont {{Nicholl}}\ \emph {et~al.}(2017)\citenamefont
  {{Nicholl}}, \citenamefont {{Williams}}, \citenamefont {{Berger}},
  \citenamefont {{Villar}}, \citenamefont {{Alexander}}, \citenamefont
  {{Eftekhari}},\ and\ \citenamefont {{Metzger}}}]{Nicholl2017}%
  \BibitemOpen
  \bibfield  {author} {\bibinfo {author} {\bibfnamefont {M.}~\bibnamefont
  {{Nicholl}}}, \bibinfo {author} {\bibfnamefont {P.~K.~G.}\ \bibnamefont
  {{Williams}}}, \bibinfo {author} {\bibfnamefont {E.}~\bibnamefont
  {{Berger}}}, \bibinfo {author} {\bibfnamefont {V.~A.}\ \bibnamefont
  {{Villar}}}, \bibinfo {author} {\bibfnamefont {K.~D.}\ \bibnamefont
  {{Alexander}}}, \bibinfo {author} {\bibfnamefont {T.}~\bibnamefont
  {{Eftekhari}}}, \ and\ \bibinfo {author} {\bibfnamefont {B.~D.}\ \bibnamefont
  {{Metzger}}},\ }\bibfield  {title} {\enquote {\bibinfo {title} {{Empirical
  Constraints on the Origin of Fast Radio Bursts: Volumetric Rates and Host
  Galaxy Demographics as a Test of Millisecond Magnetar Connection}},}\ }\href
  {\doibase 10.3847/1538-4357/aa794d} {\bibfield  {journal} {\bibinfo
  {journal} {\apj}\ }\textbf {\bibinfo {volume} {843}},\ \bibinfo {eid} {84}
  (\bibinfo {year} {2017})},\ \Eprint {http://arxiv.org/abs/1704.00022}
  {arXiv:1704.00022 [astro-ph.HE]} \BibitemShut {NoStop}%
\bibitem [{\citenamefont {Katz}(2021)}]{Katz:2020jpv}%
  \BibitemOpen
  \bibfield  {author} {\bibinfo {author} {\bibfnamefont {J.~I.}\ \bibnamefont
  {Katz}},\ }\bibfield  {title} {\enquote {\bibinfo {title} {{Testing Models of
  Periodically Modulated FRB Activity}},}\ }\href {\doibase
  10.1093/mnras/stab399} {\bibfield  {journal} {\bibinfo  {journal} {Mon. Not.
  Roy. Astron. Soc.}\ }\textbf {\bibinfo {volume} {502}},\ \bibinfo {pages}
  {4664--4668} (\bibinfo {year} {2021})},\ \Eprint
  {http://arxiv.org/abs/2012.15354} {arXiv:2012.15354 [astro-ph.HE]}
  \BibitemShut {NoStop}%
\bibitem [{\citenamefont {Wei}\ \emph {et~al.}(2022)\citenamefont {Wei},
  \citenamefont {Zhao},\ and\ \citenamefont {Wang}}]{Wei:2021vco}%
  \BibitemOpen
  \bibfield  {author} {\bibinfo {author} {\bibfnamefont {Yu-Jia}\ \bibnamefont
  {Wei}}, \bibinfo {author} {\bibfnamefont {Zhen-Yin}\ \bibnamefont {Zhao}}, \
  and\ \bibinfo {author} {\bibfnamefont {Fa-Yin}\ \bibnamefont {Wang}},\
  }\bibfield  {title} {\enquote {\bibinfo {title} {{The periodic origin of fast
  radio bursts}},}\ }\href {\doibase 10.1051/0004-6361/202142321} {\bibfield
  {journal} {\bibinfo  {journal} {Astron. Astrophys.}\ }\textbf {\bibinfo
  {volume} {658}},\ \bibinfo {pages} {A163} (\bibinfo {year} {2022})},\ \Eprint
  {http://arxiv.org/abs/2112.09292} {arXiv:2112.09292 [astro-ph.HE]}
  \BibitemShut {NoStop}%
\bibitem [{\citenamefont {Li}\ and\ \citenamefont
  {Zanazzi}(2021)}]{Li:2021rno}%
  \BibitemOpen
  \bibfield  {author} {\bibinfo {author} {\bibfnamefont {Dongzi}\ \bibnamefont
  {Li}}\ and\ \bibinfo {author} {\bibfnamefont {J.~J.}\ \bibnamefont
  {Zanazzi}},\ }\bibfield  {title} {\enquote {\bibinfo {title} {{Emission
  Properties of Periodic Fast Radio Bursts from the Motion of Magnetars:
  Testing Dynamical Models}},}\ }\href {\doibase 10.3847/2041-8213/abeaa4}
  {\bibfield  {journal} {\bibinfo  {journal} {Astrophys. J. Lett.}\ }\textbf
  {\bibinfo {volume} {909}},\ \bibinfo {pages} {L25} (\bibinfo {year}
  {2021})},\ \Eprint {http://arxiv.org/abs/2101.05836} {arXiv:2101.05836
  [astro-ph.HE]} \BibitemShut {NoStop}%
\bibitem [{\citenamefont {{Reitze}}\ \emph {et~al.}(2019)\citenamefont
  {{Reitze}}, \citenamefont {{LIGO Laboratory: California Institute of
  Technology}}, \citenamefont {{LIGO Laboratory: Massachusetts Institute of
  Technology}}, \citenamefont {{LIGO Hanford Observatory}},\ and\ \citenamefont
  {{LIGO Livingston Observatory}}}]{Reitze2019}%
  \BibitemOpen
  \bibfield  {author} {\bibinfo {author} {\bibfnamefont {David}\ \bibnamefont
  {{Reitze}}}, \bibinfo {author} {\bibnamefont {{LIGO Laboratory: California
  Institute of Technology}}}, \bibinfo {author} {\bibnamefont {{LIGO
  Laboratory: Massachusetts Institute of Technology}}}, \bibinfo {author}
  {\bibnamefont {{LIGO Hanford Observatory}}}, \ and\ \bibinfo {author}
  {\bibnamefont {{LIGO Livingston Observatory}}},\ }\bibfield  {title}
  {\enquote {\bibinfo {title} {{The US Program in Ground-Based Gravitational
  Wave Science: Contribution from the LIGO Laboratory}},}\ }\href@noop {}
  {\bibfield  {journal} {\bibinfo  {journal} {\baas}\ }\textbf {\bibinfo
  {volume} {51}},\ \bibinfo {eid} {141} (\bibinfo {year} {2019})},\ \Eprint
  {http://arxiv.org/abs/1903.04615} {arXiv:1903.04615 [astro-ph.IM]}
  \BibitemShut {NoStop}%
\bibitem [{\citenamefont {{Lu}}\ \emph {et~al.}(2022)\citenamefont {{Lu}},
  \citenamefont {{Beniamini}},\ and\ \citenamefont {{Kumar}}}]{Lu2022}%
  \BibitemOpen
  \bibfield  {author} {\bibinfo {author} {\bibfnamefont {Wenbin}\ \bibnamefont
  {{Lu}}}, \bibinfo {author} {\bibfnamefont {Paz}\ \bibnamefont {{Beniamini}}},
  \ and\ \bibinfo {author} {\bibfnamefont {Pawan}\ \bibnamefont {{Kumar}}},\
  }\bibfield  {title} {\enquote {\bibinfo {title} {{Implications of a rapidly
  varying FRB in a globular cluster of M81}},}\ }\href {\doibase
  10.1093/mnras/stab3500} {\bibfield  {journal} {\bibinfo  {journal} {\mnras}\
  }\textbf {\bibinfo {volume} {510}},\ \bibinfo {pages} {1867--1879} (\bibinfo
  {year} {2022})},\ \Eprint {http://arxiv.org/abs/2107.04059} {arXiv:2107.04059
  [astro-ph.HE]} \BibitemShut {NoStop}%
\bibitem [{\citenamefont {{Tendulkar}}\ \emph {et~al.}(2021)\citenamefont
  {{Tendulkar}}, \citenamefont {{Gil de Paz}}, \citenamefont {{Kirichenko}},
  \citenamefont {{Hessels}}, \citenamefont {{Bhardwaj}}, \citenamefont
  {{{\'A}vila}}, \citenamefont {{Bassa}}, \citenamefont {{Chawla}},
  \citenamefont {{Fonseca}}, \citenamefont {{Kaspi}}, \citenamefont
  {{Keimpema}}, \citenamefont {{Kirsten}}, \citenamefont {{Lazio}},
  \citenamefont {{Marcote}}, \citenamefont {{Masui}}, \citenamefont {{Nimmo}},
  \citenamefont {{Paragi}}, \citenamefont {{Rahman}}, \citenamefont
  {{Pay{\'a}}}, \citenamefont {{Scholz}},\ and\ \citenamefont
  {{Stairs}}}]{Tendulkar2021}%
  \BibitemOpen
  \bibfield  {author} {\bibinfo {author} {\bibfnamefont {Shriharsh~P.}\
  \bibnamefont {{Tendulkar}}}, \bibinfo {author} {\bibfnamefont {Armando}\
  \bibnamefont {{Gil de Paz}}}, \bibinfo {author} {\bibfnamefont {Aida~Yu.}\
  \bibnamefont {{Kirichenko}}}, \bibinfo {author} {\bibfnamefont {Jason W.~T.}\
  \bibnamefont {{Hessels}}}, \bibinfo {author} {\bibfnamefont {Mohit}\
  \bibnamefont {{Bhardwaj}}}, \bibinfo {author} {\bibfnamefont {Fernando}\
  \bibnamefont {{{\'A}vila}}}, \bibinfo {author} {\bibfnamefont {Cees}\
  \bibnamefont {{Bassa}}}, \bibinfo {author} {\bibfnamefont {Pragya}\
  \bibnamefont {{Chawla}}}, \bibinfo {author} {\bibfnamefont {Emmanuel}\
  \bibnamefont {{Fonseca}}}, \bibinfo {author} {\bibfnamefont {Victoria~M.}\
  \bibnamefont {{Kaspi}}}, \bibinfo {author} {\bibfnamefont {Aard}\
  \bibnamefont {{Keimpema}}}, \bibinfo {author} {\bibfnamefont {Franz}\
  \bibnamefont {{Kirsten}}}, \bibinfo {author} {\bibfnamefont {T.~Joseph~W.}\
  \bibnamefont {{Lazio}}}, \bibinfo {author} {\bibfnamefont {Benito}\
  \bibnamefont {{Marcote}}}, \bibinfo {author} {\bibfnamefont {Kiyoshi}\
  \bibnamefont {{Masui}}}, \bibinfo {author} {\bibfnamefont {Kenzie}\
  \bibnamefont {{Nimmo}}}, \bibinfo {author} {\bibfnamefont {Zsolt}\
  \bibnamefont {{Paragi}}}, \bibinfo {author} {\bibfnamefont {Mubdi}\
  \bibnamefont {{Rahman}}}, \bibinfo {author} {\bibfnamefont {Daniel~Reverte}\
  \bibnamefont {{Pay{\'a}}}}, \bibinfo {author} {\bibfnamefont {Paul}\
  \bibnamefont {{Scholz}}}, \ and\ \bibinfo {author} {\bibfnamefont {Ingrid}\
  \bibnamefont {{Stairs}}},\ }\bibfield  {title} {\enquote {\bibinfo {title}
  {{The 60 pc Environment of FRB 20180916B}},}\ }\href {\doibase
  10.3847/2041-8213/abdb38} {\bibfield  {journal} {\bibinfo  {journal} {\apjl}\
  }\textbf {\bibinfo {volume} {908}},\ \bibinfo {eid} {L12} (\bibinfo {year}
  {2021})},\ \Eprint {http://arxiv.org/abs/2011.03257} {arXiv:2011.03257
  [astro-ph.HE]} \BibitemShut {NoStop}%
\bibitem [{\citenamefont {{Curtin}}\ \emph {et~al.}(2022)\citenamefont
  {{Curtin}}, \citenamefont {{Tendulkar}}, \citenamefont {{Josephy}},
  \citenamefont {{Chawla}}, \citenamefont {{Andersen}}, \citenamefont
  {{Kaspi}}, \citenamefont {{Bhardwaj}}, \citenamefont {{Cassanelli}},
  \citenamefont {{Cook}}, \citenamefont {{Dong}}, \citenamefont {{Fonseca}},
  \citenamefont {{Gaensler}}, \citenamefont {{Kaczmarek}}, \citenamefont
  {{Lanmnan}}, \citenamefont {{Leung}}, \citenamefont {{Pearlman}},
  \citenamefont {{Petroff}}, \citenamefont {{Pleunis}}, \citenamefont
  {{Rafiei-Ravandi}}, \citenamefont {{Ransom}}, \citenamefont {{Shin}},
  \citenamefont {{Scholz}}, \citenamefont {{Smith}},\ and\ \citenamefont
  {{Stairs}}}]{Curtin2022}%
  \BibitemOpen
  \bibfield  {author} {\bibinfo {author} {\bibfnamefont {Alice~P.}\
  \bibnamefont {{Curtin}}}, \bibinfo {author} {\bibfnamefont {Shriharsh~P.}\
  \bibnamefont {{Tendulkar}}}, \bibinfo {author} {\bibfnamefont {Alexander}\
  \bibnamefont {{Josephy}}}, \bibinfo {author} {\bibfnamefont {Pragya}\
  \bibnamefont {{Chawla}}}, \bibinfo {author} {\bibfnamefont {Bridget}\
  \bibnamefont {{Andersen}}}, \bibinfo {author} {\bibfnamefont {Victoria~M.}\
  \bibnamefont {{Kaspi}}}, \bibinfo {author} {\bibfnamefont {Mohit}\
  \bibnamefont {{Bhardwaj}}}, \bibinfo {author} {\bibfnamefont {Tomas}\
  \bibnamefont {{Cassanelli}}}, \bibinfo {author} {\bibfnamefont {Amanda}\
  \bibnamefont {{Cook}}}, \bibinfo {author} {\bibfnamefont {Fengqiu~Adam}\
  \bibnamefont {{Dong}}}, \bibinfo {author} {\bibfnamefont {Emmanuel}\
  \bibnamefont {{Fonseca}}}, \bibinfo {author} {\bibfnamefont {B.~M.}\
  \bibnamefont {{Gaensler}}}, \bibinfo {author} {\bibfnamefont {Jane~F.}\
  \bibnamefont {{Kaczmarek}}}, \bibinfo {author} {\bibfnamefont {Adam~E.}\
  \bibnamefont {{Lanmnan}}}, \bibinfo {author} {\bibfnamefont {Calvin}\
  \bibnamefont {{Leung}}}, \bibinfo {author} {\bibfnamefont {Aaron~B.}\
  \bibnamefont {{Pearlman}}}, \bibinfo {author} {\bibfnamefont {Emily}\
  \bibnamefont {{Petroff}}}, \bibinfo {author} {\bibfnamefont {Ziggy}\
  \bibnamefont {{Pleunis}}}, \bibinfo {author} {\bibfnamefont {Masoud}\
  \bibnamefont {{Rafiei-Ravandi}}}, \bibinfo {author} {\bibfnamefont
  {Scott~M.}\ \bibnamefont {{Ransom}}}, \bibinfo {author} {\bibfnamefont
  {Kaitlyn}\ \bibnamefont {{Shin}}}, \bibinfo {author} {\bibfnamefont {Paul}\
  \bibnamefont {{Scholz}}}, \bibinfo {author} {\bibfnamefont {Kendrick}\
  \bibnamefont {{Smith}}}, \ and\ \bibinfo {author} {\bibfnamefont {Ingrid}\
  \bibnamefont {{Stairs}}},\ }\bibfield  {title} {\enquote {\bibinfo {title}
  {{Limits on Fast Radio Burst-like Counterparts to Gamma-ray Bursts using
  CHIME/FRB}},}\ }\href {\doibase 10.48550/arXiv.2208.00803} {\bibfield
  {journal} {\bibinfo  {journal} {arXiv e-prints}\ ,\ \bibinfo {eid}
  {arXiv:2208.00803}} (\bibinfo {year} {2022})},\ \Eprint
  {http://arxiv.org/abs/2208.00803} {arXiv:2208.00803 [astro-ph.HE]}
  \BibitemShut {NoStop}%
\bibitem [{\citenamefont {{Piro}}(2012)}]{Piro2012}%
  \BibitemOpen
  \bibfield  {author} {\bibinfo {author} {\bibfnamefont {Anthony~L.}\
  \bibnamefont {{Piro}}},\ }\bibfield  {title} {\enquote {\bibinfo {title}
  {{Magnetic Interactions in Coalescing Neutron Star Binaries}},}\ }\href
  {\doibase 10.1088/0004-637X/755/1/80} {\bibfield  {journal} {\bibinfo
  {journal} {\apj}\ }\textbf {\bibinfo {volume} {755}},\ \bibinfo {eid} {80}
  (\bibinfo {year} {2012})},\ \Eprint {http://arxiv.org/abs/1205.6482}
  {arXiv:1205.6482 [astro-ph.HE]} \BibitemShut {NoStop}%
\bibitem [{\citenamefont {{Totani}}(2013)}]{Totani2013}%
  \BibitemOpen
  \bibfield  {author} {\bibinfo {author} {\bibfnamefont {Tomonori}\
  \bibnamefont {{Totani}}},\ }\bibfield  {title} {\enquote {\bibinfo {title}
  {{Cosmological Fast Radio Bursts from Binary Neutron Star Mergers}},}\ }\href
  {\doibase 10.1093/pasj/65.5.L12} {\bibfield  {journal} {\bibinfo  {journal}
  {PASJ}\ }\textbf {\bibinfo {volume} {65}},\ \bibinfo {eid} {L12} (\bibinfo
  {year} {2013})},\ \Eprint {http://arxiv.org/abs/1307.4985} {arXiv:1307.4985
  [astro-ph.HE]} \BibitemShut {NoStop}%
\bibitem [{\citenamefont {{Wang}}\ \emph {et~al.}(2016)\citenamefont {{Wang}},
  \citenamefont {{Yang}}, \citenamefont {{Wu}}, \citenamefont {{Dai}},\ and\
  \citenamefont {{Wang}}}]{Wang2016}%
  \BibitemOpen
  \bibfield  {author} {\bibinfo {author} {\bibfnamefont {Jie-Shuang}\
  \bibnamefont {{Wang}}}, \bibinfo {author} {\bibfnamefont {Yuan-Pei}\
  \bibnamefont {{Yang}}}, \bibinfo {author} {\bibfnamefont {Xue-Feng}\
  \bibnamefont {{Wu}}}, \bibinfo {author} {\bibfnamefont {Zi-Gao}\ \bibnamefont
  {{Dai}}}, \ and\ \bibinfo {author} {\bibfnamefont {Fa-Yin}\ \bibnamefont
  {{Wang}}},\ }\bibfield  {title} {\enquote {\bibinfo {title} {{Fast Radio
  Bursts from the Inspiral of Double Neutron Stars}},}\ }\href {\doibase
  10.3847/2041-8205/822/1/L7} {\bibfield  {journal} {\bibinfo  {journal}
  {\apjl}\ }\textbf {\bibinfo {volume} {822}},\ \bibinfo {eid} {L7} (\bibinfo
  {year} {2016})},\ \Eprint {http://arxiv.org/abs/1603.02014} {arXiv:1603.02014
  [astro-ph.HE]} \BibitemShut {NoStop}%
\bibitem [{\citenamefont {{Sridhar}}\ \emph {et~al.}(2021)\citenamefont
  {{Sridhar}}, \citenamefont {{Zrake}}, \citenamefont {{Metzger}},
  \citenamefont {{Sironi}},\ and\ \citenamefont {{Giannios}}}]{Sridhar2021}%
  \BibitemOpen
  \bibfield  {author} {\bibinfo {author} {\bibfnamefont {Navin}\ \bibnamefont
  {{Sridhar}}}, \bibinfo {author} {\bibfnamefont {Jonathan}\ \bibnamefont
  {{Zrake}}}, \bibinfo {author} {\bibfnamefont {Brian~D.}\ \bibnamefont
  {{Metzger}}}, \bibinfo {author} {\bibfnamefont {Lorenzo}\ \bibnamefont
  {{Sironi}}}, \ and\ \bibinfo {author} {\bibfnamefont {Dimitrios}\
  \bibnamefont {{Giannios}}},\ }\bibfield  {title} {\enquote {\bibinfo {title}
  {{Shock-powered radio precursors of neutron star mergers from accelerating
  relativistic binary winds}},}\ }\href {\doibase 10.1093/mnras/staa3794}
  {\bibfield  {journal} {\bibinfo  {journal} {\mnras}\ }\textbf {\bibinfo
  {volume} {501}},\ \bibinfo {pages} {3184--3202} (\bibinfo {year} {2021})},\
  \Eprint {http://arxiv.org/abs/2010.09214} {arXiv:2010.09214 [astro-ph.HE]}
  \BibitemShut {NoStop}%
\bibitem [{\citenamefont {{Most}}\ and\ \citenamefont
  {{Philippov}}(2022{\natexlab{a}})}]{Most2022}%
  \BibitemOpen
  \bibfield  {author} {\bibinfo {author} {\bibfnamefont {Elias~R.}\
  \bibnamefont {{Most}}}\ and\ \bibinfo {author} {\bibfnamefont {Alexander~A.}\
  \bibnamefont {{Philippov}}},\ }\bibfield  {title} {\enquote {\bibinfo {title}
  {{Electromagnetic precursor flares from the late inspiral of neutron star
  binaries}},}\ }\href {\doibase 10.1093/mnras/stac1909} {\bibfield  {journal}
  {\bibinfo  {journal} {\mnras}\ }\textbf {\bibinfo {volume} {515}},\ \bibinfo
  {pages} {2710--2724} (\bibinfo {year} {2022}{\natexlab{a}})},\ \Eprint
  {http://arxiv.org/abs/2205.09643} {arXiv:2205.09643 [astro-ph.HE]}
  \BibitemShut {NoStop}%
\bibitem [{\citenamefont {{Most}}\ and\ \citenamefont
  {{Philippov}}(2022{\natexlab{b}})}]{Most2022b}%
  \BibitemOpen
  \bibfield  {author} {\bibinfo {author} {\bibfnamefont {Elias~R.}\
  \bibnamefont {{Most}}}\ and\ \bibinfo {author} {\bibfnamefont {Alexander~A.}\
  \bibnamefont {{Philippov}}},\ }\bibfield  {title} {\enquote {\bibinfo {title}
  {{Reconnection-powered fast radio transients from coalescing neutron star
  binaries}},}\ }\href@noop {} {\bibfield  {journal} {\bibinfo  {journal}
  {arXiv e-prints}\ ,\ \bibinfo {eid} {arXiv:2207.14435}} (\bibinfo {year}
  {2022}{\natexlab{b}})},\ \Eprint {http://arxiv.org/abs/2207.14435}
  {arXiv:2207.14435 [astro-ph.HE]} \BibitemShut {NoStop}%
\bibitem [{\citenamefont {{Higson}}\ \emph {et~al.}(2019)\citenamefont
  {{Higson}}, \citenamefont {{Handley}}, \citenamefont {{Hobson}},\ and\
  \citenamefont {{Lasenby}}}]{Higson2019}%
  \BibitemOpen
  \bibfield  {author} {\bibinfo {author} {\bibfnamefont {Edward}\ \bibnamefont
  {{Higson}}}, \bibinfo {author} {\bibfnamefont {Will}\ \bibnamefont
  {{Handley}}}, \bibinfo {author} {\bibfnamefont {Mike}\ \bibnamefont
  {{Hobson}}}, \ and\ \bibinfo {author} {\bibfnamefont {Anthony}\ \bibnamefont
  {{Lasenby}}},\ }\bibfield  {title} {\enquote {\bibinfo {title} {{Dynamic
  nested sampling: an improved algorithm for parameter estimation and evidence
  calculation}},}\ }\href {\doibase 10.1007/s11222-018-9844-0} {\bibfield
  {journal} {\bibinfo  {journal} {Statistics and Computing}\ }\textbf {\bibinfo
  {volume} {29}},\ \bibinfo {pages} {891--913} (\bibinfo {year} {2019})},\
  \Eprint {http://arxiv.org/abs/1704.03459} {arXiv:1704.03459 [stat.CO]}
  \BibitemShut {NoStop}%
\bibitem [{\citenamefont {{Speagle}}(2020)}]{Speagle2020}%
  \BibitemOpen
  \bibfield  {author} {\bibinfo {author} {\bibfnamefont {Joshua~S.}\
  \bibnamefont {{Speagle}}},\ }\bibfield  {title} {\enquote {\bibinfo {title}
  {{DYNESTY: a dynamic nested sampling package for estimating Bayesian
  posteriors and evidences}},}\ }\href {\doibase 10.1093/mnras/staa278}
  {\bibfield  {journal} {\bibinfo  {journal} {\mnras}\ }\textbf {\bibinfo
  {volume} {493}},\ \bibinfo {pages} {3132--3158} (\bibinfo {year} {2020})},\
  \Eprint {http://arxiv.org/abs/1904.02180} {arXiv:1904.02180 [astro-ph.IM]}
  \BibitemShut {NoStop}%
\end{thebibliography}%

\end{document}